\documentclass[prl,twocolumn,floatfix,nofootinbib]{revtex4}
\usepackage{amsmath}
\usepackage{amssymb}
\usepackage{ifpdf}
\usepackage[pdftex]{graphicx}
\usepackage{fancyhdr}
\usepackage{extramarks}
\usepackage{wasysym}
\usepackage{pstricks}
\usepackage{bbm}
\usepackage{epsfig}
\usepackage{natbib}
\usepackage{chapterbib}
\usepackage{psfrag}
\usepackage[percent]{overpic}
\usepackage{multirow}
\usepackage{ifthen}
\usepackage{hyperref}
\usepackage{calc}
\usepackage{rotating}
\usepackage{latexsym}
\usepackage{subfigure,rotating,rotate}
\usepackage{color}

\input{babarsym}

\newcommand{\EE}[1] {\ensuremath{\times 10^{#1}}\xspace}

\def\ze#1   {\ensuremath{\zeta_{#1}}\xspace}

\def\superb   {Super$B$\xspace}

\def\Bz   {\ensuremath{B_{z}}\xspace}

\def\CO2  {$\mathrm{CO}_2$\xspace}

\definecolor{light-gray}{gray}{0.7}
\def\gray{\color{light-gray}}

\newcommand{\xd}{\ensuremath{x_D}\xspace} 
\newcommand{\yd}{\ensuremath{y_D}\xspace} 
\newcommand{\pd}{\ensuremath{p_D}\xspace}
\newcommand{\qd}{\ensuremath{q_D}\xspace}
\newcommand{\ycp}{\ensuremath{y_{C\!P}}\xspace}
\newcommand{\xp}{\ensuremath{x^{\prime}}\xspace} 
\newcommand{\yp}{\ensuremath{y^{\prime}}\xspace} 
\newcommand{\xpp}{\ensuremath{x^{\prime\prime}}\xspace}
\newcommand{\ypp}{\ensuremath{y^{\prime\prime}}\xspace}
\newcommand{\xpsq}{\ensuremath{x^{\prime 2}}\xspace} 
 
\newcommand{\deltakpi}{\ensuremath{\delta_{\sst K\!\pi}}\xspace}
\newcommand{\deltaktwopi}{\ensuremath{\delta_{\sst K\!\pi\!\pi}}\xspace}
\newcommand{\sst}{\scriptscriptstyle}
\newcommand{\CPV}{\ensuremath{C\!PV}\xspace}

\long\def\inst#1{\par\nobreak\kern 4pt\nobreak
    {\it #1}\par\vskip 10pt plus 3pt minus 3pt}

\def\NS {\ensuremath{\Upsilon{(nS)}}\xspace}
\def\qq{\mathbbmss{q}}

\newcommand{\beq}{\begin{equation}}
\newcommand{\eeq}{\end{equation}}
\def\be{\begin{equation}}
\def\bea{\begin{eqnarray}}
\def\eea{\end{eqnarray}}
\def\nnb{\nonumber}

\def\eg{\ensuremath {e.g.}}
\def\ie{\ensuremath {i.e.}}

\makeatletter
\DeclareRobustCommand\bfseries{%
  \not@math@alphabet\bfseries\mathbf
  \fontseries\bfdefault\selectfont\boldmath}

\makeatother

\makeatletter
\newcommand{\rulesandwich}[1]{%
  {\centerline{\rule[0.5ex]{0.9\linewidth}{2pt}}}\par\nopagebreak\vskip 0.5ex%
  \centering #1\par\nopagebreak\vskip 0.5ex%
  {\centerline{\rule[0.5ex]{0.9\linewidth}{2pt}}}\smallskip
}
\def\before@@par#1#2\@@par{#2#1\@@par}
\@latex@info{//// style change: modifying section title style}
\def\@hangfrom@section#1#2{#1#2}%
\def\section{%
  \@startsection%
  {section}%
  {1}%
  {\z@}%
  {0.8cm \@plus1ex \@minus .2ex}%
  {0.5cm}%
  {%
    \normalfont\small\larger\larger\bfseries\rulesandwich
  }%
}%
\def\@part[#1]#2{%
 \@ifnum{\c@secnumdepth >\m@ne}{%
        \refstepcounter{part}%
        \addcontentsline{toc}{part}{\thepart\hspace{1em}#1}%
 }{%
      \addcontentsline{toc}{part}{#1}%
 }%
 \begingroup
    \parindent \z@ \raggedright
    \interlinepenalty\@M
    \@ifnum{\c@secnumdepth >\m@ne}{%
      \Large \bf \partname~\thepart%
      \par\nobreak
    }{}%
    \normalfont\small\larger\larger\larger\larger\bfseries
    {\centerline{\rule[0.5ex]{0.9\linewidth}{2pt}}}\par\nopagebreak\vskip 1ex%
    \centering #2\par\nopagebreak\vskip 1ex%
    {\centerline{\rule[0.5ex]{0.9\linewidth}{2pt}}}%
    \markboth{}{}\par
 \endgroup
   \nobreak
   \vskip 3ex
   \@afterheading
}%
\def\@spart#1{{\parindent \z@ \raggedright
    \interlinepenalty\@M
    \normalfont\small\larger\larger\larger\larger\bfseries
    {\centerline{\rule[0.5ex]{0.9\linewidth}{2pt}}}\par\nopagebreak\vskip 0.5ex%
    \centering #1\par\nopagebreak\vskip 0.5ex%
    {\centerline{\rule[0.5ex]{0.9\linewidth}{2pt}}}\par
  }
  \nobreak
  \vskip 3ex
  \@afterheading}
\makeatother

\def\LALRNUM{LAL-110}
\def\INFNRNUM{INFN/AE\_10/2}
\def\SLACRNUM{SLAC-R-952}

\begin{document}

\pagestyle{empty}

\onecolumngrid
\begin{flushright} 
\INFNRNUM, \LALRNUM, \SLACRNUM
\end{flushright} 
\begin{center}
{\boldmath{
\phantom{\Huge{I}}
\phantom{\Huge{I}}

{\huge\bf
{Super$B$\\
\vskip 12pt
Progress Reports}}\\
 \ \phantom{\Huge{I}}\\%
{\huge\bf{
\vskip 12pt
Physics}}\\
{\huge
{\vskip 12pt
{\gray{Accelerator}}
} \\
{\huge
{\vskip 12pt
\gray{Detector}}
} \\
{\huge
{\vskip 12pt\gray{Computing}}} \\
\phantom{\Huge{I}}
\phantom{\Huge{I}}
{\Large\bf August 7, 2010}\\
}
}
}
\end{center}
\bigskip
\begin{center}
{\large \bf Abstract}
\end{center}
\superb is a high luminosity $e^+e^-$ collider that will be able to indirectly 
probe new physics at energy scales far beyond the reach of any man made accelerator
planned or in existence.  Just as detailed understanding of the Standard Model of particle 
physics was developed from stringent constraints imposed by flavour changing 
processes between quarks, the detailed structure of any new physics is
severely constrained by flavour processes.
In order to elucidate this structure it is necessary to perform a 
number of complementary studies of a set of golden channels.  With these
measurements in hand, the pattern of deviations from the Standard Model behavior
can be used as a test of the structure of new physics.  If new physics is found at the 
LHC, then the many golden measurements from \superb will help decode the subtle nature
of the new physics.  However if no new particles are found at the LHC, \superb will be 
able to search for new physics at energy scales up to $10-100$~TeV.  In either scenario,
flavour physics measurements that can be made at \superb play a pivotal role in understanding
the nature of physics beyond the Standard Model.  Examples for using the interplay 
between measurements to discriminate New Physics models are discussed in this document.

\superb is a \sff, in addition to studying large samples of $B_{u,d,s}$, $D$ and $\tau$ decays, 
\superb has a broad physics programme that includes spectroscopy both in terms of the 
Standard Model and exotica, and precision measurements of $\sin^2\theta_W$.  In addition
to performing \CP violation measurements at the $\FourS$ and $\phi(3770)$, \superb will 
test \CPT in these systems, and lepton universality in a number of different processes.  The 
multitude of rare decay measurements possible at \superb can be used to constrain scenarios
of physics beyond the Standard Model.  In terms of other precision tests of the Standard Model, 
this experiment will be able to perform precision over-constraints of the unitarity triangle
through multiple measurements of all angles and sides.

This report extends and updates the studies presented in both the \superb Conceptual
Design Report in 2007 and the Proceedings of \superb Workshop VI in Valencia in 2008.
Together, these three documents detail the Physics case of the \superb Project.
\vfill\clearpage

{
%
\author{{\phantom{            }}}
\affiliation{{\phantom{       }\\ }}

\author{B.~O'Leary}\altaffiliation{Also affiliated with Universit\"at Bonn, D-53115 Bonn, Germany}
\affiliation{RWTH Aachen, 52056 Aachen, Germany}

\author{J.~Matias}
\author{M.~Ramon}
\affiliation{IFAE-Departament de Fisica, Universitat Autonoma de Barcelona, E-08193 Bellaterra, Barcelona}

\author{E.~Pous}
\affiliation{CC-UB and departament ECM, Universitat de Barcelona, Av. Diagonal 647, E-08028 Barcelona, Catalonia, Spain}

\author{F.~De Fazio}
\author{A.~Palano}
\affiliation{INFN Sezione di Bari; Dipartimento di Fisica, Universit\`a di Bari, I-70126 Bari, Italy}

\author{G.~Eigen}
\affiliation{University of Bergen, Institute of Physics, N-5007 Bergen, Norway}

\author{D.~Asgeirsson}
\affiliation{University of British Colombia, Vancouver, British Columbia, Canada V6T 1Z1}

\author{C.~H.~Cheng}
\author{A.~Chivukula}
\author{B.~Echenard}
\author{D.~G.~Hitlin}
\author{F.~Porter}
\author{A.~Rakitin}
\affiliation{California Institute of Technology, Pasadena, California 91125, USA}

% ...

\maketitle
}

\clearpage

\onecolumngrid\twocolumngrid\hbox{}

\begin{widetext}
This report is the result of the joint effort between the named authors and from the following contributing 
institutions who are working on the \superb project:
Universitat De Barcelona; 
INFN Bari and Universit\`a di Bari; 
INFN Bergamo and Universit\`a di Bergamo; 
University of Bergen;
INFN Bologna and Universit\`a di Bologna; 
California Institute of Technology;
Carleton University;
University of Cincinnati;
INFN CNAF;
INFN Ferrara and Universit\`a di Ferrara;
University of California, Irvine;
Taras Shevchenko National University Kyiv;
Laboratoire de l'Accelerateur Lin\'eaire;
Lawrence Berkeley National Laboratory;
Laboratori Nazionali di Frascati dell'INFN;
Laboratoire de Physique Nucl\'eaire et de Hautes Energies'
University of Maryland;
McGill University
INFN Milano and Universit\`a di Milano;
INFN Napoli and Universit\`a di Napoli Federico II;
Budker Institute of Nuclear Physics;
INFN Padova and Universit\`a di Padova;
INFN Pavia and Universit\`a di Pavia;
INFN Perugia and Universit\`a di Perugia
INFN Pisa, Universit\`a di Pisa and Scuola Normale Superiore Pisa;
Pacific Northwest National Laboratory;
Queen Mary, University of London;
Rutherford Appleton Laboratory;
INFN Roma and Universit\`a di Roma La Sapienza;
INFN Roma Tor Vergata and Universit\`a di Roma Tor Vergata;
INFN Roma Tre and Universit\`a di Roma Tre;
SLAC National Accelerator Laboratory;
Tel Aviv University;
INFN Torino and Universit\`a di Torino;
Universit\`a di Trento;
INFN Trieste and Universit\`a di Trieste;
TRIUMF;
University of British Columbia;
Universit\'e Montre\'al;
and University of Victoria.
\end{widetext}

$\,$
\vfil
\clearpage

\tableofcontents
\onecolumngrid\twocolumngrid\hbox{}

\clearpage

\pagestyle{fancyplain}

\fancyfoot{} 
\fancyfoot[LE,RO]{\it{SuperB Progress Report - The Physics - August 2010}}
\fancyhead{} 
\renewcommand{\sectionmark}[1]%
                  {\markright{#1}}
\rhead[\fancyplain{}{\bf Introduction}]%
      {\fancyplain{}{\bf\thepage}}

\graphicspath{{Introduction/}{Introduction/}}
%
%
\section{Introduction}\label{sec:intro}

The Standard Model of High Energy Physics (SM) has been developed as a result 
of decades of theoretical and experimental activity.  While all predictions of this model
are consistent with data, the SM is known to be incomplete as gravity is not included
at the very least.  The Large Hadron Collider (LHC) at CERN has recently started a decade
long programme to search for the Higgs particle and evidence for new physics (NP) at high
energies that it is hoped will lead to the discovery of a number of new heavy particles.
The Higgs boson is the physical remnant of the spontaneous breaking of the electroweak
symmetry needed to allow for mass terms for the SM particles. Besides, its presence
is also needed to unitarize the WW/ZZ scattering cross sections.
However on introducing the Higgs in the SM one creates theoretical problems that require
the inclusion of yet more particles. In particular, quantum corrections to the Higgs mass
introduce in the theory a power-like dependence on the ultraviolet cut-off which
destabilize the electroweak scale. A popular way to alleviate this problem in a ``natural''
way, namely without fine tuning the parameters of the theory, is to introduce supersymmetry
(SUSY), although other viable possibilities (compositeness, extra dimensions, etc.)
are available.  The role of the LHC is to search for the direct production of as yet
undiscovered  particles in order to help elucidate some of the limitations of the SM and
any possible extensions.  Yet this is not the only way to detect evidence for high energy
phenomena.  It is also possible  to indirectly search for new physics through precise
measurements of rare processes involving SM particles only. 
Observables measured from such processes can be significantly modified in the presence of
NP without the need to directly produce high energy particles in the laboratory.
This indirect route relies on the virtual production of particles from the 
high energy sector, studying how these may affect the behavior of SM particle 
interactions at lower energy via loop effects.  These NP effects are therefore expected
to be more easily observed in SM forbidden processes. 
Thus elucidating NP via the indirect approach is the realm of precision
measurements in clean environments.

It is possible to place constraints on the energy scale of new physics
using naturalness arguments. The energy scale required to solve the hierarchy
problem introduced by the SM Higgs sector is $\Lambda_{NP} \lesssim 1$ TeV.
Were new particles present at this scale, no fine tuning would be needed for the
SM Higgs mass.
A second estimate of the new physics energy scale can be obtained
from precision measurements of flavour related observables such as mixing 
in $K$, $B$, and $D$ decays.  Depending on the flavour observables used
and on the assumptions on the relevant couplings, one finds a lower bound
on $\Lambda_{NP}^\mathrm{flavour}$. This can exceed thousands of TeV and is driven
by constraints coming from the kaon sector. As far as $B$ physics is
concerned, considering for example the minimal SUSY extension of the SM
with $O(1)$ flavour couplings, one finds $\Lambda_{NP}^\mathrm{flavour}\gtrsim 10$
TeV. 

The inconsistency of the bounds on the NP scales $\Lambda_{NP}$ and
$\Lambda_{NP}^\mathrm{flavour}$ related to the Higgs and the flavour sectors of
the theory is usually referred to as the ``flavour problem''.
While one could still think of reconciling $\Lambda_{NP}^\mathrm{flavour}$ from
the $B$ sector with $\Lambda_{NP}$ by relaxing the naturalness requirement
(which in any case has to be relaxed to be compatible with electroweak
precision data), it is apparent that new particles with sub-TeV masses and
generic flavour couplings are not compatible with the whole set of flavour
physics measurements.
A popular solution of the flavour problem that does not rely on fine tuning
is Minimal Flavour Violation models (MFV).  MFV models assume that the
only non trivial flavour couplings are the SM Yukawa couplings and
restrict the possible NP contributions following a flavour symmetry principle.
In this way one brings $\Lambda_{NP}^{flavour}$ below the TeV and reconcile
it with $\Lambda_{NP}$.
The simultaneous solution of the hierarchy and flavour problems implies an
intriguing connection between masses and flavour properties of the NP
particles. Therefore it is imperative that both the direct and indirect searches
for NP are performed  to obtain as complete a picture of nature as possible. 
If the LHC is unable to find evidence for new physics, then detailed exploration
of the flavour sector is the only recourse without embarking on an experiment at 
significantly higher energies. On the other hand if new physics is observed 
at the LHC, then once again the flavour sector will be instrumental in 
understanding its structure.

The D0 collaboration recently claimed evidence for a di-muon asymmetry
that was incompatible with the Standard Model~\cite{Abazov:2010hv}.  
While not yet statistically significant, if this claim is confirmed 
as a first measurement of new physics, then this would mean that the 
flavour structure of physics beyond the 
Standard Model is really very interesting, and \superb will be able to discover a 
number of new-physics related effects through the many measurements described in 
this document.  Irrespective of an independent confirmation of this effect, the 
process of reconstructing the new physics Lagrangian using \superb and the 
measurements discussed in Section~\ref{sec:interplay} will be a very 
rewarding endeavor.

In addition to being able to perform measurements that in turn can be used
to constrain the flavour related properties of new physics, this experiment
will be able to perform precision tests of the SM using decays of $B$,
$D$, $\Upsilon$ mesons and $\tau$ leptons.
The studies presented here extend and update some of those found
in the \superb Conceptual Design Report~\cite{Bona:2007qt} and the Valencia
Physics Workshop  proceedings~\cite{Hitlin:2008gf}. These three documents 
together detail the \superb physics programme.  
An up-to-date description of the \superb detector can be found in Refs.~\cite{DETECTORWP}.
For comparison the physics potential of 
the proposed Belle-II experimental programme can be found in 
Ref.~\cite{Aushev:2010bq}.

\superb is a proposed high luminosity $e^+e^-$ collider operating at energies from
open charm threshold (the $\psi(3770)$ resonance) to above the $\Upsilon(5S)$
resonance~\cite{Bona:2007qt,Hitlin:2008gf}.  The aim of \superb is to accumulate 75\invab\ of data in five years
of nominal running at the 
$\Upsilon(4S)$ resonance corresponding to the world's largest samples of $B$, $D$, and 
$\tau$ pairs, as well as large samples of data at other center of mass 
energies corresponding to the $\Upsilon$ resonances.  These data are 75 
times the statistics available
at any existing experiment and can be used to probe a wide range of 
observables related
to new physics in the flavour sector at a precision level.  There is 
a proposal to construct a similar project at KEK as an upgrade
of the KEKB/Belle infrastructure referred to as 
SuperKEKB/Belle-II~\cite{Akeroyd:2004mj}. 
The Belle-II experiment ultimately aims to record only 50\invab of data.
There are two unique features of the \superb facility that provide
significant opportunities to perform or improve the precision of 
a number of important measurements: (i) the electron
beam will be longitudinally polarized, and (ii) it is envisaged that 
there will be an extended period of running at the $\psi(3770)$ resonance
in order to utilize quantum correlations at charm threshold in analogy
to the work done by the \B-factories at the $\Upsilon(4S)$.  Together 
\superb and Belle-II are referred to as Super Flavour Factories
to highlight the wide range of phenomena that can be studied at these
facilities. The new physics sensitive observables that a \sff will 
measure are complementary to those accessible at the CERN based 
flavour physics experiment LHCb. Only by measuring the full set of observables at
$e^+e^-$ and hadron colliders will we be able to optimally elucidate
details of the flavour structure at high energy.

In addition to requiring the existence of new particles at high energy, some
extensions of the SM require the existence of one or more low
mass particles.  These low mass particles may be related to the Higgs sector,
and to postulated forms of Dark Matter.  So while many of the NP sensitive 
measurements at \superb are indirect, it is also possible to directly search
for new particles that could not be detected at the LHC, but could be required
to obtain a full understanding of any discoveries at the LHC.  The potential
of such direct searches are discussed in Section~\ref{sec:directsearches}.

A long standing issue that is unresolved in physics is the reconciliation of the 
creation of equal amounts of matter and antimatter in the Big Bang, and 
today's observation of a matter dominated universe.  The
level of CP violation present in the SM is orders of magnitude smaller
than that required for our universe to have evolved into the 
state that we observe it in.  CP violation measurements in charm decays and a
discovery of Lepton Flavour  Violation in $\tau$ decays may provide new input
to help resolve this conflict.

The remainder of this document discusses the physics potential 
of \superb, firstly in terms of the physics topics, and finally in terms
of the interplay between these measurements and those of other experiments
in constraining the Standard Model and constructing features of the new physics Lagrangian:
Lepton Flavour Violation, CP violation, and
Standard Model studies of $\tau$ decays are treated in Section~\ref{sec:tau};
Sections~\ref{sec:bphysics} and~\ref{sec:bsphysics} discuss at length the 
physics potential for measurements of $B_{u,d,s}$ decays at \superb; Section~\ref{sec:charm}
discusses physics reach in the charm sector, both at charm threshold and 
at the $\FourS$ resonance;
Precision electroweak measurements facilitated by the inclusion of 
longitudinal polarization of the electron beam at \superb are reviewed in
Section~\ref{sec:ew}; The potential for spectroscopy and direct searches at
low mass are discussed in Sections~\ref{sec:spectroscopy} 
and~\ref{sec:directsearches}, respectively.  In order to appreciate the 
potential of the measurements it is necessary
to understand the impact that improvements in theory, particularly Lattice 
QCD will have on the interpretation of results.  Section~\ref{sec:lattice} 
discusses this in detail, and compares predictions made in 
Ref.~\cite{Bona:2007qt} with
the current state of the art in Lattice QCD. Finally before concluding we 
review the main physics highlights of the \superb programme in the context 
of constraining new physics scenarios and elucidating the structure of the 
new physics Lagrangian.  One of the issues here is that there are many 
postulated scenarios of new physics, so it is necessary to determine which of
these (if any) may be manifest in nature.  This task requires a detailed analysis
of many new physics sensitive observables both from direct and indirect searches.
By understanding interplay between measurements and models of new physics we 
outline a strategy for elucidating new physics in the LHC era with a \sff.

\graphicspath{{taudecays/}{taudecays/}}
\ifx\undefined\usebibtextrue
\newif\ifusebibtex
\fi
\providecommand{\usebibtex}{false}
\ifthenelse{\equal{\usebibtex}{true}}{
  \usebibtextrue
}{
  \usebibtexfalse
}
\newcommand\superkekb{Belle II\xspace}
\section{\texorpdfstring{$\tau$ physics}{tau physics}}\label{sec:tau}

Searching for lepton-flavor-violating (LFV) $\tau$ decays constitutes one
of the most clean and powerful tools to discover and
characterize NP scenarios. Although the SM
 when complemented with the experimentally observed
neutrino-mixing phenomenology does include LFV $\tau$ decays, the rates
are extremely low and experimentally
unobservable,
making the discovery of LFV an unambiguous signal for physics beyond
the SM.

Experimental investigations on
\CP\ violation in $\tau$ decay and on the $\tau$ EDM and $g\! -\!\! 2$ provide \superb with
additional experimentally clean tools to advance our knowledge on unexplored
territories, with the ability to test some specific NP scenarios.

With an integrated luminosity of 75\invab, \superb will be able to
explore a significant portion of the parameter space of most New
Physics scenarios by searching for LFV in $\tau$ decays.  While the MEG
experiment{} will search for $\mu \to
e\gamma$ with great sensitivity, \superb will uniquely explore
transitions between the third and first or second generations,
providing crucial information to determine the specific New Physics
model that produces LFV.  The LHC experiments are, in general, not
competitive in LFV searches.  Furthermore, \superb includes
features that make it superior to \superkekb for LFV searches: a
larger planned luminosity and a polarized electron beam, which
is equivalent to a substantial boost in effective luminosity, and
smaller beam currents, leading to smaller machine backgrounds.
\superb can have a 80\% longitudinally polarized electron
beam, which will provide means to improve the selection of LFV final
states, given a specific LFV interaction, or to better determine the
features of the LFV interaction, once they are found.

Experimental studies on \CP\ violation in $\tau$ decay and on the $\tau$ EDM
and $g\! - \!\! 2$ are especially clean tools, because they rely on
measurement of asymmetries with relatively small systematic
uncertainties from the experiment. The beam polarization also improves
the experimental sensitivity for $\tau$ EDM and $g\! - \!\! 2$
determinations, by allowing measurements of the polarization of a
single $\tau$, rather than measurements of correlations between two $\tau$ leptons
produced in the same event. With this technique, \superb can test
whether supersymmetry is a viable explanation for the present
discrepancy on the muon $g\! - \!\! 2$.  Although the most plausible
NP models constrained with the available experimental results
predict \CP\ violation in $\tau$ decay and the $\tau$ EDM in a range that is
not measurable, \superb can test specific models that enhance those
effects to measurable levels.

\subsection{\texorpdfstring{%
    Lepton Flavor Violation in $\tau$ decay}{%
    Lepton Flavor Violation in $\tau$ decay}\label{sec:tau:lfv}}

\subsection*{Predictions from New Physics  models}

\subsubsection*{LFV in the MSSM}

In the following, we discuss the size of $\tau$ LFV effects on decays and correlations
that are expected in supersymmetric extensions of the SM and, in particular,
in the so-called constrained MSSM,
The flavor-conserving phenomenology of this framework is
characterized by five parameters: $M_{1/2}$, $M_0$, $A_0$, $\tan\beta$, $\text{sgn}\ \mu$.
We will discuss a subset of the ``Snowmass Points and Slopes'' (SPS)~\cite{Allanach:2002nj},
which we consider adequate to illustrate the variety of predictions and the features of the model
on lepton flavor violation processes (see Table~\ref{tab:SPS:def:15}).

\begin{table}[!htb]
\caption{\label{tab:SPS:def:15}
Values of $M_{1/2}$, $M_0$, $A_0$, $\tan \beta$,
and sign of $\mu$ for the SPS points considered in the analysis.}
\begin{tabular}{cccccc}
\hline
SPS & $M_{1/2}$ (GeV) & $M_0$ (GeV) & $A_0$ (GeV) & $\tan \beta$ &
 $\mu$ \\\hline
 1\,a & 250 & 100 & -100 & 10 &  $>\,0 $ \\
 1\,b & 400 & 200 & 0 & 30 &   $>\,0 $ \\
 2 &  300 & 1450 & 0 & 10 &  $>\,0 $ \\
 3 &  400 & 90 & 0 & 10 &    $>\,0 $\\
 4 &  300 & 400 & 0 & 50 &   $>\,0 $ \\
 5 &  300 & 150 & -1000 & 5 &   $>\,0 $\\\hline
\end{tabular}
\end{table}

At all the SPS points, LFV decays are dominated by the contribution
of dipole-type effective operators of the form ($\bar l_i \sigma_{\mu\nu} l_j F^{\mu\nu}$).
Defining $\mathcal{R}^{(a)}_{(b)}=\BR(\tau \to a)/\BR(\tau \to b)$,
The dipole dominance allows us to establish the
following relations:
\begin{eqnarray}
\mathcal{R}^{(\mu ee)}_{(\mu \gamma)}\approx
1.0 \times 10^{-2} & \to & \BR(\tau \to \mu e^+ e^-) <  5 \times 10^{-10}, \nonumber \\
\mathcal{R}^{(\mu \rho^0)}_{(\mu\gamma)}
\approx  2.5 \times 10^{-3} & \to &  \BR(\tau \to \mu \rho^0) < 10^{-10}, \nonumber \\
\mathcal{R}^{(3\mu)}_{(\mu \gamma)}
\approx  2.2 \times 10^{-3}  & \to &  \BR(\tau \to 3\mu) < 10^{-10}, \nonumber \\
\mathcal{R}^{(\mu \eta)}_{(\mu\gamma)} <  10^{-3} & \to &  \BR(\tau \to \mu \eta)
< 5 \times 10^{-11}, \nonumber
\end{eqnarray}
where the bounds correspond to the present limit
\hbox{$\BR(\taumg)< 4.5 \times 10^{-8}$.} Similar relations hold for
\hbox{$\tau \to e$} transitions.  Assuming an experimental reach at
\superb at the level of $10^{-9}$ only $\tau \to \mu \gamma$ and
$\tau \to e \gamma$ decays would be within experimental reach in this
list. However, it is interesting to notice that some processes as
$\tau \to \mu \rho$ ($\rho \to \pi^+ \pi^-$) can reach branching
ratios of $10^{-10}$ for special values of the parameters
\cite{Arganda:2008jj}. Taking into account that these modes are
cleaner from the experimental point of view, they could still be
interesting processes in a \superb.

To estimate the overall scale of  $\tau \to (\mu,e) \gamma$ rates,
we must specify the value of the LFV couplings, since they are not
determined by the SPS conditions.
In the mass-insertion and leading-log approximation, assuming
that the leading LFV couplings appear in the left-handed
slepton sector, we can write:
\begin{equation}
\frac{{\rm \BR}(l_j\to l_i\gamma)}{{\rm \BR}(l_j\to l_i \bar{\nu}_i\nu_{j})}
\approx
\frac{\alpha^3}{G_{F}^2}
\frac{\left|\left(m^2_{\widetilde L} \right)_{ji}\right|^2}{M_{S}^8}
\tan^2\beta \nonumber, \label{eq:min}
\end{equation}
where, to a good approximation,
$M_{S}^8\simeq 0.5 M_0^2 M_{1/2}^2 \times (M_0^2 + 0.6 M_{1/2}^2 )^2$.
In a Grand Unified Theory (GUT) with heavy right-handed neutrinos,
the off-diagonal entries of the slepton mass matrix $m^2_{\widetilde L}$
are likely to be dominated by the flavor mixing in
the (s)neutrino sector.
These terms can be expressed as:
\begin{equation}
\label{eq:leadinglog}
\left(m^2_{\widetilde {L}}\right)_{ji} \approx
-\frac{6M_0^2+2A_0^2}{16\pi^2}\ \delta_{ij},
\end{equation}
where $\delta_{ij} = \left(Y^\dagger_\nu Y_\nu\right)_{ji}
\log(M_{GUT}/M_{R})$ in terms of the neutrino
Yukawa couplings ($Y_\nu$), the average
heavy right-handed neutrino mass ($M_R$)
and the GUT scale ($M_{GUT} \sim 10^{15}$--$10^{16}$~GeV).
The experimental information on neutrino masses and mixings is not 
sufficient to fix completely the structure in the neutrino Yukawa matrix,
even assuming some kind of quark-lepton unification. 
We can take two limiting situations that are called ``CKM-like'' and 
``PMNS-like''  \cite{Masiero:2002jn}. Taking the ``PMNS-like'' case and given 
the large phenomenological value of the 2--3 mixing in the neutrino sector
(and the corresponding suppression of the 1--3 mixing)
we expect $|\delta_{32}| \gg |\delta_{31}|$
hence $\BR(\tau \to \mu \gamma) \gg \BR(\tau \to e \gamma)$.
For sufficiently heavy right-handed neutrinos, the
normalization of $Y_\nu$ is such that
$\BR(\tau \to \mu \gamma)$ can reach
values in the $10^{-9}$ range.
In particular, $\BR(\tau \to \mu \gamma) \gsim 10^{-9}$
if at least one heavy right-handed neutrino has a mass around or above
$10^{13}$~GeV (in SPS 4) or $10^{14}$~GeV (in SPS 1a,1b,2,3,5).

A key issue that must be addressed is the role of
$\BR(\mu \to e \gamma)$ in
constraining  the LFV couplings and,
more generally, the correlations between
$\BR(\tau \to (\mu,e) \gamma)$ and $\BR(\mu \to e \gamma)$
in this framework. An extensive analysis of such questions
has been presented in Ref.~\cite{Antusch:2006vw,Arganda:2005ji},
under the hypothesis of a hierarchical spectrum
for the heavy right-handed neutrinos.

\begin{figure}[!htb]
\includegraphics[width=\linewidth,angle=0]{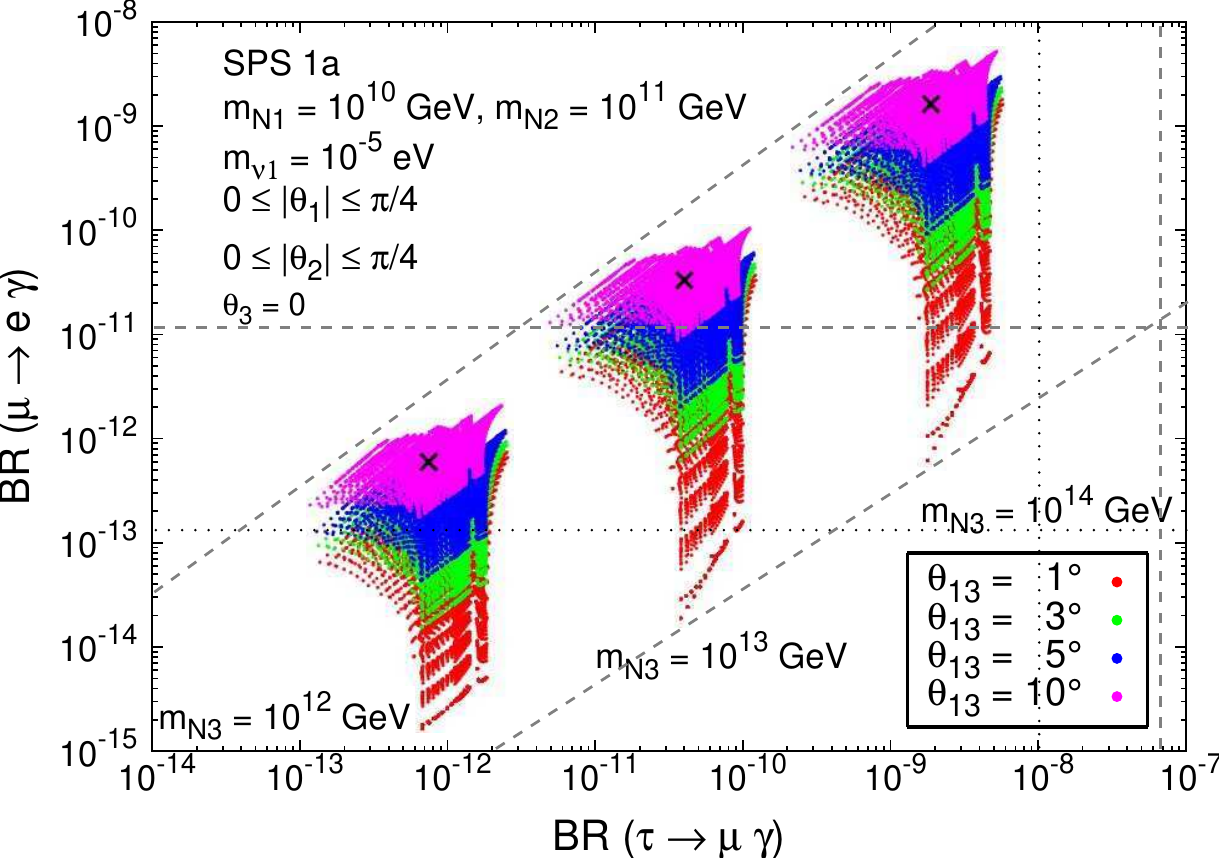}
\caption{\label{fig:herrero1}
$\BR(\tau \to \mu \gamma)$ vs.~$\BR(\mu \to e \gamma)$
in SPS~1a,
for three reference values of the heavy right-handed neutrino
mass and several values of $\theta_{13}$.
The horizontal dashed (dotted) line denotes the present experimental
    bound (future sensitivity) on $\BR(\mu \to e \gamma)$.
    All other relevant parameters are set to the values specified in
    Ref.~\cite{Antusch:2006vw}.
     }
\end{figure}

\begin{figure}[!htb]
\includegraphics[angle=-90,width=\linewidth]{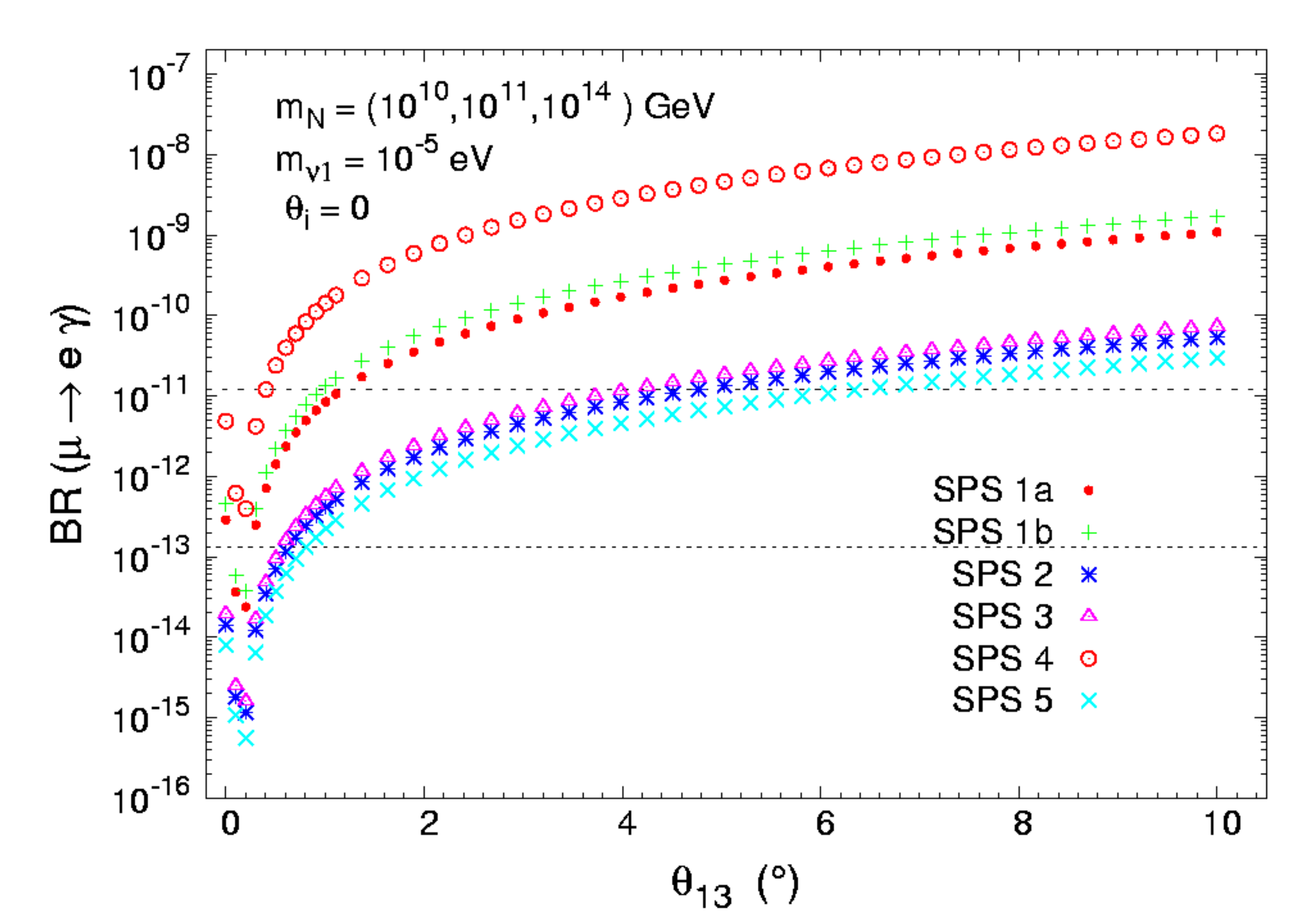}
\caption{\label{fig:SPS:t13:ad}
$\BR(\mu \to e\, \gamma$) as a function of $\theta_{13}$ (in degrees)
for various SPS points. The dashed (dotted) horizontal line denotes the present experimental
    bound (future sensitivity). All other relevant parameters are
    set to the values specified in
    Ref.~\cite{Antusch:2006vw}.
           }
\end{figure}

The overall structure of the
$\BR(\tau \to \mu \gamma)$ vs.~$\BR(\mu \to e \gamma)$
correlation in SPS~1a is shown in~Fig.~\ref{fig:herrero1}.
As anticipated, $\BR(\tau \to \mu \gamma)\sim 10^{-9}$
requires a heavy right-handed neutrino around or
above  $10^{14}$~GeV. This possibility is not excluded by
$\BR(\mu \to e \gamma)$ only if the 1--3 mixing in the
lepton sector (the $\theta_{13}$ angle of the neutrino
mixing matrix) is sufficiently small. This is a general
feature, valid at all SPS points, as illustrated in
Fig.~\ref{fig:SPS:t13:ad}.
In Table \ref{tab:SPS:BR} we show the predictions for $\BR(\taumg)$
and $\BR(\taummm)$ corresponding to the neutrino
mass parameters chosen in Fig.~\ref{fig:SPS:t13:ad}
(in particular $M_{N_3} = 10^{14}$~GeV), for the various SPS points.
Note that this case contains points that
are within the \superb sensitivity range,
yet are not excluded by $\BR(\mu \to e \gamma)$
(as illustrated in Fig.~\ref{fig:SPS:t13:ad}).
It is also interesting to notice the possible correlations with other processes.
For instance, in $SU(5)$ GUT models a large \CP phase in the $B_s$ system would
imply a large $\BR(\taumg)$ due to the unification of the squark and
slepton mass matrices at $M_{\rm GUT}$
\cite{Hisano:2003bd,Ciuchini:2003rg,Parry:2007fe,Ciuchini:2007ha}.

\begin{table}[!htb]
\caption{\label{tab:SPS:BR}Predictions for  $\BR(\tau \to \mu\, \gamma$) and \hbox{$\BR(\taummm$)}
         corresponding to the SPS points. The values of $m_{N_i}$ and $m_{\nu_1}$
         are as specified in Fig.~\ref{fig:SPS:t13:ad}~\cite{Antusch:2006vw}.}

\begin{tabular}{ccccccc}
\hline
SPS               &  1\,a   &  1\,b   &  2  &  3  &  4  &  5   \\ \hline
$\BR(\taumg)\times 10^{-9}$ & 4.2 & 7.9 & 0.18 & 0.26 & 97 & 0.019 \\
$\BR(\taummm)\times 10^{-12}$ & 9.4 & 18 & 0.41 & 0.59 & 220 & 0.043 \\ \hline
\end{tabular}
\end{table}

It is unlikely that MSSM would be realised in nature with an entirely
flavor blind soft sector while the Yukawa sector presents a highly
nontrivial structure. Thus, we must explore other ``flavored MSSM'' realizations to be
able to analyze the host of new results that will arrive from \superb and LHC
experiments. The use of flavor symmetries can explain the complicated
Yukawa structures and at the same time predict a non-trivial structure
in the soft-breaking terms. In such flavor models, we can have a
large variety of predictions with different flavor symmetries.
However, LFV processes are always the most interesting observables in
these models and it is relatively easy to obtain $\BR(\tau \to \mu
\gamma)\sim 10^{-9}$ as shown in
Ref.~\cite{Calibbi:2008qt,Calibbi:2009ja} for an $SU(3)$ flavor
symmetry. We have to emphasize here that this process can even compete
in sizable regions of the parameter space with the future bound at
MEG for the process $\mu \to e \gamma$.

\subsubsection*{LFV in other scenarios}

At large $\tan\beta$ and not too heavy Higgs masses,
another class of LFV interactions is relevant,
the effective coupling between a  $\mu$--$\tau$ pair
and the heavy (scalar and pseudoscalar) Higgs bosons.
This coupling can
overcome the constraints on $\BR(\taummm)$
and $\BR(\tau\to\mu\eta)$ dictated by $\BR(\taumg)$
in the dipole-dominance scenario.
Such a configuration cannot be realized in the CMSSM, but
it could be realized in the so-called Non Universal Higgs Masses (NUHM)
SUSY scenario. In such a framework, there are specific regions of the
parameter space in which processes like $\tau\to\mu\eta$ and
$\tau\to\mu f_1(980),\; f_1(980)\to\pi^+\pi^-$ could have
a branching ratio in the range $10^{-9}$ to $10^{-10}$,
comparable or even slightly larger than
$\BR(\taumg)$~\cite{Paradisi:2005tk,Arganda:2008jj,Herrero:2009tm}.

Another interesting set of possibilities is MSSM with R-parity
violation (RPV) \cite{Dreiner:2006gu}. In these models several of the bounds
on RPV couplings are obtained from $B$ and $\tau$
processes. \superb can improve the bounds on couplings or might even discover a
signal for RPV MSSM. The processes $\tau \to \mu \eta$ and $\tau \to \mu \mu \mu$ are
especially interesting and can be used at \superb to improve existing bounds by more than
an order of magnitude.

Finally, in some non-SUSY NP frameworks, such as Little Higgs Models 
with T parity (LHT) or $Z^{'}$ models with non-vanishing LFV couplings ($Z^{'}\ell_{i}\ell_{j}$),
the $\taummm$ rate could be significantly enhanced to a level that matches or exceeds
the rate of $\taumg$ (see {\it e.g.}~\cite{Raidal:2008jk}).
In this respect, a measurement of $\BR(\taummm)$
to the level of $10^{-10}$ would be interesting test of NP even for $\BR(\taumg)\lsim 10^{-9}$.

\subsection*{\texorpdfstring{\superb experimental reach}{\superb experimental reach}}

The vast experience accumulated on the B-factories offers a reliable
base for estimating the reach of \superb on $\tau$ LFV searches. To a
first approximation, the \superb detector is expected to have
performances comparable to or better than \babar for electron
identification and for electromagnetic energy
resolution and hermiticity. The \superb project on the other hand has
an improved momentum resolution, thanks to silicon layers closer to
the beams, and improved muon identification.

The typical $\tau$ LFV decay search consists in counting
candidate events and measuring if there is an excess against the
expected background. By running a \babar analysis unchanged
on a larger statistical sample, all expected upper limits scale with 
at least as the square root of the luminosity increase (${\propto}1/\sqrt{\cal L}$).
This extrapolation poses a lower limit for the \superb reach, which
will be ameliorated by detector improvements and only moderately
worsened by a small expected increase of beam backgrounds.
If it is possible to maintain the B factory efficiencies while keeping
the expected amount of background events negligible, then
\superb will deliver upper limits that will scale linearly with
the integrated luminosity (${\propto}1/{\cal L}$). In the first
approximation, scaling by $1/{\cal L}$ is
possible for $\tau$ LFV decays into three leptons, or into a lepton and
two hadrons in the final state (where the two hadrons may come through
a hadron resonance). On the other hand, searches for
$\taulg$ suffer higher backgrounds and tend to scale more
like ${\propto}1/\sqrt{\cal L}$.

\babar $\tau$ LFV searches are optimized for the best expected upper
limits, which typically corresponds to maximizing the signal
efficiency while keeping the expected background events of the order
one or less, when the analysis is not background dominated. Since the
analysis optimization depends on the size of the analyzed
sample and on the amount of expected backgrounds, one must re-optimize
the B-factory analyses for the \superb luminosity, especially for the
low background searches. In the following, we extrapolate from the 
most recent results from \babar by re-optimizing the analysis for
$\taulll$, and assuming a conservative $1/\sqrt{\cal L}$
scaling for $\taulg$.
The experimental reach is expressed in terms of ``the expected
$90\%$ CL upper limit'' assuming no signal, as well as in terms of a
``3$\sigma$ evidence branching fraction'' in the presence of projected
backgrounds; furthermore a minimum of 5 expected signal events is
required to establish evidence for a signal.
In the absence of signal, for large numbers of expected background
events $N_{\text{bkg}}$, the expected $90\%$ CL upper limit for the
number of signal events can be approximated as
$N^{UL}_{90} \sim 1.28(1/2 + \sqrt{1/2+N_{\text{bkg}}})$\footnote{This
  formula has been obtained by approximating the Poisson distributed
  number of background and signal events with Gaussian distributions
  and the value $1.28$ satisfies the relation
  $$90\% \approx\int_{-\infty}^{\mu+1.28\sigma}G(\mu,\sigma),$$ where
  $G(\mu,\sigma)$ is a Gaussian with mean $\mu$ and variance
  $\sigma^2$. For order 100 expected background events, the formula
  approximates toy Monte Carlo simulations within better than 5\%.}
whereas for small $N_{\text{bkg}}$ a value for
$N^{UL}_{90}$ is obtained using the method described
in~\cite{Cousins:1991qz}, which gives, for $N_{\text{bkg}} \sim 0$,
$N^{UL}_{90} \sim 2.4$.  If a signal is determined from counting
events within a signal region, the $90\%$ CL branching ratio upper
limit is:
\begin{equation}
  \B^{UL}_{90} =
  \frac{N^{UL}_{90}}{2 N_{\tau\tau} \epsilon} =
  \frac{N^{UL}_{90}}{2 \cal{L}\sigma_{\tau\tau} \epsilon}\,,
\end{equation}
where  $N_{\tau\tau}=\cal{L}\sigma_{\tau\tau}$
is the number of $\tau$-pairs produced in $e^+e^-$ collisions;
$\cal{L}$ is the integrated luminosity,
$\sigma_{\tau\tau}$=0.919\,nb~\cite{Banerjee:2007is}
 is the $\tau$-pair production cross section, and
$\epsilon$ is the signal efficiency.

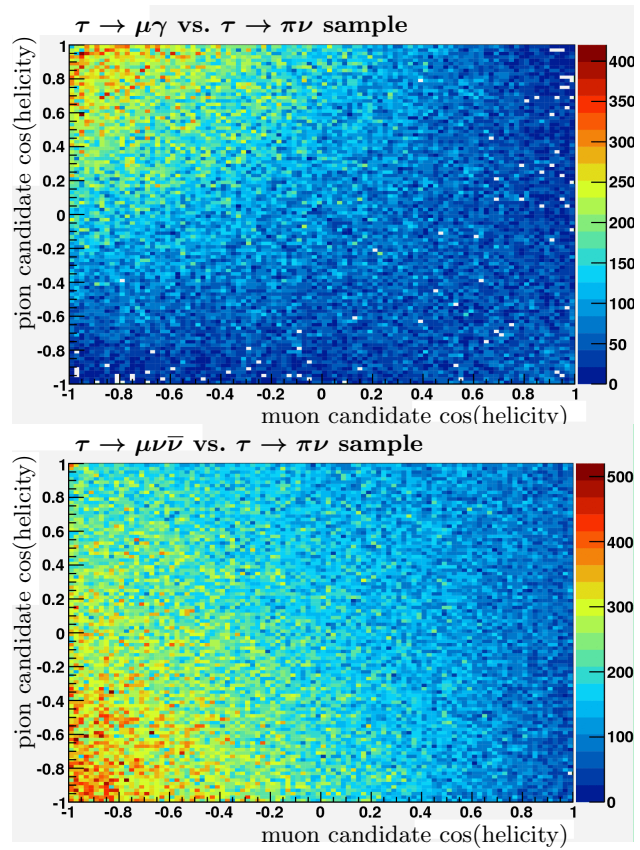
\begin{figure}[tb]
  \begin{center}\fboxrule=0pt\fboxsep=0pt
    \fbox{\begin{overpic}[trim=10 0 0 0,width=\linewidth-2\fboxsep-2\fboxrule,clip]{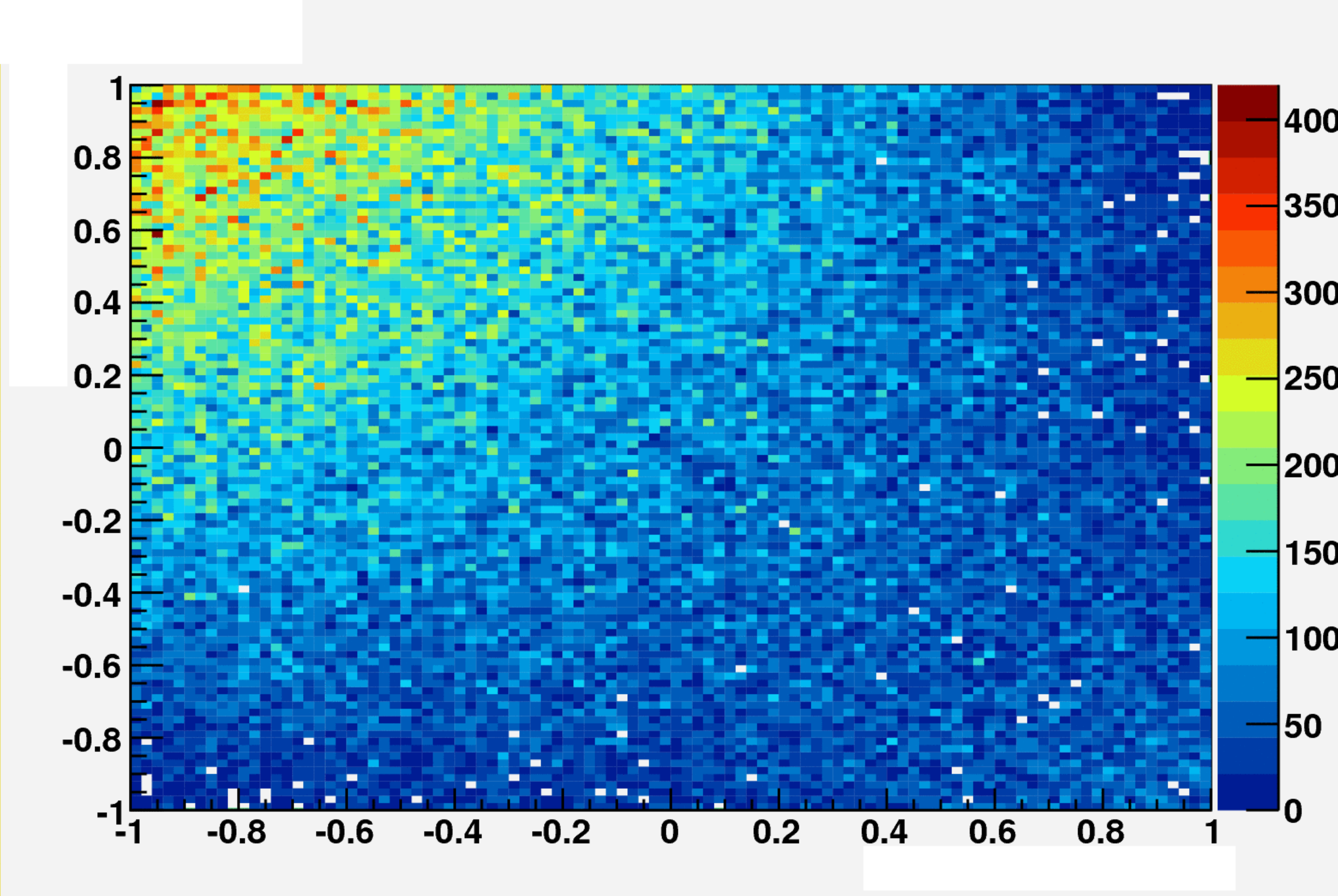}
        \put(10,63){\bfseries $\tau\to\mu\gamma$ vs.\ $\tau\to\pi\nu$ sample}
        \put(40,0){muon candidate $\cos(\text{helicity})$}
        \put(0,15){\begin{sideways}pion candidate $\cos(\text{helicity})$\end{sideways}}
        \end{overpic}}
      \vspace{1ex}
    \fbox{\begin{overpic}[trim=10 0 0 0,width=\linewidth-2\fboxsep-2\fboxrule,clip]{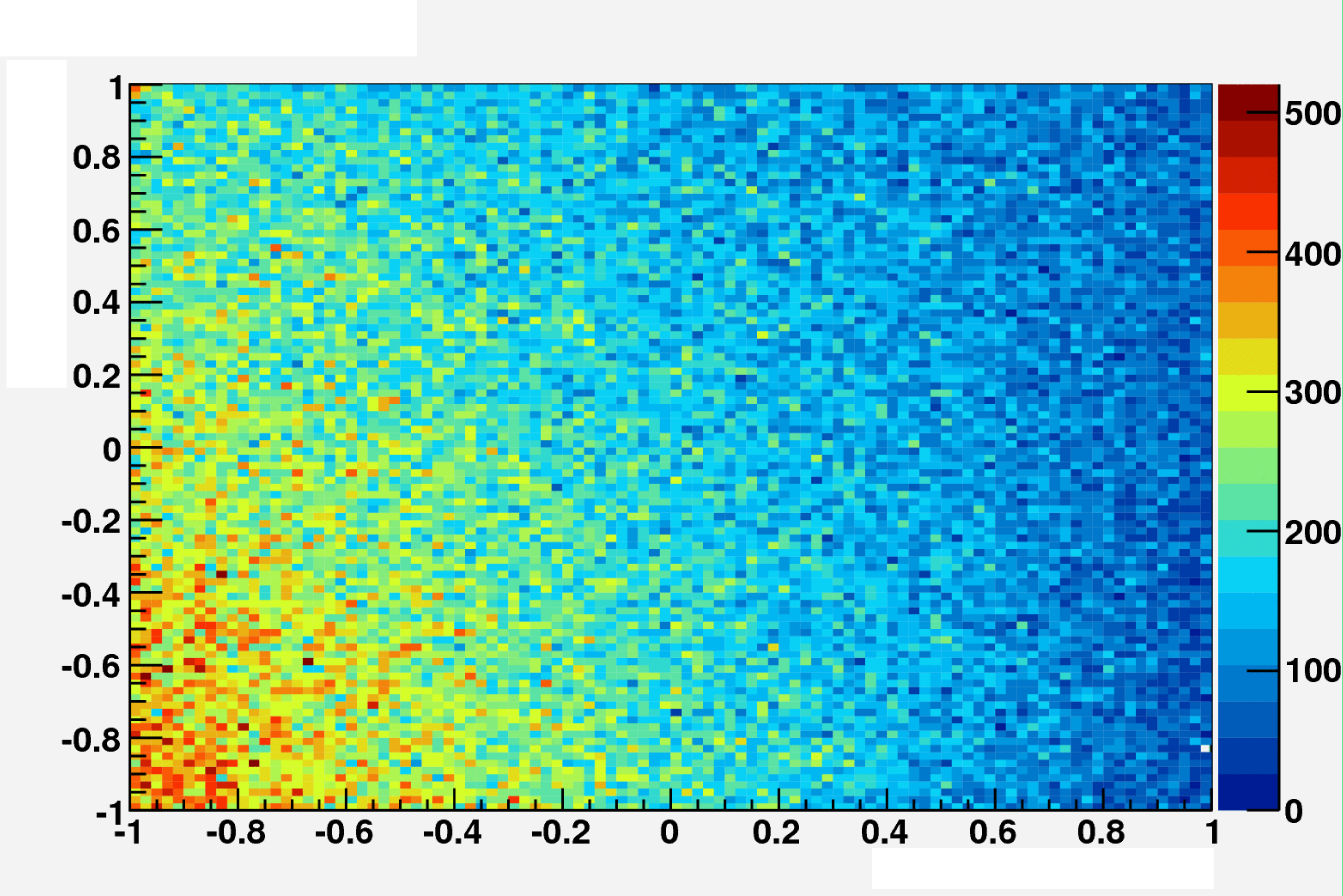}
        \put(10,63){\bfseries $\tau\to\mu\nu\nub$ vs.\ $\tau\to\pi\nu$ sample}
        \put(40,0){muon candidate $\cos(\text{helicity})$}
        \put(0,15){\begin{sideways}pion candidate $\cos(\text{helicity})$\end{sideways}}
      \end{overpic}}
  \end{center}
  \caption{\label{fig:taumug-helicities}%
    Distribution of the cosine of the helicity angle of the muon and
    pion candidates when selecting $\tau$ pairs events decaying
    to $\tau\to\mu\gamma$ and $\tau\to\pi\nu$.  The top plot shows
    simulated signal events, the bottom plot shows simulated
    $\tau\to\mu\nu\nub$ background events. In both cases, the electron
    beam is 80\% longitudinally polarized.}
\end{figure}

The \taumg projected sensitivity is based on the most recent
\babar result~\cite{Aubert:2009tk}. Some \superb
improvements with respect to \babar are taken into account:
\begin{itemize}

\item
  The smaller beam size and (to a minor extent) the improved momentum
  resolution will improve the invariant mass and energy resolution of
  the $\tau$ candidates and are expected to reduce the signal region area by
  35\%.
  
\item
  The improved coverage for photons is expected to increase the
  acceptance by 20\%.
  
\end{itemize}
Further gains are possible by re-optimizing the analysis for the
\superb detector and exploiting beam polarization effects. The high
energy electrons beam at \superb can be ${\sim}80\%$ longitudinally
polarized, influencing the angular distribution of the $\tau$ decay products
in a way that depends on the interaction that causes
LFV. Figure~\ref{fig:taumug-helicities} shows that with beam
polarization the helicity angles of the $\tau$ pair decay products
can be used to significantly suppress the background when one $\tau$
decays to $\mu\gamma$ and the other one to $\pi\nu$. Similar
background suppressions are also obtained with $\rho\nu$ and
$e\nu\nub$ decays on the other side. While further investigations are
ongoing to quantify analysis improvements, we provide here a more
conservative estimate of the \superb reach on $\tau$ LFV, which does
not consider beam polarization and other possible improvements on the
analysis.
For \taumg, we expect the final efficiency to be
$\sim 7.3\%$ and the final background to be $\sim 260$ events.
This leads to an expected 90\%CL upper limit of $2.4\EE{-9}$ and a
3$\sigma$ evidence reach of $5.4\EE{-9}$. One additional benefit of
beam polarization is the possibility to determine the helicity
structure of the LFV coupling from the final state momenta
distributions (see for instance Ref.~\cite{Matsuzaki:2007hh} for the
$\taummm$ process).
The extrapolation of the \taueg search receives benefits from similar
improvements, and has a projected 90\% CL upper limit of
$3.0\EE{-9}$ and a 3$\sigma$ evidence reach of $6.8\EE{-9}$.

\begin{figure}[tb]
  \begin{center}\fboxrule=0pt\fboxsep=0pt
    \fbox{\begin{overpic}[width=\linewidth-2\fboxsep-2\fboxrule]{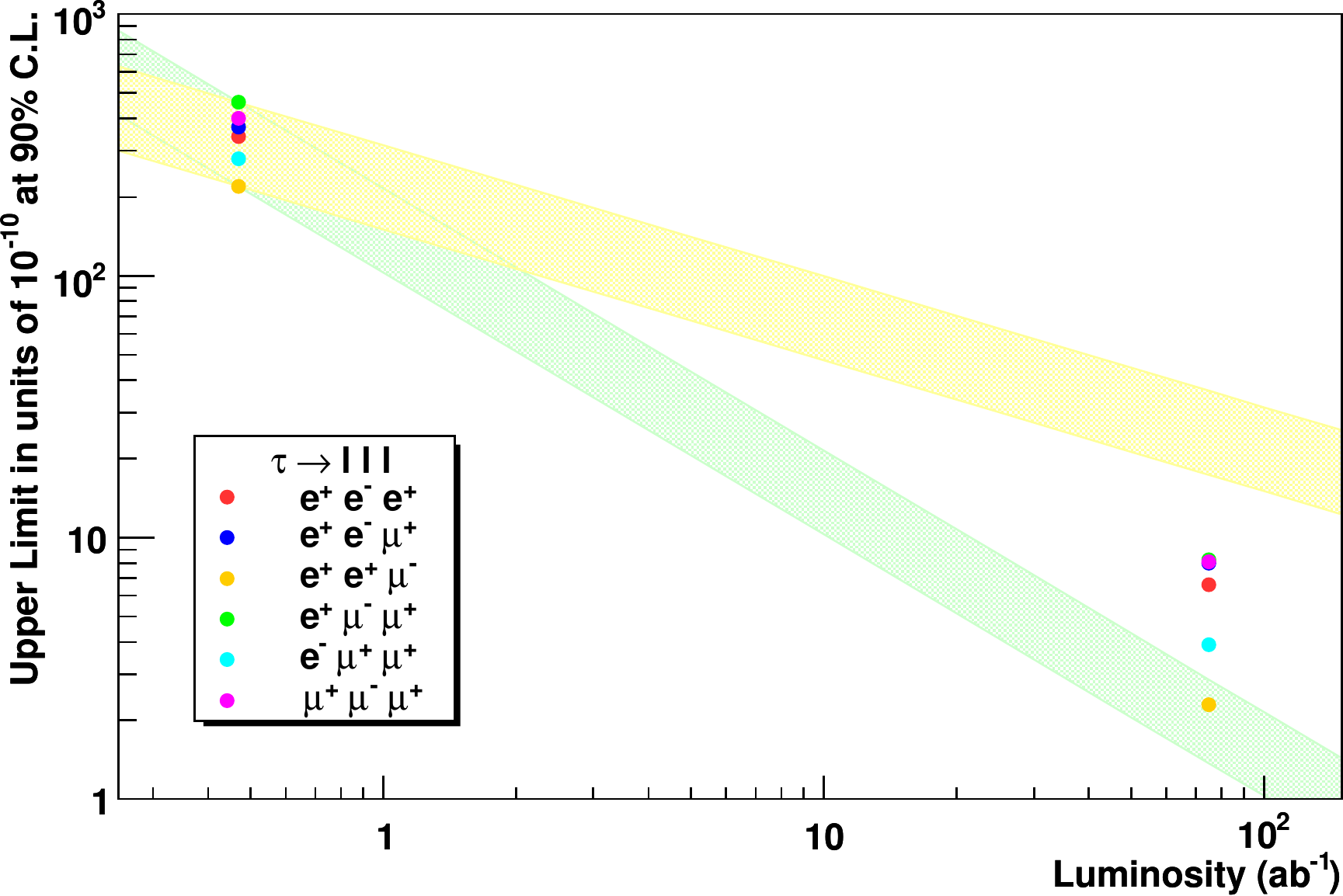}
      \end{overpic}}
  \end{center}
  \caption{\label{fig:taulll-extrapolations}%
    Expected \superb 90\% CL upper limits for \taulll LFV decays compared
    with the most recent \babar expected upper limits. The upper and
    lower bands indicate the $1/\sqrt{\cal L}$ and $1/\cal L$
    extrapolations, respectively.}
\end{figure}

By re-optimizing the \babar analyses for 75\invfb of data, we obtained
refined projected upper limits for LFV searches for $\tau$ into three
leptons~\cite{Oberhof:Thesis:2009}, which lie between the ${\propto}1/{\cal L}$ and the
${\propto}1/\sqrt{\cal L}$ extrapolations (see
Fig.~\ref{fig:taulll-extrapolations}).
\superb detector
improvements are expected to have a minor impact for these channels,
and they are conservatively neglected. After optimization
the expected backgrounds are small so beam polarization has a minor
impact (which we neglect here) on the expected reach of the search.
The re-optimization has been performed by using the \babar data and
the simulation of the \babar detector. The expected 90\% confidence
upper limits are in the range $2.3\EE{-10}$ to $8.2\EE{-10}$, depending
on the channel, and the $3\sigma$ evidence branching fractions are
$1.2\EE{-9}$ to $4.0\EE{-9}$.
For technical reasons, the amount of simulated data that has been used
(equivalent to about twice the \babar collected luminosity) permits
only a crude estimate of some specific backgrounds that have exactly
the same particle content of the signal as a result of rare
and accidental combination of SM processes. For instance, the
\babar simulated samples only contain a few events where a
$\tau\to\mu\nu\bar{\nu}$ decay combines with an \epem pair
from an ISR photon to accidentally match the $\tau$ mass and energy,
therefore the extrapolation to the \superb integrated luminosity
has some uncertainty. To improve these extrapolations, one needs 
to simulate large samples of generic background that can be used to 
study the \superb environment.  Work is ongoing within \superb to 
this end.

We consider the projected results for $\taulll$ as indicative
of the \superb sensitivity for hadronic LFV final states containing a
lepton (either a muon or electron) and a hadronic system such as a
pseudoscalar or vector meson ($\pi^0$, $\eta$, $\eta'$, $K^0_S$,
$\omega$, $\phi$, $K^{*}$, $f_1$, {\it etc.}) or a non-resonant two body 
system comprising a combination of pions or kaons.

The LFV searches $\tau\ra\ell\pi^0$ and  $\tau\ra\ell\eta \,
(\eta\ra\gamma\gamma)$, will suffer from accidental backgrounds
similar to $\taulg$.  These backgrounds arise when combinations of 
two hard photons from an initial state radiation (ISR) event accidentally 
are reconstruct with an invariant mass $m_{\gamma\gamma}$ compatible with the $\pi^0$ or $\eta$
mass.  However the rate for two hard-photon ISR emission will be about 100 
times lower than that of single hard photon emission.  By requiring $m_{\gamma\gamma}$ 
to be consistent with that of a $\pi^0$ or $\eta$ we are able to suppress
much of this background.
Consequently background from ISR events are not expected to be an issue 
at \superb luminosities.

Compared with \superkekb, the \superb project is expected to have a significantly
better reach on $\tau$ LFV as a result of: (i) a larger design instantaneous and integrated
luminosity and (ii) the availability of a highly polarized electron beam in the 
baseline design of the collider.
Table~\ref{tab:LFVExptSensitivities} summarizes the expected sensitivities
at \superb for golden LFV decays, which do not yet include possible
analysis optimization and the full exploitation of beam polarization
effects.

\begin{table}[!htb]
  \caption{\label{tab:LFVExptSensitivities}
    Expected $90\%$ CL upper limits and 3$\sigma$ evidence reach on
    LFV decays with $75 \ {\rm ab}^{-1}$ with a polarized electron
    beam.
  }
  \begin{center}
    \begin{tabular}{lcc}
      \hline 
      \multicolumn{1}{c}{\multirow{2}*{Process}} &
      \multicolumn{1}{c}{Expected} &
      \multicolumn{1}{c}{3$\sigma$ evidence} \\
      &
      \multicolumn{1}{c}{90\%\,CL upper limit} &
      \multicolumn{1}{c}{reach}  \\
      \hline
      $\BR(\tau \to \mu\,\gamma)$          &  $2.4\EE{-9}$ &  $5.4\EE{-9}$  \\
      $\BR(\tau \to e\,\gamma)$            &  $3.0\EE{-9}$ &  $6.8\EE{-9}$  \\
      $\BR(\tau \to \ell\ell\ell)$         &  $2.3{-}8.2\EE{-10}$ & $1.2{-}4.0\EE{-9}$  \\
      \hline
    \end{tabular}
  \end{center}
\end{table}

\subsection{\texorpdfstring{%
    \CP Violation in $\tau$ decay}{%
    CP Violation in $\tau$ decay}\label{sec:tau:cpv}}

\CP violation in the quark sector has been observed both
in the $K$ and in the $B$ systems. All experimental results thus far can
be explained by the complex phase of the CKM matrix.
On the contrary $\CP$ violation in the lepton sector has not been observed 
yet. $\CP$-violating effects in charged-lepton
decays within the SM are predicted to be vanishingly small.
For instance, the $\CP$ asymmetry rate of $\tau^{\pm}\to K^{\pm}\pi^0\nu$
is estimated to be of order ${\cal O}(10^{-12})$~\cite{Delepine:2005tw}.
For the decay $\tau^{\pm}\to K_S \pi^\pm \nu$ a small \CP\ asymmetry of $3.3\EE{-3}$
is induced by the known \CP-violating phase of the $\KzKzb$
mixing amplitude~\cite{Bigi:2005ts}. This asymmetry is known to a precision of 2\%.
Hence the \CP violating asymmetry in this mode can serve as a calibration measurement
for searches for effects in other $\tau$ decays.  Any observed deviation from 
expected asymmetries in $\tau$ decays would be a clear sign of NP.

Most of the known NP models cannot generate observable
\CP-violating effects in $\tau$ decays (see {\it e.g.},~\cite{Raidal:2008jk}).
The only known exceptions are RPV SUSY~\cite{Delepine:2007qg,Dreiner:2006gu}
or specific non-supersymmetric multi-Higgs models~\cite{Datta:2006kd,Kiers:2008mv,Kimura:2008gh}.
In such frameworks NP contributes at tree level, and if the sfermions or charged Higgs particles 
are relatively light with sizable couplings to the light quarks, then the NP contributions
can be significant. In some cases the $\CP$ asymmetries of various $\tau$ decay channels or 
$T$-odd  \CP-violating asymmetries in the angular distribution can be enhanced up to the 
level of $10^{-1}$.  Such enhancements from NP are compatible with limits from other observables, 
and saturate at the experimental limits obtained by CLEO~\cite{Bonvicini:2001xz,Kiers:2008mv,Kimura:2008gh}.
In particular, these models have been shown to be able to produce sizable asymmetries in the decays 
$\tau \to K \pi  \nu_\tau$,  $\tau \to K \eta^{(\prime)}  \nu_\tau$,  and 
$\tau \to K \pi \pi  \nu_\tau$~\cite{Delepine:2007qg,Datta:2006kd,Kiers:2008mv,Kimura:2008gh}.

A first search for \CP violation in $\tau$ decay has been conducted by the CLEO
collaboration~\cite{Bonvicini:2001xz}, looking for a
tau-charge-dependent asymmetry of the angular distribution of the
hadronic system produced in $\tau \to K_S\pi\nu$.  In multi-Higgs
doublet NP scenarios the \CP-violating asymmetry arises from the Higgs coupling
and the interference between $S$ wave scalar exchange and $P$ wave
vector exchange.  The Cabibbo-suppressed decay mode into $K_S\pi\nu$
has a larger mass-dependent Higgs coupling.  Furthermore, events in the
sidebands of the $K_S$ mass distributions can be used to calibrate the
detector response.  Using a data sample of 13.3\invfb ($12.2\EE{6}$ $\tau$
pairs) CLEO obtains the mean of the optimal asymmetry observable $\left<\xi\right> =
(-2.0 \pm 1.8)\EE{-3}$.
As this measurement relies on detector calibration using data side-band
events it is conceivable that \superb with 75\invab would not be
limited by systematics and could reach an experimental sensitivity of $\sigma_{\left<\xi\right>} \approx 2.4\EE{-5}$.

\subsection{\texorpdfstring{%
    Measurement of the $\tau$ electric dipole moment}{%
    Measurement of the $\tau$ electric dipole moment}\label{sec:pol:edm}}

In minimal SUSY frameworks with flavor-independent \CP-violating phases, 
like the constrained MSSM, lepton EDMs ($d_{\ell}$) scale linearly with the
lepton mass. As a result, the existing limits on the electron EDM generally
preclude any visible effect in the $\tau$ and $\mu$ cases.
In more general MSSM models, however, the strength of \CP violation may 
be different for different flavors and this simple linear scaling does not
apply \cite{Calibbi:2008qt}. A very simple example is given by models where the
 \CP-violation phases are associated with the third generation, in our case, 
to the stau trilinear coupling, $A_\tau$ \cite{Pilaftsis:1999qt}. In this 
case the $\tau$ EDM will be large and EDMs for the first two generations 
will be suppressed by small mixings. Unfortunately, there are also situations 
where the additional flavor dependence can generate a further suppression 
in the $\tau$ EDM~\cite{Calibbi:2008qt}. Thus, it is necessary 
to measure all three lepton EDMs independently in order to determine the 
flavor dependence of \CP phases.
Furthermore in multi-Higgs models the EDMs scale with the cube of the lepton
mass~\cite{Barger:1996jc}, thus $d_{\tau}$ can be enhanced significantly.
However, in this case the $e$ and $\mu$ EDMs receive sizeable two-loop 
contributions from Barr-Zee diagrams.  Again these scale linearly
with the lepton masses.  As a result, one can derive an approximate
bound
 $d_{\tau}\lsim 0.1\times (m_{\tau}/m_{\mu})^3 (m_{\mu}/m_e)d_{e}$
which is still very strong.  From the
present experimental upper bound on the electron EDM,
$d_{e}\lsim 10^{-27} e\,{\rm cm}$ it follows that $d_{\tau}\lsim 10^{-22} e\,{\rm cm}$.

The $\tau$ EDM influences both the angular
distributions and the polarization of the $\tau$ produced in \epem
annihilation. With a polarized electron beam it is possible to
reconstruct observables from the angular distribution of the products of
a single $\tau$ decay that unambiguously discriminates between the
contribution due to the $\tau$ EDM and other
effects~\cite{GonzalezSprinberg:2007qj,Bernabeu:2006wf}. 
Recent studies have provided an estimate of the \superb
upper limit sensitivity for the real part of the $\tau$ EDM
$\left|\text{Re}\{d_\tau^\gamma\}\right| \le
7.2\EE{-20}\,e\,\text{cm}$ with
75\invab~\cite{GonzalezSprinberg:2007qj}. The result
assumes a $100\%$ polarized electron beam colliding with unpolarized
positrons at the \FourS resonance. Uncertainty on the polarization is neglected, and
a perfect reconstruction of the decay $\tau \to \pi\nu$ is assumed. Studies
have been done assuming more realistic conditions:
\begin{itemize}
\item An electron beam with a linear polarization of $80\% \pm 1\%$.
\item $80\%$ geometric acceptance.
\item Track reconstruction efficiency $97.5\%\pm 0.1\%$ (similarly to
  that achieved in LEP analyses~\cite{Schael:2005am} and
  \babar ISR analyses~\cite{Davier:2008:babarIsrEffSyst}.
\end{itemize}
The process $\epem \to \tautau$ is simulated with the KK
generator~\cite{Jadach:1999vf} and the Tauola package for $\tau$
decay~\cite{Jadach:1999vf}; the simulation includes the
complete spin correlation density matrix of the initial-state beams
and  the final state $\tau$ leptons. $\tau$ EDM effects are simulated by
weighting the $\tau$ decay product angular distributions.
These studies are not complete, and do not yet include
uncertainties in reconstructing the $\tau$ direction. Preliminary
indications are that the $\tau$ EDM experimental resolution is $\approx
10\EE{-20} e\,\text{cm}$, corresponding to an angular asymmetry of
$3\EE{-5}$. Uncertainties in track reconstruction give a systematic 
contribution of $\approx 1\EE{-20}\,e\,\text{cm}$.  Asymmetries
proportional to the $\tau$ EDM depend on events that go into the same
detector regions but arise from $\tau$ leptons produced at different angles,
minimizing the impact of efficiency uncertainties. It must be noted
that all the hadronic $\tau$ channels theoretically have the same
statistical power as the $\tau \to \pi\nu$ mode in measuring the $\tau$
polarization~\cite{Kuhn:1995nn}, and can therefore be used to improve
the experimental resolution.

A search for the $\tau$ EDM using unpolarized beams has been completed at
Belle~\cite{Inami:2002ah}.  In this case, one
must measure correlations of the angular distributions of both $\tau$
leptons in the same events thereby losing both reconstruction efficiency and
statistical precision.  The analysis shows the impact of inefficiency
and uncertainties in the $\tau$ direction reconstruction, and also
demonstrates that all $\tau$ decays, including leptonic decays with two
neutrinos, provide statistically useful information for measurement of the $\tau$
EDM.  With $29.5\invfb$ of data, the experimental resolution on the
real and imaginary parts of the $\tau$ EDM is
between $0.9\EE{-17}\,e\,\text{cm}$ and $1.7\EE{-17}\,e\,\text{cm}$, including systematic effects.  
An extrapolation to \superb with a data sample of $75\invab$ (assuming systematic
effects can be reduced according to statistics) corresponds to an
experimental sensitivity of between $17\EE{-20}\,e\,\text{cm}$ and $34\EE{-20}\,e\,\text{cm}$.

\subsection{\texorpdfstring{%
    Measurement of the $\tau$ $g-2$}{%
    Measurement of the $\tau$ g-2}\label{sec:pol:gm2}}

The Standard Model prediction for the muon anomalous magnetic
moment is not in perfect agreement with recent experimental results. In particular,
 $\Delta a_{\mu} = a_{\mu}^{\rm exp} - a_{\mu}^{\rm SM}
\approx(3 \pm 1)\times 10^{-9}$.
Within the MSSM, this discrepancy can naturally
be accommodated if $\tan\beta \gsim 10$ and $\mu >0$.

A measurement of the $\tau$ anomalous magnetic moment
could be used to confirm or disprove the
possibility that the discrepancy in $\Delta a_{\mu}$ is the result of NP.
The natural scaling of heavy-particle effects
on lepton magnetic dipole moments,
implies $\Delta a_\tau/\Delta a_\mu \sim m^{2}_{\tau}/m^{2}_{\mu}$.
Thus, if we interpret the present muon discrepancy
as a signal of NP we would expect $\Delta a_{\tau}\approx 10^{-6}$.

In the supersymmetric case, such an estimate holds for all
the SPS points (see Table~\ref{tab:SPS:gm2}) and, more generally,
in the limit of almost
degenerate slepton masses. If $m_{\tilde{\nu}_{\tau}}^2\ll m_{\tilde{\nu}_{\mu}}^2$
(as happens, for instance, in the so-called effective-SUSY scenario),
$\Delta a_{\tau}$ could be enhanced up to the $10^{-5}$ level.

\begin{table}[!htb]
\caption{\label{tab:SPS:gm2}
Values of $\Delta a_{\mu}$ and $\Delta a_{\tau}$ for various SPS points.}
\begin{tabular}{ccccccc}
\hline
SPS               &  1\,a   &  1\,b   &  2  &  3  &  4  &  5   \\ \hline
$\Delta a_{\mu}\times 10^{-9}$ &  3.1 & 3.2 & 1.6  & 1.4  & 4.8  & 1.1 \\
$\Delta a_{\tau}\times 10^{-6}$ &  0.9 & 0.9 & 0.5  & 0.4  & 1.4  & 0.3 \\ \hline
\end{tabular}
\end{table}

In a manner similar to an EDM, the $\tau$ anomalous moment ($g - 
2$) influences both the angular distribution and the polarization of
the $\tau$ produced in \epem annihilation.
Polarized beams allow the measurement of the real part of the
$g - 2$ form factor by measuring the $\tau$ polarization with just the
$\tau$ polar angle distribution, i.e.\ without looking at the angular
distribution of the $\tau$ decay products in the $\tau$ rest frame.
Bernab\'eu {\it et al.}~\cite{Bernabeu:2007rr} estimate that \superb
with 75\invab will measure the real and imaginary part of the $g - 2$
form factor at the \FourS with a resolution in the range
$[0.75 - 1.7]\EE{-6}$. Two measurements of the real part of
$g - 2$ are proposed: one fitting just the polar angle distribution of the $\tau$ leptons,
and one based on the measurement of the transverse and
longitudinal polarization of the $\tau$ from the angular distribution of its decay
products.  All events with $\tau$ leptons decaying either in $\pi\nu$ or
$\rho\nu$ are considered, but no detector 
effects are accounted for. For the $\tau$ polarization measurements,
electron beams with 100\% polarization are assumed.
Studies simulating more realistic experimental conditions are ongoing.
While the polar angle distribution measurement will conceivably suffer
from uncertainties in the $\tau$ direction reconstruction, the
preliminary results on the $\tau$ EDM measurement mentioned above
indicate that reconstruction systematic effects are small for asymmetries using the $\tau$ polarization.
Using the estimated precision on the $\tau$ EDM measurement, we
expect that \superb can measure the real part
of the $g - 2$ form factor  with a statistical error of $2.4\EE{-6}$.
With such resolution, \superb will be able to measure the SM-predicted $\tau$ 
magnetic anomaly at the percent level.

\subsection{Search for second-class currents}

In the SM, approximate conservation of isospin symmetry
implies that hadronic currents corresponding to $J^{PG}= 0^{+-},
0^{-+}, 1^{++}, 1^{--}$, known as second-class currents
(SCC)~\cite{Weinberg:1958ut}, are suppressed
by the difference between the down- and up-quark masses.
This suppression makes the search for decays mediated by SCC a test of the
SM, a way to shed light on hadronic states, and a way to
search for new physics contributions to SM-suppressed decays.

For example, the SM branching fraction of the SCC decay $\tau^- \to
\eta \pi^-\nu_\tau$ is predicted to be in the range $(2{-}4)\times
10^{-6}$\cite{Nussinov:2008gx}, assuming, as commonly believed,
that the $a_0^-(980)$ is a four-quark state. A
measured branching fraction around $10^{-5}$ would mean
that $a_0^-(980)$ is actually a $d\bar u$ meson, and a value greater
than about $3\times 10^{-5}$ could indicate the possible existence of
a new-physics scalar component in the weak interaction.

There is no published search for $\tau^- \to \eta \pi^-\nu_\tau$, but
BaBar has searched for the related decay $\tau^- \to \eta'(958)
\pi^-\nu_\tau$ using $384$~fb$^{-1}$, and has determined its branching
fraction to be smaller than $7.2\EE{-6}$~\cite{Aubert:2008nj}.
With the full data set of $75$~ab$^{-1}$, SuperB could push the limit
to about a third of the theoretical upper limit of $1.4\EE{-6}$~\cite{Nussinov:2009sn}.  Similarly, the experimental reach
for $\tau^- \to \eta \pi^-\nu_\tau$ is expected to place limits on
scalar new-physics contributions and may be sensitive enough to elucidate
the nature of the $a_0^-(980)$.
BaBar has also studied $\tau^- \to \omega \pi^-\nu_\tau$ and has
set a 0.69\% upper limit on the SCC fraction in this
decay~\cite{Aubert:2009cc}. 

\graphicspath{{Bphysics/}{Bphysics/}}
\section{\B Physics at the $\Upsilon(4{\rm S})$}\label{sec:bphysics}

%
%

This section contains highlights of the $B$ physics programme from \superb.
The focus of much of the material presented here is the search for physics
beyond the standard model.  The following sub-sections discuss time-dependent 
\CP measurements, theoretical and experimental aspects of a number of rare decay
golden channels including $B\to K^{(*)}\nu\overline{\nu}$, $B\to X_{s,d}\gamma$, 
$B\to X_{s,d} \ell^+\ell^-$, $B\to \ell \nu (\gamma)$, measurements of \Vub
and \Vcb, $\Delta m_d$ and CPT tests.  In addition to these processes that can 
shed light on new physics there are standard model control measurements
comprising precision CKM determination.  One area of $B$ decays that 
is a rich test bed for theoretical understanding and testing of new tools
is that of charmless hadronic decays.  This area will remain interesting
in the era of \superb.

\subsection{New Physics in \CP violation\label{sec:bdecays:cpv}}

\subsubsection{$\Delta S$ measurements}

It is possible to use time-dependent CP (TDCP) asymmetry measurements to search for 
signs of new physics (NP) in the form of heavy particles contributing to loop 
topologies and additional contributions from NP for tree level processes.
In order for such a search to have a reasonable chance of seeing NP one 
has to study a mode, or set of modes, that are loop (or penguin) dominated.  The golden
channels for this type of measurement fall into the categories of penguin-dominated
TDCP measurements of $b\to s$ transitions and tree level
$b\to c\overline{c}s$ transitions.  These measurements can 
trigger the observation of NP if the value of 
$S^f = \stwobeff$ measured in one of these decays deviates significantly from that
measured in the tree dominated $c\overline{c}s$ decays like $J/\psi K^0$ 
($S=\sin 2\beta$), or from that predicted by the Standard Model ($S^{SM}$).
The current level of such deviations $\Delta S^{tree} 
= S^{SM} - S$ and $\Delta S^{penguin} = S^{f} - S_{SM}$ from
the theoretically clean penguin (tree) modes are $2.7\sigma$ ($2.1\sigma$) from the SM 
prediction~\cite{Buras:2008nn,Lunghi:2009sm}, and the deviation 
$\Delta S^{f} = S^{f} - S$ is small using current data.  Such tantalizing hints
of a deviation beckons us to study this area further 
to see if these deviations are indications of NP, or 
if these effects are merely statistical fluctuations. 

The interpretation of the precise data on the TDCP asymmetries
in terms of the CKM parameters requires a reasonable control over the hadronic 
matrix elements. In particular, the ratio of the penguin versus the tree contribution
has to be known from the theoretical side in order to turn the measurements of 
CP asymmetries into a test of the Standard Model (SM).  

Explicitly, the typical amplitude for a non-leptonic two-body decay can be written as: 
\begin{equation} \label{Amp} 
A(B^0\to f) = {\cal A}\left[1+ r_f \, e^{i\delta_f}  \,  e^{i\theta_f}\right],
\end{equation}
where usually ${\cal A}$ is the tree amplitude and $r_f$ denotes the modulus of the 
penguin-over-tree ratio, which has a strong phase $\theta_f$. The weak phase $\delta_f$ 
is in the cases at hand the CKM angle $\gamma$, while the modulus of the CKM factors 
is absorbed into $r_f$.  The observables $C^f$ and $S^f$  can be expressed as:
\begin{eqnarray} 
 C^f &=&  
-2  r_f \sin\theta_f \sin\delta_f,   \\   
 S^f &=&   
\sin\phi+2 r_f \cos\theta_f\sin(\phi+\delta_f) \\ 
\nonumber &&  +r_f^2\sin(\phi+2\delta_f),  
 \end{eqnarray} 
where $\phi$ is the mixing phase stemming from the $\Delta B = 2$ interaction. 

The key issue in the theoretical understanding of CP asymmetries is the ratio $r_f$, 
which for the ``gold-plated modes'' is doubly Cabibbo suppressed. However, at the precision 
of a \sff even a small $r_f$ will be observable and hence relevant for 
the analysis. 

There are basically two ways to assess the ratio $r_f$. From the theoretical 
side one may compute $r_f$ with a non-perturbative method, namely a variant
of factorization (when applicable)~\cite{Beneke:2005pu,Li:2006jv,Williamson:2006hb} or
try a phenomenological estimates otherwise~\cite{Boos:2004xp,Gronau:2008cc}.
The second possibility is to rely more on data, making use of approximate flavour symmetry
relations between matrix
elements~\cite{Ciuchini:2005mg,Faller:2008zc,Jung:2009pb,Fleischer:1999nz}~\footnote{In practice,
these two ways are not well separated: theoretical approaches often use phenomenological constraints
and data-drive approaches relies also on theory to some extent.}. 

Theoretical calculations of $\Delta S^f = S^f - \sin \phi$ typically produce results in the range
$\Delta S^f \sim 10^{-3}$ up to few$\times 10^{-2}$. 
Based on data there are indications that $r_f$ in fact can be sizable~\cite{Ciuchini:2005mg}. 
As an example, based  on the data of $B \to J/\psi \pi^0$ one extracts values for $r_f$ which can be
as large as 0.8, yielding  shifts as large as $\Delta S^f \sim - 7\%$~\cite{Faller:2008zc}. 
In fact the data on  $B \to J/\psi \pi^0$  indicate a negative shift, which would soften the
currently existing tension between $\sin 2 \beta$ and $V_{ub}$. 

Table~\ref{tbl:bdecays:spenguin} summarizes the precision of current measurements~\cite{hfag} and
\superb extrapolations, together with a reference set of theoretical predictions for
$\Delta S^f$~\cite{Beneke:2005pu,Cheng:2005bg,Cheng:2005ug} in the SM.
The NP discovery potential deviations required at a \superb factory to observe NP are also shown.
Where appropriate, reducible systematic uncertainties, and data driven bounds on the SM uncertainties have 
been scaled by luminosity from current measurements in making these 
extrapolations.  

With the theoretical predictions for $\Delta S^f$ used in Table~\ref{tbl:bdecays:spenguin},
the golden $b\to s$ penguin modes for this NP search are $\B^0\to \eta^\prime K^0$ and
$\B^0\to K^0_SK^0_SK^0_S$, together with $\B^0\to f^0_0 K^0_S$ for which the calculation of
the SM uncertainty is however less accurate. Some interesting three-body modes, notably
$\B^0\to\phi K^0_S \pi^0$ and $\B^0\to\pi^0\pi^0 K^0_S$, presently lack an assessment of the
theoretical uncertainty.

One can see from the table that it is possible to discover NP if there is a deviation of 0.02
from SM expectations of \stwob\ as measured in tree decays.  It is possible to observe a deviation of $5\sigma$
or more of about 0.1 in \stwobeff\ from $b\to s$ transitions in the golden modes.
It is worth noting however that these conclusions may change depending on the models used for
computing $\Delta S^f$. Indeed not all sources of theoretical error are under control in these
estimates and in some case even the sign of the correction can be model dependent. On the other hand,
theoretical estimates not explicitly data-driven also rely on experimental information to some extent and
and could benefit from the \superb large data set. This improvement has not be taken into account in 
Table~\ref{tbl:bdecays:spenguin}. Clearly, if \superb will find significant deviations in these measurements,
further theoretical and phenomenological work will be required to pin down the SM value of $\Delta S^f$
and firmly establish the presence of NP. In the absence a theoretical leap in the understanding of
non-leptonic decays, data-driven methods are expected to play a prominent role. In this respect, the opportunity
of measuring several modes with different theoretical uncertainties, but possibly correlated NP contributions, is
a unique advantage of \superb.

The golden $b\to d$ process is $\B^0 \to J/\psi \pi^0$ from an experimental perspective. Yet current theoretical
understanding indicates that the measurements of $S^f$ for $b\to d$ modes are theoretically limited.

\begin{table*}[!ht]
\caption{Current experimental precision of $S^f$~\cite{hfag}, and that expected at a \superb experiment with 75\invab of data.  
The $5\sigma$ discovery limit deviations at 75\invab are also listed.  The first entry in the table
corresponds to the tree level calibration mode, and the next two sections of the table refer to $b\to s$
and $b\to d$ transitions. Theoretical estimates of $\Delta S^f$ are taken from Refs.~\cite{Beneke:2005pu,Cheng:2005bg,Cheng:2005ug}.
A long dash `$-$'denotes that there is no theoretical estimate of $\Delta S^f$ computed yet for a given mode,
thus the corresponding discovery limits are not evaluated.}\label{tbl:bdecays:spenguin}
\begin{center}
\small
\begin{tabular}{l|ccc|ccc|cc}\hline
Mode                & \multicolumn{3}{c}{Current Precision} & \multicolumn{3}{|c|}{Predicted Precision (75\invab)} &
\multicolumn{2}{c}{Discovery Potential} \\ 
                    & Stat.        & Syst.       & $\Delta S^f$(Th.) & Stat.        & Syst.       & $\Delta S^f$(Th.)  & $3\sigma$ & $5\sigma$ \\ \hline
$J/\psi K^0_S$      & 0.022        & 0.010       & $0\pm0.01$        & 0.002        & 0.005       & $0\pm0.001$        & 0.02  & 0.03 \\ \hline
$\eta^\prime K^0_S$ & 0.08         & 0.02        & $0.015\pm0.015$   & 0.006        & 0.005       & $0.015\pm0.015$    & 0.05  & 0.08\\
$\phi K^0_S \pi^0$  & 0.28         & 0.01        & $-$               & 0.020        & 0.010       & $-$                & $-$   & $-$\\
$f_0 K^0_S$         & 0.18         & 0.04        & $0\pm0.02$        & 0.012        & 0.003       & $0\pm0.02$         & 0.07  & 0.12\\
$K^0_SK^0_SK^0_S$   & 0.19         & 0.03        & $0.02\pm0.01$     & 0.015        & 0.020       & $0.02\pm0.01$      & 0.08  & 0.14\\
$\phi K^0_S$        & 0.26         & 0.03        & $0.03\pm0.02$     & 0.020        & 0.005       & $0.03\pm0.02$      & 0.09  & 0.14\\
$\pi^0 K^0_S$       & 0.20         & 0.03        & $0.09\pm0.07$     & 0.015        & 0.015       & $0.09\pm0.07$      & 0.21  & 0.34\\
$\omega K^0_S$      & 0.28         & 0.02        & $0.1\pm0.1$       & 0.020        & 0.005       & $0.1\pm0.1$        & 0.31  & 0.51\\
$K^+K^-K^0_S$       & 0.08         & 0.03        & $0.05\pm0.05$     & 0.006        & 0.005       & $0.05\pm0.05$      & 0.15  & 0.26\\
$\pi^0\pi^0 K^0_S$  & 0.71         & 0.08        & $-$               & 0.038        & 0.045       & $-$                & $-$   & $-$\\ 
$\rho K^0_S$        & 0.28         & 0.07        & $-0.13\pm0.16$    & 0.020        & 0.017       & $-0.13\pm0.16$     & 0.41  & 0.69\\\hline
$J/\psi \pi^0$      & 0.21         & 0.04        & $-$               & 0.016        & 0.005       & $-$                & $-$   & $-$\\
$D^{*+}D^{*-}$      & 0.16         & 0.03        & $-$               & 0.012        & 0.017       & $-$                & $-$   & $-$\\
$D^{+}D^{-}$        & 0.36         & 0.05        & $-$               & 0.027        & 0.008       & $-$                & $-$   & $-$\\ \hline
\end{tabular}
\end{center}
\end{table*}

\subsection{Theoretical aspects of rare decays}
\label{sec:bphysics:rare}

\subsubsection{New physics in $B\to K^{(*)}\nu\bar\nu$ decays}
\label{sec:bphysics:rare:snunubar}

Rare $B$ decays with a $\nu\bar\nu$ pair in the final state are interesting probes of new 
physics, since they allow one to transparently study $Z$ penguin and other electroweak 
penguin effects in the absence of dipole operator and Higgs penguin contributions, which 
are often more important than $Z$ contributions in $b\to s\ell^+\ell^-$ decays. Moreover, 
since the neutrinos escape the detector unmeasured, the $B\to K^{(*)}+E_{\rm miss}$ channel 
can also contain contributions from other light SM-singlet particles substituting the neutrinos in the decay.

The inclusive decay $\bar B\to X_s\nu\bar\nu$ is the theoretically cleanest $b\to s\nu\bar\nu$ decay due to the 
absence of form factor uncertainties, but is experimentally very challenging to measure. The exclusive decay 
$B\to K\nu\bar\nu$ currently provides most stringent constraints on NP with an experimental upper bound 
only a factor of three above the SM prediction. The $B\to K^*\nu\bar\nu$ decay has the advantage that, in 
addition to its differential decay rate,  it in principle provides access to an additional observable via the 
angular distribution of the $K^*$ decay products $K^\pm\pi^\mp$: the $K^*$ longitudinal polarization fraction 
$F_L(q^2)$, which is theoretically very clean since form factor uncertainties cancel to a large 
extent~\cite{Altmannshofer:2009ma}.

The SM predictions and current experimental upper bounds are summarized in table~\ref{tab:bsnn-SMexp}. However, for the modes
involving a charged $B$ in the initial state, it should be noted that the bounds in the rightmost column do not take into account
an important background from $B\to\tau\nu$ decays with the $\tau$ subsequently decaying to a $K$ or $K^*$ and a (anti-)neutrino,
which has been recently pointed out in~\cite{Kamenik:2009kc}. This contribution is expected to be small at \superb (roughly
15--30\% of the SM value for $B^+\to K^+\nu\bar\nu$). With data available at \superb it will be possible to accurately determine
the background contribution from $\mathcal B(B\to\tau\nu)$  decays and on doing so increase the precision with which we can extract
the signal. The sensitivities quoted in the table are conservative for this reason.

\begin{table}
\caption{SM predictions and experimental 90\% C.L. upper bounds for the four $b\to s\nu\bar\nu$ observables.}
\label{tab:bsnn-SMexp}
\centering
\renewcommand{\arraystretch}{1.5}
\begin{tabular}{lll}
\hline
Observable  & SM prediction &  Experiment \\
\hline
$\mathcal B (B^0 \to K^{*0} \nu\bar\nu)$ & $( 6.8^{+1.0}_{-1.1} ) \times 10^{-6}$~\cite{Altmannshofer:2009ma} & $< 80 \times 10^{-6}$ \cite{:2008fr} \\
$\mathcal B(B^+ \to K^+ \nu\bar\nu)$   & $( 3.6 \pm 0.5 ) \times 10^{-6}$~\cite{Bartsch:2009qp} & $< 14 \times 10^{-6}$ \cite{:2007zk} \\
$\mathcal B(\bar B \to X_s \nu\bar\nu)$ & $( 2.7\pm0.2 ) \times 10^{-5}$ \cite{Altmannshofer:2009ma} & $< 64 \times 10^{-5}$ \cite{Barate:2000rc} \\
$\langle F_L(B \to K^* \nu\bar\nu) \rangle$ & $0.54 \pm 0.01$ \cite{Altmannshofer:2009ma} & --  \\
\hline
\end{tabular}
\renewcommand{\arraystretch}{1}
\end{table}

The $b\to s\nu\bar\nu$ transition is governed by the effective Hamiltonian
\begin{equation}
{\mathcal H}_{\rm eff} = - \frac{4\,G_F}{\sqrt{2}}V_{tb}V_{ts}^*\left(C^\nu_L \mathcal O^\nu_L +C^\nu_R \mathcal O^\nu_R  \right) ~+~ {\rm h.c.} ~,
\end{equation}
where the operators are $\mathcal{O}^\nu_{L,R} =\frac{e^2}{8\pi^2}(\bar{s}  \gamma_{\mu} P_{L,R} b)(  \bar{\nu} P_L \nu)$,
and the $C^\nu_{L,R}$ are the corresponding Wilson coefficients. In the SM, $C^\nu_L \approx - 6.38$ and the right-handed Wilson
coefficient vanishes. In models beyond the SM, both $C^\nu_L$ and $C^\nu_R$ can be non-zero and complex; however, the two exclusive
and the inclusive decay rates as well as $F_L$ only depend on two independent combinations of these Wilson coefficients, which can
be written as
\begin{equation}  \label{eq:epsetadef}
 \epsilon = \frac{\sqrt{ |C^\nu_L|^2 + |C^\nu_R|^2}}{|(C^\nu_L)^{\rm SM}|}~, \qquad
 \eta = \frac{-{\rm Re}\left(C^\nu_L C_R^{\nu *}\right)}{|C^\nu_L|^2 + |C^\nu_R|^2}~,
\end{equation}
implying $(\epsilon,\eta)_{\rm SM}=(1,0)$.
This allows one to express the observables of $b\to s\nu\bar\nu$ decays in a general NP model as
\begin{eqnarray}
\label{eq:epseta-BKsnn}
 R(B \to K^* \nu\bar\nu) & =&  (1 + 1.31 \,\eta)\epsilon^2, \\
\label{eq:epseta-BKnn}
 R(B \to K \nu\bar\nu)   & =& (1 - 2\,\eta)\epsilon^2, \\
\label{eq:epseta-BXsnn}
 R(\bar B \to X_s \nu\bar\nu) & =& (1 + 0.09 \,\eta)\epsilon^2, \\
\label{eq:epseta-FL}
 \langle F_L \rangle/\langle F_L \rangle_{\rm SM}             & =&  \frac{(1 + 2 \,\eta)}{(1 + 1.31 \,\eta)},
\end{eqnarray}
where $R(X)=\mathcal B(X)/\mathcal B(X)_{\rm SM}$ and $\langle F_L \rangle$ refers to $F_L$ 
appropriately integrated over the neutrino invariant mass~\cite{Altmannshofer:2009ma}. 
Eq.~(\ref{eq:epseta-FL}) highlights an important feature of the observable $F_L$: 
it only depends on $\eta$ and not on $\epsilon$.  Any experimentally observed 
deviation from the SM prediction of $F_L$ would unambiguously imply the presence of right-handed currents.

\begin{figure}[tb]
\centering
\includegraphics[width=0.47\textwidth]{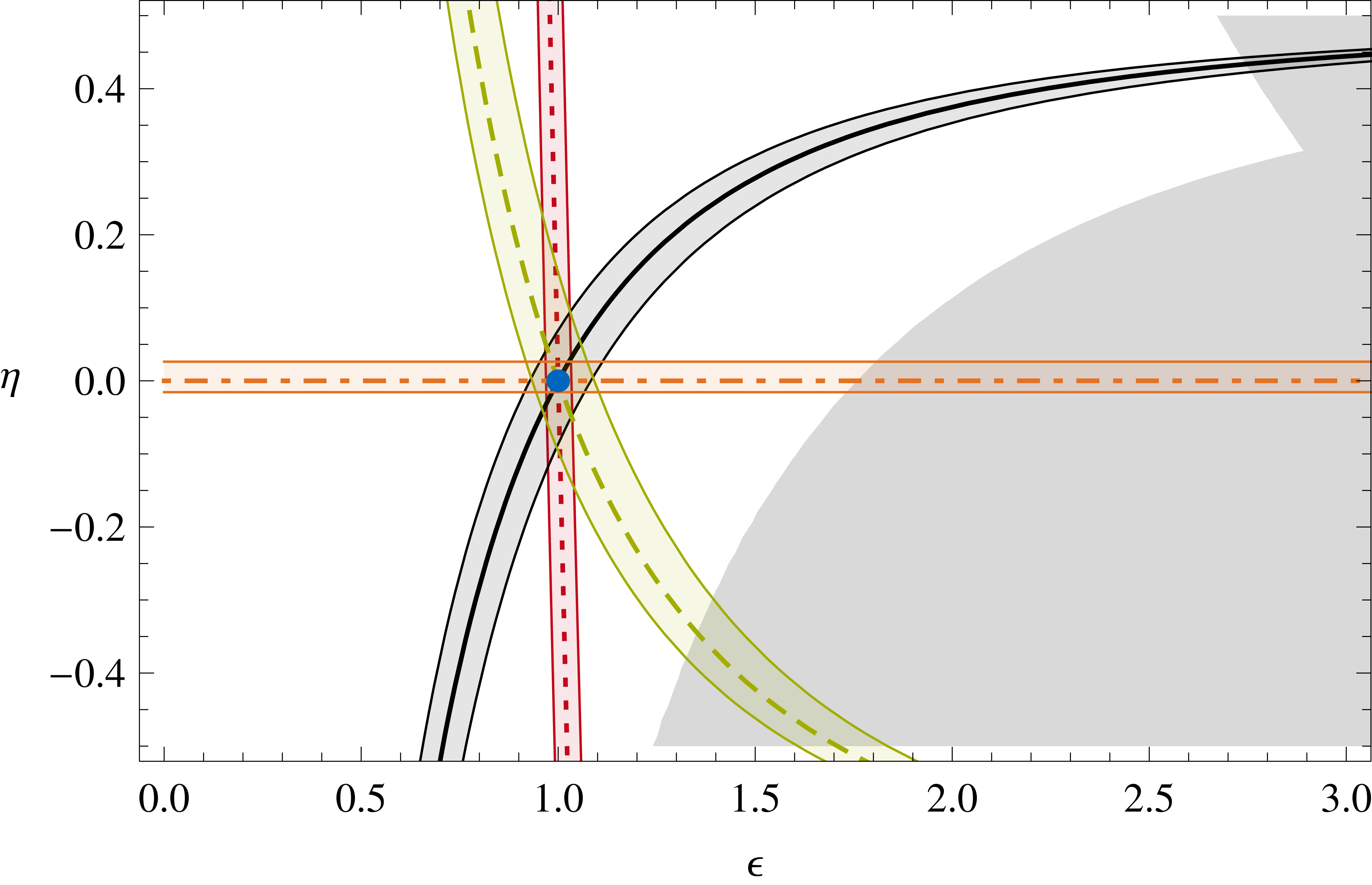}
\caption{Hypothetical constraints on the $\epsilon$-$\eta$-plane, assuming all four observables have been measured with
infinite precision. The error bands refer to theory uncertainties only. The green band (dashed line) represents
$\mathcal B(B \to K^* \nu\bar\nu)$, the black band (solid line) $\mathcal B(B \to K \nu\bar\nu)$, the red band (dotted line)
$\mathcal B(B \to X_s \nu\bar\nu)$ and the orange band (dot-dashed line) $\langle F_L \rangle$. The shaded area is ruled out
experimentally at the 90\% confidence level.}
\label{fig:epseta}
\end{figure}

In Fig.~\ref{fig:epseta}, the existing constraints on the $\epsilon$-$\eta$ plane are shown 
in combination with the hypothetical constraints arising from a measurement of all 
four observables with infinite precision. It is self-evident that the complementarity 
between the different modes allows us to over-constrain the point $(\epsilon, \eta)$.

Concerning the size of possible NP effects in $b\to s\nu\bar\nu$ decays, it is instructive to parameterize the dominance of
$Z$ penguin contributions in many models by a modified $bsZ$ coupling~\cite{Buchalla:2000sk}. In this way, the NP contributions
to $b\to s\nu\bar\nu$ transitions are automatically correlated to other $b\to s$ transitions sensitive to this coupling.
A particularly stringent constraint in this respect turns out to be the branching ratio of the inclusive decay
$\bar B\to X_s\ell^+\ell^-$. Assuming no NP contributions apart from the modified $bsZ$ couplings, the measurement of this
branching ratio implies that the $b\to s\nu\bar\nu$ branching ratios cannot be enhanced by more than a factor of two above the SM.
However, this bound can be weakened substantially by assuming other NP contributions to $\bar B\to X_s\ell^+\ell^-$, such as
photon penguins.

Very large effects can in principle be obtained in family non-universal $Z'$ models. If 
the $Z'$ couples more strongly to neutrinos than to charged leptons, the constraints 
on the flavour-changing couplings from $\bar B\to X_s\ell^+\ell^-$ and $B_s\to\mu^+\mu^-$ 
can be weakened or entirely absent.

In the Minimal Supersymmetric Standard Model (MSSM), NP effects in the $b\to s\nu\bar\nu$ 
observables turn out to be quite limited, even in the general, non-minimal flavour violating 
case~\cite{Yamada:2007me,Altmannshofer:2009ma}. While gluino contributions to $C_{L,R}^\nu$ 
are strongly constrained by the $b\to s\gamma$ decay, $\tan\beta$-enhanced Higgs-mediated 
contributions to $C_R^\nu$ are also negligible once the stringent bound from $B_s\to\mu^+\mu^-$ 
is taken into account. Visible effects can thus only be generated in presence of a 
sizable $(\delta_u^{RL})_{32}$ mass insertion by means of up-squark-chargino loops. 
Consequently, while $C_R^\nu$ (and thus $F_L$) is SM-like, the branching ratios can be 
enhanced or suppressed by at most 35\%~\cite{Altmannshofer:2009ma}.

Spectacular NP effects can be obtained in models with light invisible particles produced
in the $b\to s$ transition, even if the $b\to s\nu\bar\nu$ amplitude is unaffected by NP,
since experiments actually measure the process $b\to s+E_{\rm miss}$. This can happen
\eg in models with light scalar dark matter~\cite{Bird:2004ts}, light
neutralinos~\cite{Adhikari:1994wh,Dreiner:2009er}, light Next to Minimal Supersymmetric
Standard Model (NMSSM) pseudoscalar Higgs~\cite{Hiller:2004ii} or light radions~\cite{Davoudiasl:2009xz}.
A crucial point in this case is that the invariant mass distributions of the $b\to s+E_{\rm miss}$
decays can be strongly modified, which has to be taken into account in the experimental searches.
In addition, the parameterization in Eqns.~(\ref{eq:epseta-BKsnn})--(\ref{eq:epseta-FL}) do not apply
in this case. Therefore, a contribution from particles other than neutrinos to the $b\to s+E_{\rm miss}$
observables would be signaled by a failure of the individual constraints on the $\epsilon$-$\eta$ plane
meeting at a single point.

Summing up,  $b\to s\nu\bar\nu$ transitions are interesting probes of NP, as one
can perform a theoretically clean study of non-standard $Z$ penguin effects. The 
experimentally accessible 
observables are the differential branching ratios of  $B\to X_s\nu\bar\nu$, $B\to K\nu\bar\nu$ 
and $B\to K^*\nu\bar\nu$ and the $K^*$ longitudinal polarization fraction $F_L(q^2)$ in  
$B\to K^*\nu\bar\nu$. In a general NP model, these observables depend on two real parameters,
$\epsilon$ and $\eta$, which can be over-constrained by the four measurements.

While the effects in models with minimal flavour violation (MFV) are quite limited and 
in the non-MFV MSSM can reach at most 35\%, well-motivated models exist where much 
larger effects are possible, \eg $Z'$ models with family non-universal couplings.

Since the neutrinos in the final state cannot be detected, the actual measurements 
probe the process $b\to s + E_{\rm miss}$, which can receive contributions from 
particles other than neutrinos in models with new light invisible particles. In 
this case, spectacular effects and strong modifications of the invariant mass 
spectra can be obtained.

\subsubsection{$\bar B \rightarrow  X_s \gamma$ and $\bar B \rightarrow X_s \ell^+ \ell^-$}

The two inclusive rare decays $\bar B \rightarrow X_s \gamma$ and $\bar B \rightarrow X_s \ell^+ \ell^-$
are both dominated by perturbative contributions. 
 The relevant Lagrangian density can be found in Refs.~\cite{Misiak:2006zs} and~\cite{Huber:2007vv}.
The SM prediction of ${\cal B}(\bar B \to X_s \gamma$) for $E_{\gamma} > 1.6 \ {\rm GeV}$ is
\[
  \label{eq:btosgamma:NNLL}
  {\cal B}( \bar B \to X_s\gamma)|_{E_\gamma >1.6 \ \rm{GeV}} =
  \left\{
    \begin{array}{ll}
      (3.15  \pm 0.23) \times 10^{-4}   & \hbox{\cite{Misiak:2006zs}}.
    \end{array}
  \right.
\]
The overall uncertainty consists of non-perturbative ($5\%$),
parametric ($3\%$), higher-order ($3\%$) and $m_c$-interpolation ($3\%$), which
have been added in quadrature. Ref.~\cite{Becher:2006pu} found a different (but compatible) result using
resummation techniques, which however has been strongly questioned in Ref.~\cite{Misiak:2008ss}.

This result is based on a global effort to calculate the perturbative corrections to the NNLL
level~\cite{Misiak:2004ew,Gorbahn:2004my,Gorbahn:2005sa,Czakon:2006ss,Blokland:2005uk,Melnikov:2005bx,Asatrian:2006ph,Asatrian:2006sm,Misiak:2006ab,Bieri:2003ue}.
There are other perturbative NNLL corrections that are not yet included
in the present NNLL estimate, but are expected to be smaller than the current
uncertainty, producing a shift of the central value of about 1.6$\%$ at maximum~\cite{Misiak:2008ss}.

While the uncertainties due to the input parameters and due to the
$m_c$~interpolation could be further reduced, the perturbative error
of $3\%$ will remain unless a new major effort to compute the NNNLO
is carried out.
However, the theoretical prediction has now reached the non-perturbative
boundaries. The largest uncertainty is presently due to non-perturbative
corrections that scale with $\alpha_s \Lambda_{\rm QCD}/m_b$.
A local expansion is not possible for these contributions.
A specific piece of these additional non-perturbative corrections
has been estimated~\cite{Lee:2006wn}, and found to be consistent with the
dimensional estimate of $5\%$.   More recently, a systematic analysis  
found an overall uncertainty of $4-5\%$ due to all power corrections of this kind~\cite{Benzke:2010js}.

Two explicit examples demonstrate the stringent constraints that
can, with these uncertainties, be derived from the measurement of the $\bar B \rightarrow X_s \gamma$
branching fractions.

\begin{figure}[!htb]
  \begin{center}
    \includegraphics[width=8cm,angle=0]{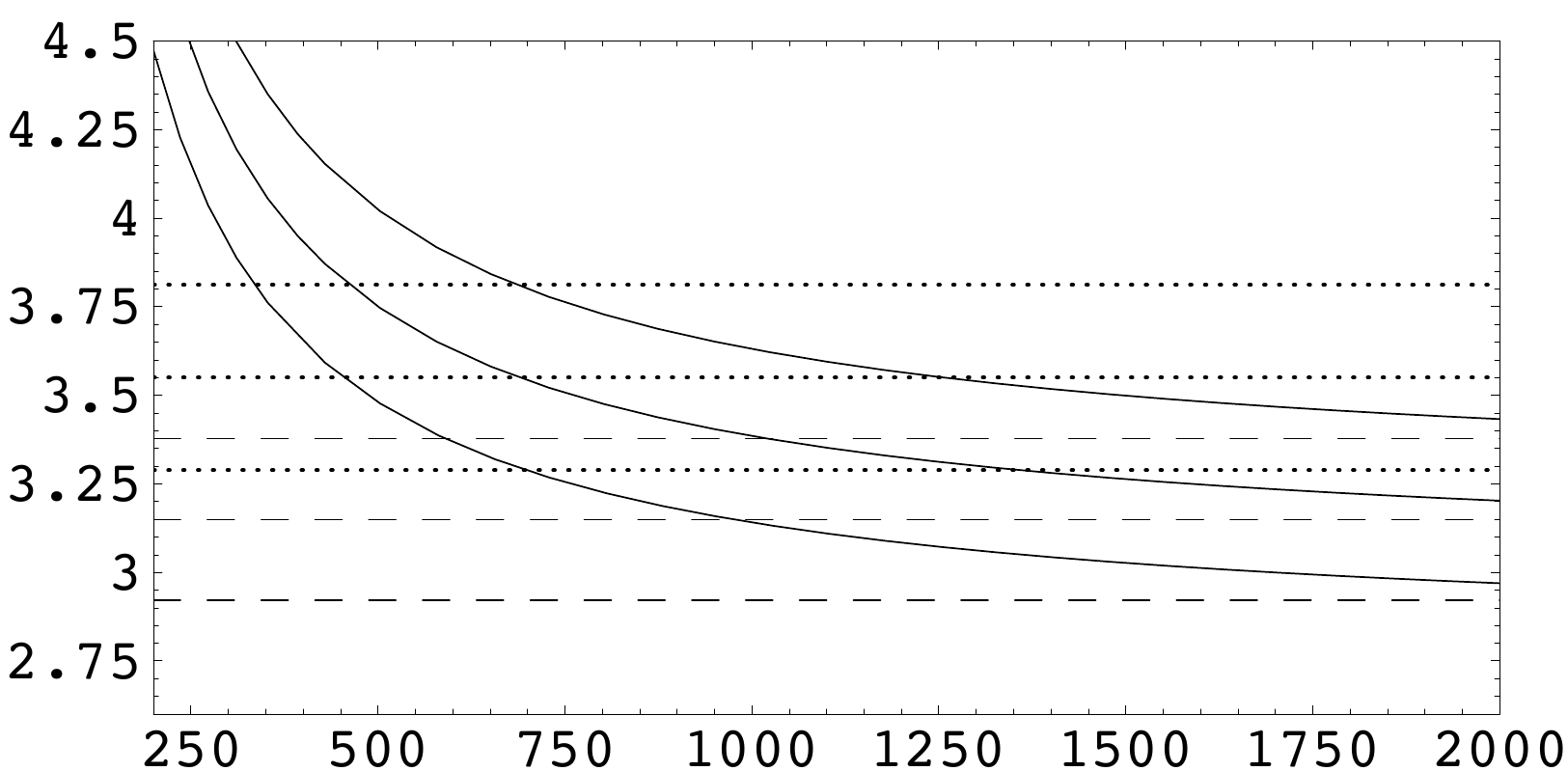}
  \end{center}
  \caption{
    ${\cal B}(\bar B \to X_s \gamma) \times 10^{-4}$ as a function of the charged
    Higgs boson mass $M_{H^{+}}$ (GeV) in the 2HDM~II for $\tan\beta=2$ (solid lines).
    Dashed and dotted lines show the SM and experimental results,
    respectively.  The central line for each of the cases corresponds to the central value,
    and the other lines have been obtained by combining errors in quadrature.
    \label{fig:MHc}
  }
\end{figure}

Fig.~\ref{fig:MHc} shows the dependence of ${\cal B}(\bar B \to X_s \gamma)$ on the
charged Higgs mass in the 2-Higgs-doublet model (2HDM-II) ~\cite{Misiak:2006zs}.
The bound on $M_{H^{+}}$ = 295 GeV at 95$\%$~CL, shown in Fig.~\ref{fig:MHc}, is currently
the strongest available lower limit on the charged Higgs mass.

\begin{figure}[!htb]
\begin{center}
\vspace{2mm}
\makebox{
\begin{psfrags}
\newcommand{\psfragtextscale}{1}
\providecommand{\psfragtextscale}{1}
\providecommand{\psfragmathscale}{\psfragtextscale}
\providecommand{\psfragnumericscale}{\psfragtextscale}
\providecommand{\psfragtextstyle}{}
\providecommand{\psfragmathstyle}{}
\providecommand{\psfragnumericstyle}{}

\psfrag{x}[cc][cc][1.1][0]{$1/R~[{\rm TeV}]$}
\psfrag{y}[bc][bc][1.1][0]{${\cal B} (\bar{B} \to X_s \gamma)~[10^{-4}]$}
\makebox{\hspace{-1.1cm} \includegraphics[width=3.5in,height=2.25in]{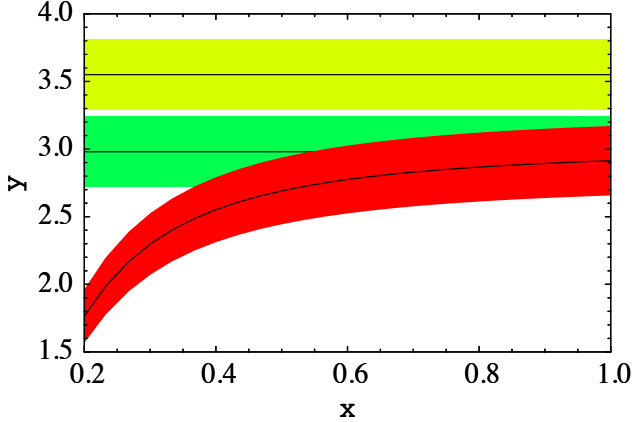}}
\end{psfrags}
}
\end{center}
\vspace{-4mm}
\caption{Branching fraction for $E_0 = 1.6 \, GeV$ as a function of $1/R$.
  The red (dark gray) band corresponds to the LO mUED result.
  The $68 \%$ CL range and central value of the
  experimental/SM result is indicated by the yellow/green
  (light/medium gray) band underlying the straight solid line.}
\label{fig:ACDLO}
\end{figure}

Similarly, the bound on the inverse compactification radius of the minimal
universal extra dimension model (mUED) derived from ${\cal B}(\bar B \rightarrow X_s \gamma)$~\cite{Haisch:2007vb}
is $1/R > 600$ GeV at $95\%$ confidence level, as shown in Fig.~\ref{fig:ACDLO}.

The angular decomposition of the $\bar B\to X_s\,\ell^+\ell^-$ decay rate provides three independent observables, $H_T$, $H_A$,
$H_L$ from which one can extract the short-distance electroweak Wilson coefficients that test for NP~\cite{Lee:2006gs}:
\begin{eqnarray}\label{eq:d3Gamma}
\frac{\d^3\Gamma}{\d q^2\,  \d z}
&=& \frac{3}{8} \bigl[(1 + z^2) H_T(q^2)
\nonumber\\ &\quad&
+ 2(1 - z^2) H_L(q^2)
\nonumber\\ &\quad&
+  2 z H_A(q^2)
\bigr]
\,.\end{eqnarray}
Here, $z=\cos\theta$, where $\theta$ is the angle between the $\ell^\pm$ and $B$ meson three 
momenta in the di-lepton rest frame, $H_A$ is equivalent to the forward-backward asymmetry, and the $q^2$ 
spectrum is given by $H_T + H_L$.  The observables  depend on the Wilson coefficients 
$C^{eff}_7, C^{eff}_9$, and $C^{eff}_{10}$ in the SM. 

In the $\bar B \rightarrow X_s \ell^+\ell^-$ system, one has to remove contributions from $c \bar c$
resonances that appear as large peaks in the dilepton invariant mass spectrum, using 
appropriate  kinematic cuts.
It is conventional to define ``perturbative windows'' with $s=q^2/m_b^2$ away from charmonium resonances,
namely the low dilepton-mass region $ 1 \ {\rm GeV} < q^2 < 6 \ {\rm GeV}$
and the high dilepton-mass region with $q^2 > 14.4 \ {\rm GeV}$.
In these windows theoretical predictions for the invariant mass spectrum
are dominated by the perturbative contributions; in principle
a theoretical precision of order $10\%$ is possible.

The calculations in $\bar B \to X_s \ell^+ \ell^-$ have achieved a very sophisticated level. The recently 
calculated NNLL  QCD contributions~\cite{Bobeth:1999mk,Asatryan:2001zw,Asatryan:2002iy,Ghinculov:2002pe,Asatrian:2002va,Ghinculov:2003bx,Ghinculov:2003qd,Bobeth:2003at,Asatrian:2003yk,Gorbahn:2004my}
have significantly improved the sensitivity of the inclusive $\bar B \rightarrow X_s \ell^+ \ell^-$ decay 
in testing extensions of the SM in the flavour dynamics sector. 
In particular, the value of the dilepton invariant mass $q^2_0$, for which the differential 
forward-backward asymmetries (FBA) vanishes is one of the most precise predictions in flavour physics 
with a theoretical uncertainty of order $5\%$. This corresponds well to the 
expected experimental sensitivity of $4-6 \%$ at \superb.

Also non-perturbative corrections scaling with $1/m_b^2$, $1/m_b^3$, or $1/m_c^2$ have to be  taken into 
account. Moreover, factorizable long-distance contributions away from the resonance peaks are 
important; here using the Kr\"uger-Sehgal (KS) approach~\cite{Kruger:1996cv} one avoids the  problem of double-counting.

In the high-$q^2$  region, one encounters the breakdown of the heavy-mass expansion at the endpoint; 
while the partonic contribution vanishes in the end-point, the $1/m_b^2$ and $1/m_b^3$ corrections 
tend towards a non-zero value. However, for an integrated high-$q^2$ spectrum an effective expansion 
is found in inverse powers of $m_b^{\rm eff} = m_b \times (1 - \sqrt{\hat s_{\rm min}})$ rather than $m_b$.

Recently, further refinements were presented such as the NLO QED two-loop corrections to the Wilson coefficients whose size
is of order $2\%$~\cite{Gambino:2004mv}. Furthermore, it was shown that in the QED one-loop corrections to matrix elements
large collinear logarithms of the form $\log(m_b^2/m^2_{\rm lepton})$ survive integration if only a restricted part of the
dilepton mass spectrum is considered. This adds another $+2\%$ contribution in the low-$q^2$ region for ${\cal B}
(\bar B\to X_s \mu^+\mu^-)$~\cite{Huber:2005ig}. This results in the following predictions 
\begin{equation}
\label{muonBR} 
{\cal B} (\bar B\to X_s \ell^+\ell^-)_{low} = \begin{cases}
(  1.59  \pm 0.11 ) \times 10^{-6}  & \ell=\mu  \cr
(  1.64  \pm 0.11 ) \times 10^{-6}   & \ell=e \, .    \cr
\end{cases}
\end{equation}
In Ref.~\cite{Huber:2007vv} the results for the high-$q^2$ region and for the FBA were derived. The result for the branching ratio
(BR) in the high-$q^2$ region reads
\begin{equation}
\label{muonBRhighs} 
{\cal B} (\bar B\to X_s \ell^+\ell^-)_{high}  = \begin{cases}
2.40 \times 10^{-7} \; (1^{+0.29}_{-0.26} ) & \ell=\mu \cr
2.09 \times 10^{-7} \; (1^{+0.32}_{-0.30} ) & \ell=e \, . \cr
\end{cases}
\end{equation}
In this case the relative impact of the collinear QED logarithm is about $-8$\% ($-20$\%) for muons (electrons) and therefore much
larger than in the low-$q^2$ region due to the steep decrease of the differential decay width at large $q^2$. The large error in
Eq.~(\ref{muonBRhighs}) is mainly due to the sizable uncertainties in the parameters that enter the $O(1/m_b^3)$ non-perturbative
corrections. As was pointed out in Ref.~\cite{Ligeti:2007sn} the error can be significantly decreased by normalizing the
$\bar B \rightarrow X_s \ell^+ \ell^-$ decay rate to the semi-leptonic $\bar B \rightarrow X_u \ell\bar\nu$ decay rate
{\textit{with the same $q^2$ cut}}. This will only be possible in the future at a \sff. For a lower cut of $q_0^2 =
14.4$~GeV$^2$ this leads to~\cite{Huber:2007vv}
\begin{equation} \label{muonR} 
{\cal R}^{\ell\ell}(\hat s_0) = 
\begin{cases}
2.29 \times 10^{-3} ( 1 \pm 0.13) & \ell=\mu \cr
1.94 \times 10^{-3} ( 1 \pm 0.16) & \ell=e \, , \cr
\end{cases}
\end{equation}
where $\hat s = q^2/m_b^2$. The uncertainties from poorly known $O(1/m_b^3)$ power corrections are now under control 
and the largest source of error is $V_{ub}$. 

The zero of the FBA is found to be at
\begin{equation}
\label{eq:muonzero}
(q_0^2)_{\ell\ell} = \begin{cases}
( 3.50 \pm 0.12) \, {\rm GeV}^2  &  \ell=\mu \cr
( 3.38 \pm 0.11) \, {\rm GeV}^2  &  \ell=e\, . \cr
\end{cases}
\end{equation}
The error is about 3\% but includes parametric and perturbative uncertainties only. 
This 3\% error is applicable only in the absence of cuts on $m_X$. 
However, unknown subleading non-perturbative corrections of order $O(\alpha_s \Lambda/m_b)$, which are estimated to give an
additional uncertainty of order 5\%. It is often argued that especially the small $\mu$ dependence at the zero is an accident
and should  be increased by hand. However, by comparing the NLO-QCD with the NNLO-QCD result one can clearly show that the
$\mu$ dependence is a reasonable reflection of the perturbative error. Moreover, the zero is stable under change of the $b$
quark mass scheme; the variation is below $2$\% when switching from 1S to $\overline{\rm MS}$ or pole scheme. There are also
predictions for the FBA integrated over bins in the low-$q^2$ region, which are usually chosen to be $q^2 \in [1,3.5]$~GeV$^2$
and $q^2 \in [3.5,6]$~GeV$^2$~\cite{Huber:2007vv}:
\begin{eqnarray}
\label{integratedAFB}
\bar{\cal A}_{\ell\ell}^{(1,3.5)} &=&  \begin{cases}( -9.09 \pm 0.91 )\%    & \ell=\mu \cr ( -8.14 \pm 0.87 )\%  & \ell= e \cr \end{cases} \; ,\\
\bar{\cal A}_{\ell\ell}^{(3.5,6)}  &=&  \begin{cases} ( 7.80 \pm 0.76 )\%  & \ell=\mu  \cr ( 8.27 \pm 0.69 )\%   & \ell=e      \cr \end{cases} \; ,
\end{eqnarray}
These quantities already allow us to discriminate between different NP 
scenarios~\cite{Lee:2006gs,Ali:2002jg,Gambino:2004mv}. 
Moreover, by also including the third independent variable in the double differential decay,
(see  Eq. (\ref{eq:d3Gamma})) in integrated form,  one can fix both the magnitude and sign of all 
relevant Wilson coefficients in the SM and to put constraints on the parameter space of NP models~\cite{Lee:2006gs}.
NP might also affect the high-scale Wilson Coefficients in such a way that they acquire additional phases.
In Refs.~\cite{Huber:2005ig,Huber:2007vv}, the results for the branching ratio and FBA in terms of generic high-scale Wilson
Coefficients are given.

Two  issues need further comments: (1) After including the NLO QED matrix elements, the electron and muon channels receive
different contributions due to terms involving $\ln(m_b^2/m_\ell^2)$. This is the only source of the difference between these
two channels. The results presented  in Eqns.~(\ref{muonBR})--(\ref{integratedAFB}) correspond to the process
$\bar B \to X_s \ell^+ \ell^-$ in which QED photons are included in the $X_s$ system and the di-lepton invariant mass does not
contain any photon, i.e. $q^2 = (p_{\ell^+} + p_{\ell^-})^2$. This would be exactly the case in a fully inclusive analysis using
the recoil technique possible at \superb. (How photons are treated in the present B-factory studies is discussed in
Ref.~\cite{Huber:2008ak}.)

(2) There is another important source of non-perturbative contributions:
Measurements in the low $q^2$ region require an experimental cut on the hadronic invariant mass, $m_X < m_X^{\rm cut}$, to
suppress the huge background from $b\to c(\to s \ell^+ \nu) \ell^-\bar\nu$ transitions. The latest \babar~\cite{Aubert:2004it}
and Belle~\cite{Iwasaki:2005sy} analyses use $m_X^{\rm cut} = 1.8\gev$ and $m_X^{\rm cut} = 2.0\gev$, respectively.
The situation is completely analogous to the inclusive determination of $\Vub$. The $m_X$ cut causes the rates to be sensitive
to the $B$ meson shape functions~\cite{Lee:2005pk, Lee:2005pwa}, introducing hadronic uncertainties that can easily spoil the
sensitivity to NP in this decay. (The large $q^2$ region is unaffected by the $m_X$ cut.)

At present, the $m_X$ cut is taken into account by extrapolating the measurements to the full $m_X$ range using MC
based on a Fermi motion model. This extrapolation is not reliable and can give at best a rough estimate of the effect
of the $m_X$ cut. At \superb, measurements of $\bar B\to X_s\ell^+\ell^-$ will reach comparable precision as the current
$\bar B\to X_s\gamma$ and $\bar B\to X_u\ell\bar\nu$ measurements and will thus require the same level of rigor in dealing
with hadronic shape function effects.

At leading order in $\Lambda_\mathrm{QCD}/m_b$, the cut on $m_X$ leads to a $10\%-30\%$ reduction in the rate,
which to a good approximation is universal among the different short distance contributions~\cite{Lee:2005pk, Lee:2005pwa}.
An accurate calculation of the cut rate requires good knowledge of $m_b$ and the shape function, which will become available
from measurements of $B\to X_s\gamma$ and $B\to X_u\ell\bar\nu$, as explained in Section.~\ref{sec:inclvub}.

At subleading order in $\Lambda_\mathrm{QCD}/m_b$ additional corrections due subleading shape functions arise~\cite{Lee:2008xc}.
Very little is known about the subleading shape functions, which at present causes an irreducible hadronic uncertainty. Their
effects were estimated in Ref.~\cite{Lee:2008xc} by scanning over a range of models (so the estimates should be taken with caution).
Depending on the observable and the value of the $m_X$ cut, the subleading shape functions induce corrections in the rates relative
to the leading-order result anywhere between $-10\%$ to $+10\%$ with equally large uncertainties. They also break the universality
in the different short distance contributions, causing a shift of about $-0.05\gev^2$ to $-0.1\gev^2$ in the zero of the
forward-backward uncertainty with an equally large uncertainty.

Hence, the current theory uncertainties from non-perturbative corrections in the low $q^2$ range are above 10\%. They can be
decreased by raising $m_X^{\rm cut}$, which however will cause an increase in the experimental uncertainties. With the full \superb
data set, it may be possible to push the non-perturbative uncertainties well below the 10\% level by constraining both the leading
and subleading shape functions using the combined $\bar B\to X_s\gamma$, $\bar B\to X_u\ell\bar\nu$, and $\bar B\to X_s \ell^+\ell^-$
data.

More details on the subjects of this section can be found for example in Ref.~\cite{Hurth:2010tk}.

\subsubsection{Angular analysis of $B \to K^* l^+ l^-$}

Few modes are able to provide such a wealth of information as the decay
mode of $B \to K^* l^+ l^-$, ranging from FBA~\cite{Alok:2009tz},
isospin
asymmetries~\cite{Feldmann:2002iw} to angular observables~\cite{Kruger:2005ep,
Lunghi:2006hc, Egede:2008uy, neverending}. Each of these observables constructed
can provide information on a different type of NP, isospin
breaking, right-handed currents, etc. In this sense exploring this mode at
a \superb machine is a worthy effort. One of the most interesting
observables are those coming from the angular distribution.

One can analyze in full detail the 4- body decay distribution
of the $B \to K^*(\to K \pi) l^+ l^-$ in the context of QCD Factorization~\cite{Beneke:2001at}
designing a new method to construct observables based on three steps.
First, use the spin amplitudes of the $K^*$ as the key ingredient. Second,
construct a quantity with these spin amplitudes that maximizes the
sensitivity to a certain type of NP (right-handed currents for
example), canceling at the same time,  the
dependence on the
poorly known soft form factors at LO. This later point is inspired in a
way in
the idea of the zero of the FBA. This particular
point of the $A_{FB}$ has attracted a lot of interest due to its
cleanliness given the cancellation of
form factor dependence at LO. The angular observables 
$A_T^2, A_T^3$ and $A_T^4$ and defined by~\cite{Kruger:2005ep,
Egede:2008uy}:

$$ A_T^{(2)}=\frac{|A_\perp|^2-|A_\||^2}{|A_\perp|^2+|A_\||^2},
\quad \quad A_T^{(3)}=\frac{|A_{0L}A_{\| L}^* + A_{0R}^* A_{\| R}|}{
\sqrt{|A_0|^2 |A_\perp|^2}}, $$
$$A_T^{(4)}=\frac{|A_{0L} A_{\perp L}^* - A_{0R}^* A_{\perp R}|}{|A_{0L}^* A_{\| L} + A_{0R} A_{\| R}^*|},
$$
exhibit this quality in the full $q^2$ region and not just at a single point.
Finally, the
third step for the constructed quantity to be considered an observable (in this context an
  observable is a quantity that can be extracted  from the angular distribution), is that this
quantity
fulfills all the symmetries of the distribution.

This last point is one of new fundamental ingredients in this procedure~\cite{neverending}. 
This requires the identification of all four symmetries~\cite{neverending} of the
4-body decay distribution in the massless case. And those are precisely the
symmetries that the quantities constructed should respect.
The identification of the symmetries~\cite{neverending}
allows one to explicitly solve the spin amplitudes in
terms of the coefficients of the distribution. An interesting byproduct of
this is a highly non trivial constraint between the coefficients of
the 4-body distribution, considered before as independent parameters, that should
be fulfilled in the SM when leptons are
taken to be massless.

It is important to remark that the new observables that can be easily obtained from a full angular
distribution analysis at \superb have a huge sensitivity to right-handed currents
driven by the operator $O_7^\prime$, to which $A_{FB}$ is blind, and they all  have only  a mild dependence on soft  form factors.
In particular,   $A_T^2$ contains all the information of the FBA and more. The zero of the $A_{FB}$ occurs also in $A_T^{2}$ at
the LO in the same position and the absence of a zero affects both observables in the same way. Moreover, $A_T^2$ exhibits a
maximal deviation from the SM prediction that is approximately zero, for certain types of models in the region
between 1 and 2 $GeV^2$. 

Theoretical uncertainties in those exclusive modes are always larger than in the corresponding 
inclusive modes due to the well-known problem that $\Lambda / m_b$ cannot be calculated within 
QCD factorization. This obviously  restricts the new-physics sensitivity of those exclusive 
modes compared to the corresponding inclusive ones~\cite{Egede:2008uy}.

\subsubsection{$\bar B \rightarrow  X_d \gamma$ and $\bar B \rightarrow X_d \ell^+ \ell^-$}

The rare decay modes $b\to d\gamma$ and $b\to d \, \ell^+ \ell^-$ offer an interesting
phenomenology which is complementary to the corresponding $b \to s$ transitions. The
corresponding effective Hamiltonian is
\begin{eqnarray}
H_{\rm eff} (b \to q)&=&
- {4 G_F\over \sqrt{2}} V_{tb}^{} V_{tq}^* \left[ \sum_{i=1}^8 C_i (\mu) \, O_i (\mu) \right.\nonumber\\
&&\left. + \epsilon_q \, \sum_{i=1}^2 C_i (\mu) (O_i (\mu)-O_i^u (\mu) )\right] \!\! ,\nonumber\\
\end{eqnarray}
where the relevant operators are
\begin{eqnarray}
O_1 &=& (\bar{q}_{L} \gamma_\mu T^a  c_{ L}\black) (\bar{c}_{ L}\black \gamma^\mu T^a b_{ L})\nonumber  \\
O_1^{u} &=& (\bar{q}_{L} \gamma_\mu T^a  u_{ L}\black) (\bar{u}_{ L}\black \gamma^\mu T^a b_{ L}) \nonumber \\
O_2 &=& (\bar{q}_{L} \gamma_\mu c_{ L} \black) (\bar{c}_{ L} \black \gamma^\mu b_{ L})\nonumber\\
O_2^{u} &=& (\bar{q}_{L} \gamma_\mu u_{ L} \black) (\bar{u}_{ L} \black \gamma^\mu b_{ L})\nonumber\\
O_7 &=& e/(16 \pi^2)\,  m_b (\mu) \,  (\bar{q}_{L} \sigma_{\mu \nu} b_{R}) F^{\mu \nu}\nonumber \\
O_8 &=& g_s /(16 \pi^2) \, m_b (\mu) \, (\bar{q}_{L} T^a \sigma_{\mu \nu} b_{R }) G^{a \mu \nu},
\end{eqnarray}
and once again the $C_i$ are Wilson Coefficients.
The decisive difference between $b \to d$ and $b \to s$ decays is the size of the respective
$\epsilon_q = (V_{ub}^{} V_{uq}^*) / (V_{tb}^{} V_{tq}^*)$, which are
\begin{eqnarray}
\epsilon_s &=& \textstyle \frac{V_{ub}^{} V_{us}^*}{V_{tb}^{} V_{ts}^*} \displaystyle = -\lambda^2 (\bar \rho -i\bar \eta) \simeq -0.01 +0.02 \, i, \nonumber \\
\epsilon_d &=& \textstyle \frac{V_{ub}^{} V_{ud}^*}{V_{tb}^{} V_{td}^*} \displaystyle =\frac{ \bar \rho -i\bar \eta}{ 1-\bar \rho -i\bar \eta} \simeq -0.02 +0.42 \, i .
\end{eqnarray}
As a consequence CP asymmetries are tiny in $b \to s$, but sizable in $b \to d$ transitions, as we will see below.
Moreover, due to the democratic pattern of the CKM elements, up-quark loops play an important r\^ole in $b \to d$ transitions. Their
QCD corrections are known at two loops~\cite{Asatrian:2003vq,Seidel:2004jh}. But also non-perturbative power-corrections
have to be taken into account.  The $1/m_c^2$ corrections are known and well under control, 
whereas the contributions to the $u$ quark loops have been shown to be order 
 ${\cal O}(\Lambda_{\rm QCD}/m_b)$~\cite{Buchalla:1997ky}.
However, a recent systematic analysis of power corrections~\cite{Benzke:2010js} 
has found  that the contribution 
due to the operator $O_1^u$ in interference with the operator $O_7$ vanishes in the total
rate; thus,  there is no additional uncertainty due to the $u$ quark loops any longer, and the decay
rate of $\bar B \rightarrow X_d \gamma$ is as theoretically clean as the decay rate 
$\bar B \rightarrow X_s \gamma$. The complete effect of power corrections on CP asymmetries is 
 not estimated yet. For this purpose  the systematic analysis in Ref.~\cite{Benzke:2010js} has to be extended.

The inclusive decay $\bar B \to X_d \gamma$ has been studied at various places in the Literature.
The theoretical predictions for the branching ratio ${\cal B}[\bar B \to X_d \gamma]$ for photon
energies $E_\gamma>1.6$~GeV are~\cite{Hurth:2003pn,Hurth:2003dk}
\bea 
{\cal B} [\bar B \to X_d \gamma]
& = & \Big[ 1.38 \left. {}^{+0.14}_{-0.21}   \right|_{m_c \over m_b}
                     \pm 0.15_{\rm CKM}  \nonumber \\
                && \pm 0.09_{\rm param.} \pm 0.05_{\rm scale} \Big] \cdot 10^{-5},  \nonumber \\
{{\cal B} [\bar B \to X_d \gamma] \over {\cal B} [\bar B \to X_s \gamma]}
& = & \Big[ 3.82 \left. {}^{+0.11}_{-0.18}   \right|_{m_c \over m_b}
                     \pm 0.42_{\rm CKM} \nonumber \\
                && \pm 0.08_{\rm param.} \pm 0.15_{\rm scale} \Big] \cdot 10^{-2}, \nonumber \\ \label{eq:BRratio}
\eea
where the errors in the ratio in Eq.~(\ref{eq:BRratio}) are dominated by CKM uncertainties.
Similar numbers
are found by a second analysis~\cite{Ali:1998rr,Battaglia:2003in}
\bea
\displaystyle \langle {\cal B}(\bar B\to X_d \gamma) \rangle &\simeq& 1.3 \cdot 10^{-5},\nonumber \\
\displaystyle \frac{\langle {\cal B}(\bar B\to X_d \gamma) \rangle}{\langle {\cal B}(\bar B\to X_s \gamma) \rangle} &=&
\frac{|\xi_t|^2}{|\lambda_t|^2} + \frac{D_u}{D_t} \, \frac{|\xi_u|^2}{|\lambda_t|^2} +  \frac{D_r}{D_t} \,
\frac{{\rm Re}(\xi_t^*\xi_u)}{|\lambda_t|^2}, \nonumber \\
&\simeq& 3.6 \cdot 10^{-2} \; .
\eea
Here $\langle {\cal B}(\bar B\to X_d \gamma) \rangle$ always denotes the charge-conjugate averaged branching ratio,
$\lambda_i=V_{ib}V^*_{is}$ and $\xi_i=V_{ib}V^*_{id}$ are combinations of CKM elements, and the $D_i$ are
functions of $m_t$, $m_b$, $m_c$, $\mu_b$, and $\alpha_s$.

The SM predictions for the direct CP asymmetries
\bea
A_{\rm CP}^{b\to q \gamma} & \equiv & { \Gamma{[\bar B \to X_{q} \gamma]}
   - \Gamma{[B \to X_{\bar q} \gamma]} \over
      \Gamma{[\bar B \to X_q \gamma]} + \Gamma{[B \to X_{\bar q} \gamma]}} \nnb
\eea
are~\cite{Hurth:2003pn,Hurth:2003dk}
\bea 
A_{\rm CP}^{b\to s \gamma}
& = & 
\Big[ 0.44 \left. {}^{+0.15}_{-0.10} \right|_{m_c \over m_b} 
   \pm 0.03_{\rm CKM}   \left. {}^{+0.19}_{-0.09} \right|_{\rm scale} \Big],
 \% \nnb\\ 
A_{\rm CP}^{b\to d \gamma} 
& = & 
\Big[ -10.2 \left. {}^{+2.4}_{-3.7} \right|_{m_c \over m_b}
  \pm 1.0_{\rm CKM} \left. {}^{+2.1}_{-4.4} \right|_{\rm scale} \Big] \% \; . \nonumber
\eea
The additional parametric uncertainties are sub-dominant. Again, a second analysis finds
similar results~\cite{Ali:1998rr,Battaglia:2003in}
\bea
 A_{\rm CP}^{b\to s \gamma} &\simeq&
\frac{{\rm Im}(\lambda_t^*\lambda_u) \, D_i}{|\lambda_t|^2 \, D_t} \sim 0.5\%, \nonumber \\
 A_{\rm CP}^{b\to d \gamma} &\simeq&
\frac{{\rm Im}(\xi_t^*\xi_u) \, D_i}{|\xi_t|^2 \, D_t} \sim -13\% \; .
\eea
The radiative FCNC processes exhibit yet another interesting quantity, which can serve as a Null 
test of the SM, 
 namely the unnormalized, untagged
$\bar B \to X_{s+d} \, \gamma$ CP asymmetry. This quantity vanishes in the U-spin limit
$m_s=m_d$~\cite{Soares:1991te}, and hence in this limit and for real Wilson coefficients one has
\bea
\left[\Gamma(\bar B \to X_s \, \gamma)-\Gamma(B \to X_{\bar s} \, \gamma)\right]&+& \nnb \\
\left[\Gamma(\bar B \to X_d \, \gamma)-\Gamma(B \to X_{\bar d} \, \gamma)\right] &=& 0.
\eea
The size of the untagged $\bar B \to X_{s+d} \, \gamma$ CP asymmetry is a measure of U-spin breaking.
In the SM within the partonic contribution one finds~\cite{Hurth:2001yb,Hurth:2001ja}
\begin{equation}
| \Delta {\cal B}(B \to X_s \gamma) + \Delta {\cal B}(B \to X_d \gamma) | \sim 1 \cdot 10^{-9} \; .
\end{equation}
Power corrections beyond the leading partonic contribution are expected to be small, since 
U-spin breaking in $1/m_{b,c}^2$ corrections bring up a factor of $m_s^2/m_b^2$.
This is also shown for  the non-perturbative corrections to the $u$ quark loops which 
scale like $\Lambda_{\rm QCD}/m_b$~\cite{Hurth:2001ja}. 
We therefore conclude that any sizable value of the untagged
$\bar B \to X_{s+d} \, \gamma$ \CP asymmetry is a direct signal for NP.

\begin{figure}[!htb]
  \begin{center}
\hspace*{5pt}\includegraphics[width=0.49\textwidth]{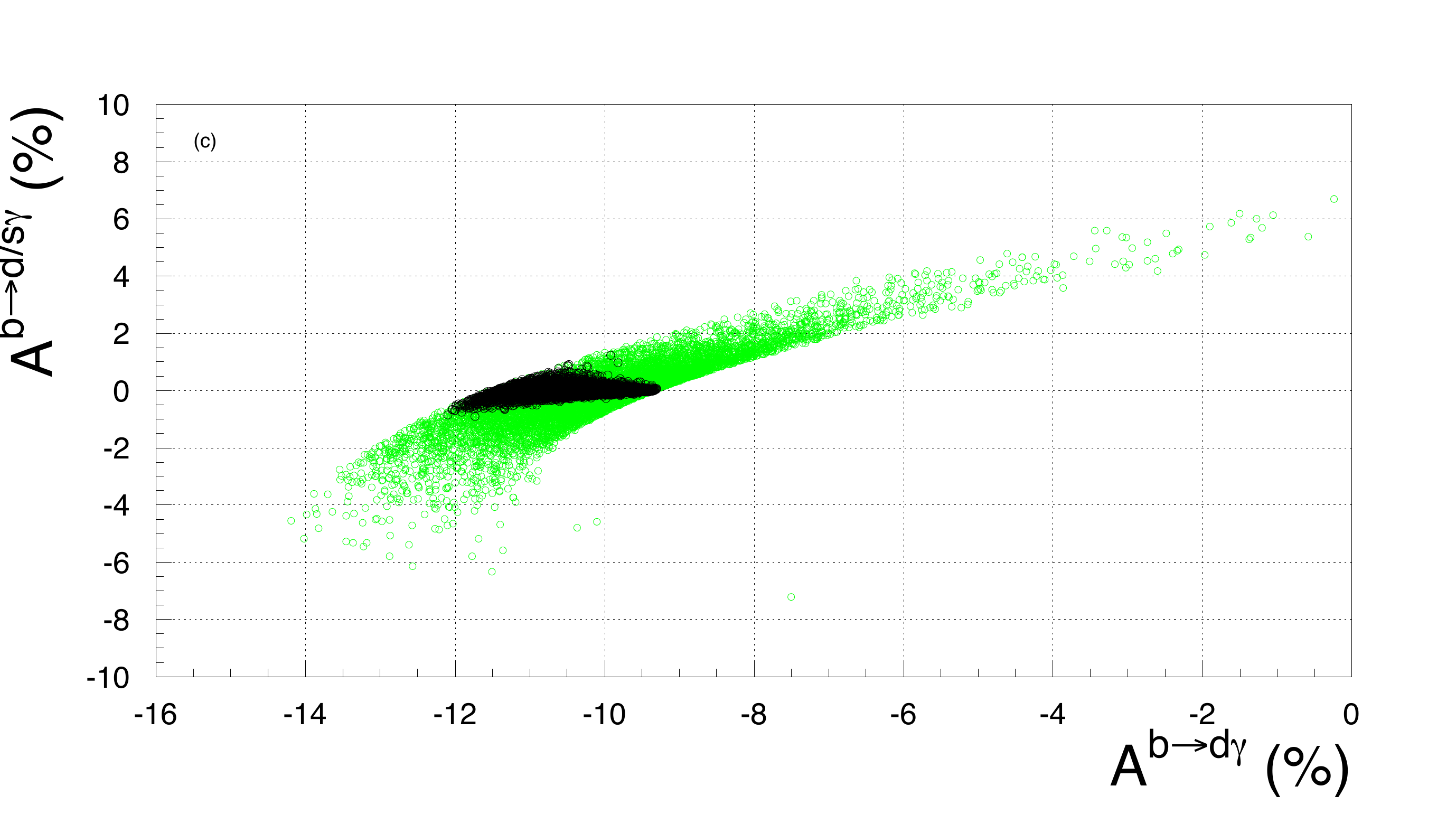}
  \caption{\label{fig:untaggedCPAinMFV}
    Untagged rate asymmetry in MFV with flavour-blind phases with (black/dark) and without (green/light) neutron and electron EDM
    constraints imposed~\cite{Hurth:2003dk}.
    }
  \end{center}
\end{figure}

The untagged $\bar B \to X_{s+d} \, \gamma$ \CP asymmetry also offers the possibility for
interesting analyses in NP models. For example, the experimental accuracy of the untagged CP asymmetry
at the current $B$-factories is $\pm 3$\%, which allows one to distinguish between MFV and more general flavour models,
where the untagged \CP asymmetry can reach $\sim 10$\%~\cite{Hurth:2003dk}. However, a distinction between MFV with and
without flavour-blind phases not possible at existing $B$-factories, but within reach of a \superb facility,
see Fig.~\ref{fig:untaggedCPAinMFV}. 

The inclusive rare decay $\bar B \to X_d \, \ell^+\ell^-$ offers, just as in the case of $b \to s$ transitions,
a complementary test of the SM compared to $\bar B \to X_d \,\gamma$, since due to the three-body final state
and the presence of the axial current the kinematic structure is richer. This results in three independent functions
of the di-lepton invariant mass $q^2$, two of which are the differential branching ratio and the 
FBA~\cite{Lee:2006gs}. Just as in the $b \to s$ case, one distinguishes several windows in $q^2$. The low-$q^2$
window $1<q^2<6$~GeV$^2$ is dominated, due to the local OPE, by the quark-level decay and its perturbative corrections.
The $\rho$, $\omega$, and $\bar cc$ ($J/\psi$, $\psi^\prime$) resonances are cut out in this window, and the effect of their
respective tails can be taken into account within the KS approach~\cite{Kruger:1996dt}. This way on-shell up-quark
loops can be avoided. A systematic analysis of power corrections would be desirable.  
The differential branching ratio integrated over the low-$q^2$ window is~\cite{Asatrian:2003vq}
\bea
{\cal R}_{\rm quark} &=& \int\limits_{0.05}^{0.25} \! d\hat s \; R_{\rm quark} =(4.75 \pm 0.25) \times 10^{-7} . \nonumber
\eea
This number is without power corrections and without taking into account resonances. A very preliminary
investigation of the feasibility of studying $\bar B \to X_d \, \ell^+\ell^-$ at \superb looks promising.

\subsection{Experimental aspects of rare decays}

\subsubsection{$B \to K^{(*)} \nu \overline \nu$}

The recoil technique has been developed in CLEO, and subsequently adopted by both \babar and Belle,
in order to search
for rare $B$ decays with undetected particles, like neutrinos, in the final states.
This technique consists of the reconstruction of one of the two $B$ mesons 
(the $B_{tag}$) in a hadronic or semi-leptonic final state, and the search for the signal decay
of the other $B$ (the $B_{sig}$) in the rest of the event. The reconstruction of the $B_{tag}$ allows
one to select a pure $B \overline B$ sample.  Having identified the $B_{tag}$, everything in the rest
of the event by default is the $B_{sig}$ candidate, and so this technique provides a clean environment 
to search for rare decays.

Since the typical efficiency of the $B_{tag}$ reconstruction is below 1\%, the use of this technique at 
the present $B$-Factories is almost limited to the search for rare decays with undetected particles like
neutrinos in the final state, where strong kinematic constraints are missing. On the other hand, the 
larger statistics available at \superb would make this technique convenient also for the search of other
rare decays, where the high purity of the $B \overline B$ sample selected with the recoil method would provide
a high level of background suppression.

We have investigated the potential of using the recoil technique at \superb and studied in particular the 
$B \to K^{(*)} \nu \overline \nu$ decays. This channel is an interesting probe for NP in $Z^0$
penguins~\cite{Buchalla:2000sk}, such as chargino-up-squark contributions in a generic supersymmetric model. Moreover,
due to the presence of undetected neutrinos in the final state, the experimental signature of these decays
is $B \to K^{(*)} + $ missing energy, so that the measured decay rate is sensitive to exotic sources of 
missing energy, such as light dark matter~\cite{Bird:2004ts} or ``unparticles''~\cite{Georgi:2007ek,Aliev:2007gr}. 

In this analysis, the $B_{tag}$ is reconstructed in the hadronic modes $B \to D^{(*)} X$, where 
$X = n\pi+ mK + pK_{s} + q\pi^0 $ and $n+m+r+q<6$, or semi-leptonic modes
$B \to D^{(*)} \ell \nu$, where $\ell = e,\,\mu$. In the search for $B \to K \nu \overline \nu$, the signal
is given by a single track identified as a kaon. In the search for $B \to K^* \nu \overline \nu$, we look for
a $K^{*}$ in the $K^{*0} \to K^+ \pi^-$, $K^{*\pm} \to K^0_S \pi^\pm$ and $K^{*\pm} \to K^\pm \pi^0$ modes.

The \superb fast simulation has been used to simulate signal events in the \superb and \babar setup. This
test showed a 20 to 30\%  increase in the efficiency at \superb relative to \babar, depending on the final state,
mainly provided by the larger acceptance due to the lower $\Upsilon(4S)$ boost.
The simulation of $B \overline B$ events, where one of the two $B$ mesons decays semi-leptonically
and the other generically, showed that a 10\% reduction in this important background source
is also possible in the semi-leptonic recoil analysis. Based on these observations, we estimate a 
20 to 30\% increase in the $S/\sqrt(B)$ ratio, where $S$ and $B$ are the signal and background yields, respectively.

\begin{figure}[!htb]
  \begin{center}
  \includegraphics[width=0.45\textwidth]{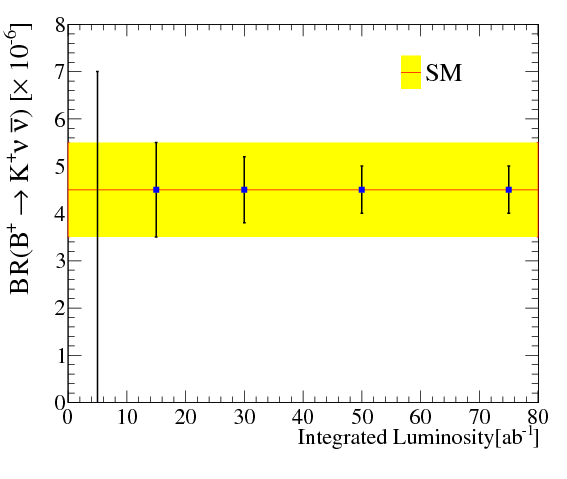}
  \includegraphics[width=0.45\textwidth]{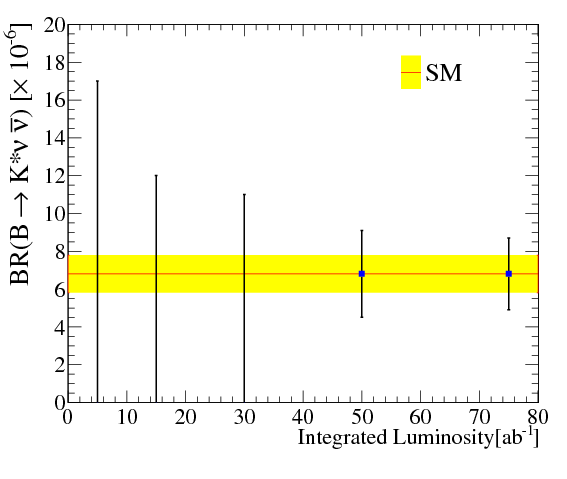}
 \caption{\label{fig:knunu_sensitivity} Expected precision of the measurements of the 
branching fractions of (top) $B^+ \to K^+ \nu \overline \nu$ and (bottom) 
$B^{*} \to K^{*} \nu \overline \nu$ evaluated as a function of the integrated luminosity.}
  \end{center}
\end{figure}

The results in Table~\ref{tab:bsnn-SMexp} have been used to update the study presented in~\cite{Hitlin:2008gf}. The expected
sensitivity to $B \to K^{(*)} \nu \overline \nu$ is shown in Fig.~\ref{fig:knunu_sensitivity}.
The $3\sigma$ observation of the $B \to K \nu \overline \nu$ decay is expected with a data sample of $10\,ab^{-1}$, while 
$50\,ab^{-1}$ will be needed to observe $B \to K^* \nu \overline \nu$ decays, assuming the branching fraction 
occurs at a rate consistent with SM-based calculations (See Table~\ref{tab:bsnn-SMexp}).

We also investigated the feasibility of an angular analysis of the
$B \to K^* \nu \overline \nu$ decay. Along with the measurement of the Branching Fractions, this analysis would
provide a constraint for the two parameters $\epsilon$ and $\rho$ given in Eq.~\ref{eq:epsetadef}.

\begin{figure}[!htb]
  \begin{center}
    \includegraphics[width=0.45\textwidth]{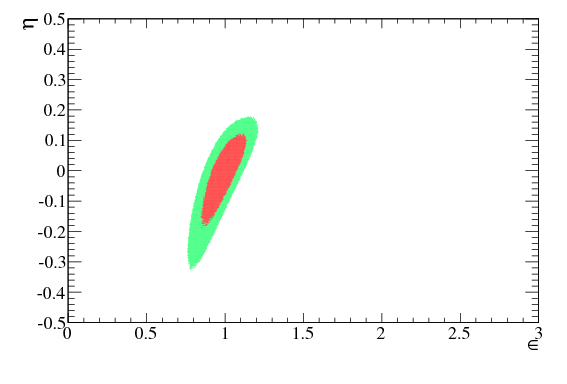}
  \caption{\label{fig:angular_constraint} Expected constraint on the $(\epsilon,\eta)$ plane discussed in Section~\ref{sec:bphysics:rare:snunubar},
from the measurement of the Branching Ratios of $B \to K^{(*)} \nu \overline \nu$ and the angular analysis of $B^0 \to K^{*0} \nu \overline \nu$ at $75\,ab^{-1}$ (See Figure~\ref{fig:knunu_sensitivity}). }
  \end{center}
\end{figure}

In the angular analysis, the distribution of the cosine of the
angle $\theta$ between the $K^*$ flight direction in the $B_{sig}$ rest frame and the $K$ flight direction 
in the $K\pi$ rest frame has to be studied.  At least in the hadronic analysis, the $B_{sig}$ rest frame can be 
deduced from the fully reconstructed $B_{tag}$, and $\theta$ can be easily determined.
From toy MC studies, neglecting $\cos(\theta)$ resolution effects and assuming a flat background
on $\cos(\theta)$, we estimated that the $B^0 \to K^{*0} \nu \overline \nu$ channel
in the hadronic recoil could provide an error of about 0.3
on the parameter $<F_L>$ with 75\invab of data.
The combination of this information with the measurement of the branching ratios 
would provide a constraint in the plane $(\epsilon,\eta)$, as shown in Fig.~\ref{fig:angular_constraint},
where NP would show up as a deviation from the SM values $(1,0)$.

In summary, in the search for rare $B$ decays at \superb, the high $B \overline B$ statistics
would enable one to consider using the recoil technique, 
consisting of the full reconstruction of one of the two $B$ mesons in a hadronic
or semi-leptonic mode, and the subsequent search for a signal in the rest of the event.
By using the recoil technique one will loose signal efficiency, but in return be able to
identify and reconstruct signal and a very clean environment.  We have
investigated the reach of \superb in the search for the $B \to K^{(*)} \nu \overline \nu$ 
decays in both semi-leptonic and hadronic recoil samples. Preliminary results based on the \superb fast
simulation have shown that a 10 to 30\% improvement in the sensitivity with respect
to the \babar setup is possible, allowing for a $3\sigma$ observation of the 
$B^{0,\pm} \to K^{(*)0,\pm} \nu \overline \nu$ decays. An angular analysis
of the decay will also be feasible.

\subsubsection{$B \to \ell \nu$ and $B \to \ell \nu \gamma$}

One of the most important applications of the recoil technique (tag reconstruction method) described in the 
previous section is in the search for $B^+ \to \tau^+ \nu$, where the
presence of two or more neutrinos in the final state effectively eliminates any meaningful kinematic information that can 
be used to identify the signal 
decay~\footnote{This is actually not quite true.  Leptonic $\tau$ decays from 
$B^+ \to \tau^+ \nu$ preferentially produce low momentum electrons and muons, 
while $\tau^+ \to \pi^+ \nu_\tau$ produces high momentum pions.  To date,
neither \babar nor Belle has explicitly incorporated this feature into their studies.}.
Searches performed to date have therefore
utilized the tag reconstruction method (both hadronic and semi-leptonic tags), relying on signal topology for the 
selection, exploiting the low particle multiplicity of the signal decays compared with background events.  After a 
cleanly-reconstructed tag $B$ has been selected, the event is required to possess either one or three additional charged tracks,
particle ID requirements are imposed to distinguish between hadronic and leptonic $\tau$ decay candidates, then mode-specific
constraints are imposed on the presence of neutral energy in the calorimeter.
The limitation on this technique is ultimately imposed by systematic uncertainties on the background resulting from knowledge of the
background track multiplicity and calorimeter ``extra energy'' distributions.  The extra energy distribution is particularly
problematic, since it relies on excellent understanding of low energy clusters resulting from diverse sources including beam
backgrounds, hadronic cluster ``split-offs'', neutral hadrons etc.  One advantage that \superb has over the existing $B$ Factories
in this respect is the ability to validate the MC modeling of the extra energy using ``signal-like'' exclusive decay control samples,
\eg\ by plotting  the extra energy distribution for events in which a clean $B \to K^* \gamma$ or similar event has been
reconstructed in addition to the tag $B$.  With the existing $B$ factory data sets, this technique does not yield sufficient
statistics for meaningful studies.  However at \superb this ability is expected to permit $B^+ \to \tau^+ \nu$ branching fraction
measurements to remain statistically limited even with the full \superb data  statistics.  

One can find a brief theoretical overview of the $B\to \ell \nu$ channels in Ref.~\cite{Bona:2007qt}.
Both in the SM and in the generic 2HDM, the ratios of leptonic branching fractions is given simply by the
ratio of final state lepton masses squared.  It is therefore extremely useful to have a good measurement of
not only $B^+ \to \tau^+ \nu$, but also $B^+ \to \mu^+ \nu$ in order to provide an internal consistency check
of these measurements.  This is particularly important given the current $\sim 2 \sigma$ discrepancy between $V_{ub}$
measurements, the CKM fit excluding $V_{ub}$ and $B^+ \to \tau^+ \nu$.  $B^+ \to \mu^+ \nu$ results have been reported
by both \babar and Belle (using un-tagged analyses) with resulting branching fraction limits which are within about a factor 
of two of SM expectations.  In addition a first study has been published by \babar using the hadronic tag reconstruction method.
The tagged and untagged approaches are complementary and both are expected to yield clear observations of $B^+ \to \mu^+ \nu$ 
with \superb luminosity.  Due to the cleaner signal signature in this mode, it is likely that the ultimate precision on 
$B^+ \to \mu^+ \nu$ will be similar to $B^+ \to \tau^+ \nu$.  \babar has recently extended the $B^+ \to \ell^+ \nu$ ($\ell = e, \mu$)
to include the radiative decay $B^+ \to \ell^+ \nu \gamma$ using a hadronic tag method.  Although this analysis suffers from limited 
statistics with the \babar dataset, the additional kinematic constraints which 
are available with this method (as well as improved over continuum backgrounds) permit an almost background free signal selection, although 
at the cost of signal efficiency.  This method is expected to have sensitivity to the SM rate well within the nominal \superb luminosity
range, and it will be important contribution toward a precision determination of $B^+ \to \ell^+ \nu$ as in the 
limit of small photon energy this radiative mode becomes an important component of the $B^+ \to \ell^+ \nu$ signal.

\subsubsection{Experimental aspects of $\bar B\to X_s \gamma$}

In the $B$ factory era, $\bar B\to X_s \gamma$ has been an extremely
important channel for searching for NP and for constraining
new models that go beyond the SM.  The inclusive branching fraction for this
decay is not small ($\sim 3 \times 10^{-4}$) and several measurements
have been made at \babar and Belle during the B-factory
era~\cite{Limosani:2009qg,Aubert:2005cua,Aubert:2006gg, Aubert:2007my}. Currently the
experimental world-average~\cite{hfag} has a total
uncertainty of 7\%:
\begin{equation}
B(\bar B \to X_s \gamma) = (3.52 \pm 0.23 \pm 0.09) \times 10^{-4}.
\end{equation}
This branching fraction is for $E_\gamma > 1.6$ GeV, where theoretical
models have been used to extrapolate from the experimental photon
energy cut, typically $\sim 1.9$ GeV, down to 1.6 GeV.  The uncertainties quoted are experimental
(statistical plus systematic) and model uncertainty inherent in the
extrapolation. 

At the same time, theorists have carried the
calculation of the branching fraction to NNLO, resulting in a quite
precise theoretical calculation of $B(\bar B \to X_s \gamma)$ within the SM~\cite{Misiak:2006zs}:
\begin{equation}
B(\bar B \to X_s \gamma) = (3.15 \pm 0.23) \times 10^{-4},
\end{equation}
which also has a total error of 7\%.  The experimental and theoretical
values are in reasonable agreement.

Part of the experimental challenge has been to make the most inclusive
measurements possible, in order to compare them to the inclusive
theoretical calculation without introducing excessive model
dependence.  The earliest  measurements used the
``sum of exclusive modes'' technique, whereby a large number of
individual exclusive decay channels are fully reconstructed.  This
approach is easier experimentally, but the large systematic uncertainties
associated with the unseen modes will make this technique obsolete at the
\superb. 

The alternative approach is the fully inclusive method, which attempts
to make no requirements whatsoever on the $X_s$ system, thereby making
the branching fraction measurement fully inclusive. The one exception
is making a cut on photon energy (which is equivalent to selecting on
$m(X_s)$), although much effort is devoted to keeping the photon
minimum energy as low as possible. 

The fully inclusive approach can adopt the recoil tagging technique
to identify the ``other'' $B$ in the event {\it i.e.} the $B_{tag}$
from the \FourS (as described above) in order to reduce the large
continuum background.  At the B-factories, the lepton tagging method,
which has an efficiency of roughly ten times the hadronic tag
efficiency, has proved more useful.  However, at the \superb factory,
where there will be a very large number of reconstructed hadronic
tags, this method will make an important contribution to the
measurement of $B(\bar B \to X_s \gamma)$.  

\begin{figure}
    \includegraphics[scale=0.4]{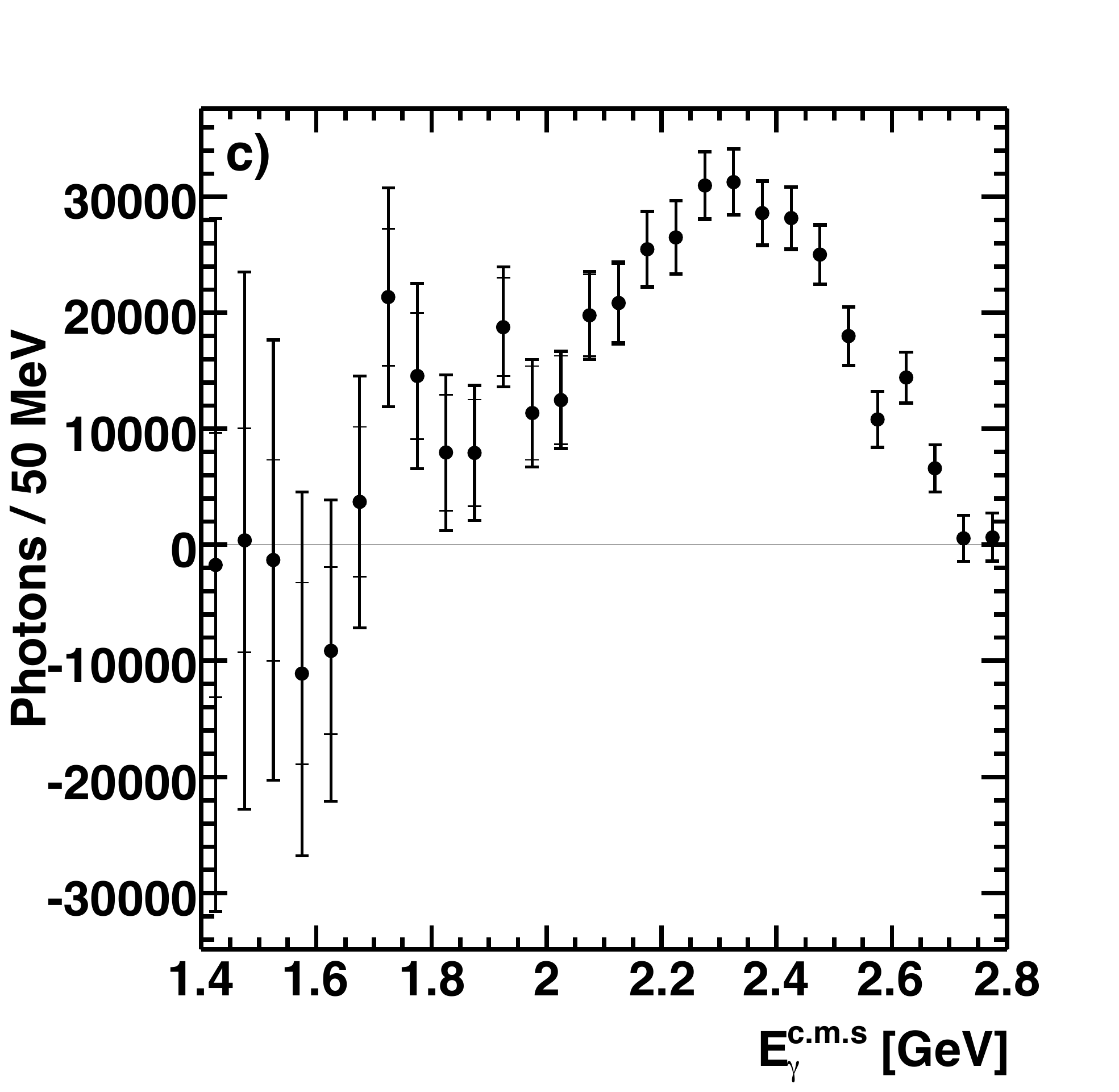} 
\caption{\label{fig:belle_bsg}
Efficiency-corrected photon spectrum for $\bar B \to X_s \gamma$ decays. Taken from reference~\cite{Limosani:2009qg}.}
\end{figure}

Figure~\ref{fig:belle_bsg} shows the efficiency-corrected photon
spectrum from Belle's recent measurement. This fully inclusive
analysis is based on combining a lepton-tagged sample and an untagged
sample. The large uncertainties visible at lower energies are caused by
the systematic uncertainty in the $B\bar{B}$ background subtraction.

At the \superb the uncertainty on the branching fraction measurement
will be dominated by the systematic error --- we estimate that a
systematic error of 3\% will be achievable at \superb.

NP can also modify the direct CP asymmetries of $\bar B \to X_s \gamma$ decays
and these asymmetries have been measured at the B-factories, although
with still large statistical errors.  The increased statistics
available at the \superb factory will make these CP asymmetries
important tools in the search for NP.  Experimentally,
different analysis approaches measure different asymmetries.  

The fully inclusive analyses do not distinguish the Cabibbo-suppressed
process $\bar B\to X_{d} \gamma$ from $\bar  B \to X_s \gamma$, and so measures the asymmetry of
the sum: $A_{CP}(\bar B\to X_{s+d}\gamma)$.  The sum of exclusive-modes
analysis does measure $A_{CP}(\bar B \to X_s \gamma)$ and with good precision,
since many of the systematic effects cancel in the asymmetry ratio.
However, since this measurement is not truly inclusive (typically
about 50\% of the inclusive rate is reconstructed), there may be
problems interpreting the result and comparing to the SM theoretical
calculation for the inclusive process.  The most precise $A_{CP}$
measurements to date have been made with the exclusive channel $B\to
K^{\ast}\gamma$.

Selected experimental results for these asymmetry measurements are reported in 
Table~\ref{tab:bsg_acp}.
\begin{table}
\begin{center}
\caption{Selected experimental results on $A_{CP}$ from the B-factories. The first uncertainty
listed is statistical, the second systematic.}\label{tab:bsg_acp}
\begin{tabular}{lc} \hline
Quantity  &  Result \\ \hline

$A_{CP}(\bar B \to X_s \gamma)$               & $-0.011\pm0.030\pm0.014$~\cite{Aubert:2008gvb} \\
                             & $0.002 \pm 0.050\pm0.030$~\cite{Nishida:2003yw}  \\ \hline
$A_{CP}(B\to X_{s+d}\gamma)$ & $-0.11\pm0.12\pm0.02$~\cite{Aubert:2006gg}   \\
                             & $0.10\pm0.18\pm0.05$~\cite{Aubert:2007my}    \\ \hline
$A_{CP}(B\to K^{\ast}\gamma)$ & $-0.003\pm0.017\pm0.007$~\cite{Aubert:2009we}   \\ \hline
\end{tabular}\end{center}
\end{table}   
The SM predictions for each of these quantities is quite close to
zero, so any measurement of a substantial asymmetry would be an
indication of NP~\cite{Hurth:2003dk,Kagan:1998bh,Greub:2005xxxx}.  We note that currently
the experimental measurements have small systematic errors,
making them very attractive for the high-statistics environment of
\superb.

Another quantity related to the $b \to s \gamma$ channel is the time-dependent
CP asymmetry in the decay $B^0\to K_s \pi^0 \gamma$.  In the SM the
photon is this decay has, to order $\sim m_s/m_b$, definite helicity: $b\to s\gamma_{L}$ and
$\bar{b}\to\bar{s}\gamma_{R}$. This results in quite small
CP-violation in this decay, with the SM expectation being
$S_{K_s\pi^0\gamma} \approx 0.02$~\cite{Atwood:1997zr,Grinstein:2005nu}. A larger
observed value would be an indication of NP.  The latest
\babar measurements has found $S_{K_s\pi^0\gamma} = -0.03\pm 0.29 \pm
0.03$~\cite{Aubert:2008gy}.  Again, we remark that at \superb we
can expect a reduction in the statistical error of roughly a factor of
10, making this measurement, with its small systematic uncertainty, 
a powerful probe of physics beyond the SM at the \superb
factory.

It is interesting to compare what LHCb will be doing with the $b \to s \gamma$
channel in the next few years. Most LHCb studies on radiative penguin
decays thus far have focused on the related channel $B\to
K^{(*)}\mu^+\mu^-$.  They will not be able to perform an
inclusive measurement, and so have focused on a few exclusive
analyses such as time-dependent CP asymmetries in $B\to K^{*}\gamma$
and its $B_s$ analog, $B_s\to \phi\gamma$.  This last is of particular
interest since it cannot be done at \superb (assuming that only a
relatively small number of $B_s$ mesons will be produced there).  An LHCb
MC based study reports that with an integrated luminosity of
2 fb$^{-1}$, some 11,000 $B_s\to \phi\gamma$ will be
reconstructed. With this sample, a precision of 0.2 on
${\cal{A}}$$^\Delta$, the relevant CP-violating parameter for this mode,
is expected~\cite{LHCB:2009ny}.

In summary, the radiative penguin process $\bar B \to X_s \gamma$ has played a crucial
role in searching for physics beyond the SM at the $B$ factories.  This
channel will continue to be important at the \superb factory, where the
increased statistics will quickly yield a branching fraction
measurement with a total uncertainty close to 3\%. Furthermore, the
large \superb data sample will make possible precision measurements of
direct CP-violating asymmetries for inclusive $\bar B \to X_s \gamma$, as well as
time-dependent CP-violation in the exclusive mode $\B\to
K_s\pi^0\gamma$.

\subsubsection{Inclusive and exclusive $b \to s  \ell^+ \ell^-$ }

The study of exclusive and inclusive rare semileptonic decays, 
$B \to X_{s, d} \ell^+ \ell^-$, where $\ell^+ \ell^-$ is either $e^+ e^-$ or $\mu^+ \mu^-$, is an important task for a
\superb factory. 
 
In the exclusive decays $B \to K \ell^+ \ell^-$ and $B \to K^* \ell^+ \ell^-$, both \babar~\cite{Aubert:2008ps,Aubert:2008ju}
and Belle~\cite{Belle:2009zv} measured branching fractions (${\cal B}$), \CP asymmetries (${\cal A}_{CP}$), isospin asymmetries
(${\cal A}_I$), $\mu^+\mu^-/e^+ e^-$ ratios (${\cal R}_{K^{(*)}}$), the $K^*$ longitudinal polarization (${\cal F}_L$) and the
lepton FBA
(${\cal A}_{FB}$) in several $q^2$ bins. \babar~\cite{Aubert:2007mm} and Belle~\cite{Wei:2008nv} also searched for
$B \to \pi \ell^+ \ell^-$ and $B \to \rho \ell^+ \ell^-$. Both experiments measured partial branching fractions in the inclusive
mode in several $q^2$ bins ~\cite{Aubert:2004it, belle3},  where $B \to X_s \ell^+ \ell^-$ was approximated by a sum of 20 (36)
exclusive final states in \babar (Belle). \babar\, is presently working on an update with the full data set using 28 final states.
Here, the decay model uncertainty is reduced to $10\%$.

Another inclusive approach consists of reconstructing one $B$ meson fully in an exclusive 
final state and looking for an $\ell^+ \ell^-$ pair in the recoil. The advantage is that 
no assumptions have to be made on the $X_s \ell^+ \ell^-$ final states, thus this is a true 
inclusive measurement. The disadvantage is 
that the $B$ tagging reduces the total selection efficiency. Using both hadronic tags and 
semileptonic tags, the tagging efficiency can be up to $1.75\%$. The $B$ reconstruction 
removes semileptonic backgrounds from the opposite $B$ and from $D \bar D$. Additional requirements 
are necessary to remove backgrounds from $B$ semileptonic cascade decays of the signal $B$. 
Residual semileptonic backgrounds can be subtracted bin-by-bin using an $e-\mu$ data sample. 
Furthermore, the $B \to X_d \ell^+ \ell^-$ contribution also needs to be subtracted. 

The best approach to $B \to K^*  \ell^+\ell^-$  consists of a measurement of the full angular distribution in several $q^2$ bins. In the present $B \to K^* \ell^+ \ell^-$ analyses, the one-dimensional angular distributions in $\cos \theta_K$ and $\cos \theta_\ell$ are used to extract ${\cal F}_L(q^2)$ and ${\cal A}_{FB} (q^2)$. In the SM ${\cal A}_{FB}$ crosses zero 
around 4.2~$\rm GeV^2/c^4$~\cite{Feldmann:2002iw}. The $\phi$ distribution allows two additional asymmetries, ${\cal A}_{Im}(q^2)$ 
an interference between transverse and longitudinal components and ${\cal A}^{(2)}_{T}(q^2)$, an asymmetry between 
transverse and parallel components~\cite{Kruger:1999xa}. At  \superb, we collect enough events to measure the full angular 
distribution and extract all its  nine coefficients, which are functions of $q^2$ ~\cite{Kruger:1999xa}. In addition, we 
have sufficient sensitivity to examine the $\cos \theta_\ell$ distribution for scalar and pseudoscalar 
contributions.
In particular $B \to K \ell^+ \ell^-$ is rather sensitive, since ${\cal A}_{FB}$ vanishes in the SM. 
In the inclusive analyses, we will explore the $\cos \theta_\ell$ angular distribution and determine the functions $H_T(q^2)$, 
$H_L(q^2)$ and $H_A(q^2)$~\cite{Lee:2006gs} in several $q^2$ bins.

Since the \superb detector will be an improvement on \babar\, detector, we focus on \babar\, measurements and 
scale statistical errors to a luminosity of 75~\invab by $\sqrt{{\cal L}_{\babar}/75~ab^{-1}}$ in order to be conservative. 
For decay rates and rate asymmetries of exclusive modes, we use the \babar\, publication~\cite{Aubert:2008ps}. 
For ${\cal F}_L(q^2)$ and ${\cal A}_{FB}(q^2)$ in $B \to K^* \ell^+ \ell^-$, we use a recent study based on the 
total \babar\, luminosity of $425 fb^{-1}$ assuming ${\cal B}(B \to K \ell^+ \ell^-) =0.48 \times 10^{-6}$ 
and ${\cal B}(B \to K^* \ell^+ \ell^-) =1.15\times 10^{-6}$. The present statistical uncertainties of decay 
rates and rate asymmetries of the sum of exclusive decays are obtained from another recent \babar\, study 
based on $425 fb^{-1}$ assuming ${\cal B}(B \to X_S \ell^+ \ell^-) =4.5 \times 10^{-6}$. The statistical 
uncertainty in the fully inclusive mode is obtained by scaling that in the sum of exclusive modes by the 
square root of the ratio of expected events in the two approaches. Since we have no studies on inclusive 
angular analyses yet, we base our estimates for $H_L(q^2)$ and $H_A(q^2)$ on the ${\cal F}_L(q^2)$ 
and ${\cal A}_{FB}(q^2)$ results measured in the $B \to K^* \ell^+ \ell^-$ angular analysis.

Table~\ref{tbl:bdecays:nsll} shows event yields for the different decay channels at the \babar luminosity of $425~fb^{-1}$ and extrapolations to 75~\invab for \superb. The statistical and systematic uncertainties of branching fractions, rate asymmetries and angular observables in  different regions of $q^2$ for $B \to K \ell^+ \ell^-$ and $B \to K^* \ell^+ \ell^-$ at \babar (425~$fb^{-1})$ and extrapolations at \superb (75~\invab) are summarized in Table~\ref{tbl:bdecays:kll} and Table~\ref{tbl:bdecays:kstll}, respectively. Table~\ref{tbl:bdecays:xsll} shows estimates of statistical and systematic uncertainties of corresponding observables in  the $B \to X_s \ell^+ \ell^-$ analyses.

The systematic errors for observables in the exclusive modes are taken from the latest \babar 
publications~\cite{Aubert:2008ps,Aubert:2008ju}, while the systematic error of the branching fraction in the sum of exclusive modes is estimated by adding in quadrature the individual contributions with updated values. For the cross feed, multivariate selection, and fitting systematics we assume $3\%, 2\% $ and $1.4\%$, respectively. These values are chosen after comparing systematic errors of the analysis of the sum of 20 exclusive modes~\cite{Aubert:2004it} with those in the exclusive modes. This amounts to a total systematic error of $5.6\%$ in the branching fraction. The systematic error of the total branching fraction of the fully inclusive mode is estimated by adding in quadrature contributions from tagging, tracking, lepton identification, event selection, background parametrization, fitting, and total number of $B$ mesons. For the present \babar data set, we estimate a systematic error of $6 \%$. The systematic errors of the partial branching fractions, 
rate asymmetries and angular observables are assumed to scale in a similar was as the total branching fraction in $B \to K^* \ell^+ \ell^-$.

At \superb, both exclusive and inclusive $b  \to s \ell^+ \ell^-$ modes will be measured with high precision. 
For example, for an integrated luminosity of 75 \invab, we expect to observe 8,200 selected $B \to K^* \ell^+ \ell^-$ signal events in the low $q^2$ region ($< 8~ \rm GeV^2/c^4$) and 5,500 selected signal events reconstructed fully inclusively with the recoil method. Thus, for both inclusive and exclusive decays we have sufficient statistics to measure the $q^2$ dependence of branching fractions and angular observables. For most observables, the statistical precision will be around one per cent or below. Thus, these measurements will be systematics limited. Reducing the $q^2$ region to $1~\rm GeV^2/c^4$ $< q^2 < 6~GeV^2/c^4$ yields an increase in the statistical uncertainty by a factor of 1.3 for rate asymmetries and by a factor of 1.38 for angular observables.

For $B \to \pi \ell^+\ell^-$, we expect about 700 events in the entire $q^2$ region at 75~\invab. We find the same result whether we scale the results of a previous analysis or whether we adjust the $K \ell^+ \ell^-$ results by $|V_{ts}/V_{td}|^2=0.206^2$ obtained from $B_s$  and $B_d$ mixing. Using the latter scaling for $K^* \ell^+ \ell^-$ results, we expect about 800 $\rho \ell^+ \ell^-$ signal events in the entire $q^2$ region at 75~\invab. Using the recoil method, we also expect a sizable sample for $B \to X_d \ell^+ \ell^-$ modes. From both exclusive and inclusive modes we can determine $|V_{td}/V_{ts}|$.
\superb should also be able to discover the exclusive modes $B \to K^+ \tau^+ \tau^-$and $B \to K^{*0} \tau^+ \tau^-$.
Thus, at \superb there is a great potential to see NP at the order of ${\cal O}(0.1)$. 

At the expected design luminosity of $2~fb^{-1}$ per year, the LHCb experiment expects to observe 4,000 
$K^{*0}(\to K^+ \pi^-) \mu^+ \mu^-$ events with a background of 1,000 events in the $q^2$ region 
$4 m_\mu^2 < q^2 < 9~\rm GeV^2/c^4$~\cite{lhcb}. This yields a large sample to perform a full 
angular analysis and determine the nine coefficients $I_i (q^2)$. In a recent study, LHCb focussed 
on the observables ${\cal A}_{FB}$, ${\cal F}_L$, ${\cal A}_{Im}$ and 
${\cal A}^{(2)}_T$. For $10~fb^{-1}$,  integrated over the $q^2$ region $1~ \rm GeV^2/c^4 <$ $q^2 < 6~ \rm GeV^2/c^4$, 
LHCb estimates statistical uncertainties of $\sigma_{{\cal A}_{FB}}=(+0.0047,-0.0050)$,  
$\sigma_{{\cal F}_L}=(0.0052, -0.0058)$,  $\sigma_{{\cal A}_{Im}}=(+0.0060, -0.0057)$, and 
$\sigma_{{\cal A}^{(2)_T}}=(+0.095, -0.094)$, respectively. The zero crossing of ${\cal A}_{FB}$ 
is determined with a relative uncertainty of $4\%$ at
$q^2_0=4.33^{+0.18}_{-0.16}~\rm GeV^2/c^4$. The LHCb errors are purely statistical. Systematic 
uncertainties have not been addressed and are expected to be larger than those in \babar. 
In a toy study based on \babar simulated events scaled to a luminosity to 75~\invab, we extract a relative statistical uncertainty of $9\%$ for the zero crossing after fitting the $q^2$ dependence of ${\cal A}_{FB}(q^2)$ in the region of $2.75~\rm GeV^2/c^4$ to $5.75~\rm GeV^2/c^4$ with a linear shape. We think that further optimization of the angular fits is possible and that statistical errors may be further reduced.   

CDF has presented a study of $B \to K^{*0} \mu^+ \mu^-$ using $4.4~fb^{-1}$ of data~\cite{cdf}. 
For $q^2 < 8.68~\rm GeV^2/c^4$ they observe $34.3\pm 6.7$ events, yielding 780 events expected at $10~fb^{-1}$, 
if the $b \bar b$ cross section at 14~TeV is a factor of ten higher than that at 2~TeV. For ${\cal A}_{FB}$ and 
${\cal F}_L$ in the $q^2$ bin  $q^2 < 4.3~\rm GeV^2/c^4$,  the statistical uncertainties of $(+0.31,-0.33)$ and 
$(+0.23,-0.24)$ at $4.4~fb^{-1}$ are reduced to $\pm 0.07$ and $\pm 0.05$, respectively at $10~fb^{-1}$. The 
systematic errors at $4.4~fb^{-1}$ are 0.05 and 0.03, respectively. Thus, the extrapolated CDF yields are a 
factor of 5 lower than the LHCb simulation, while those in ${\cal A}_{FB}$ and ${\cal F}_L$ are factors of 14 
and 8.3 higher, respectively.

\begin{table*}[!ht]
\caption{Number of events for $B \to K \ell^+ \ell^-$, $B \to K^* \ell^+ \ell^-$, $B \to X_s \ell^+ \ell^-$  via the sum of exclusive modes (SE) and $B \to X_s\ell^+ \ell^-$ via the recoil method (RM) for luminosities of $\rm 425~fb^{-1}$ and 75~\invab. The signal yields are shown for the entire $q^2$ region,  $0.1~\rm GeV^2/c^4 $ $ < q^2 <   7.84~\rm GeV^2/c^4 $  and  $ 1~\rm GeV^2/c^4$ $ < q^2 <   6~\rm GeV^2/c^4 $. Uncertainties in the yields are of the order of $20\%$.}
\label{tbl:bdecays:nsll}
\begin{center}
\small
\begin{tabular}{|l|ccc|ccc|} \hline
  \multicolumn{1}{l}{} & \multicolumn{3}{l}{Number of events in $\rm 425~fb^{-1}$}  & \multicolumn{3}{l}{Expected number of events in 75~\invab} \\  \hline
Mode       & all $q^2$       & 0.1--7.84  & 1--6  & all $q^2$       &0.1--7.84  &1--6 \\ \hline\hline
$K \ell^+ \ell^-$ &   90 & 42 & 26& 15,900 & 7,340 & 4,600  \\ 
$K^* \ell^+ \ell^-$&  110 & 46 & 24 & 19,400 & 8,200 & 4300    \\ 
$X_s \ell^+ \ell^-$ SL   & 270       &171 &101    & 47,500   &30,000 &17,900   \\ 
$X_s \ell^+ \ell^-$ RM   & 49       &31  &18   &8,600   &5,500   &3250 \\ 
\hline 
\end{tabular}
\end{center}
\end{table*}

\begin{table*}[!ht]
\caption{Present and extrapolated statistical and systematic uncertainties of the total branching fraction, partial branching fractions, \CP asymmetries, isospin asymmetries, lepton flavor ratio for $B \to K \ell^+ \ell^-$ after combining 
$e^+ e^-$ and $\mu^+\mu^-$ modes as well as $K^+$ and $K^0_S$ modes.}
\label{tbl:bdecays:kll}
\begin{center}
\small
\begin{tabular}{|lc|cc|cc|}
\hline 
 \multicolumn{2}{l}{}  & \multicolumn{2}{c}{$\babar\, (425~fb^{-1})$}  & \multicolumn{2}{c}{\superb (75~\invab)} \\ \hline
Observable & $q^2$ region $\rm [GeV^2/c^4]$         & Stat.        & Sys.          & Stat.        & Sys.    \\ \hline
$\sigma{\cal B}/ {\cal B}  $& all     & 0.175      & 0.05      & 0.011   & 0.025-0.035        \\ 
$\sigma{\cal B}/ {\cal B}   $ & 0.1--7.02   &0.20       & 0.044     & 0.012   & 0.022-0.035    \\ 
$\sigma{\cal B}/ {\cal B}  $ & 10.24--12.96 and $>14.06$  & 0.27       &0.052     & 0.017   & 0.026-0.039    \\ \hline
${\cal R}_{K} $   & all &  0.34     & 0.05      & 0.021  &      0.025-0.038   \\ 
${\cal A}_{CP} $ & all    & 0.18      & 0.01        & 0.012       & 0.008-0.01      \\ 
 ${\cal A}_I   $ &  0.1--7.02 & 0.56      & 0.05    & 0.034        & 0.025-0.035   \\ 
\hline 
\end{tabular}
\end{center}
\end{table*}

\begin{table*}[!ht]
\caption{Present and extrapolated statistical and systematic uncertainties of the total branching fraction, partial branching fractions, \CP asymmetries, isospin asymmetries, lepton flavor ratio, longitudinal polarization and lepton FBA for $B \to K^* \ell^+ \ell^-$ after  combining $e^+ e^-$ and $\mu^+\mu^-$ modes as well as $K^{*+}$ and $K^{*0}$ modes.}
\label{tbl:bdecays:kstll}
\begin{center}
\small
\begin{tabular}{|lc|cc|cc|}\hline 
 \multicolumn{2}{l}{}  & \multicolumn{2}{c}{$\babar\, (425~fb^{-1})$} & \multicolumn{2}{c}{\superb (75~\invab)} \\ \hline
 Observable & $q^2$ region $\rm [GeV^2/c^4]$   & Stat.        & Sys.          & Stat.        & Sys. \\         \hline
$\sigma{\cal B}/ {\cal B} $& all     & 0.162     & 0.063      &0.01   & 0.032-0.048     \\ 
$ \sigma{\cal B}/ {\cal B} $ & 0.1--7.02   & 0.23       &0.070     & 0.014   & 0.035-0.053   \\ 
$ \sigma{\cal B}/ {\cal B}  $ & 10.24--12.96 and $>14.06$    &0.24       & 0.071      & 0.015  & 0.036-0.054  \\ \hline
${\cal R}_{K*}  $& all     & 0.34       & 0.07     & 0.02  & 0.035-0.048       \\ 
$ {\cal A}_{CP} $& all     & 0.15       & 0.01     & 0.009  & 0.008-0.01      \\ 
${\cal A}_I   $&  0.1--7.02      & 0.17      &0.03     & 0.01 &0.015-0.023       \\ \hline
$ {\cal F}_L   $&  0.1--4      & 0.15       &0.04      & 0.011  & 0.02-0.03      \\ 
$ {\cal F}_L   $&  4--7.84      & 0.14       &0.04      & 0.011  & 0.02-0.03      \\ 
$ {\cal  A}_{FB}  $&0.1--4       &0.14    &0.05     &0.011   &0.025-0.038      \\ 
$ {\cal  A}_{FB}  $& 4--7.84       &0.14    &0.05     &0.011   &0.025-0.038      \\ 
\hline
\end{tabular}
\end{center}
\end{table*}

\begin{table*}[!ht]
\caption{Present and extrapolated statistical and systematic uncertainties of the total branching fraction, partial
branching fractions, \CP asymmetries, isospin asymmetries, lepton flavor ratio, and angular observables for
$B \to X_s \ell^+ \ell^-$. The first two columns show the results for the sum of exclusive modes (SE), the second two
columns those for the recoil method (RM), respectively. The sum of exclusive modes including 28 final states has an additional
uncertainty of $\sim 10\%$ from the decay model.
}
\label{tbl:bdecays:xsll}
\begin{center}
\small
\begin{tabular}{|lc|cccc|cccc|}\hline 
 \multicolumn{2}{l}{}  & \multicolumn{4}{c}{$\babar\, (425~fb^{-1})$} & \multicolumn{4}{c}{\superb (75~\invab)} \\  \hline
 Observable & $q^2$ region  & Stat.        & Sys.  & Stat. & Sys.       & Stat.        & Sys.  & Stat. & Sys.     \\ 
 &  $\rm [GeV^2/c^4]$  &  SE     &  SE &RM  & RM   & SE      &SE &RM &  RM  \\ 
 \hline
$\sigma{\cal B}/{\cal B}$  & all          &0.11  & 0.056  &0.26 &0.06  &0.008   &0.03-0.05   &0.019   &0.03-0.05 \\ 
$\sigma{\cal B}/ {\cal B}$   &0.1--1 & 0.29   & 0.07  &0.69 &0.07  &0.022   &0.04-0.06  &0.052    & 0.04-0.06\\ 
$\sigma{\cal B}/ {\cal B}$   & 1--4 & 0.23   & 0.06  &0.53  &0.06  &0.017   &0.03-0.05  &0.040   &0.03-0.05\\ 
$\sigma{\cal B}/ {\cal B}$   & 4--7.84 & 0.18   & 0.06  &0.43  &0.06  &0.014   &0.03-0.05  &0.032   & 0.03-0.05\\ 
$\sigma{\cal B}/ {\cal B}$   & 10.24--12.96 & 0.31   & 0.07  &0.73  &0.07  &0.024   &0.04-0.06  & 0.055  &0.04-0.06\\ 
$\sigma{\cal B}/ {\cal B}$   & $>$14.06 & 0.29   & 0.07  &0.69  &0.07  &0.022   &0.04-0.06  &0.052  &0.04-0.06\\ \hline
${\cal R}_{X_s} $& all                       & 0.21    & 0.06  &0.50 &0.06  &0.016   &0.03-0.05   &0.038   &0.03-0.05\\  
${\cal R}_{X_s}  $ &0.1--7.84 & 0.25      & 0.06&0.58    &0.06   &   0.019  & 0.03-0.05      &0.044 &0.03-0.05\\ 
${\cal A}_{CP}  $ & all  & 0.06       & 0.01    &0.14   &0.01  & 0.004  & 0.005-0.008      &0.011  &0.005-0.008 \\ 
${\cal A}_{CP}  $ & 0.1--7,84   & 0.07       & 0.01    &0.16  &0.01   &   0.005  & 0.005-0.008   &0.012  &0.005-0.008 \\ 
${\cal A}_I  $    & all  & 0.05       & 0.06    &0.12 &0.06   & 0.004        & 0.03-0.05      &0.009 &0.03-0.05 \\ 
${\cal A}_I  $    & 0.1--7.84& 0.06       & 0.06    &0.14 &0.06   & 0.005        & 0.03-0.05      &0.011 &0.03-0.05 \\ \hline
${\cal H}_L $ & 0.1--1  & 0.17       & 0.04   &0.40  & 0.04     & 0.013        & 0.02-0.03      &0.030  &0.02-0.03  \\ 
${\cal H}_L $ &  1--4  & 0.17      & 0.04   &0.40  &0.04    & 0.013        & 0.02-0.03      &0.030  & 0.02-0.03\\ 
${\cal H}_L $ &  4--7.84   & 0.13       & 0.04   &0.27  &0.04     & 0.009        & 0.02-0.03      &0.021  &0.02-0.03 \\ 
${\cal H}_{A}  $  &  0.1--1   & 0.22       & 0.06  &0.51 &0.06    & 0.016        & 0.03-0.05     &0.039  &0.03-0.05 \\ 
${\cal H}_{A} $  &1--4   & 0.22       & 0.06  &0.51 &0.06     & 0.016        & 0.03-0.05     &0.039  &0.03-0.05  \\ 
${\cal H}_{A} $  &  4--7.84   & 0.15       & 0.06  &0.35 &0.06     & 0.011        & 0.03-0.05     &0.026 &0.03-0.05 \\ 
\hline
\end{tabular}
\end{center}
\end{table*}

\subsubsection{More on  $B \to X_{s/d}  \ell^+ \ell^-$ with a hadronic tag}

\begin{figure}[!htb]
  \begin{center}
    \includegraphics[width=0.5\textwidth]{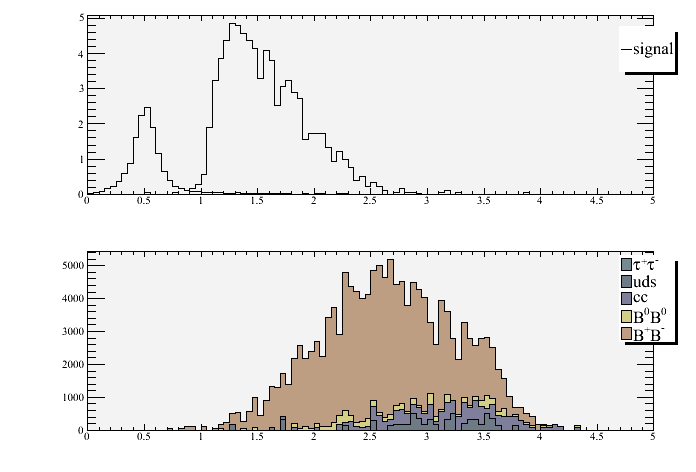}
  \caption{\label{fig:myplot}
The (top) reconstructed hadronic mass for exclusive $B^+ \to K^+ \mu^+ \mu^-$ (low mass region) and
inclusive $B^+ \to X_s \mu^+ \mu^-$ (region above $m_{X_s} = 1$GeV) signal MC.  The 
expected background is shown in the bottom plot.}
  \end{center}
\end{figure}

It may be possible to study the process $B \to X_s \ell^+ \ell^-$ (and maybe also $B \to X_d \ell^+ \ell^-$, albeit with limited
statistics) fully inclusively at \superb using the hadronic tag reconstruction method.  Although this technique has very low
efficiency and hence can statistically limit the sensitivity of rare decay studies, it has a number of advantages.  First, by
selecting signal events using primarily information from the reconstructed $B_{tag}$ and the di-lepton system, the $X_s$ system
can potentially be selected without substantial biases to the hadronic mass distribution.  Even $K_L$ modes are potentially
accessible, regardless of whether the $K_L$ interacts in the detector. Secondly, the tag reconstruction effectively eliminates the
``irreducible'' (in the un-tagged analysis) backgrounds from double-semileptonic $B\bar{B}$ decays and cascade $b \to c \to s$
semileptonic decays.  The first of these is eliminated by the requirement of an exclusively reconstructed hadronic $B$ meson
accompanying the signal candidate, while the second is reduced based on missing energy considerations.  Although this method still
needs to be studied carefully with large background statistics in the context of the \superb fast simulation, an initial study based
on \babar MC appears promising.  Hadronic $B$ decays are first reconstructed using the usual tag reconstruction method.
These events are additionally required to possess a $\ell^+\ell^-$ ($\ell = e, \mu$) system, and all remaining detector activity
in the event is defined to comprise the $X_s$ hadronic system, and the combination of $X_s$ with the di-lepton system is required
to be kinematically consistent with a $B$ meson recoiling against the reconstructed $B_{tag}$.  Note that in this case we have not
only the usual $\Delta E$ and $m_{ES}$ variables at our disposal, but also the angle between the 3-momenta of the reconstructed tag
and signal $B$ candidates.  Vetoes are imposed on the $\ell^+\ell^-$ invariant mass to remove long distance contributions from
$J/\Psi$ and $\Psi(2S)$, and the mass of the hadronic system $X_s$ is computed directly from the combination of the $B_{tag}$ 
4-vector and the $\ell^+\ell^-$ 4-vector (i.e. without any direct reconstruction of the hadronic system itself).
Fig.~\ref{fig:myplot} (top) shows the reconstructed hadronic mass for exclusive $B^+ \to K^+ \mu^+ \mu^-$ (low mass region) and 
inclusive $B^+ \to X_s \mu^+ \mu^-$ (region above $m_{X_s} = 1$GeV) signal MC.  The expected background is shown in the 
lower plot.  Both signal and background plots are normalized to 75 ab$^{-1}$, however the background statistics represent 
only about 1 ab$^{-1}$.  The low mass region is essentially background free, while in the higher mass region the background 
is predominantly from $B^+B^-$ events in which the two muon candidates originate from the same $B$.  This remaining background 
can then be further reduced through an appropriate choice of muon PID (and vertexing, in the case of the corresponding $e^+e^-$ 
modes), and more stringent missing energy requirements on the signal $B$ candidate. More detailed studies within the \superb 
simulation framework are needed to establish an expected signal significance, but the method shows promise for a fully 
inclusive determination of $B \to X_s \ell^+\ell^-$ branching fractions and angular asymmetries.  This method can also 
in principle be extended to include $B \to X_d \ell^+\ell^-$ simply by applying a kaon tag/veto to the hadronic system.

\subsection{Determination of \Vub and \Vcb}

\subsubsection{Inclusive Determination of \Vub}
\label{sec:inclvub}

The precise determination of \Vub is an essential ingredient in the determination of the CKM matrix parameters. As the precise
study of inclusive decays is a unique feature of \superb, it will be important to make maximal use of the data to extract \Vub with
small but robust uncertainties. The main experimental and theoretical challenge in the inclusive \Vub determination is the
background from $B\to X_c\ell\bar\nu$ decays which is roughly $50$ times larger than the signal. At \superb the experimental
uncertainties will be reduced compared to the existing measurements at the $B$ factories. In addition to reduced statistical
uncertainties, the much larger statistics will also lead to reduced systematic uncertainties by allowing for cleaner data samples
and a better understanding of the $\bar B\to X_c\ell\bar\nu$ background. For the theoretical uncertainties, one must distinguish
between uncertainties due to higher-order perturbative and power corrections and parametric uncertainties due to input parameters.
Currently, the dominant theoretical uncertainties are parametric due to $m_b$ and the leading $B$-meson shape function, and these
can be reduced by more precise measurements as explained below. The current
approaches~\cite{Lange:2005yw,Andersen:2005mj,Gambino:2007rp} heavily rely on modeling the shape function, and as a result the
present uncertainty in the inclusive \Vub determination~\cite{Barberio:2008fa} has been the subject of intense debate.

Measurements near the phase space boundary are experimentally cleaner with less contamination from $\bar B\to X_c\ell\bar\nu$,
thus allowing for reduced systematic uncertainties, but are more sensitive to $m_b$ and shape-function effects, leading to
increased theoretical uncertainties. Hence, the choice of kinematic cuts is a trade-off between experimental and theoretical
uncertainties, and so there is no unique optimal region of phase space from which to extract \Vub. Even if the kinematic cuts
can be relaxed, at the expense of increased systematic uncertainties, to the point where there is no shape-function sensitivity
in the theory, the most sensitivity to \Vub in the data still comes from the shape-function region near the endpoint. In addition,
one should recall that the Monte Carlo (MC) signal model requires the knowledge of the shape function as well.

Ultimately, the best determination of \Vub with the smallest uncertainty might be achieved by performing a combined fit to all
available measurements that simultaneously determines \Vub and the required inputs, such as $m_b$ and the leading shape function,
as proposed in Ref.~\cite{Ligeti:2008ac}. This follows the same strategy successfully employed in the inclusive determination of
\Vcb. It allows for the combination of different measurements with a consistent treatment of correlated uncertainties in
measurements and input parameters. Moreover, it is straightforward to consistently include additional constraints on $m_b$ (\eg
from the $\bar B\to X_c\ell\bar\nu$ fits) and the shape function, \eg from the measured $\bar B\to X_s\gamma$ photon energy
spectrum. Finally, the goodness of the fit itself provides an important test of the underlying theory.

Experimentally this strategy requires measurements of $\bar B\to X_u\ell\bar\nu$ decay spectra, which are already possible with
the current data sets~\cite{Bizjak:2005hn, Aubert:2007rb, Tackmann:2008qa}, and will be possible with increased precision at \superb.
In the fit, the normalization of the spectra determines the value of \Vub, while their shape determines $m_b$ and the shape function.
Hence, in this way the increased statistics at \superb is effectively utilized to also reduce the dominant theoretical uncertainties.
Theoretically, the implementation of such a fit is complicated by the fact that one is fitting a function (rather than a few numbers
as is the case for \Vcb). To obtain small but reliable shape function uncertainties, it is desirable to include as much perturbative
information on the shape function as possible, but at the same time have a model-independent treatment of its nonperturbative content.
This can be achieved for instance using the approach of Ref.~\cite{Ligeti:2008ac} which we now brielfy outline.

\begin{figure}[!htb]
  \begin{center}
    \includegraphics[width=0.45\textwidth]{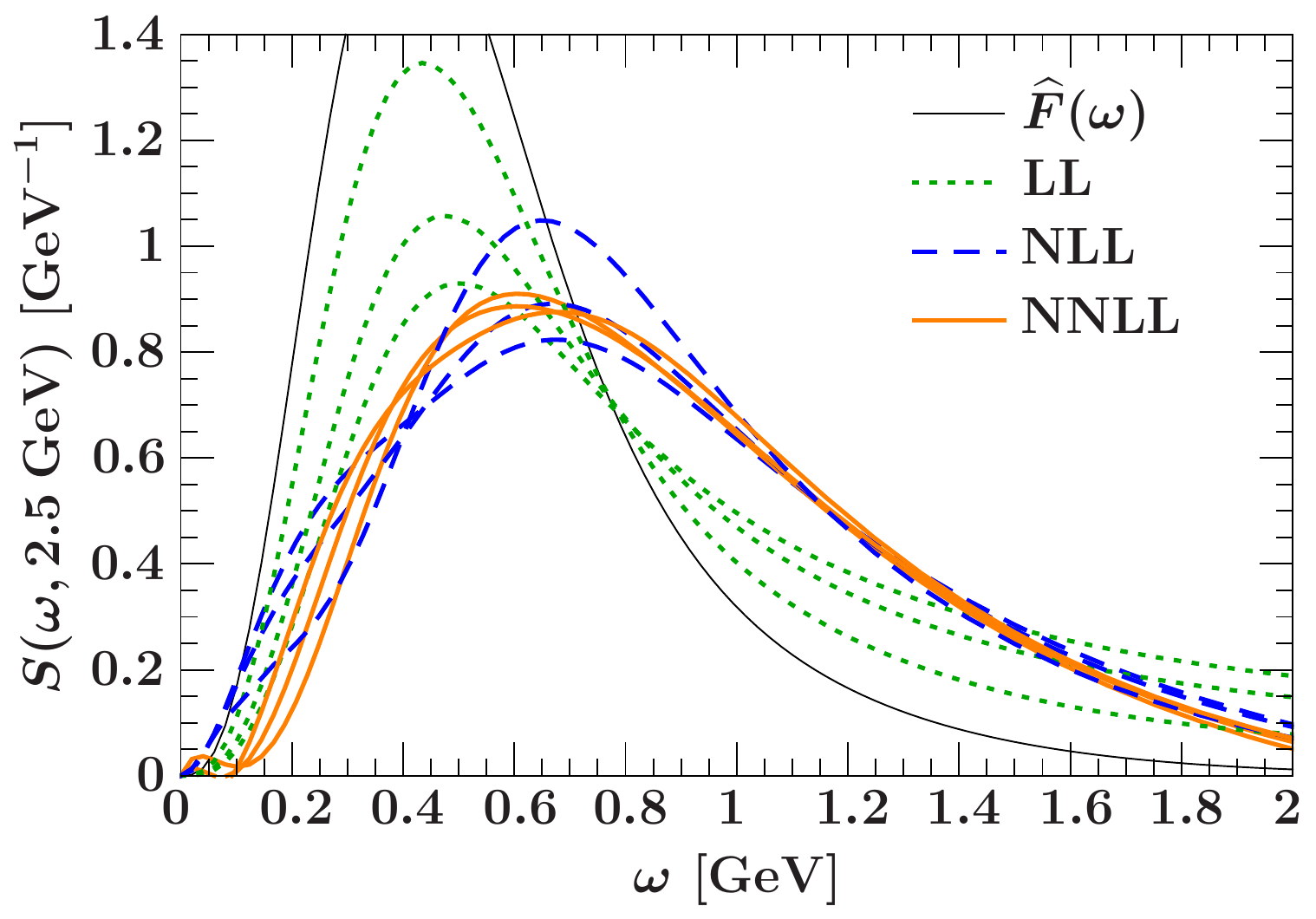}
  \caption{\label{fig:shapefunction}
   The shape function $S(\omega, \mu)$ obtained by evaluating Eq.~(\ref{eq:shapefunction}) at a low scale $\mu_\Lambda$
   and RG evolving the result to the common scale $\mu = 2.5\gev$. The three curves at each order correspond to the three
   values $\mu_\Lambda = \{1.1, 1.3, 1.8\}\gev$. The solid black curve shows the assumed input function $\widehat F(k)$.}
  \end{center}
\end{figure}

First, the shape function $S(\omega, \mu)$ is factorized as
\begin{equation} \label{eq:shapefunction}
S(\omega, \mu_\Lambda) = \int\! dk\, \widehat C(\omega - k, \mu_\Lambda) \widehat F(k)
\,,\end{equation}
where $\widehat C$ is computed in perturbation theory and is known at the two loop level~\cite{Bauer:2003pi, Becher:2005pd}, while
$\widehat F(k)$ is a purely non-perturbative function to be extracted from data. Given $\widehat F(k)$, the shape function can be
obtained from Eq.~(\ref{eq:shapefunction}) order by order in perturbation theory and the perturbative uncertainty can be estimated
by varying the scale $\mu$, as illustrated in Fig.~\ref{fig:shapefunction}.

Next, to extract $\widehat F(k)$ from data in a model-independent way, it is expanded in a complete set of basis functions
$\{f_n(k)\}$ that are designed to converge quickly for functions consistent with confinement,
\begin{equation} \label{eq:SFbasisexpansion}
\widehat F(k) = \biggl|\sum_{n = 0}^\infty c_n f_n(k) \biggr|^2
\,.\end{equation}
Here, the basis coefficients $c_n$ are the unknown fit parameters. In practice, the series has to be truncated after a certain
number of terms depending on the precision of the data. This results in a small residual model uncertainty, which can be studied
systematically, \eg by varying the number of coefficients, and decreases with increasing experimental precision.

Given Eq.~(\ref{eq:SFbasisexpansion}), the $i^{th}$ experimentally measured bin, $\Gamma^i$, is calculated as
\begin{equation}
\Gamma^i  = \Vub^2 \sum_{m,n} c_m\, c_n\,\Gamma_{mn}^i
\,,\end{equation}
where $\Gamma_{mn}^i$ is the contribution to $\Gamma^i$ from the product of basis functions
$f_m(k) f_n(k)$ in Eq.~(\ref{eq:SFbasisexpansion}), which can be computed in advance. Hence,
one simply has to fit polynomials of the basis coefficients $c_n$ to the experimental measurements.
Measurements of the $\bar B\to X_s\gamma$ photon energy spectrum are included in an analogous way.
In addition, known constraints on the moments of $\widehat F(k)$ are included as additional constraints
on the basis coefficients. For example the $0$th and $1$st moments are
\begin{equation} \label{eq:MomentsFit}
1 = \sum_n c_n^2
\,,\text{ and }
m_B - \widehat m_b = \sum_{m,n} M^1_{mn} c_m c_n
\,,\end{equation}
where $M^1_{mn}$ is the first moment of $f_m(k) f_n(k)$. In this way, existing information on
$\widehat m_b$ can be included in the fit (the hat indicates a suitable short-distance scheme).
The above combined fit approach will be key to push the precision of the inclusive \Vub
determination at \superb below the 5\% level. It makes maximal use of the increased data set
by providing a rigorous treatment of uncertainties, provides tests for the theory, and yields
an improved determination of $m_b$ and the leading shape function. With sufficient data and
measurements of different spectra, the same methods can also be used to reduce uncertainties
due to subleading shape functions. A precise knowledge of leading and subleading shape functions
is also crucial for a clean theoretical interpretation of measurements of the inclusive radiative
decays $\bar B\to X_s\gamma$ and $\bar B\to X_s\ell^+\ell^-$.

\subsubsection{Inclusive Determination of \Vcb} 
\noindent
The determination of \Vcb from inclusive semi-leptonic $b \to c$ transitions relies on the heavy 
quark expansion (HQE), resulting in an expansion in inverse powers of $m_b$ and 
$m_c$. In the meantime, the  methodology is in a very mature state, leading finally to a theoretical 
uncertainty as low as $1\%$ in $V_{cb}$~\cite{Benson:2003kp}. 
At \superb this will be the limiting factor, since the large 
number of available events will reduce the experimental uncertainty to a negligible level. 

The extraction of \Vcb relies on the HQE for the total rate, which takes the schematic form
\begin{eqnarray}
&& \Gamma = \frac{G_F^2 m_b^5}{192 \pi^3} \Vcb^2 \\  \nonumber
&& \left[ f_0 + \left(\frac{\Lambda_{\rm QCD}}{m_b}\right)^2   f_2 
+ \left(\frac{\Lambda_{\rm QCD}}{m_b}\right)^3   f_3 + 
\left(\frac{\Lambda_{\rm QCD}}{m_b}\right)^4   f_4 \right. 
\\ \nonumber 
&& + f_5 \left( a_0  \left(\frac{\Lambda_{\rm QCD}}{m_b}\right)^5 
+ a_2 \left(\frac{\Lambda_{\rm QCD}}{m_b}\right)^3  
   \left(\frac{\Lambda_{\rm QCD}}{m_c}\right)^2 \right) \\
&& \left. + ... +  f_7  \left(\frac{\Lambda_{\rm QCD}}{m_b}\right)^3 \left(\frac{\Lambda_{\rm QCD}}{m_c}\right)^4 \nonumber \cdots \right],
\end{eqnarray} 
where the $f_i$ and $a_i$ 
are functions of $m_b / m_c$ which are - aside from logarithms  $\ln (m_b / m_c)$
- regular as $m_c \to 0$. 

In order to obtain a precise determination of \Vcb one has to control the following 
inputs:
\begin{enumerate} 
\item On the first sight there is a strong dependence on the $b$ quark mass due to the 
$m_b^5$ factor. However, the real dependence is weaker due to the presence of the charm 
mass $m_c$. In fact, the semi-leptonic decays depend roughly only on $m_b - 0.6\, m_c$, where 
this difference can be determined from the moments of the decay spectra of $b \to c$ 
semi-leptonic transitions with a sufficient accuracy.
\item QCD radiative corrections are taken into account up to and including $\alpha_s^2$ terms
\cite{Pak:2008cp,Biswas:2009rb}. 
These corrections are under reasonable control provided a suitable scheme for the 
quark masses  is used, in which 
the bulk part of the radiative corrections has been absorbed into the quark masses. 
Mainly used are the ``kinetic scheme''~\cite{Bigi:1994ga} and the
 ``$1S$ scheme''~\cite{Hoang:1998hm}, both of 
which yield comparably small uncertainties.  
\item The non-perturbative inputs are given by local forward matrix elements of operators, which 
are calculated from the HQE. The quantity $f_0$ does not have any hadronic matrix element 
and is simply the result of the parton model calculation;  $f_0$ is calculated including the complete 
$\alpha_s$ corrections.  The current fits include $f_2$ and $f_3$, where $f_2$ depends on 
\begin{eqnarray} 
2 M_b \mu_\pi^2 &=& \langle B | \bar{b} (i \vec{D})^2  b | B \rangle,  \\
2 M_b \mu_G^2 &=& \langle B | \bar{b} (\vec{\sigma} \cdot \vec{B})  b | B \rangle, 
 \end{eqnarray}
where $\vec{B}$ denotes the chromomagnetic field inside the $B$ meson, while $f_3$ 
depends on 
\begin{eqnarray} 
2 M_b \rho_D^3 &=& \langle B | \bar{b}  \vec{D} \cdot \vec{E}   b | B \rangle,  \\
2 M_b \rho_{LS}^3 &=& \langle B | \bar{b} (\vec{\sigma} \cdot (\vec{D} \times  \vec{E}))  b | B \rangle.
 \end{eqnarray}
These matrix elements have to be fitted from the data. In particular, suitably chosen moments 
of the lepton energy, the hadronic energy and the hadronic mass spectra are sensitive to these 
matrix elements~\cite{hfag,Gambino:2004qm}.  

\item QCD corrections to the subleading terms $f_2$, $f_3$, $\ldots$ are only partially known. 
Up to now only corrections of order $\alpha_s \mu_\pi^2 / m_b^2$ are calculated 
\cite{Becher:2007tk} and not yet included in 
the fit. 
\end{enumerate} 

The perspectives for a further theoretical improvement are quite limited and depend on the 
following points
\begin{itemize}
\item Currently the leading element missing in the theoretical analysis
is the calculation of the corrections of order
$\alpha_s \mu_G^2 / m_b^2$, which would complete our knowledge of the $\alpha_s / m_b^2$ 
terms.  
\item A detailed consideration of the various quark mass determinations has to be performed 
in order to improve the control over the uncertainties related to the quark-mass input. Currently 
the quark masses are taken from $B$ decays, which turn out to be consistent with determinations 
from $e^+ e^-$ threshold data. However, the uncertainties have to be scrutinized carefully.
\item The number of non-perturbative parameters proliferates drastically, i.e.\ $f_4$ already 
depends on 9, $f_5$ on 18 and $f_6$ on 72 unknown matrix elements. The ability to extract 
this information from data remains limited even at \superb, in which case a calculation of these 
matrix elements would be needed; however, the perspectives to get a reliable calculation is also very limited. 
\end{itemize}

In conclusion, with the current theoretical technology a relative theoretical uncertainty of 
$1\%$ may be reached in this determination on \Vcb. The current total uncertainty is roughly 
twice as large due to the experimental error; at a \superb the experimental error will shrink 
considerably, but from todays perspective there is a  brick wall at the level of 1\%.   

\subsection{Studies in Mixing and \CP Violation in Mixing}
\label{sec:cpvmixing}

\subsubsection{Measurements of the mixing frequency and \CP asymmetries}

The measurement of the mixing frequency $\deltamd$ at \superb is of interest as this physics 
parameter will come to be a significant systematic uncertainty in many of the time-dependent
\CP asymmetry studies and other NP searches.  The current precision on this 
parameter is dominated by early measurements of the $B$-Factories, so there is the potential 
to improve knowledge of this parameter sufficiently so that it no longer plays an important
role in error evaluation of other more important observables.  It is anticipated that 
one will be able to measure \deltamd with a precision of better than $\pm 0.006$ at 
\superb and will be systematically limited.  This level of precision is comparable with
the current PDG average value~\cite{pdg}.

With the discoveries of \CP violation in decay and indirect \CP violation at the $B$-Factories, it is 
natural to continue to search for \CP violation in mixing.  The test for this phenomenon is a part
of generic time-dependent \CP violation measurements where one searches for $|\lambda| = q / p \neq 0$, where
$q/p$.  The cosine coefficient measured at the same time as the $\Delta S$ parameters discussed previously
are related to $|\lambda|$, and the Charmonium decays are a good place to search for CP violation in mixing
where \superb will be able to achieve a precision on $C= (1-|\lambda|^2)/(1+|\lambda|^2)$ of $0.005$ with $J/\psi K^0$.
It is possible to perform a precision measurement of $|q / p|$ using di-lepton events with a precision of a
few per mille as discussed in Section~\ref{sec:cpt}.

\subsubsection{New Physics in mixing\label{sec:bdecays:mixing}}

\begin{figure}[htb!]
  \begin{center}
    \includegraphics[width=0.45\textwidth]{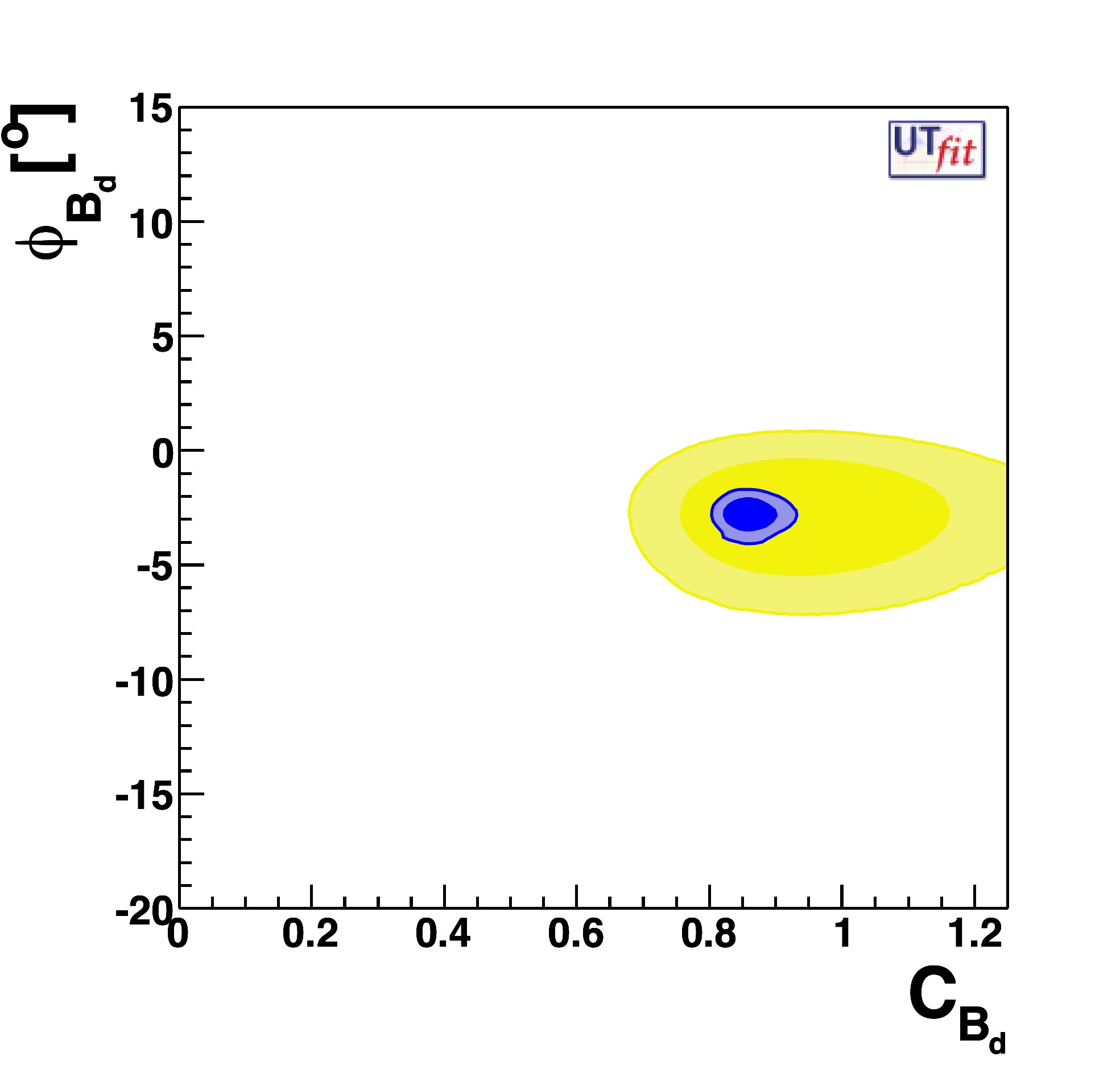}
    \caption{
Current constraints on possible NP amplitude and phase contributions to $B_d$ mixing~\cite{Bona:2005eu}.  The SM solution is $(\phi_{B_d}, C_{B_d}) = (0, 1)$, and any deviation would be the result of NP. The light shaded regions correspond to the current 68, 95 and 99\% CL constraints on $C_{B_d}$ and $\phi_{B_d}$, while the dark shaded regions correspond to the expected result from \superb.
    }    \label{fig:bphysics:npinmixing}
  \end{center}\end{figure}

It is possible to search for signs of NP in mixing in a model independent way.  This is done, starting from 
a tree level determination of the apex of the unitarity triangle $(\rhobar, \etabar)$, and searching for any perturbation
from the SM solution using a generic parameterization of NP with an amplitude and phase, in addition to the 
SM contribution.  The ratio of NP and SM amplitudes can be parameterized simply in terms of 
an amplitude ratio $C_d$ and phase difference $\phi_{B_d}$:
\begin{eqnarray}
C_d e^{2i\phi_{B_d}} &=& \frac{\bra{B_d}{\cal H}_{eff}^{NP+SM}\ket{\overline{B}_d}}{\bra{B_d}{\cal H}_{eff}^{SM}\ket{\overline{B}_d}} \nonumber \\
  &=&  \frac{ A_d^{SM} e^{2i \phi_d^{SM}} + A_d^{NP} e^{2i (\phi_d^{SM} + \phi_d^{NP})}} { A_d^{SM} e^{2i \phi_d^{SM}} } \nonumber
\end{eqnarray}
where the SM phase for $B_d$ mixing $\phi_d^{SM} = \beta$.  The corresponding constraints on the NP phase
and amplitude ratios are given in Fig.~\ref{fig:bphysics:npinmixing}.  Current data is consistent with small values of the 
amplitude for NP, and a large NP phase.  Using data from \superb we would be able to make a precision
search for NP in $B_d$ mixing.

\subsubsection{Tests of \CPT}
\label{sec:cpt}

The combined symmetry of \C, \P, and \T otherwise written as \CPT is conserved in locally gauge invariant quantum field
theory.  The role of \CPT in our understanding of physics is described in more detail in
Refs.~\cite{luders_1954,jost_1957,pauli_1957,dyson_1958} and an observation of \CPT violation would be a sign of new
physics.  \CPT violation could be manifest in neutral meson mixing, so a \sff is 
well suited to test this symmetry. The text-book description of neutral meson mixing in terms of the complex 
parameters $p$ and $q$ can be extended to allow for possible \CPT violation.  On doing so the heavy and light 
mass eigenstates of the \Bz meson \BH and \BL become 
\begin{equation}
\ket{\BLH} = p \sqrt{1 \mp z} \ket{\Bz} \pm q  \sqrt{1 \pm z} \ket{\Bzb},\nonumber
\end{equation}
where \Bz and \Bzb are the strong eigenstates of the neutral \B meson. The \CPT conserving solution is recovered 
when $z=0$ and if \CP and \CPT are conserved in mixing then $|q|^2+|p|^2=1$.

There are two types of \CPT test that have been performed at the current \B-factories.  The more powerful
of these methods is the analysis of di-lepton events where both \B mesons in an event decay into an
$X^\mp\ell^\pm \nu$ final state.  Di-lepton events can be categorized by lepton charge into three types: $++$,
$+-$ and $--$ where the numbers of such events $N^{++}$, $N^{+-}$ and $N^{--}$ are related to \Deltagamma and $z$ as a
function of \deltat as described in Ref.~\cite{Aubert:2007bp}. Using these distributions one can construct
two asymmetries: the first is a \T/\CP asymmetry:
\begin{eqnarray}
{\cal A}_{\T/\CP} &=&\frac{P(\Bzb \to \Bz) - P(\Bz \to \Bzb)} {P(\Bzb \to \Bz) + P(\Bz \to \Bzb)}\nonumber\\
                  &=&\frac{N^{++} - N^{--}}{N^{++} - N^{--}} \nonumber\\
                  &=& \frac{1 - \modqovp^4}{1 + \modqovp^4}, \nonumber
\end{eqnarray}
and the second is a \CPT asymmetry:
\begin{eqnarray}
{\cal A}_{\CPT}(\deltat)  &=& \frac{N^{+-}(\deltat > 0) - N^{+-}(\deltat < 0)}{N^{+-}(\deltat > 0) + N^{+-}(\deltat <
0) } \nonumber \\
                  &\simeq& 2 \frac{\mathrm{Im} z \sin (\deltamd \deltat) - \mathrm{Re} z \sinh \left( \frac{\Deltagamma \deltat }{2}\right)}
                             {\cosh \left( \frac{\Deltagamma \deltat }{2}\right) + \cos(\deltamd \deltat)}, \nonumber
\end{eqnarray}
where ${\cal A}_{\CPT}(\deltat)$ is sensitive to $\Deltagamma \times \mathrm{Re} z$.
In the SM ${\cal A}_{\T/\CP}\sim 10^{-3}$ and ${\cal A}_{\CPT} = 0$~\cite{beneke_cpt_2003,Ciuchini:2003ww}.
\babar\ measures~\cite{Aubert:2006nf}:
\begin{eqnarray}
\modqovp - 1  &=& (-0.8 \pm 2.7 \stat \pm 1.9 \syst) \times 10^{-3}, \nonumber \\
\mathrm{Im} z &=& (-13.9 \pm 7.3 \stat \pm 3.2 \syst) \times 10^{-3}, \nonumber \\
\Deltagamma \times \mathrm{Re} z &=& (-7.1 \pm 3.9 \stat \pm 2.0 \syst ) \times 10^{-3}, \nonumber
\end{eqnarray}
which is compatible with no \CP violation in $\Bz-\Bzb$ mixing and \CPT conservation. It is possible to study
variations as a function of sidereal time, where 1 sidereal day is approximately $0.99727$ solar
days~\cite{Kostelecky:1997mh} where $z$ depends on the four momentum of the \B candidate.  \babar\ re-analyzed their data and find that it is consistent with $z=0$ at 2.8 standard deviations~\cite{Aubert:2007bp}. 
With data from the first few years of operation at \superb it would be possible
to perform a more precise test of \CPT than performed by the current experiments and on doing so
continue the search for \CPT violation.  These measurements would become limited by systematic 
uncertainties after the first few years of running at \superb.  The precision on \CPT violating 
observables that could be reached with \superb is:
\begin{eqnarray}
\sigma(\mathrm{Im}) z &=& 0.6 \times 10^{-3}, \nonumber \\
\sigma(\Deltagamma \times \mathrm{Re} z) &=& 0.3 \times 10^{-3}, \nonumber
\end{eqnarray}
However with such a measurement
it would be possible to test if the $2.8\sigma$ hint for \CPT violation were a real effect or the result 
of a statistical fluctuation.

\subsection{Why measure $\gamma$ precisely (and how)?}
\label{sec:bphysics:gamma}

The measurement of the unitarity triangle angle $\gamma$ is a standard candle measurement that can be 
used to perform precision tests of the SM. In this section we discuss why it is important to 
measure $\gamma$ precisely, whether such a measurement is ``safe'' from NP, and also how we can perform 
such a measurement.

The main idea behind the measurement of $\gamma$ dates back to almost two decades ago, and 
in retrospect this is a straight forward concept. One uses the interference between $b\to c \bar{u} s$ 
and $b\to u \bar{c} s$~\cite{Gronau:1991dp,Gronau:1990ra} transitions. The sensitivity to the 
weak phase $\gamma$ comes from the interference between $B^-\to D K^-$ decay followed by $D\to f$ and the 
$B^-\to \bar{D} K^-$ decay followed by $\bar{D}\to f$, where $f$ can be any common final state of $D$ and $\bar{D}$. 

Depending on the choice of the final state $f$ in $D$ decay there have been many variations of this 
central idea proposed in the literature.  With $f$ for instance taken to be a  CP odd ($-1$) 
eigenstate (\eg $K_S \pi^0$)~\cite{Gronau:1991dp,Gronau:1990ra}, a flavor state (\eg 
$K^+\pi^-$)~\cite{Atwood:1996ci,Atwood:2000ck}, singly Cabibbo suppressed (\eg 
$K^{*+} K^-$)~\cite{Grossman:2002aq}, and many-body final state (\eg 
$K_S\pi^+\pi^-$)~\cite{Giri:2003ty,Poluektov:2004mf}. Other extensions include many-body 
$B$ final states $B^+\to D K^+\pi^0, K_S \pi^+$, $B^0\to D\pi^-K^+$~\cite{Aleksan:2002mh,Gershon:2008pe, Gershon:2009qr,Gershon:2009qc}, 
where $D^{0*}$ is used in addition to $D^0$~\cite{Bondar:2004bi}, the use of self tagging $D^{0**}, D_2^{*-}$~\cite{Sinha:2004ct,Gershon:2008pe} as well as neutral $B$ and $B_s$ decays (time dependent~\cite{Aleksan:1991nh,Dunietz:1995cp,Fleischer:1996ai,Fleischer:2003aj,Fleischer:2003ai}, time-integrated~\cite{Gronau:2004gt,Gronau:2007bh}, and self-tagging~\cite{Dunietz:1991yd}).

For $N_B$ different $B$ decay channels and $N_D$ different $D$ decay channels one has  $3N_B+N_D$ unknowns that are fit from data (in addition to $\gamma$, which is common to all the decays). On the other hand, one has $\sim 4 N_B N_D$ observables -- different branching ratios and CP asymmetries. It then immediately follows that the best strategy is to 
combine as many channels as possible in order to perform a precision measurement of $\gamma$. 

An interesting question is whether the weak phase $\gamma$ measured from $B\to DK$ decays is the SM weak phase. In other words 
-- to what extent is $B\to DK$ a SM reference point for the value of $\gamma$? One would expect that since  $B\to DK$ decays 
are mediated exclusively by tree diagrams, the contributions from NP are negligible. Schematically we have
\beq
 A(B^-\to DK^-)\sim \underbrace{V_{cb}}_{\lambda^2} \underbrace{V_{us}}_{\lambda}T,\quad \underbrace{V_{ub}}_{\lambda^3} \underbrace{V_{cs}}_{1}C,
\eeq
where we have also denoted explicitly the CKM scaling in the Wolfenstein parameterization. 
How could this amplitude be modified by NP? If it is modified due to non-SM
charged currents then the effect of NP would be likely to show up elsewhere, 
\eg in semi-leptonic $B$ decays (if NP is leptophobic, this may be harder to see, though). 
If the modification is due to non-SM neutral currents then two insertions of NP flavor 
violation (FV) are needed, \eg to generate a correction to $(\bar c u)(\bar s b)$ operator. 
In this case the NP effects are doubly suppressed. Compare this scenario with for instance 
the decay $B\to \pi\pi$, where one can have a single NP FV insertion, \eg to generate a correction to
$(\bar u u)(\bar d b)$ operator. Thus, if we have TeV NP,  then  FV is small and the effect is
suppressed in $B\to DK$. If, however we have general FV at high scale then the deviation is as likely
to show in either of the two,  $B\to DK$ and $B\to \pi\pi$ (but in this case the effect is also going
to be very small given constraints from the $\bar K-K$ mixing). 

Another interesting question is why measure the SM weak phase  $\gamma$ from $B\to DK$ and not from 
$B\to \rho\rho$, $\rho\pi$, $\pi\pi$ decays, which are also tree dominated. Schematically we have
\beq
A(B\to \rho\rho, \rho\pi, \pi\pi)\sim \underbrace{V_{ub}}_{\lambda^3} \underbrace{V_{ud}}_{1} T, \quad
\underbrace{V_{cb}}_{\lambda^2} \underbrace{V_{cd}}_{\lambda} P.
\eeq
The weak phase $\gamma$ ($\alpha=\pi-\beta-\gamma$) determined from these decays is the SM one unless
one has isospin breaking NP. This means that the $\gamma$ measured would be the SM one even if there is NP in
$B-\bar B$ mixing or in QCD penguins. NP in electro-weak penguins would, however, shift the measured value of
$\gamma$ from the SM one.  Still, the NP has to compete with the tree transitions!

The extraction of $\gamma$ from $B\to DK$ has one major advantage over extraction from $B\to \rho\rho$, $\rho\pi$, $\pi\pi$. 
In the latter case there are irreducible theoretical errors in extraction of $\gamma$ (via $\alpha$):  due to 
(i) isospin breaking, which is hard to estimate to better than factors of few (i.e. that it is at a few 
percent level~\cite{Gronau:2005pq}) and (ii) due to the uncertainties in resonance shapes that are used 
in the extraction from $B\to \rho\pi$~\cite{Gronau:2004tm}. There are no such errors in extraction of $\gamma$ 
from $B\to DK$.  Isospin is not used at any point, while reference to resonance parameterization can be 
avoided even in the case of multi-body $D$ decays~\cite{Giri:2003ty}. The remaining theoretical errors 
are much smaller and negligible at the precision level achievable at a \sff as we will 
see in more details below. 

Charm factories (CLEO-c or BES-III) can have a big impact by measuring strong phases in $D$ decays. Then 
only parameters of the $B$ system need to be measured in $B$ decays. Let us take as 
an example: $B^\pm\to [K_S\pi^+\pi^-]_D K^\pm$ where the $D$ decay amplitude varies over the Dalitz plot
\begin{eqnarray}
A_D(s_{12},s_{13}) &\equiv& {A_{12,13}}\,e^{i{\delta_{12,13}}},
\\
&\equiv & A(D^0 \to K_S(p_1) \pi^-(p_2) \pi^+(p_3)), \nonumber
\end{eqnarray}
where $s_{12(13)}=(p_1-p_{2(3)})^2$. The variation of the strong phase $\delta_{12,13}$ over the 
Dalitz plot can be measured by fitting the data with a resonance model, which leads to a hard 
to reduce $\sim 10^\circ$ systematic error on $\gamma$. A model independent treatment is possible 
using input from the CLEO-c and BES-III experiments as discussed in Ref.~\cite{Giri:2003ty}. 

In the model independent method one partitions the Dalitz plot in bins~\cite{Giri:2003ty} and 
introduces the sine and cosine of the strong phase differences averaged over the bins as new 
variables
\begin{eqnarray}
{c_i} &\equiv &\int_i dp\; 
A_{12,13}\,A_{13,12}\cos(\delta_{12,13}-\delta_{13,12}), \nonumber
\\
{s_i} &\equiv &\int_i dp\; A_{12,13}\,A_{13,12}\sin(\delta_{12,13}-\delta_{13,12}).
\label{si}
\end{eqnarray}
Since $D$ decays do not violate $CP$ one can relate $c_i$ and $s_i$ in $i^{th}$ bin with the CP 
conjugated one, $\bar i$, hence $c_{\,\bar{i}}=c_i$, and $s_{\,\bar{i}}=-s_i$. A MC based 
feasibility study showed that the optimal strategy using this model independent approach is 
statistically only $30\%$ worse than the model dependent method~\cite{Bondar:2005ki,Bondar:2008hh}.
The prior measurement of $c_i$, and $s_i$ at charm factory (or at \superb via running at charm threshold) 
is a crucial ingredient to measure $\gamma$ in this way. The measurement of $c_i$, and $s_i$ is possible 
because in the $\Psi(3770)\to D\bar D$ mesons in the final state are entangled. Let us assume that 
the $D$ meson flying to the right decays to bin $i$ of $K_S \pi^+\pi^-$ final state. Then by choosing 
different decay channels of the $D$ meson flying to the left, one has enough information to measure 
both the phase and the amplitude of the $D\to K_S \pi^+\pi^-$ decay~\cite{Giri:2003ty}. 

This was done by CLEO-c~\cite{Briere:2009aa}. Using optimal binning with the least variation in strong
phase difference across a bin $\Delta \delta_f$~\cite{Bondar:2005ki,Bondar:2008hh} for 
$D\to K_{S/L}\pi^+\pi^-$ decays and using entangled decays of $\Psi(3770)\to D\bar D$ with $D$'s 
decaying into flavor tagged, CP-tagged and $D\to$  $K_{S/L}\pi^+\pi^-$ decay modes the collaboration 
estimates that the decay model uncertainty on $\gamma$ is reduced to $1.7^\circ$ (from toy MC based studies with 
$\gamma=60^\circ$, $\delta_B=130^\circ$ and $r_B=0.1$). 
$D$ Decays involving more than three final-state particles can also
be used to measure $\gamma$. In this case, it is simpler and in some
modes sufficient to measure the total $B\to DK$ decay rates rather
than to analyze the full phase-space distribution~\cite{Atwood:2003mj,Lowery:2009id}. This is
also the case for some three-body modes~\cite{Soffer:2008md}.

There are several ways that would improve constraints on $\gamma$: include as many $D$ decay modes as possible and
include more $B$ decay modes. To include $B^\pm\to D^*K^{*\pm}$ would be very hard, since one would need to 
perform an angular analysis of the decay. It is much easier to include neutral modes $B^0\to D K_S$, $B_s^0\to D\phi,D\eta^{(')}$.
At first glance neutral $B$ decays are less attractive since in this case both $b\to u s \bar c$ and $b\to c s \bar u$ are color
suppressed, while in charged $B$ decays $b\to c s \bar u$ is color allowed. The neutral decays therefore have smaller decay rates
(${A_n}\sim \frac{1}{3}{A_c} $). However, the statistical error on $\gamma$ scales  roughly as the smallest of the two interfering
amplitudes, which  both in $B^+$ and $B^0$ decays are color suppressed. Furthermore, using isospin (and neglecting annihilation)
one gains an extra constraint~\cite{Gronau:2004gt}
\beq
A(B^+\to D^0K^+)\simeq \sqrt{2}A(B^0\to D^0K_S),
\eeq
which reduces the number of independent unknowns to be determined.  §Time dependent measurements of $B^0$ decays contain full
information~\cite{Gronau:1990ra,Kayser:1999bu}, but already time integrated  $B^0$ rates (untagged rates) alone suffice to determine
$\gamma$~\cite{Gronau:2004gt,Gronau:2007bh}. The analysis can also be simplified by the use of self-tagging modes:
$B^0\to DK^{*0}\to D \pi^-K^+$~\cite{Dunietz:1991yd}.

In the above discussion we have neglected $D-\bar D$ mixing, direct CP violation in $D\to f$ decays and 
$\Delta \Gamma$ in $B_d$ and $B_s$ time integrated decays. We now move to the estimate of the related theory errors.

The $D-\bar D$ mixing parameters $x_D ={\Delta m_D \over \Gamma_D} , y_D = {\Delta \Gamma_D \over 2\Gamma_D} $ are 
both measured to be $O(10^{-2})$.  Furthermore in the SM, $D-\bar D$ mixing is CP conserving to a very good 
approximation, with the mixing phase $\theta \sim O(10^{-4})$. For CP conserving $D-\bar D$ mixing the only 
important change in equations used to extract $\gamma$ from $B\to DK$ are in the interference terms, where one 
changes the relative strong phase $\delta_f$ with a time averaged one, $\langle \delta _f\rangle$. $D-\bar D$ 
mixing also dilutes the interference so that it gets multiplied by $e^{-\epsilon_f}$. The effect on 
$\gamma$ is small and is of second order in small parameters $\epsilon_f\sim O(x_D^2, y_D^2)$~\cite{Grossman:2005rp}. 
These terms can still be enhanced, if the suppressed term is multiplied by a large amplitude, which happens for doubly 
Cabibbo suppressed $D$ decays. Even in this case, however, the shift is small,  $\Delta \gamma\lesssim 1^\circ$, 
while otherwise it is much smaller~\cite{Grossman:2005rp}.

To recapitulate, the effect of CP conserving $D-\bar D$ mixing on the measurement of $\gamma$ is small. It can 
also be included for precisely measured $x_D$ and $y_D$. For the model independent analysis the 
news are even better -- the way the model-independent analysis is set up, the inclusion of $D-\bar D$ 
mixing actually does not require any change in the analysis. Since both the average of the sine 
and of the cosine of the strong phase are measured independently from experiment, this means that 
the dilution due to time averaging is already determined experimentally~\cite{Grossman:2005rp}. The 
last  approximation we used was neglecting $\Delta \Gamma_{d,s}$.  The inclusion of $\Delta \Gamma_s$ 
dependence is important only in untagged $B_s\to D\phi$  so that $\Delta \Gamma_s$ needs to be well 
measured and kept in the analysis~\cite{Gronau:2007bh}. Once all these reducible theoretical errors 
are taken into account the theory error would come from
higher electroweak corrections and CP violation in $D$ decays. The resulting theoretical error is conservatively
$\Delta \gamma<10^{-5}$, so the measurement will be statistics dominated for a long time.

We conclude  that measurement of $\gamma$ from $B\to DK$ decays is the theoretically cleanest measurement 
of the SM weak phase and thus represents a standard candle with which to test the SM. This can for 
instance be contrasted with  the measurements of $\alpha$ that may start to become theory limited at a \sff
due to poorly known isospin breaking effects (for example see Ref.~\cite{Bona:2007qt}). 
Under quite general assumptions the measurement 
of $\gamma$ is also safe from NP contamination. An important input is provided by charm factories that 
can measure the strong phase differences in the $D$ decays from entangled $\Psi(3770)$ decays. The 
estimated systematic uncertainty on $\gamma$ is below $2^\circ$ and can be reduced with increased 
statistics. This is true also of other systematic uncertainties in the present analyses, including the 
neglect of $D-\bar D$ mixing and $\Delta \Gamma$. They stem from simplifications made for 
convenience when analyzing currently available data. While in the future more complicated analyses will 
be required that incorporate these effects directly. Thus the irreducible theoretical error 
on $\gamma$ is well below \sff sensitivities. 

\subsection{Charmless hadronic $B$ decays}

Charmless hadronic $B$ decays can be used to test the SM and CKM theory in
detail. In principle, such decays can be used to measure all of the angles of the unitarity
triangle, however SM uncertainties are a concern for all such measurements
as existing calculations are either computing the hadronic amplitudes using
factorization, or invoking flavour SU(2) or SU(3) symmetries in order to achieve
their goals.
Without further improvements in the theoretical tools available, in many
cases the potential of using charmless hadronic $B$ data, both branching fractions
and direct \CP asymmetries, to test the CKM picture will be limited by theoretical uncertainties,
coming either by unknown power-suppressed terms or by flavour symmetry breaking.
On the other hand, a \sff provides a full set of high precision
data to test and possibly improve the theoretical tools developed so far to
describe these hadronic $B$ decays.

\subsection{Precision CKM\label{sec:bdecays:phenomenology:ckm}}
By the time \superb starts to take data it is expected that the knowledge
of the CKM matrix parameters (sides and angles) will be dominated by a
combination of measurements from the B-factories and LHCb.  These will
include measurements of $\beta$ and $\gamma$ with a precision of the order
of $1^\circ$, and a measurement of $\alpha$ with a precision of $5-6^\circ$.
LHCb will not be able to improve upon the existing measurements of 
\Vub and \Vcb, which have uncertainties of 8\% and 2\%, respectively.
\superb will be able to perform precision measurements of the angles
of the unitarity triangle as well as \Vub and \Vcb.  The anticipated
precision attainable for these observables is given in 
Table~\ref{tbl:bphysics:ckm}. Together with hadronic parameters computed
mainly using lattice QCD (see Section~\ref{sec:lattice}), this 
set of information will play a vital role in defining a 
model-independent determination of quark mixing in the Standard 
Model, thus providing a precision test of the CKM anzatz.  Precision 
knowledge of the CKM matrix itself facilitates several NP
search opportunities available to \superb and other experiments. 

\begin{table}[!ht]
\caption{The expected precision on CKM observables from \superb.  The third column indicates if the measurement
is theoretically clean, or dominated by theory uncertainties.}\label{tbl:bphysics:ckm}
\begin{center}
\small
\begin{tabular}{l|cc}\hline
CKM observable & Precision (75\invab) & Theory uncertainty \\ \hline
$\beta$ (\ccbars)   & $0.1^\circ$ & clean \\
$\alpha$            & $1-2^\circ$ & dominant \\
$\gamma$            & $1-2^\circ$ & clean \\
\Vcb (inclusive)    & 1.0\% & dominant \\
\Vcb (exclusive)    & 1.0\% & dominant \\
\Vub (inclusive)    & 2.0\% & dominant \\
\Vub (exclusive)    & 3.0\% & dominant \\ \hline
\end{tabular}
\end{center}
\end{table}

\graphicspath{{Bphysics/}{Bphysics/}}
\section{\B Physics at the $\Upsilon(5{\rm S})$}\label{sec:bsphysics}

\bigskip
\smallskip

Measurement of CKM- and New Physics-related
quantities in the $B_s$ sector is a natural extension of the traditional 
$B$~Factory program.
In some cases, studies of $B_s$ mesons allow the extraction of
the same fundamental quantities accessible at a $B$~Factory
operating at the $\Upsilon(4{\rm S})$ resonance,
but with reduced theoretical uncertainty.
Experiments running at hadronic machines are expected
to be the main source of $B_s$-related measurements.
In particular, in the near future,
the increased dataset of the Tevatron experiments and the start
of the \lhcb, ATLAS, and CMS programs will surely yield important new results.

It is worth noting, however, that despite the rapid $\BsBsb$ oscillation frequency, it is 
feasible to carry out $\Bs$ studies in the very clean environment of
$e^+e^-$ annihilation machines by running at the $\FiveS$ resonance,
where it is possible to perform measurements involving neutral particles
({\it e.g.}, $\pi^0$, $\eta$ and $\eta^\prime$ mesons,
radiative photons, {\it etc.}).
CLEO~\cite{Artuso:2005xw,Bonvicini:2005ci,Huang:2006em}
and \belle~\cite{Drutskoy:2006fg,Abe:2006xc}
have had short runs at the $\FiveS$,
measuring the main features of this resonance.
The results clearly indicate the potential for an $e^+e^-$ machine
to contribute to this area of $B$ physics,
and have inspired the work in this section,
and elsewhere~\cite{Hou:2006mx,Hou:2007ps,Baracchini:2007ei}.
Note that, in contrast to much of the remainder of this chapter, there are
no experimental analyses for many of the measurements of interest,
and therefore our studies are based on Monte Carlo simulations.

A detailed study of the physics capability of \superb at the $\FiveS$
can be found in the Conceptual Design Report~\cite{Bona:2007qt}.  The main 
conclusions of that study are summarized here.

The production of $B_s$ mesons at the $\FiveS$ allows comprehensive
studies of the decay rates of the $B_s$ with a completeness
and accuracy comparable to that currently available for $B_d$ and $B_u$ mesons,
thereby improving our understanding of $B$ physics and helping to reduce
the theoretical uncertainties related to New Physics-sensitive $B_d$ quantities.
Moreover, $B_s$ physics provides additional methods and observables to probe New Physics 
effects in $b \to s$ transitions. In the following, we concentrate on this second point,
providing examples of some of the highlight measurements that could be performed by 
\superb\ operating at the $\FiveS$ resonance.

The $\FiveS$ resonance is a $J^{PC} = 1^{--}$ state
of a $b\bar{b}$ quark pair with an invariant mass of $m_{\FiveS} = (10.865 \pm 0.008) \ {\rm GeV}/c^2$~\cite{Besson:1984bd,Lovelock:1985nb,Yao:2006px}.
The cross section $\sigma(e^+e^- \to  \FiveS)$ is 
$0.301 \pm 0.002 \pm 0.039 \ {\rm nb}$~\cite{Huang:2006mf}, which is about three
times smaller than $\sigma(e^+e^- \to  \FourS)$. 
Unlike the $\FourS$ state, this resonance is sufficiently massive to
decay into several $B$ meson states:
vector-vector ($B^* \bar{B}^*$),
pseudoscalar-vector ($B \bar{B}^*$),
and pseudoscalar-pseudoscalar ($BB$) combinations of charged $B$ mesons,
as well as neutral $B_d$ and $B_s$ mesons, as well as into
$B^{(*)}\bar B^{(*)} \pi$ states.  The $B$ pair production rates
at the $\FiveS$ resonance are summarized in Ref.~\cite{Bona:2007qt}.
As with reconstructing $B$ decays at the $\FourS$, one can use the
precisely determined initial state kinematics to compute the usual 
discriminating variables $m_{\rm ES}$ and $\Delta E$.  There is a 
small complication that the different $B$ pairs produced occupy
slightly different regions in the $m_{\rm ES} - \Delta E$ plane
and this can be used to study fine details of the decay properties
of these $B$ mesons.
With the small beam energy spread of \superb,
the resolution of $m_{\rm ES}$ will be comparable to that of the current $B$~Factories,
resulting in almost negligible crossover between
$\BsBsb$ and $B\Bbar \pi$ states.

\mysubsubsection{Measurement of $B_s$ Mixing Parameters}
\label{sec:U5s_timeind}
In analogy with the $B_d$ system, the absolute value and the 
phase of the $\BsBsb$ mixing amplitude can be used to test for 
the presence of New Physics in $\Delta B=2$ $b \to s$ transitions.
These measurements can be made at hadronic colliders~\cite{Bona:2006sa}.
The recent measurement of
$\Delta m_s$~\cite{Abulencia:2006ze,Abulencia:2006mq,Abazov:2006dm} 
provides the first milestone in this physics program.
Similar tests for New Physics effects can be made by measuring
quantities such as $\Delta \Gamma_s$ and the $\CP$ asymmetry in
semi-leptonic decays $A^s_{\rm SL}$.  These observables can be 
measured using the large statistics, and high reconstruction 
efficiency available in the clean environment of \superb.
It is not necessary to resolve $B_s$ oscillations to make these measurements.

In a generic New Physics scenario, the effect of $\Delta B = 2$ New Physics contributions
can be parameterized in terms of an amplitude and phase, $C_{B_s}$ and $\phi_{B_s}$,
(in analogy with Section~\ref{sec:bdecays:mixing}).
In the absence of New Physics effects, $C_{B_s}=1$ and $\phi_{B_s}=0$.
The measured values of $\Delta m_s$ and $\sin 2\beta_s$
are related to Standard Model quantities through the relations :
\begin{eqnarray}
  \Delta m_s^{\rm exp} &=& C_{B_s} \cdot \Delta m_s^{\rm SM},\\
  \sin 2\beta_s^{\rm exp} &=& \sin (2\beta_s^{\rm SM} + 2\phi_{B_s}) \,.
\end{eqnarray}
The semi-leptonic $\CP$ asymmetry~\cite{Laplace:2002ik}
and the value of $\Delta \Gamma_s/\Gamma_s$~\cite{Dunietz:2000cr}
are sensitive to New Physics contributions to the $\Delta B=2$ effective Hamiltonian,
and can be expressed in terms of the parameters $C_{B_s}$ and $\phi_{B_s}$.

Different experimental methods have been proposed to extract the
lifetime difference $\Delta \Gamma_s$~\cite{Dighe:1998vk}.
For instance, $\Delta \Gamma_s$ can be obtained from the angular distribution
of untagged $B_s \to J/\psi \phi$ decays.
This angular analysis allows separation of the $\CP$ odd and $\CP$ even
components of the final state, which have a distinct time
 evolution,
given by different combinations of the
two exponential factors $e^{-\Gamma_{L,H}t}$.
This allows the extraction of the two parameters $\Gamma_{L,H}$
or, equivalently, $\Gamma_s$ and $\Delta \Gamma_s$.
The weak phase of the mixing amplitude, $\beta_s$, also
appears in this parameterization, and a constraint on this phase
can be extracted along with the other two parameters
(see Eq.~\ref{eq:BsPDFuntagged} below).
Measurements of $\Delta \Gamma_s$ and $\beta_s$ have been performed
by CDF~\cite{Aaltonen:2007he} and D\O~\cite{Abazov:2008fj}.
With a few \invab of data at the $\FiveS$ \superb will be able to 
improve upon the current experimental precision, and provide a 
useful second measurement to cross check any results from \lhcb in this area.

We have also studied the performance of two different experimental techniques
that can be used to to extract the semi-leptonic asymmetry $A^s_{\rm SL}$,
defined as:
\begin{eqnarray}
  A_{\rm SL}^s = \frac{1-|q/p|^4}{1+|q/p|^4} \, .
\end{eqnarray}

The first technique consists of exclusively reconstructing one of the two $B$
mesons into a self-tagging hadronic final state
(such as $B_s \to D_s^{(*)} \pi$)
and looking for the signature of a semi-leptonic decay (high momentum lepton) in
the rest of the event.
The second approach is more inclusive,
using all events with two high momentum leptons.
In this case, contributions from $B_s$ and $B_d$ decays
cannot be separated, and a combined asymmetry, $A_{\rm CH}$ is measured.
Results from this type of analysis are available from D\O~\cite{Abazov:2007nw}.
We expect to be able to reach precisions of 0.006 and 0.004 on
$A^s_{\rm SL}$ and $A_{\rm CH}$, respectively, with 1\invab of data.
These measurements quickly become systematically limited at \superb, however
the achievable precision would be a clear improvement over the current experimental
situation. The cleaner experimental environment at \superb suggests that this
experiment is better suited at making precision  measurements
of the semi-leptonic asymmetries than experiments at a hadron collider.
For example, the measurement of $A_\mathrm{SL}^s$ (and, to a lesser extent, also to $A^d_\mathrm{SL}$), can be used 
to test the Littlest Higgs Model with T-parity as discussed in Ref.~\cite{Bona:2007qt}.

\mysubsubsection{Time Dependent $\CP$ Asymmetries at the \FiveS}
\label{sec:U5s_timedep}
Let us consider a $B_s$ pair produced at the $\FiveS$ resonance,
through a $B_s^* \overline {B}_s^*$ state.
If one of the two $B_s$ mesons decays into a $\CP$ eigenstate $f$
and the other to a flavour-tagging final state,
the untagged time-dependent decay rate $R(\Delta t)$ as a function of the proper 
time difference $\Delta t$ can be written in terms of the parameter
$\lambda_f = \frac{q}{p}\frac{\bar{A}_f}{A_f}$ as~\cite{Dunietz:2000cr}:

\begin{eqnarray}
R(\Delta t)  =
  {\cal N} 
  \frac{e^{-| \Delta t | / \tau(\Bs)}}{2\tau(\Bs)}
  \Big[
  \cosh(\frac{\Delta \Gamma_s \Delta t}{2}) - \nonumber \\
  \frac{2\, \Re(\lambda_f)}{1 + |\lambda_f|^2} \sinh(\frac{\Delta \Gamma_s \Delta t}{2})
  \Big] \, ,
  \label{eq:BsPDFuntagged}
\end{eqnarray}
where the normalization factor ${\cal N}$
is fixed to $1 - (\frac{\Delta \Gamma_s}{2\Gamma_s})^2$.
Here we have neglected $\CP$ violation in mixing.

It is not possible to perform a a similar
time-dependent analysis to that for the case of $B_d \to J/\psi\Kz$ decays,
at \superb as the detector would be unable to resolve the very fast $B_s$ oscillations.
However, since $\Delta \Gamma_s \neq 0$,
the untagged time-dependent decay rate also allows $\lambda_f$ to be probed,
through the $\Re(\lambda_f)$-dependence of the coefficient of the
$\Delta t$-odd $\sinh(\frac{\Delta \Gamma_s \Delta t}{2})$ term.
Such an analysis has been performed by D\O~\cite{Abazov:2007tx,Abazov:2007zj}.
A ``two-bin'' time-dependent analysis using this approach is possible at
\superb.

If one considers the decay $B_s \to J/\psi \phi$ decay, and for simplicity 
assumes that this is a pure $\CP$-even eigenstate (more generally a full 
angular analysis can be used to isolate $\CP$-even and $\CP$-odd contributions),
it is possible to measure the weak phase of \Bs\ mixing $2\beta_s$.
A precision of $\sim 10^\circ$ and $\sim 3^\circ$ can be achieved on $\beta_s$,
with $1 \ {\rm ab}^{-1}$ and $30 \ {\rm ab}^{-1}$ of integrated luminosity,
respectively. There is a two-fold ambiguity resulting from the sign of $\beta_s$
that can produce almost twice the resolution in the measurement, when 
$\beta_s$ has a value close to zero as in the SM.  Such a measurement 
as this is not limited by systematics and the precision can be 
improved by collecting more data.

While \lhcb is expected to achieve a better precision on the measurement
of $\beta_s$ using a tagged analysis of $B_s \to J/\psi \phi$, the
strength of \superb lies in the ability to make measurements that are not
possible in a hadronic environment, in analogy with the $\Delta S$ measurements
discussed for $B_d$ decays (Section~\ref{sec:bdecays:cpv})
there is an effective $\beta_s$ (denoted $\beta_{s, \rm eff}$) that will form a 
secondary basis for new physics 
searches. As with the $B_d$ case it will be necessary to compare 
the SM expectations of $\beta_s$ with the measurements from tree decays and
with $\beta_{s, \rm eff}$ from penguin-dominated rare decays. Among the 
interesting final states \superb\ can study are
$B_s \to J/\psi \eta$, $B_s \to J/\psi \eta^\prime$,
$B_s \to D^{(*)+}_s D^{(*)-}_s$, $B_s \to D^{(*)} \KS$,
$B_s \to D^{(*)} \phi$, 
and $B_s \to \phi \eta^\prime$.
Studies on the measurement of the effective $\beta_s$ 
using the pure $b \to s$ penguin transition $B_s \to \Kz \bar {K}^0$,
indicate that \superb will be able to measure this phase with a precision
of $11^\circ$ given $30 \ {\rm ab}^{-1}$ of data.

\mysubsubsection{Rare Radiative \Bs\ Decays}
It is possible to search for possible NP effects by comparing
measurements of $\Delta B=1$ $b \to s$ transitions,
measurements of $|V_{td}/V_{ts}|$, and $\Delta m_s$.
\superb will be able to perform a precision measurement 
of  $|V_{td}/V_{ts}|$ using the ratio
$R = \BR(B_d^0 \to \rho^0 \gamma)/\BR(B_d \to K^{*0}\gamma)$
to a precision that is expected to be ultimately limited by the
presence of a power-suppressed correction term.
The ratio
 $R_s = \BR(B_s^0 \to K^{*0} \gamma)/\BR(B_d^0 \to K^{*0} \gamma)$
has the advantage that there is no $W$ exchange diagram contribution to 
hinder interpretation of results.
Assuming that $\BR(B_s^0 \to K^{*0} \gamma) = 1.54 \times 10^{-6}$,
and taking reasonable estimates from lattice QCD
for the form factor ratio $\xi$
to extract $|V_{td}/V_{ts}|$ with a precision of a few percent
with a multi-\invab sample of data,
as shown in Table~\ref{tab:inputsbs}.

\mysubsubsection{Measurement of $B_s \to \gamma \gamma$}
In analogy with the $B_d$ decay $b \to s \gamma$, the decay
$B_s \to \gamma \gamma$ is considered a promising golden channel to 
search for new physics at \superb.
The final state contains both $\CP$-odd and $\CP$-even components,
allowing for the study of $\CP$-violating effects with $B$~Factory 
tagging techniques.
The Standard Model expectation for the branching ratio is
$\BR(B_s \to \gamma \gamma) \sim (2-8) \times 10^ {-7}$~\cite{Reina:1997my}.
New Physics effects are expected to give sizable contributions to the decay rate
in certain scenarios~\cite{Aliev:1993ea,Devidze:1998cf}.
For instance, in R-parity-violating SUSY models,
neutralino exchange can enhance the branching ratio up to
$\BR(B_s \to \gamma \gamma)\simeq 5 \times 10^ {-6}$~\cite{Gemintern:2004bw}.
On the other hand, in R-parity-conserving SUSY models,
in particular in softly broken supersymmetry,
$\BR(B_s \to \gamma \gamma)$ is found to be highly correlated
with $\BR(b \to s \gamma)$~\cite{Bertolini:1998hp}.

Experimentally the measurement of $B_s \to \gamma \gamma$ will be
much less demanding at \superb than the well established
measurement of final states such as $B_d^0 \to \pi^0 \pi^0$.
The presence of two high-energy photons in the final state is a 
clear signature for the signal, particularly with a
recoil technique. Both \babar~\cite{Aubert:2001fm} and \belle~\cite{Abe:2005bs}
have published results of searches for $B_d^0 \to \gamma \gamma$,
setting the current experiment upper limit at
$\BR(B_d \to \gamma \gamma) < 6.2 \times 10^{-7}$
which is a proof of principle that one can measure the corresponding \Bs decay
at \superb. We anticipate that it will be possible to observe $14$ signal 
events and $20$ background events in a sample of $1 \ {\rm ab}^{-1}$ assuming
a Standard Model branching fraction. With $30 \ {\rm ab}^{-1}$, one can achieve 
a statistical error of $7\%$ and a systematic error smaller than $5\%$ from
a straight forward analysis.  It would be possible to improve
upon this precision using tagging information, which would also facilitate the 
measurement of a direct CP asymmetry in this mode.

\mysubsubsection{Phenomenological Implications}
The experimental measurements of $\Delta \Gamma$, $A^{s}_{\rm SL}$, $A_{\rm CH}$
and $\CP$ violation parameters described in the previous sections
can be used to determined the $\Delta B = 2$ New Physics contributions
in the $B_s$ sector.
The knowledge of $\overline{\rho}$ and $\overline{\eta}$ is assumed to come
from studies at the $\FourS$.

\begin{figure}[t]
  \begin{center}
    \includegraphics[width=0.35\textwidth]{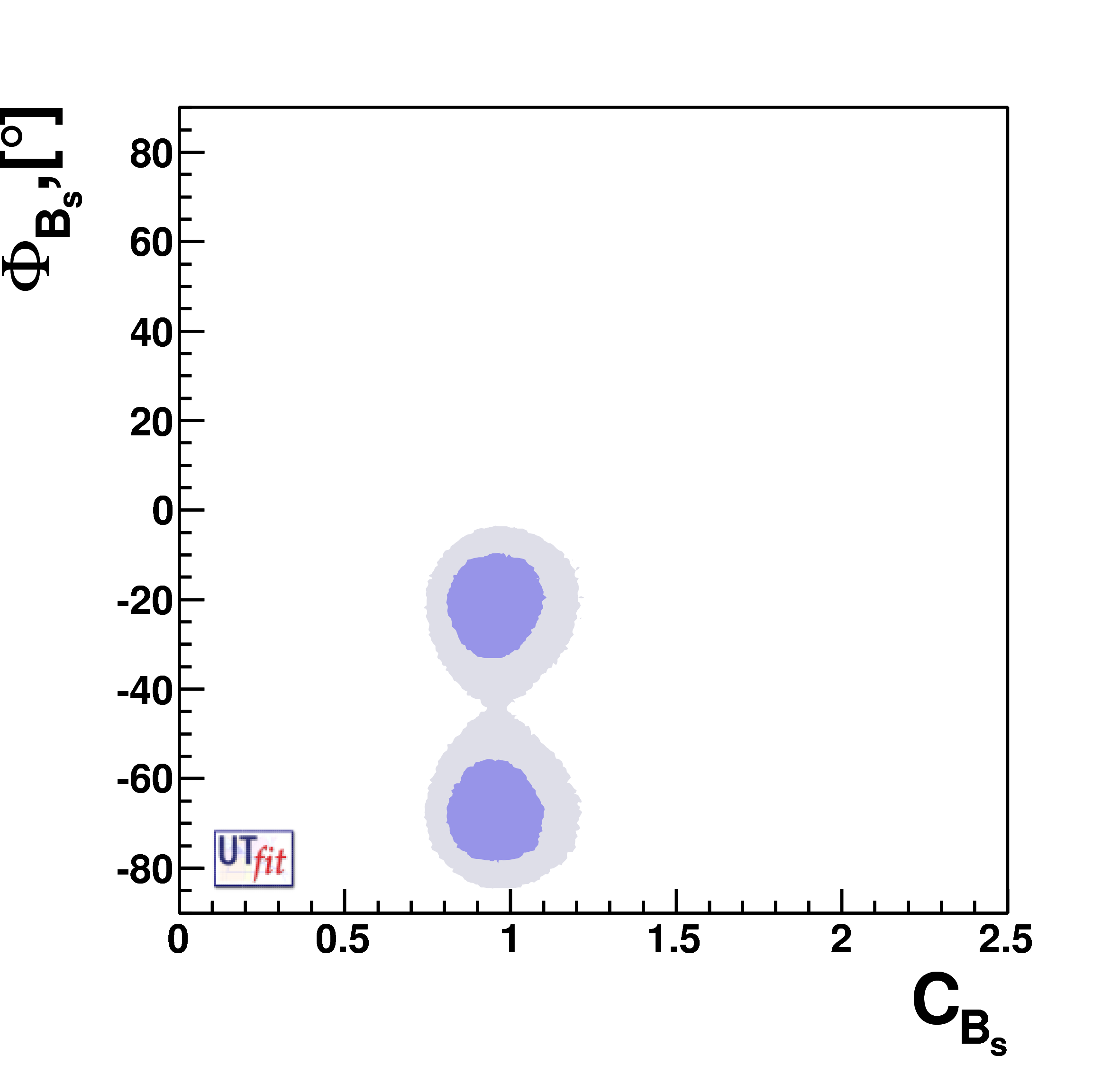}
    \includegraphics[width=0.35\textwidth]{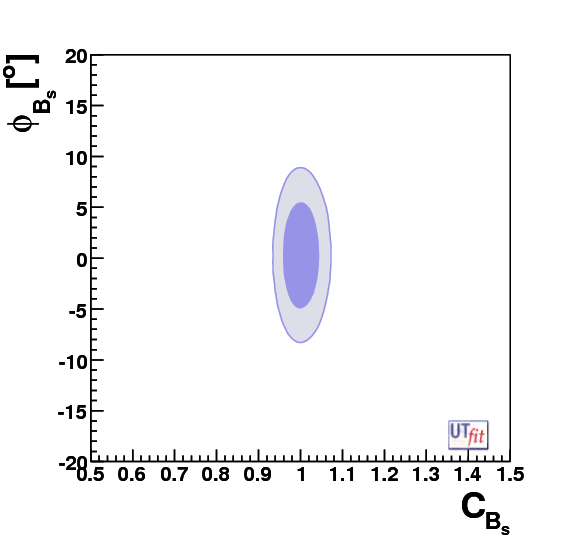}
    \caption{
      Allowed regions in the $C_{B_s}$--$\phi_{B_s}$ plane
      given by the current data (top) and at the time of \superb\ (bottom).
      Note that the scales for the axes are different in the two cases.  See 
      Table \ref{tab:deltaf2sfit} for the corresponding values anticipated for 
      these measurements.
    }
    \label{fig:cvsphi}
  \end{center}
\end{figure}

To illustrate the impact of the measurement at \superb at the $\FiveS$,
we show in Fig.~\ref{fig:cvsphi} selected regions in the
$\phi_{B_s}$--$C_{B_s}$ plane (right),
compared to the current situation (left).
Corresponding numerical results are given in Table~\ref{tab:deltaf2sfit}.

\begin{table}[ht]
  \caption{
    Uncertainty of New Physics parameters $\phi_{B_s}$ and $C_{B_s}$
    using the experimental and theoretical information available
    at the time of \superb\ and given in Tables~\ref{tab:inputsbs}
    ($30 \ {\rm ab}^{-1}$) and~\ref{tab:lattice}.
    These uncertainties are compared to the present determination.
  }
  \label{tab:deltaf2sfit}
  \begin{center}
    \begin{tabular}{ccc}
      \hline
      Parameter    & Today        & At \superb\ ($30 \ {\rm ab}^{-1}$) \\
      \hline
      $\phi_{B_s}$ & $(-3 \pm 19)^\circ \cup (94 \pm 19)^\circ$ & $\pm 1.9^{\circ}$ \\
      $C_{B_s}$    & $1.15 \pm 0.36$     & $\pm 0.026$ \\
      \hline
    \end{tabular}
  \end{center}
\end{table}

It is important to note that the uncertainty on the parameter $C_{B_s}$
is dominated by the uncertainty on $f_{B_s}$ and bag parameters.
The error on $\phi_{B_s}$ is not limited by systematics and theory,
and can be improved to $1$--$2^\circ$ with a
longer dedicated run at the $\FiveS$.  \lhcb\ will also measure the 
New Physics phase $\phi_{B_s}$ and is expected to achieve a comparable
sensitivity with full statistics ($\sim 10 \ {\rm fb}^{-1}$) of
$\sim 1^\circ$.

\mysubsection{Summary\label{sec:bdecays:summary}}

The recently reported evidence from the D0 collaboration for a di-muon asymmetry $A_{\rm CH}$
that is incompatible with the Standard Model~\cite{Abazov:2010hv} has been followed by theoretical
attempts to interpret the result, for 
example~\cite{Eberhardt:2010bm,Dighe:2010nj,Dobrescu:2010rh,Chen:2010wv,Buras:2010mh,Ligeti:2010ia}.
If this effect is confirmed, then we can expect to observe new physics
in meson decays in the $B$ and $D$ sector.  One prime example
of an observable that would be expected to manifest new physics in $B_s$ 
decays is the semi-leptonic asymmetry $A^s_{\rm SL}$.  
In some models the phases postulated to be responsible for large semi-leptonic
asymmetries are flavour blind, and thus we would expect effects to also
be manifest in other meson decays as well as in the $B_s$ sector.
Other models have postulated a richer texture of new physics that may be related
to this D0 result.  In both cases, and even if the D0 result turns out to 
be a statistical fluctuation rather than evidence for new physics, \superb 
will be able to test a variety of new physics scenarios using a wide 
array of measurements as outlined in Section~\ref{sec:interplay}.

The results presented in this section section are summarized
in Table~\ref{tab:inputsbs} for two scenarios (i) a short ($1 \ {\rm ab}^{-1}$)
and (ii) a long ($30 \ {\rm ab}^{-1}$) run at the $\FiveS$ resonance.
Collecting $1 \ {\rm ab}^{-1}$ will take less than one month at the
\superb design luminosity of $10^{36} \ {\rm cm}^{-2} \ {\rm sec}^{-1}$.

\begin{table}[htb]
  \caption{
    Summary of the expected precision of some of the most important
    measurements that can be performed at \superb\
    operating at the $\FiveS$ resonance,
    with an integrated luminosity of
    $1 \ {\rm ab}^{-1}$ and $30 \ {\rm ab}^{-1}$.
  }
  \label{tab:inputsbs}
  \begin{center}
    \begin{tabular}{lcc}
      \hline
      Observable    &  $1 \ {\rm ab}^{-1}$ & $30 \ {\rm ab}^{-1}$ \\
      \hline
      $\Delta \Gamma$                 & $0.16 \ {\rm ps}^{-1}$ & $0.03 \ {\rm ps}^{-1}$ \\
      $\Gamma$                        & $0.07 \ {\rm ps}^{-1}$ & $0.01 \ {\rm ps}^{-1}$	\\
      $A^s_{\rm SL}$                  & 0.006                  & 0.004 \\
      $A_{\rm CH}$                    & 0.004                  & 0.004 \\
      $\BR(B_s \to \mu^+ \mu^-)$      &  -                     & $<8 \times 10^{-9}$ \\
      $|V_{td}/V_{ts}|$               & 0.08                   & 0.017 \\
      $\BR(B_s \to \gamma \gamma)$    & $38 \%$                & $7 \%$ \\
      $\beta_s$ (angular analysis)    & $20^\circ$             & $8^\circ$ \\
      $\beta_s$ ($J/\psi \phi$)       & $10^\circ$             & $3^\circ$ \\
      $\beta_s$ ($K^0 \bar{K}^0$)     & $24^\circ$      & $11^\circ$ \\
      \hline
    \end{tabular}
  \end{center}
\end{table}

While it is clear that \superb\ cannot compete with hadronic experiments
on modes such as $B_s \to \mu^+\mu^-$ and $B_s \to J/\psi \phi$,
it is also evident that many important channels that are not easily accessible
at hadronic experiments such as \lhcb will be measurable at \superb.
Besides the flagship measurement of the semi-leptonic asymmetry $A^s_\mathrm{SL}$,
the channels $B_s \to \gamma \gamma$ and $B_s \to K^0 \bar{K}^0$ will also be measurable
at \superb. Therefore \superb\ will complement the results from \lhcb and enrich 
the search for new physics in flavour decays by accumulating several 
\invab of data at the $\FiveS$ resonance~\cite{Baracchini:2007ei}.

Measuring an absolute branching fraction
in a hadronic environment is limited by ones determination of luminosity
and the production mechanisms at play.
So in addition to being able to study these $B_s$ golden modes, it is  
anticipated that there will be benefits to the field when interpreting
some \lhcb analyses if one can obtain precision measurement of 
at least one absolute branching fraction from \superb.  In order to measure
an absolute branching fraction with precision it will be necessary to study 
a number of $B_s$ decays at the \FiveS. In time and with an understanding 
of the performance of \lhcb, it will be possible to identify a full list 
of useful branching fractions to measure at \superb and thus understand 
better how much data to record at the \FiveS resonance.


\graphicspath{{Dphysics/}{Dphysics/}}
\section{Charm Physics}\label{sec:charm}
\newcommand{\note}[1]{{\bf NOTE}#1}

The SM projects a rather mundane weak phenomenology for charm transitions; yet as
has been stated since the early discussions about a Tau-Charm Factory in the late 
1980's, this fact can be turned to our advantage: detailed studies 
in particular of CP invariance in 
charm decays can act as (almost) zero-background searches 
for physics beyond the SM. While no clear signal for the intervention of NP has been uncovered yet 
in charm transitions, the situation has changed qualitatively in the last two years: 
\begin{itemize}
\item 
$D^0 - \bar D^0$ oscillations have been resolved experimentally with $x_D$, $y_D$ 
$\sim 0.5 - 1 \%$. 
\item 
This breakthrough has lead to `new thinking' among theorists. They have 
begun to realize that scenarios of NP motivated by considerations  
{\em outside} of flavour dynamics can produce an observable footprint in charm decays; i.e., 
one is no longer forced to invoke the old `stand-by' of NP scenarios, namely 
SUSY models with {\em broken} $R$ parity, to produce observable effects in an ad-hoc 
fashion. There is every reason to think that this emerging renaissance of creative thinking 
about charm dynamics will continue and bear novel fruits. 

\end{itemize}
The \sff allows comprehensive charm studies in 
two different environments: 
\begin{enumerate}
\item 
One has the large production rate of charm mesons and baryons at (or
close to) the
$\Upsilon (4S)$ and can benefit greatly from the Lorentz boost imparted onto the charm hadrons. 
\item 
The \sff design discussed here allows running at the charm threshold 
region, where one can make use of quantum correlations.  With data collected
at charm threshold one will also be able to use a $D$ recoil technique to search 
for rare decays that may be otherwise background dominated. Such recoil analyses
in $B$ decays provide useful constraints on scenarios of physics beyond the SM,
and the $D$ analogues of these will, in general, also be interesting.
The anticipated ultra-high luminosity is again crucial, since high statistics can be achieved 
with relatively limited running. 
This will provide a raw sample of $1.8\times 10^9$ \Dz\Dzb and $1.5\times 10^9$
\Dp\Dm pairs.
We will demonstrate that valuable information relevant to interpretation of the role of
NP in charm decays can be made from a sample of $\sim 500~\invfb$ obtained from a \superb run at the 
$\psi(3770)$.  This could be accumulated in a few months' running.
\end{enumerate}

\subsection{On the Uniqueness of Charm}

In general NP will induce flavour changing neutral currents (FCNC). The SM had to be crafted judiciously to
have them greatly suppressed for strangeness; the weight of FCNC is then even more reduced for the up-type quarks $u$,
$c$ and $t$. Yet NP scenarios could exhibit a very different pattern with FCNC being significantly more relevant for
up-type quarks. 

Among those it is only the charm quark that allows the full range of probes for FCNC in general and for 
CP violation in particular \cite{CICERONE}. For top quarks do not hadronize \cite{RAPALLO} thus eliminating the
occurrence of $T^0 - \bar T^0$ oscillations. Neutral pions etc. cannot oscillate, since they are their own
antiparticles; furthermore CPT constraints are such that they rule out most CP asymmetries. 

In general, particles and couplings that enhance FCNC in charm above
the SM expectation are distinct from those that contributed to FCNC
in the B and K sectors. Thus, charm sector provides a unique window
to observe or constrain BSM physics.


\subsection{$D^0 - \bar D^0$ Oscillations}

\subsubsection{Experimental Status}

While the existence of $D^0 - \bar D^0$ oscillations is considered as 
established - $(\xd,\yd) \neq (0,0)$ - the size of $\xd$, $\yd$ and even their 
relative strengths are not known with sufficient accuracy to know if \CPV\ is 
manifest in mixing. Their accurate values will hardly shed light on their 
theoretical interpretation; yet having them is not merely a `noble goal' (G. 
Wilkinson), but a practical one: for knowing their values with some accuracy 
will help validate measurements of the presumably small CP asymmetries, as 
discussed later.  


So far, almost all the information on mixing parameters has come from decays 
where the final state $f$ is accessible to either \Dz or \Dzb.  
In such cases,
deviations from exponential behavior in the number of \Dz(\Dzb)'s, $N(\bar N)$,
at time $t$ have been exploited.  To second order in $x$ and $y$,
\begin{align}
\label{eq:nonexponential}
  \begin{split}
    N(t)      &= N(0) e^{-\Gamma t}\times [1+ 
                 \frac{x^2+y^2}{4}|\lambda_f|^{2}(\Gamma t)^2    \\
              &+ |\lambda_f|(y\cos\overline{\delta_f\!+\!\phi_f}
                          -x\sin\overline{\delta_f\!+\!\phi_f})(\Gamma t)], \\
    \bar N(t) &= \bar N(0) e^{-\Gamma t}\times [1+
                 \frac{x^2+y^2}{4}|\lambda_f|^{-2}(\Gamma t)^2   \\
              &+ |\lambda_f|^{-1}(y\cos\overline{\delta_f\!-\!\phi_f}
                               -x\sin\overline{\delta_f\!-\!\phi_f})(\Gamma t)],
  \end{split}
\end{align}
where $\lambda_f = \left(q\bar{\cal A}_f\right)/\left(p{\cal A}_f\right)$, $
\phi_f    = \psi_f+\phi_m$, and $\phi_m    = \arg{\qd/\pd}$.
The first and second terms in Eq.~(\ref{eq:nonexponential}) correspond,
respectively, to direct decay ($\Dz\!\to\! f$)
\footnote{Charge-conjugate modes are implicitly included unless noted 
otherwise.}, 
and to decay after mixing ($\Dz\!\to\!\Dzb\!\to\! f$).  
The third term, linear in $t$, is due to the interference between these two.

The decay amplitudes ${\cal A}_f$ and $\bar{\cal A}_f$ describe, 
respectively, the processes $\Dz\!\!\to\!\!f$ and $\Dzb\!\!\to\!\!f$
with relative strong (weak) phases $\delta_f$ ($\psi_f$).
This phase is generally unknown, and this limits the
measurability of \xd and \yd to quantities rotated by $\delta_f$.  However, 
in decays to self-conjugate multi-body states (the ``golden channels") such 
as $\KS\, h^+\!h^-$, $h=\pi$ or $K$, where $f$ is expressible as a 
combination 
of \CP odd and even eigenstates, $\delta_f$ is zero (or $\pi$), making it 
possible to measure \xd, \yd, (and $|\qd/\pd|$ and $\phi_m$) directly, with a 
time-dependent Dalitz plot (TDDP) analysis of the final, 3-body system.

Three kinds of successful mixing parameter measurements have exploited the
linear dependence of the interference term in Eq.~(\ref{eq:nonexponential})
on \xd and \yd (both $\ll 1$):  Wrong-Sign (WS) decays
$\Dz\to\Kp\pim$; decays to \CP\ eigen-states $h^-h^+$ ($h=K$ and $h=\pi$); 
and decays to 3-body states ($\Kp\pim\piz$, $\KS\pip\pim$ and $\KS\Kp\Km$).

WS semi-leptonic decays $\Dz\to X^+\ell^-\bar\nu_{\ell}$ have 
also been examined for mixing.  Such decays can only arise from mixing
($\Dz\!\to\!\Dzb$) followed by decay, so their time-dependence is
described by the second term alone in Eq.~(\ref{eq:nonexponential}).
The rates, proportional to $(x^2+y^2)/4\sim 5\times 10^{-5}$, are very small,
however, and only upper limits have been found so far.

%
Evidence for \DzDzb oscillations was found by \babar%
\cite{Aubert:2007wf}
and confirmed by CDF
\cite{Aaltonen:2007uc}
from WS decays $\Dz\!\to\!\Kp\!\pim$ by comparing their time-dependence with
that for decays to the Right-Sign (RS) final state, $f=\Km\!\pip$.  In 
the WS case, direct decays are doubly Cabibbo-suppressed (DCS), so 
$|\lambda_f|\gg 1$ and deviations from exponential are quite large.  By 
contrast, such deviations for RS decays are negligible.  Even assuming 
\CP conservation ($\phi_m=\phi_f=0$), the strong phase difference 
$\deltakpi$ between $\Dz$ and $\Dzb$ decays to $\Kp\!\pim$ is virtually 
unknown, making it possible only to measure $\xpsq$ and $\yp$, where \xp, 
and \yp\ are $(\xd, \yd)$, rotated by angle \deltakpi
\begin{align}\label{eq:xpyp}
  \begin{split}
  \xp &= \xd\cos\deltakpi + \yd\sin\deltakpi \\
  \yp &= \yd\cos\deltakpi - \xd\sin\deltakpi \\
  \end{split}
\end{align}
and not $x$ and $y$ directly.

%
Mean lifetimes, $\tau_{hh}$, of decays to \CP-even states $f=h^+h^-$ (where
$h=\pi$ or $K$) are related to \ycp, the value of \yd\ if \CP\ is conserved.  
With \CP conservation, \ycp is given by
\beq
  \label{eq:ycp}
  \ycp \approx \frac{\tau_{\Km\!\pip}}{\tau_{hh}}-1,
\eeq
where $\tau_{\Km\!\pip}$ is the lifetime for the mixed-\CP\ state 
$f=\Km\!\pip$.

Measurements of \ycp\ by Belle 
\cite{Staric:2007dt}
and \babar%
\cite{Aubert:2007en,Aubert:2009ck}
show evidence for mixing ($\ycp\neq 0$) at a level of at least $3\sigma$
in each case, and are in good agreement.  The world average for all 
measurements is $1.107\pm 0.217\%$
\cite{Schwartz:2009jv}.

%
WS decays to three-body final states $\Kp\!\pim\!\piz$ have been studied 
by \babar
\cite{Aubert:2008zh}.  
In these decays, the final state $f$ is specified by
its position $(s_1,s_2)$ in the Dalitz plot (DP) representing the phase space
available to the three-body system.  The coordinates are two of the three
squared invariant masses.
Eq.~(\ref{eq:nonexponential}) applies for each point in the DP so, with a model
for the variation of strong phase $\delta(s_1,s_2)$ over the DP due to final 
state interactions, both \xd and \yd are measurable through the interference
term in which they are linear.  However, since there is an unknown, strong 
phase \deltaktwopi, arising from the decay, only \xpp and \ypp (\xd and \yd, 
respectively, rotated by the \deltaktwopi) are measurable.  However, unlike
the 2-body decay to $\Kp\!\pim$, these rotated parameters are both linear,
not quadratic as in WS $\Kp\pim$ decays - a distinct advantage.

Golden channel decays (to 3-body self-conjugate states) do not suffer 
from the unknown strong decay phase, so measurement of \xd\ and \yd\ are, 
possible.  Measurements of the $\KS\,\pip\!\pim$ final state
carried out by Belle 
\cite{Abe:2007rd}
and \babar
\cite{delAmoSanchez:2010xz}
have uncertainties in \xd\ and \yd\ of $\sim 3\times 10^{-3}$.  In
each case, uncertainties in the assumptions made in the decay models used to 
describe the strong phase variations in the DP introduce irreducible systematic
uncertainties of $\sim 1\times 10^{-3}$.

\subsubsection{Combination of measurements and \CPV}
Asymmetries between \Dz and \Dzb
event samples have also been measured, providing information on the 
\CP\ mixing parameters $|\qd/\pd|$ and $\arg(\qd/\pd)$.  In 3-body decays to self-conjugate 
final states, these parameters can be determined directly from time-dependent effects on the Dalitz plot population.  Asymmetries in direct decay rates (either allowing ${\cal A}_{\bar f}\neq\bar{\cal A}_f$ or not) have also provided information on direct \CPV.  However, all these asymmetries are, so far, consistent with zero.

In all, 28 mixing observables have been measured.  The Heavy Flavor Averaging 
Group (HFAG) has included these, with their covariances, in a $\chi^2$ fit to 
obtain mixing parameter values
\cite{Schwartz:2009jv},
both allowing for \CPV\ and requiring \CP\ conservation.  The values obtained
from this \CPV\ fit are
\begin{center}
\begin{tabular}{rclrcl}
   $x$ &=& $(0.98^{+0.24}_{-0.26})~\%$ &
   $y$ &=& $(0.83\pm 0.16)~\%$         \\
   $|\qd/\pd|$ &=& $0.87^{+0.17}_{-0.15}$    &
   $\phi_{\sst M}$ &=& $(-8.5^{+7.4}_{-7.0})^{\circ}$ \\
   $\delta_{\sst K\pi}$   &=& $(26.4^{+9.6}_{-9.9})^{\circ}$ &
   $\delta_{\sst K\pi\pi}$&=& $(14.8^{+20.2}_{-22.1})^{\circ}$ \\
\end{tabular}
\end{center}
To summarize, mixing has clearly been established, but so far there is no evidence for \CPV
in charm decays. As shown later, measurements of \xd depend heavily on the
golden channels, and of \yd on the \ycp results.  Uncertainties in \xd and
\yd are of order $2\times 10^{-3}$, too large to detect \CPV differences
between $\Dz$ and $\Dzb$.

\subsubsection{Measurements of strong phases}
Data taken at the $\psi(3770)$ ($D\bar D$ threshold), allow independent
determination of the strong phases $\delta_f$
\cite{Asner:2005wf,Rosner:2008fq}.
Using the coherence of $\Dz$ and $\Dzb$ pairs from $\psi(3770)$ decays, 
values of the strong phases $\delta_f$ have been obtained by the CLEO-c collaboration
\cite{Asner:2008ft,Lowery:2009id}.
An $818~\invpb$ sample of such decays for 
$\delta_{\sst K\!\pi}=(50^{+38}_{-28})^{\circ}$ (for $\Dz\to\Kp\!\pim$) 
and $\delta_{\sst K\!\pi\!\pi}=(59^{+32}_{-28})^{\circ}$ (for $\Dz\to\Kp\!\pim\!\piz$)
\footnote{We take the results from the fit made without input from other mixing
measurements, and adjust for a $180^{\circ}$ difference in phase definition 
used in this paper.}.
We note that the overall phase for $\Kp\pim\piz$ decays is not quite what is 
required to convert $x''$ and $y''$ to \xd and \yd since the population of the 
quantum-correlated DP's for events from
the 818~\invpb sample is not quite the same as that observed in freely
decaying $\Dz$'s.  A re-averaging process would be required.  In the
projections presented here, however, we simply shift the central values
as outlined to agree with the \babar central values for \xd and \yd.

These results are less precise than the indirectly determined values from
the HFAG averages.  More $\psi(3770)$ data is forthcoming, however, from
BES~III which should improve this estimate by a factor $\sim 6$.  A 
$500~\invfb$ \superb run (approximately the same integrated luminosity 
as that accumulated by the \babar experiment) at threshold would improve 
on the CLEO measurement by a factor $\sim 25$, so should add much to the 
precision of the mixing parameters.  We examine this possibility in
Section~\ref{sec:mixcando}.

More interestingly, threshold data also provides a measurement of point to 
point variations in strong phase over the $\Kp\!\pim\!\piz$ and 
$\KS\,h^+h^-$ Dalitz plots that is independent of any model.  The 
precision and granularity required of such measurements, {\em per se}, 
is hard to assess without detailed simulation, but it should add a 
valuable reality check to any of the decay amplitude models presently 
in use.  Such information will clearly, therefore, reduce the model 
uncertainties inherent in both TDDP mixing measurements as well as in 
those of CKM $\gamma$ that also requires such phase information
\cite{Bondar:2005ki}.

\subsubsection{Theoretical Interpretation}

Most authors have concluded that effects even as `high' as $x_D \simeq 1\% \simeq y_D$ could conceivably be generated by SM dynamics alone 
(see, e.g., \cite{FALK1,FALK2,CICERONE,URIDOSC}).
Some, however, think that $x_D$ in particular might contain a sizable or even large 
contribution from NP \cite{CICERONE}. Short of a breakthrough in our computational powers -- one that 
lattice QCD seems unlikely to achieve -- this issue cannot be decided by theoretical means. 

The current world average value of \xd, appears to lie at the 
tantalizingly high end of SM expectations.  The new TDDP analysis by the \babar collaboration
\cite{delAmoSanchez:2010xz}
of the golden channels, however, indicates a value that is consistent with zero.  This will reduce the average considerably, to little more than $3\sigma$ from zero.  
To make progress in understanding what role (if any) new physics
beyond the SM plays in the charm sector, we not only need to know how 
large \xd is, but also to understand at what level \CPV occurs in either 
mixing or in decay.  Such effects will show up in asymmetries in effective values for \xd (or \yd) obtained from separated samples of \Dz and \Dzb decays.  To answer either of these questions will require measurements of \xd with a precision at least of the order of $10^{-4}$.

\subsubsection{Measuring $x_D$ and $y_D$ at \superb}

At \superb, we will address this situation in several ways.
First, sample sizes at the $\Upsilon(4S)$ will be much larger, thereby
improving statistical precision on all current measurements by a 
factor $\sim 12$.  Simulation studies have shown that \superb also
provides improved $\Dz$ decay time resolution that effectively enhances 
this statistical significance by a further factor $\sim 1.5$ in event yield
\cite{Andreassen:4Ssim}
though, to be conservative, we do not include this in our estimates.  

Secondly, in addition to \xd and \yd, TDDP analyses of golden decays 
can provide direct measurement of the \CPV parameters $|\qd/\pd|$ and 
$\phi_m$ that should now approach a level where \CPV could be identified.

\superb will use data from $\psi(3770)$, anticipated to come
on the \superb time-scale from BES~III, to overcome the limitation 
imposed by uncertainty in the models used to define the strong phase 
structure in the golden channel DP's.  Such data can also add information 
on $\delta_f$ required for mixing measurements in other channels.
  
Further improvements would also come from \superb data from a 
dedicated $500~\invfb$ run at the $\psi(3770)$.  We estimate that this 
should improve precision in \xd (and \yd) by a factor two and to come 
close to the goal of $\sim 10^{-4}$
\footnote{We might also speculate on possibilities
for time-dependent measurements at threshold utilizing the boost unique to 
\superb, but these studies still have to be made.}.

\subsubsection{Projections for mixing measurements at \superb}
\label{sec:mixcando}

Realistic estimates for the \superb mixing reach can be made on the basis 
of what has been achieved with \babar's accumulation of $482~\invfb$ 
running at the $\Upsilon(4S)$.  Also, by projecting 
results already obtained by CLEO-c, we can estimate the gain we might 
expect from the measurement of strong phases from $D\bar D$ threshold 
data either from BES~III, or from a $500~\invfb$ \superb sample
\footnote{In using BES~III results, we ignore some differences wrt CLEO-c.  BES~III
has no RICH which will affect the precision of measurements of channels
($\Dz\to K\pi$, for example) where $K/\pi$ separation is important.  We
also observe that systematic uncertainties in BES~III may take some time
to be as well understood as those in CLEO-c.}.

\begin{figure*}[!htbp]
\begin{center}
\begin{tabular} {cc}
\resizebox{16cm}{!}{
  \includegraphics[width=0.44\textwidth]{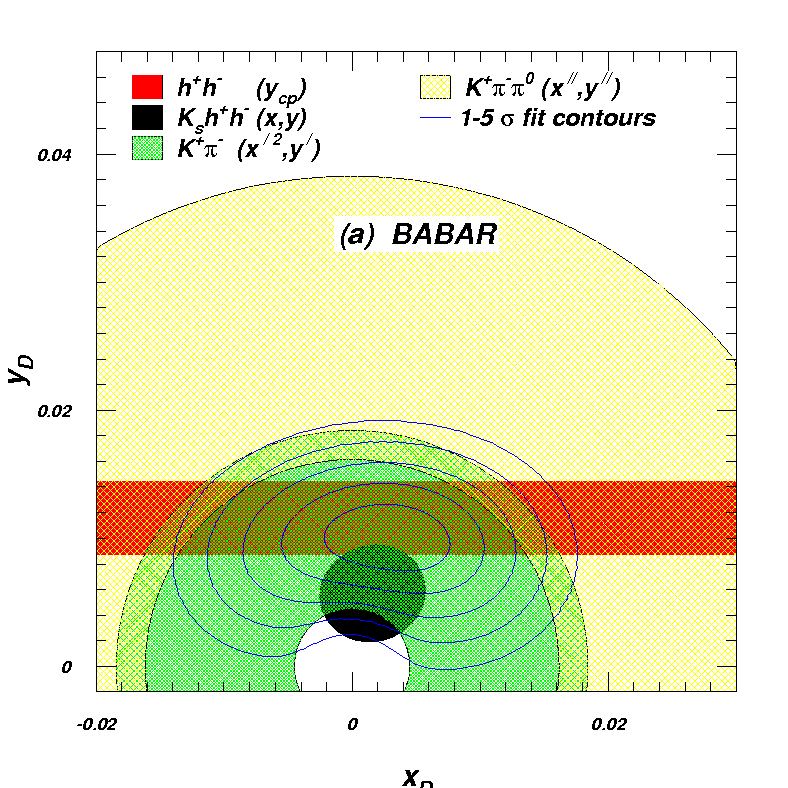}
  \includegraphics[width=0.44\textwidth]{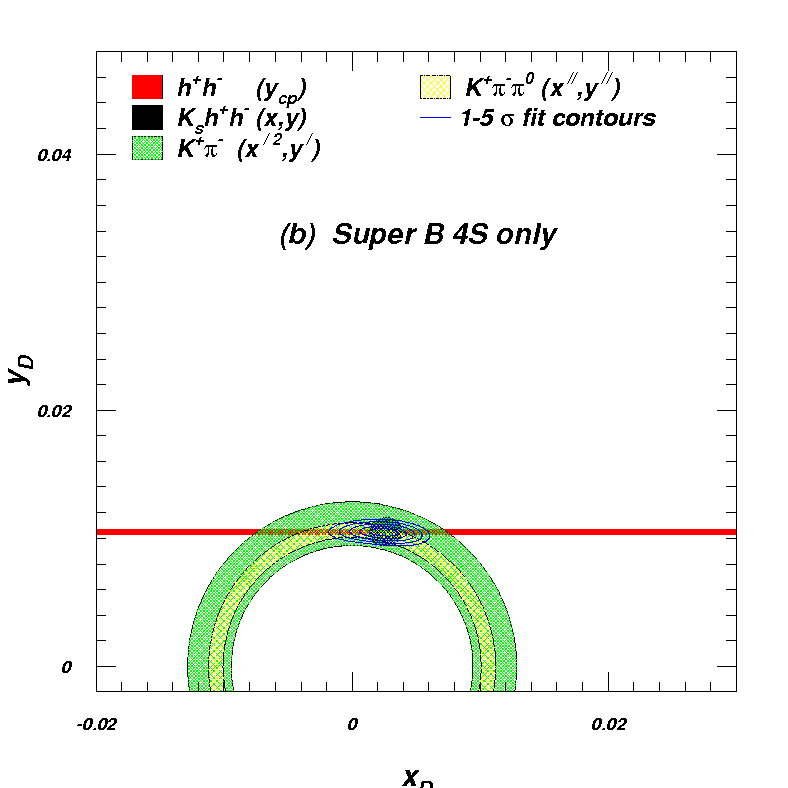}
}
\end{tabular}
  \caption{Mixing observables projected into the (\xd,\yd) plane. 
   Shaded areas indicate the coverage of measured observables lying 
   within their 68.3\% confidence region.  Contours enclosing 68.3\% ($1\sigma$), 95.45\% ($2\sigma$), 
   99.73\% ($3\sigma$), 99.994\% ($4\sigma$) and $1-5.7\times 10^{-7}$ 
   two-dimensional confidence regions from the $\chi^2$ fit to these 
   results are drawn as solid lines.
   (a) Shows current results from \babar alone.  (The $\Kp\!\pim\!\piz$ 
   projection is omitted since it obscures much of the area shown);  
   (b) includes results anticipated from a $75~\invab$ \superb 
   run at the $\Upsilon(4S)$ only (no data from $\psi(3770)$ running.  
  }
  \label{fig:babarmix}
  \end{center}
\end{figure*}

In Fig.~\ref{fig:babarmix}(a) the four main mixing results from \babar are 
shown combined into average values for (\xd, \yd).  To compute this average 
we used a $\chi^2$ minimization technique, similar to that employed by HFAG
\cite{Schwartz:2009jv},
which includes effects from correlations between the measured observables, 
$(\xpsq,\yp)$ from WS $\Dz\to\Kp\!\pim$ decays
\cite{Aubert:2007wf},
$(\xpp,\ypp)$ from TDDP analysis of $\Kp\!\pim\!\piz$
\cite{Aubert:2008zh},
$\ycp$ from both tagged
\cite{Aubert:2007en}
and untagged
\cite{Aubert:2009ck}
samples of $\Dz\to h^-\!h^+$ decays and $(\xd,\yd)$ from the combined
$\KS\, h^+\!h^-$ golden channel samples
\cite{Jordi:moriond}
($h=\pi, K$).  In each case, results are based upon the assumption of
no \CPV
 
We are omitting other mixing results anticipated to come from \babar 
data that should also be projected to \superb.  These include
three further golden channels - $\Dz\to h^+\!h^-\piz$, 
$\Dz\to\Kp\!\!\Km\!\pip\!\pim$, $\Dz\to\KS\pip\!\pim\!\piz$.  A joint
analysis of the channels $\Dz\to\KS\!K^{\mp}\!\pi^{\pm}$ is also anticipated.

The averages obtained are reported in Table~\ref{tab:babarmix}.
The figure indicates the 68.3\% confidence region that each of these 
measurements covers in the (\xd,\yd) plane.  It is evident that the 
most precise information on \xd comes
from the golden channels (black ellipse) and, for \yd, from the $\ycp$ 
results (horizontal band).

\begin{table}[!htbp]
  \begin{center}
  \caption{Mixing parameters (\xd,\yd) and strong phases 
   $\deltakpi$ and $\delta_{\Kp\pim\piz}$ obtained from 
   $\chi^2$ fits to observables obtained either from \babar 
   or from their projections to \superb.  Fit a) is for $482~\invfb$
   from \babar alone and this is scaled up in b) to $75~\invab$ at
   $\Upsilon(4S)$ for \superb.  In each case, no input from measurements
   of strong phase is included.  Fit c) includes strong phase information
   projected to come from a BES~III run at $D\bar D$ threshold, and
   d) is what would be possible from a $500~\invfb$ $D\bar D$
   threshold run at \superb.  For each of these scenarios, the 
   uncertainties due to statistical limitation alone are entered, in 
   parentheses, on the line below the results for the corresponding 
   fit.  In all but fit a) (\babar results) the central values 
   have no meaning.
  }
  \label{tab:babarmix}
\begin{tabular}{lccccc}
\hline
Fit   & 
$ x \times 10^{3}$                          &
$ y \times 10^{3}$                          &
$\delta_{\sst K^+\!\pi^-}^{\circ}$          &
$\delta_{\sst K^+\!\pi^-\!\pi^0}^{\circ}$   \\ [3pt]
\hline
(a) 
  & $  3.01^{+  3.12}_{  -3.39} $
  & $ 10.10^{+  1.69}_{  -1.72} $
  & $ 41.3^{+ 22.0}_{ -24.0}    $
  & $ 43.8  \pm 26.4  $
  \\[3pt]
Stat.
  & $  (2.76) $
  & $  (1.36) $
  & $  (18.8) $
  & $  (22.4) $
  \\[3pt]
(b) 
  & $  xxx ^{+  0.72}_{  -0.75} $
  & $  xxx   \pm 0.19  $
  & $  xxx^{+  3.7}_{  -3.4} $
  & $  xxx^{+  4.6}_{  -4.5} $
\\
Stat.
  & $  (0.18) $
  & $  (0.11) $
  & $  (1.3)  $
  & $  (2.9)  $
  \\[3pt]
(c) 
  & $  xxx  \pm 0.42 $
  & $  xxx   \pm 0.17  $
  & $  xxx  \pm 2.2  $
  & $  xxx^{+  3.3}_{  -3.4} $
\\
Stat.
  & $  (0.18)  $
  & $  (0.11)  $
  & $  (1.3)   $
  & $  (2.7)   $
  \\[3pt]
(d) 
  & $  xxx   \pm 0.20  $
  & $  xxx   \pm 0.12  $
  & $  xxx  \pm 1.0    $
  & $  xxx  \pm 1.1    $
\\
Stat.
  & $  (0.17)  $
  & $  (0.10)  $
  & $  (0.9)   $
  & $  (1.1)   $
 \\
 \hline
\end{tabular}

  \end{center}
\end{table}

This fit procedure is repeated on projections to various \superb scenarios.
The first of these, in Fig.~\ref{fig:babarmix}(b), shows expectations
of measurements solely from a $75~\invab$ run at the $\Upsilon(4S)$.
To compare with \babar results, the various observables are shifted
to correspond to values expected for the \xd and \yd averages for \babar,
and are then smeared with correlated uncertainties based on those
reported by \babar, and projected to the \superb sample.

In making this projection, it is assumed that statistical and systematic
uncertainties will shrink in accordance with the square root of 
the luminosities - a reasonable assumption since major systematic 
uncertainties are estimated from data and simulated studies that should 
scale in this way.  The Dalitz plot amplitude model uncertainty in the 
TDDP of the golden channels is, however, left unchanged.  While a better
understanding of such models could develop on the \superb time-scale, our 
conservative assumption is that it will not reduce this uncertainty.

\begin{figure*}[!htbp]
\begin{center}
\begin{tabular} {ccc}
\resizebox{16cm}{!}{
  \includegraphics[width=0.32\textwidth]{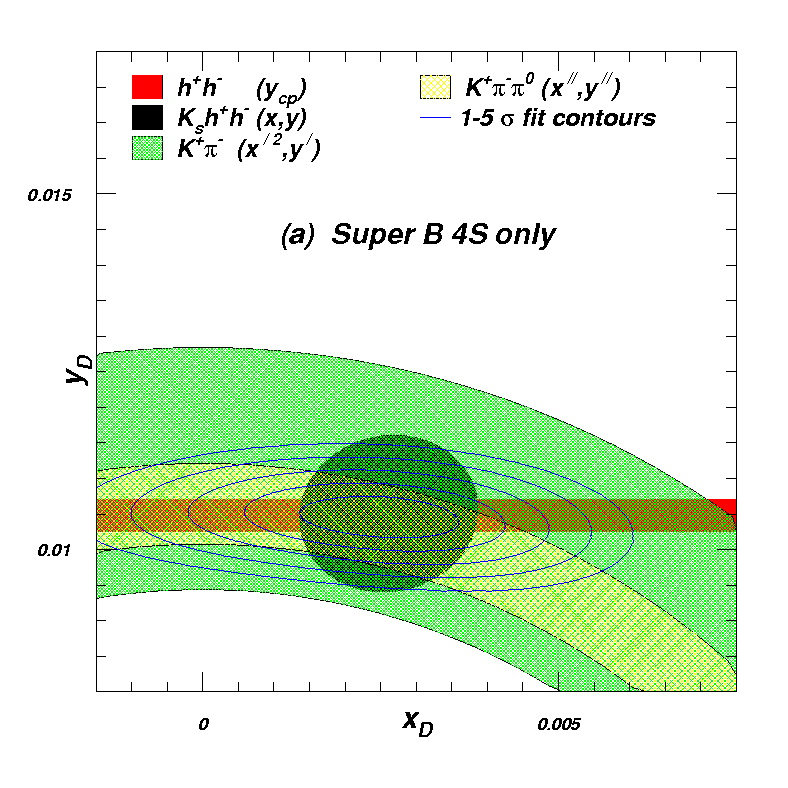}
  \includegraphics[width=0.32\textwidth]{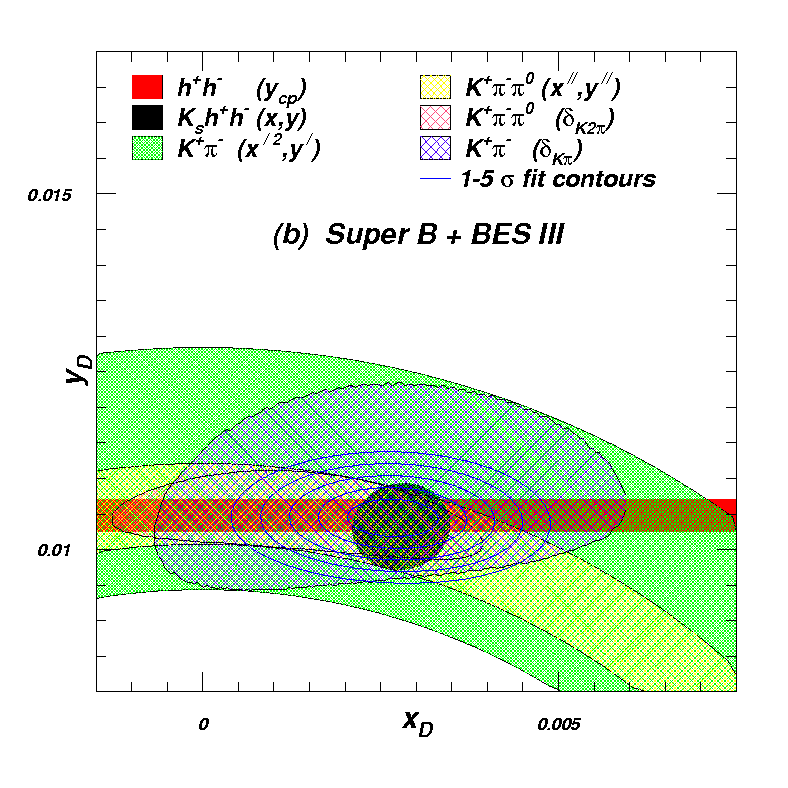}
  \includegraphics[width=0.32\textwidth]{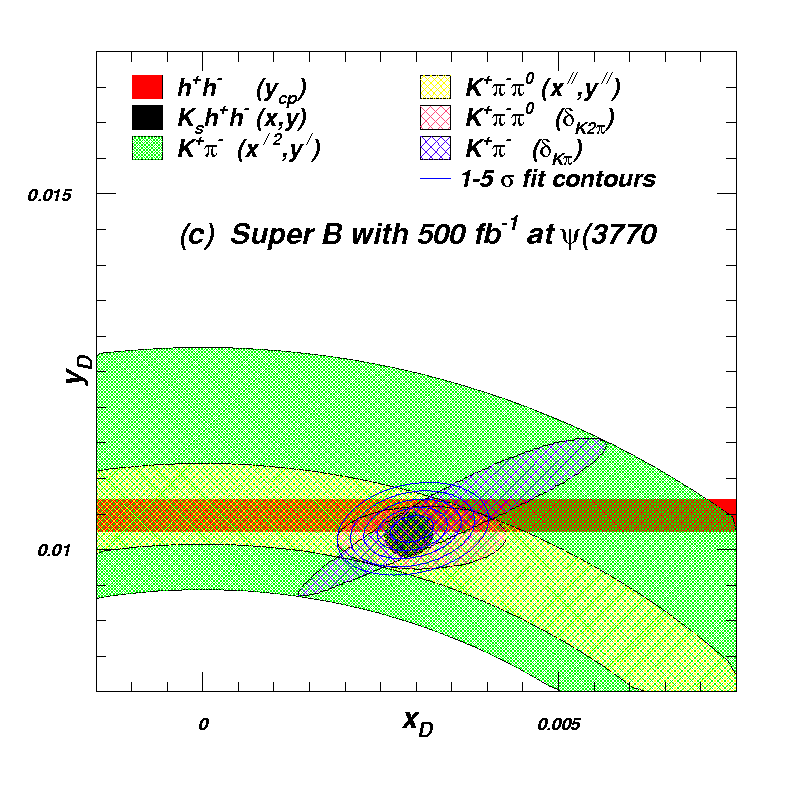}
}
\end{tabular}
  \caption{Mixing observables projected into the (\xd,\yd) plane. 
   Shaded areas indicate the coverage of measured observables lying 
   within their 68.3\% confidence region.  In (b) and (c), the
   projections of $\deltakpi$ and $\deltaktwopi$ measurements are also shown.
   Contours enclosing 68.3\% ($1\sigma$), 95.45\% ($2\sigma$), 
   99.73\% ($3\sigma$), 99.994\% ($4\sigma$) and $1-5.7\times 10^{-7}$ 
   two-dimensional confidence regions from the $\chi^2$ fit to these 
   results are drawn as solid lines.
   (a) includes results anticipated from a $75~\invab$ \superb
   run at the $\Upsilon(4S)$ only (no data from $\psi(3770)$ running.  
   In (b) a $10~\invfb$ threshold run by BES~III
   is estimated to provide a factor three improvement in the 
   uncertainties arising from the Dalitz plot amplitude model, and 
   a factor six in the measurement of the strong phase for 
   $\Dz\to\Kp\!\pim$ and $\Kp\!\pim\!\piz$ decays.  In (c) a $75~\invab$
   \superb run at the $\Upsilon(4S)$ is combined with a $500~\invfb$ run at    the
   $\psi(3770)$, assumed to reduce DP model uncertainty by a
   factor 10 and in the strong phases by factor given by the ratio of 
   luminosities wrt CLEO-c.  Note that the figures use the SAME vertical
   and horizontal scales.
  }
  \label{fig:babarmix2}
  \end{center}
\end{figure*}

This assumption is treated differently in the three \superb scenarios 
illustrated in Figs.~\ref{fig:babarmix2}(a)-(c).  In these, measurement 
of strong phases from data taken at $D\bar D$ threshold by CLEO-c
\cite{Lowery:2009id},
from data anticipated to come from BES~III, and from a dedicated $500~\invfb$
run at $\psi(3770)$ by \superb, are examined.  In Fig.~\ref{fig:babarmix2}(a),
no such data are included (in fact, CLEO-c results make little difference).
BES~III measurements are included in Fig.~\ref{fig:babarmix2}(b) and the
putative \superb measurements in Fig.~\ref{fig:babarmix2}(c).

Our assumption is that such measurements will not only provide the average 
strong phases to be used in various channels, but will also improve our DP 
model.  Without detailed modeling of ways to include this, we
estimate, a factor 3 improvement in DP model uncertainty for BES~III data 
and a factor 10 for the \superb threshold run.

Averages that result from these assumptions for all scenarios are reported 
in Table~\ref{tab:babarmix}.

The most precise localization of \xd and \yd is seen to
come from the golden channels, and it is clear that uncertainties in the 
DP amplitude models represent a limiting factor in this precision.
It is also noteworthy that the TDDP measurements for $\Kp\pim\piz$ decays, 
with linear dependence on $x^{\prime\prime}$, provide a better localization 
of \xd and \yd than do those from the WS $\Kp\pim$ mode, with dependence 
only on $x^{\prime 2}$.

The most dramatic improvements in precision of \xd and \yd result from a 
better understanding of DP amplitude models.

LHCb has projected the statistical uncertainties expected in $\Dz$ mixing observables corresponding to a $10~\invfb$ run
\cite{Spradlin:2007pi}.  
In Table~\ref{tab:dmixyields}, we compare these projections with the same observables measured in \babar and also with their projections to a $75~\invfb$ \superb sample at $\Upsilon(4S)$.
\begin{table}[!hbt]
\caption{Event yields and projected statistical uncertainties for various
observables for the final \babar sample, a projected $10~\invfb$
(approximately five year) LHCb run
and for a $75~\invab$ \superb run at $\Upsilon(4S)$.  For 
\babar, the yields for published mixing results using both $D^*$-tagged 
and untagged $\Km\!\Kp$ and for WS $\Kp\pim$ events are scaled up from 
published results to the final integrated luminosity of $482~\invfb$.  
LHCb estimates come from Ref.~\cite{Spradlin:2007pi}.}
\label{tab:dmixyields}
\begin{tabular}{lccc}
\hline
Decay Mode               
 & \babar           & \superb           & LHCB \\[3pt]
\hline
\boldmath\bf$\Kp\!\Km$ ($D^*$-tag): \\
$N$ (Events)             
 &  $88\times 10^3$ & $13.7\times 10^6$ &   $8\times 10^6$ \\
$\Delta\ycp$ (stat)
 & $\pm 3.9\times 10^{-3}$ & $0.28\times 10^{-3}$ & $0.5\times 10^{-3} $ \\
\boldmath\bf$\Kp\!\Km$ (no tag):    \\
$N$ (Events)             
 & $330\times 10^3$ & $51.4\times 10^6$ &   -- \\
$\Delta\ycp$ (stat)             
 & $\pm 2.3\times 10^{-3}$ & $0.19\times 10^{-3}$ & -- \\
\boldmath\bf$\Kp\!\pim$ (WS):    \\
$N$ (Events)
 & $5.1\times 10^3$ & $0.79\times 10^6$ & $0.23\times 10^{6}$ \\
$\Delta y'$ (stat)
 & $\pm 4.4\times 10^{-3}$ & $0.31\times 10^{-3}$ & $0.87\times 10^{-3}$ \\
$\Delta x'^2$ (stat)
 & $\pm 3.0\times 10^{-4}$ & $0.21\times 10^{-4}$ & $0.64\times 10^{-4}$ \\
\hline
\end{tabular}
\end{table}

Table~\ref{tab:babarmix} indicates the statistical uncertainties in the values of \xd and \yd expected in the various averaging scenarios.  The LHCb collaboration also plans to use 3-body golden channels and decays to the 4-body, self-conjugate state $\Kp\!\Km\!\pip\!\pim$.  These channels will, we can assume, benefit from BES~III results from $\psi(3770)$ in a way similar to that discussed above.  No projections for these modes are yet available, nor are systematic uncertainties.

In summarizing these projections,  it is clear that interesting levels of sensitivity in mixing measurements in the precision range of $10^{-4}$, are achievable both by \superb and LHCb, though \superb can be expected to do better.  It is observed that uncertainties in \xd are typically twice those of \yd.  This is probably due to the precise \ycp measurements.  It is also clear that the TDDP analyses for golden channels are most important, and that the main limiting factor for all experiments is the DP model uncertainty.  This can be largely mitigated using BES~III $\psi(3770)$ data.  A further factor two improvement is also possible with a $500~\invfb$ $\psi(3770)$ run of \superb.

\subsubsection{Estimated sensitivity to \CPV from mixing measurements}

\CPV in mixing, or in its interference with decay can reveal information
on the underlying parameters in the mass matrix
\cite{Kagan:2009gb},
and would have an important bearing on the role of NP.
A simple strategy for studying \CPV is to measure asymmetries in
{\em effective} values $(\xd^+,\yd^+)$ for \Dz and $(\xd^-,\yd^-)$ for \Dzb.  
Systematic uncertainties will be almost identical for \Dz as for \Dzb, so 
their contribution to uncertainties in these differences can be neglected.  
Statistical uncertainties are listed in Table~\ref{tab:babarmix} and it is 
seen that \superb will be sensitive, at the $3\sigma$ level, to a 
difference $\xd^+-\xd^-~(\yd^+-\yd^-)$ of $5(3)\times 10^{-4}$ in the 
average $x (y)$ values.

If observed, and if they were due to \CPV in mixing, they would provide a 
measurement of $|\qd/\pd|$.  Neglecting direct \CPV, $\xd^+\simeq |\qd/\pd|\xd$ 
and $\xd^-\simeq |p/q|\xd$, with similar relations for $\yd^+$ and $\yd^-$.  
Asymmetries are, therefore, given by
\begin{equation}
\label{eq:xydiff}
  a_{z} = \frac{z^+ - z^-}{z^+ + z^-}  \approx\frac{1-\left|\frac{q}{p}\right|^2}{1+\left|\frac{q}{p}\right|^2},
\end{equation}
where $z$ can be $\xd, \yd, \ycp, \yp, \xpp$ or $\ypp$.

This test can be made in different decay modes.  If \CPV originates in the 
decay, rather than in mixing, then the asymmetries will depend on the mode.  

These asymmetries are largely independent of systematic uncertainty.  
Statistical uncertainties for various \babar analyses are projected 
to \superb to obtain the precisions in $|\qd/\pd|$ listed in Table~\ref{tab:d0cpv}.  If \CPV originates in mixing, these asymmetries should be the same in all modes.  
\begin{table*}[!hbt]
\caption{Estimates for uncertainties in \CPV mixing parameters $|\qd/\pd|$ and
$\phi_{\sst M}$ obtainable at \superb using various methods.  Asymmetries 
$a_z$ and $a_{SL}$ are as defined in
the text, and are determined for the observables and channels indicated.
Time-dependent Dalitz plot (TDDP) analyses, allowing \CPV, include
scenarios where uncertainties from the decay model are reduced from Belle
estimates \cite{Abe:2007rd} by either a factor 3 (``BES~III DP model") or 
a factor 10 (``\superb DP model'').}
\label{tab:d0cpv}
\begin{tabular}{lccc}
\hline
{\bf Strategy}      &  {\bf Decay}       
& {\bf\boldmath $\sigma(|\qd/\pd|)\times 10^2$} 
& {\bf\boldmath $\sigma(\phi_{\sst M})^{\circ}$} \\[3pt]
\hline
\multicolumn{3}{l}{
\bf HFAG (direct \CPV allowed):}  \\[4pt]
Global $\chi^2$ fit
 & $<$All modes$>$   
 & $\pm 18$     
 & $\pm 9$   \\[8pt]
\hline
\multicolumn{3}{l}{\bf Asymmetries $a_z$:}  \\[4pt]
$    \xd $          & $<$All modes$>$    & $\pm 1.8$     &    --      \\
$    \yd $          & $<$All modes$>$    & $\pm 1.1$     &    --      \\
$    \ycp $         & $\Kp\!\Km$         & $\pm 3.8$     &    --      \\
$    \yp $          & $\Kp\!\pim$        & $\pm 4.9$     &    --      \\
$    \xpsq $        & $\Kp\!\pim$        & $\pm 4.9$     &    --      \\
$    \xpp $         & $\Kp\!\pim\!\piz$  & $\pm 5.4$     &    --      \\
$    \ypp $         & $\Kp\!\pim\!\piz$  & $\pm 5.0$     &    --      \\[4pt]
\hline
\multicolumn{3}{l}{\bf TDDP (\CPV allowed):}  \\[4pt]
Model-dependent     & $\KS h^+\!h^-$     & $\pm 8.4$     &  $\pm 3.3$ \\
BES~III DP model    & $\KS h^+\!h^-$     & $\pm 3.7$     &  $\pm 1.9$ \\
\superb DP model    & $\KS h^+\!h^-$     & $\pm 2.7$     &  $\pm 1.4$ \\[4pt]
\hline
\multicolumn{3}{l}{\bf SL Asymmetries $a_{SL}$:}  \\[4pt]
$75~\invab$ at $\Upsilon(4S)$
                    & $X\ell\nu_{\ell}$  & $\pm 10$      &            \\
$500~\invfb$ at $\psi(3770)$
                    & $K\pi$             & $\pm 10$      &            \\
$500~\invfb$ at $\psi(3770)$
                    & $X\ell\nu_{\ell}$  &   TBD         &            \\
\hline
\end{tabular}
\end{table*}

For the golden channels, a direct measurement of \CPV parameters $|\qd/\pd|$ 
and $\phi_{\sst M}$, is possible.  \babar has yet to make this measurement, 
so the statistical and systematic uncertainties obtained from the Belle 
analysis
\cite{Abe:2007rd}
are projected, in Table~\ref{tab:d0cpv} to the \superb luminosity at 
$\Upsilon(4S)$.  Uncertainties arising from the Dalitz plot model will 
be important, and a \superb run at threshold will increase the \CPV reach, 
as indicated in the Table.

A third metric for \CPV also comes from measurement of the asymmetry:
\begin{equation}
\label{eq:asldef}
  a_{SL} =   \frac{\Gamma_{\ell^-}-\bar\Gamma_{\ell^+}}
                  {\Gamma_{\ell^-}+\bar\Gamma_{\ell^+}}
         =   \frac{|q|^4-|p|^4}{|q|^4+|p|^4},
\end{equation}
where $\Gamma_{\ell^-}~(\bar\Gamma_{\ell^+})$ are decay rates for
``wrong-sign" semi-leptonic (SL) $D~(\bar D)$ decays.  Such decays can 
only occur, without NP contributions, through mixing, and have a 
time-dependence $\propto t^2e^{-\Gamma t}$.  Though difficult to measure, 
this asymmetry can be large.  For the current world average value for $|\qd/\pd|$
\cite{Schwartz:2009jv},
it is in the 90\% confidence range $a_{SL}\in\{+0.3,-0.75\}$.  If its
measured value is not zero, then it is clear evidence for \CPV in 
mixing and, therefore, for NP.

Measurement of $a_{SL}$ using $\Upsilon(4S)$ data can only come from WS,
SL decays, so far unseen by any experiment.  The precision
achievable, $\sigma(a_{SL}\sim\pm\sqrt{S+B}/S$, is limited by the number 
of background events, $B$, beneath a WS signal of $S$ events in the sample
(of both polarities combined) selected for the measurement.
 
An estimate of $S$ at \superb can be obtained from RS signals that have
been observed by \babar, using the reasonably well-known mixing rate 
$R_M\simeq 5\times 10^{-5}$.  Background $B$ under a WS signal can be 
expected at roughly the same level as under the corresponding RS signal.

With these assumptions, we can estimate $S(B)$ from two \babar observations.  
The first, a
conventional, singly-tagged RS $D^{*+}\to \Dz\pi_s^+~(\Dz\to \Km\!e^+\nu_e$ 
sample
Ref.~\cite{Aubert:2004bn} 
predicts $S (B)\sim 2140 (\sim~3M)$ events at \superb, resulting in an
uncertainty, $\sigma(a_{SL})\sim 0.8$ that is not useful.
The second, a double-tagged sample
\cite{Aubert:2007aa},
with a reconstructed $D$ or $D^*$ of known flavour on the recoil side
is much cleaner and, when projected to \superb, predicts a WS signal 
of $S=50\pm 16$ events on a background $B=195$ leading to
$\sigma(a_{SL})\simeq 0.3$.  Improvements in the selection of the latter
signal are known, in retrospect, to be possible and can lead to a 
precision of $\sigma(a_{SL})\simeq 0.20$.  This would result in a 
measurement of $|\qd/\pd|$ with precision $0.10$, listed in 
Table~\ref{tab:d0cpv}.

This asymmetry can also be measured in $\psi(3770)\to\Dz\Dzb$ decays from
the $500~\invfb$ sample.  In this case, three types of event can be used:
\begin{eqnarray*}
  \begin{array}{lclclcl}
    \Dz &\to& K\pi            &;& \Dzb &\to& K\pi,            \\
    \Dz &\to& X\ell\nu_{\ell} &;& \Dzb &\to& X\ell\nu_{\ell}, \\
    \Dz &\to& X\ell\nu_{\ell} &;& \Dzb &\to& K\pi,           
  \end{array}
\end{eqnarray*}
in which both $\Dz$ and $\Dzb$ decay to the $K^{\pm}\!\pi^{\mp}$ mode with the
same sign $K$ mesons, where both $\Dz$ and $\Dzb$ decay semi-leptonically with 
leptons of the same sign, or where one of the $D$'s decays semi-leptonically 
and the other to $K\pi$ where the lepton on one side has the same sign as 
the $K$ meson on the other.  In the first case, Bose symmetry prevents
DCS decays to a WS $K\pi$ state so that mixing has to occur in either 
the \Dz or the \Dzb decay.

CLEO-c has published data with 630 events with a (RS-RS)
$K\pi$ on each side, with virtually no background events from their 280~\invpb
$\psi(3770)$ sample.  Guessing that the background for a (RS-WS) combination
(not published) is 0.1 events, then this projects to 56 signal and 178 
background events in our putative 500~\invfb sample.  This would allow a 
precision of approximately $26\%$ in $a_{SL}$ (13\% in $|\qd/\pd|$).  No data on
the second mode are published, though studies are underway.  

The third mode has, however, been observed by CLEO-c
\cite{Asner:private} 
and the electron events are virtually background free.  In these cases, however,
an ambiguity arises over which decay is WS.  The probability for the $K\pi$
decay to be DCS is $\sim 3\times 10^{-3}$ and for the SL decay to have been
preceded by mixing the probability is $\sim 5\times 10^{-5}$.  It is possible 
that the different (and coherent) time-dependencies can be used to distinguish 
these, but a simulation study must be made to estimate how well this could work.

We can, perhaps, imagine a combined result from all three modes with 
$\sigma(a_{SL}\sim~20$\% (10\% uncertainty in $|\qd/\pd|$).  This estimate, with 
other estimates for \CPV reach, are included in Table~\ref{tab:d0cpv}.  Asymmetries
in other mixing parameters come closer to challenging SM estimates than do those 
from $a_{SL}$ measurements.

\subsection{CP Violation}

\subsubsection{Generalities}

On the phenomenological level one differentiates between two classes of CP violation, namely 
{\em indirect} CP violation residing in $\Delta C =2$ dynamics driving oscillations and 
{\em direct} CP violation affecting $\Delta C=1$ decays. These two sources can produce three 
classes of effects \cite{CPBOOK}: 
\begin{enumerate}
\item 
`CP violation {\em in} $D^0 - \bar D^0$ oscillations': due to the SM's selection rules this is most 
cleanly expressed through a difference in the transitions to `wrong-sign' leptons: 
\begin{eqnarray}
a_{SL}(D^0) &\equiv& \frac{\Gamma (D^0(t) \to \ell^-\bar\nu K^+) - 
\Gamma (\bar D^0 \to \ell^+\nu K^-)}
{\Gamma (D^0(t) \to \ell^-\bar\nu K^+) + \Gamma (\bar D^0 \to \ell^+\nu K^-)},\nonumber\\ 
&=& \frac{|q_D|^4 - |p_D|^4}{|q_D|^4 + |p_D|^4} \; . 
\end{eqnarray}
While the fraction of wrong-sign leptons oscillates with the time of decay, the fractional 
asymmetry does not. Data tell us that the production rate of `wrong-sign' leptons in $D$ 
decays is very low. Yet as illustrated below their CP asymmetry could be rather large. 

It should be noted that also non-leptonic modes of neutral $D$ mesons 
depend on the quantity $|q_D/p_D|$, see Eq.~(\ref{SMALLT}).  
\item 
`CP violation {\em involving} $D^0 - \bar D^0$ oscillations': it can emerge in non-leptonic 
final states common to $D^0$ and $\bar D^0$ decays in qualitative, though of course 
not quantitative analogy to $B_d \to \psi K_S$. Relevant channels are 
$D^0 \to K_S\phi/\eta$, $K^+K^-/\pi^+\pi^-$, $K^+\pi^-$ on the Cabibbo allowed, once 
and twice forbidden levels, respectively. CP asymmetries are driven by 
$|q_D/p_D| \neq 1$ as well as Im$\frac{q_D}{p_D}\bar \rho (f) \neq 0$ with 
$\bar \rho (f) = T(\bar D^0 \to f)/T(D^0 \to f)$ denoting the ratio of decay amplitudes.  
Such asymmetries depend on the time of decay in a characteristic way, which can be 
well approximated by a linear dependence due to $x_D$, $y_D \ll 1$: 
\beq
\frac{\Gamma (D^0(t) \to f) - \Gamma (\bar D^0(t) \to f)}{\Gamma (D^0(t) \to f) + \Gamma (\bar D^0(t) \to f)}
\equiv S_f \frac{t}{2\bar \tau},
\label{SFDEF}
\eeq 
with 
\begin{eqnarray}  
S_f&=& - \eta_f y_D \left(\left|\frac{q_D}{p_D}\right|- \left|\frac{p_D}{q_D}\right|\right)\cos2\varphi + \nonumber \\
 && -\eta_f x_D \left(\left|\frac{q_D}{p_D}\right|+ \left|\frac{p_D}{q_D}\right|\right)\sin2\varphi,
\label{SMALLT}
\end{eqnarray}  
in the {\em absence} of {\em direct} CP violation. In that case one has a useful connection between 
the two asymmetries listed so far \cite{BBBR,NIR2}: 
\beq
S_f = - \eta_f \frac{x_D^2 + y_D^2}{y_D} a_{\rm SL}(D^0).
\eeq

\item 
`Direct CP violation' characterized by a difference in the moduli of the decay amplitudes 
describing CP conjugate transitions: 
\beq 
|T(D \to f)| \neq |T(\bar D \to \bar f)|. 
\eeq
For two-body final states it requires the presence of two coherent amplitudes differing 
in both their weak as well as strong phases. 

\end{enumerate}
Three-body final states with their much richer dynamical structure can provide us with 
more detailed information about the operators driving these decays 
\cite{miranda}. Accordingly they require a more involved analysis. Fortunately a great deal of experience 
exists on how to deal with it through Dalitz plot studies. A \sff provides a 
particularly suitable environment, since it allows one to study not only all charged particle final 
states like $D^{\pm} \to \pi^{\pm}\pi^+\pi^-$ but also ones with neutrals like 
$D^0 \to \pi^+\pi^-\pi^0$ and $D^{\pm} \to \pi^{\pm}\pi^0\pi^0$. Comparing transitions 
with different charge combinations provides insight into the impact of the strong interactions. 
A working group of theorists and experimentalists has been formed under the 
name `Les Nabis' \cite{NABIS} to refine the theoretical tools for Dalitz plot studies 
to a degree that the huge statistics anticipated from a \sff can be exploited.  
While a full Dalitz plot description has to be the ultimate goal, achieving it represents a long term 
task. A model independent method has been proposed in Ref.\cite{miranda} as an intermediate 
step at least.  

\subsubsection{SM Expectations}

As far as {\em direct} CP violation in the SM is concerned, it can occur only in singly Cabibbo 
suppressed channels, but not in Cabibbo allowed and doubly suppressed ones, where one has 
only a single weak amplitude. Thus any observation of a CP asymmetry in the latter establishes 
the intervention of NP -- except for final states containing $K_S$ mesons, where the CP odd 
component in the $K_S$ wave function induces an asymmetry \cite{CPBOOK}. 
Cabibbo suppressed modes like $D^0 \to K^+K^-$, $\pi^+\pi^-$ are expected to show 
direct CP violation within the SM, yet only on the ${\cal O}(10^{-4})$ level. 

While $D^0 - \bar D^0$ oscillations are dominated by long distance dynamics within the SM, 
CP violation can arise there through $|q_D/p_D| \neq 1$ via a deficit in weak universality, 
albeit only on less than the $10^{-3}$ level~\cite{URIDOSC}. Time dependent CP asymmetries involving 
oscillations can arise also in the SM. Since, however, they are driven by terms of the form 
$x_D$ or $y_D$ $\times$ Im$\frac{q_D}{p_D}\bar \rho (f)$, they cannot exceed 
the $10^{-5}$ level. 

{\em In summary}: Due to the impact of non-perturbative dynamics that are beyond firm theoretical control one cannot 
make accurate predictions on SM CP asymmetries in charm decays. Nevertheless one can make 
highly non-trivial ones, as sketched above, namely that they are at best tiny. One 
can{\em not count} on NP creating large CP asymmetries in $D$ transitions, but its manifestations might be clearer here than in $B$ decays;
for the SM creates much smaller  ``backgrounds"; i.e., SM \CP effects are much larger in $B$ decays than in $D$ decays,
thus:
\beq 
\left[ \frac{\rm exp. \; NP\; signal}{\rm SM \; CP~``backgr."}\right]_{{\bf D}} > 
\left[ \frac{\rm exp. \; NP\; signal}{\rm SM \; CP~``backgr."}\right]_{{\bf B}}.
\eeq

\subsubsection{Experimental Landscape}

While it is an experimental fact that no evidence for CP violation has emerged 
in charm transitions so far, one should not over-interpret this statement.  In 
particular, CP asymmetries involving oscillations depend on expressions of the 
form $x_D$ or $y_D$ $\times$ weak phases and with $x_D$ and $y_D$ $\leq 1\%$ 
one can hardly exceed the 1\% level. To put it differently: only recently has 
one entered a regime where NP has a chance to induce an observable asymmetry, 
yet now any improvement in experimental sensitivity could reveal an effect. 

\CPV in a decay $D\to f$ results in a time-integrated asymmetry
\begin{equation}
\label{eq:ti_asymm}
  A_{f} = \frac{\Gamma-\bar\Gamma}{\Gamma+\bar\Gamma},
\end{equation}
where $\Gamma$ is the decay rate, and $\bar\Gamma$ is that for the conjugate 
decay $\bar D\to\bar f$.  For charged $D$ or $D_s$ any asymmetry would arise
from direct \CPV, while for \Dz, it could also result from \CPV in the mixing
or the interference between mixing and decay (indirect \CPV).  In this case, for a small asymmetry,
\begin{equation}
\label{eq:tizero_asymm}
  A_{f} = a_d+a_m+a_i,
\end{equation}
where $a_d$, $a_m$ and $a_i$ are, respectively, asymmetries resulting from direct CPV in the decay, mixing and from the interference between these.

Until recently, measurements of $A_f$ were limited to a precision of a few times $10^{-2}$ by two experimental uncertainties.  The first arose from an asymmetry of order $10^{-2}$ in detection and reconstruction efficiency between positively and negatively charged particles that used to be measured using samples of events generated in simulations.  These simulations are always limited in the precision with which they can mimic differences in interactions between positive and negative particles and the various detector components.  The second uncertainty arose from the poorly-known production asymmetry inherent in $\epem\to c\bar c$ interactions resulting from $Z^0-\gamma$ interference, and from higher order effects (ISR, FSR, box diagram, etc.)  Together with the built-in forward-backward asymmetry in detector efficiency, this led to an unknown apparent asymmetry, also of order $10^{-2}$.   

A way to overcome these uncertainties has recently been found by the \babar collaboration
\cite{Aubert:2007if}
in their measurement of $A_{\Kp\!\Km}$ from $D^*$-tagged $D^*\to\Dz(\to\Kp\!\Km)\pi_s$ decays.  They use data rather than simulations to estimate the charge asymmetry in efficiency for the $\pi_s$'s, measuring the ratios of $\pi_s$ to the corresponding $\Dz\to\Km\!\pip$ decay for each charge separately. These ratios should be the same, assuming only that these Cabibbo favored decays exhibit no \CPV.  The uncertainty in the charge asymmetry, once a subject of simulation-limited systematic uncertainty, is thereby limited only by the data sample size.  They also eliminated the effect of the production asymmetry simply by
evaluating $A_f$ in slices of production angle.  

With these innovations, a precision in $A_f^{\Kp\!\Km}$ of $3.6\times 10^{-3}$ (upper limit of $\sim~1\%$) was obtained.  $A_f$ for $\Dz\to\pi^+\!\pi^-$ and for $\Dz\to\pi^+\!\pi^-\!\pi^0$ modes were also measured with precisions in the range $4-5\times 10^{-3}$
\cite{Aubert:2007if,Aubert:2008yd}.

The Belle collaboration
\cite{belle:2008rx}
used these technique to obtain a similar precision in the $\Kp\!\Km$ mode,
and have recently adapted it to decays of charged mesons in the modes
$D^+_{(s)} \to K_S^0\pi^+$ and $D^+_{(s)} \to K_S^0K^+$
\cite{Ko:2010ng},
obtaining comparable precisions in the few times $10^{-3}$ range.
These decays are expected to have an asymmetry, induced by the \CP-odd
component in $\KS$, of $-2\hbox{Re}\{\epsilon_K\}\simeq -3.3\times 10^{-3}$
\cite{Bigi:2001sg}.

Since systematic uncertainties can now be estimated from data, they should scale in the same way as statistical uncertainties.  \superb, therefore, should be able to
achieve a precision of a few parts in $10^{-4}$ in similar measurements.  
For these singly Cabibbo-suppressed decays, this is close to SM expectations.
Furthermore, the SM limit for $D^+_{(s)} \to K_S^0\pi^+$ and $D^+_{(s)} \to K_S^0K^+$ decays of $\simeq -3.3\times 10^{-3}$ can be confronted with a truly precise measurement at \superb.  It is conceivable, therefore, that effects from NP should be observable!

The effects of \CPV in multi-body final states such as $\Dz\to h^+\! h^-\!\piz$ 
will ultimately be more likely to appear in the sub processes of which it is made.
For instance, $A_f$ may be expected to be different for $\Dz\to\rho^0\piz$ from 
that for $\Dz\to f_0(980)\piz$.  Asymmetry in the total $h^+\! h^-\!\piz$ system 
would, therefore, probably be diluted.  \babar examined this possibility in 
several ways for these channels
\cite{Aubert:2008yd}.  
Differences in the (normalized) \Dz and \Dzb Dalitz plot distributions, and 
their Legendre polynomial moments in each of the three channels were examined 
for structure.  None was found, to a precision of a few parts in $10^{-2}$.

Such model-independent tests are virtually free from PID and other experimental
asymmetries, since rates are normalized to the total number of events in
\Dz and \Dzb Dalitz plots.  A $75~\invfb$ at the $\Upsilon(4S)$ sample from 
\superb should, therefore, be capable of observing a \CPV effect, for instance,
in $\Dz\to\rho^0\piz$ at a few parts per $10^{-3}$.

Other tests of \CPV can also be projected to \superb performance levels.
Measurements of the ``$T$-odd" quantity
\cite{Bigi:2001sg}
\begin{equation}
\label{eq:xydiff}
  C_T = \vec p_{\Kp}\cdot\left(\vec p_{\pip}\times\vec p_{\pim}\right),
\end{equation}
can provide a sensitive test for $\Dz\to\Kp\!\Km\!\pip\!\pim$ decays.
In practice, it is necessary, to eliminate the effects of final state
interactions, to measure the difference in asymmetries $A_T-\bar A_T$,
where
\begin{equation}
\label{eq:xydiff}
  A_T = \frac{\Gamma(\Dz, C_T>0)-\Gamma(\Dz, C_T<0)}
             {\Gamma(\Dz, C_T>0)+\Gamma(\Dz, C_T<0)},
\end{equation}
and $\bar A_T$ is the corresponding quantity for $\Dzb$ decays.  This has
recently been measured by the \babar collaboration
\cite{Aubert:2010xj}
who find a value $(1.0\pm 6.7)\times 10^{-3}$.  The systematic uncertainty
includes a contribution of $3.5\times 10^{-3}$ from PID.  Improvements in PID
are contemplated at \superb, so the limiting precision could be somewhat
better than this - perhaps $2\times 10^{-3}$.

\subsubsection{Littlest Higgs Models with T Parity -- A Viable Non-ad-hoc Scenario}

What has changed over the last two years -- and is likely to produce further `fruits' in the future -- 
is that theorists have developed {\em non-}ad-hoc scenarios for NP -- i.e. ones {\em not} motivated by considerations of flavour dynamics -- that are {\em not} minimal flavour violating  
\cite{BBBR,NIR2}. 

`Little Higgs' models are motivated by the desire to `delay the day of reckoning'; i.e.,  
to reconcile the {\em non}-observation of NP effects in the electroweak parameters even on the 
quantum level with the possibility to discover NP quanta via their direct production in LHC 
collisions. A sub-class of them -- Little Higgs models with T parity -- are {\em not} minimal flavour violating in general and in particular can generate observable CP violation in charm decays 
\cite{BBBR}. Since they are relatively `frugal' in introducing extra parameters, observing 
their quanta in high $p_{\mathrm t}$ collisions would allow one to significantly tighten predictions 
of their impact on $K$, $D$ and $B$ decays.  

While these models are hard pressed 
to generate values for $|q_D/p_D|$ {\em out}side its present experimental range of 
$0.86^{+0.17}_{-0.15}$, they can well induce it {\em in}side it; i.e., they could move 
$|q_D/p_D|$ much further away from unity than the less than $10^{-3}$ amount expected 
for the SM. Likewise they 
could produce CP asymmetries in $D^0 \to K_S\phi$, $K^+K^-$, $\pi^+\pi^-$ up to the 
1\% level; i.e., much larger than the $10^{-5}$ SM expectation. It should also be noted that in some parts of the parameter space of these models 
their impact could not be identified in $B$ decays: in particular the CP asymmetry 
in $B_s \to \psi \phi$ would still remain below 5\% as predicted in the SM. Their strongest 
correlation exists with the branching ratio for the ultra-rare mode 
$K_L \to \pi^0 \nu \bar \nu$ \cite{BBBR}. 

\subsection{Rare Decays}

\subsubsection{$D^0 \to \mu^+\mu^-$, $\gamma \gamma$}

$D^0 \to \mu^+\mu^-$ has, potentially, the cleanest experimental signature
if seen at rates greater than $\Dz\to\gamma\gamma$. 
However, its rate suffers greatly from helicity suppression and the
need for weak annihilation -- two effects that are basically model independent. In the SM the rate is estimated to be greatly dominated by long-distance dynamics -- yet on a very tiny level 
\cite{Burdman:2003rs}:
\begin{eqnarray}
{\rm BR}(D^0 \to \mu^+\mu^-)_{\rm SM} &\simeq& {\rm BR}(D^0 \to \mu^+\mu^-)_{\rm LD}  \nonumber \\
&\simeq  &
3\cdot 10^{-5} \times {\rm BR}(D^0 \to \gamma\gamma )_{\rm SM}. \nonumber\\
\end{eqnarray}
With the SM contribution to $D^0 \to \gamma \gamma$ again being dominated by long-distance 
forces \cite{Burdman:2003rs}
\beq
{\rm BR}(D^0 \to \gamma\gamma )_{\rm SM} \simeq {\rm BR}(D^0 \to \gamma\gamma )_{\rm LD} 
\sim (1 \pm 0.5)\cdot 10^{-8} \; , 
\eeq
one infers 
\beq 
{\rm BR}(D^0 \to \mu^+\mu^-)_{\rm SM} \sim 3 \cdot 10^{-13}, 
\eeq
to be compared with the present bounds 
\bea 
\label{MUMU}
{\rm BR}(D^0 \to \mu^+\mu^-)_{exp} &\leq& 5.3 \cdot 10^{-7},  \\ 
\label{2GAMMA}
{\rm BR}(D^0 \to \gamma \gamma)_{exp} &\leq& 2.7 \cdot 10^{-5}.  
\eea
The bound of Eq.(\ref{2GAMMA}) implies a bound of $10^{-9}$ in Eq.(\ref{MUMU}) -- i.e., a 
much tighter one. 
In either case there is a rather wide window of opportunity for discovering 
NP in $D^0 \to \mu^+\mu^-$. As pointed out in \cite{PETROV2} in several NP models there is actually a relatively tight 
connection between the NP contributions to BR$(D^0 \to \mu^+\mu^-)$ and 
\xd.

Specifically, LHT makes short-distance contributions to $D^0 \to \mu^+\mu^-$ and 
$D^0 \to \gamma \gamma$ that can be calculated in a straightforward way as 
a function of viable LHT parameters. Their size is under active study now \cite{AYAN}. No matter what 
drives $D^0 \to \gamma \gamma$ - whether it is from short or long distance dynamics -- it 
provides a long distance contribution to $D^0 \to \mu^+\mu^-$. For a proper interpretation of these 
rare $D$ decays it is thus important to search for $D^0 \to \gamma \gamma$ with as high a 
sensitivity as possible.

\subsubsection{$D \to l^+l^- X$}

It has been suggested that studying $D \to \gamma X$ etc. is very unlikely to 
allow establishing the presence of NP because of uncertainties due to long 
distance dynamics 
\cite{Burdman:2003rs}. 
The same strong caveat probably 
applies also to $D \to l^+l^- X$, unless
a CP asymmetry is observed there, in particular in the lepton spectra. 
However, BaBar's experimental limits on the $D^+\to\pi^+ l^+l^-$ 
branching fractions~\cite{Aubert:2006aka}, obtained with $288~\invfb$,
are about an order of magnitude above the theoretical
calculations based on long-distance effects~\cite{Fajfer:2007qb}.
Therefore, the high SuperB luminosity should enable 
probing of these effects, as well as provide hadronic-mode
measurements needed to improve the calculations.
For many $D^0\to Xl^+l^-$ decays, the only existing upper limits are 
from searches by CLEO~\cite{Freyberger:1996it}, performed with 
$3.85~\invfb$.  
There is clearly much room for improvement in these modes.

\subsection{Experimental possibilities for rare decay searches at \superb}

The scale for rare decay rates at \superb is set by the numbers of $D$ mesons
that will be produced.  From a $75~\invab$ sample at $\Upsilon(4S)$ 
$\sim~7.5\times 10^{10}$ are expected.  A $500~\invfb$ sample at $\psi(3770)$ 
produces $\sim~2\times 10^9$.  A special advantage in rare decay searches from
the use of events at threshold is that backgrounds are extremely low, in
most instances.

The \superb reach can be estimated for the $\Dz\to\mu^+\mu^-$ rate from current 
measurements from \babar (and Belle).  The best published limit on the 
$\Dz\to\mu^+\mu^-$ rate so far is from \babar ($13\times 10^{-7}$) 
\cite{Aubert:2004bs}.
Lower, unpublished limits are also available now: $4.3\times 10^{-7}$ from CDF
\cite{Casey:2008hcp} 
(360 $\invpb$, unpublished) and $1.4\times 10^{-7}$ from the Belle collaboration 
\cite{Won:2009eps}.
A further, similar result from \babar is also imminent.

Results from neither of these experiments are yet limited by systematic 
uncertainty.  The major source of background in each case is from 
$\Dz\to\pip\pim$ decays, where the $\pi$'s either decay in flight to $\mu$ or 
are mis-identified by the PID devices.  This background peaks at a mass below, 
but has a significant tail in the $\mu^+\mu^-$ invariant mass signal region.  
There is also a flat, combinatorial background from semi-leptonic $B$ decays 
that is hard to eliminate.  These two sources account for 90\% of the background.

In the significantly larger \superb samples from $\Upsilon(4S)$ running, 
these backgrounds should be relatively simple to parameterize, and the major 
limiting factor should be the uncertainty in their shapes.  It is reasonable 
to assume that limits in the lower $10^{-8}$ level should be achievable.  
For comparison, the LHCb experiment can also reach a level of about 
$2,5\times 10^{-8}$ before reaching a systematic limit.  

Prospects for searches for these decays in a \superb run at threshold could 
provide an interesting opportunity in the search for NP.  Further study is, 
however, still required and is ongoing.  BES~III estimates are for a limit of
$1.7\times 10^{-6}$ per $\invfb$, but this estimate cannot be easily scaled
up to the $500~\invfb$ anticipated at \superb.

At the $\psi(3770)$, there will be no background from
$B$ decays of course, but the most serious background will come from
$\Dz\to\pip\pim$ decays.  The \Dz's are produced with virtually no transverse 
momentum so, in this view, the muons have equal and opposite momenta - an 
excellent kinematic signature that should help reduce the number of 
$\pi^{\pm}\to\mu^{\pm}\nu_{\mu}$ decays in flight, which confuse PID selectors.  
The muons from $\Dz\to\mu^+\mu^-$ decays will have laboratory momenta of 
$\sim 0.9-1.0$~\gevc, a range where none of the \babar PID devices would
work well in separating them from pions.  \superb PID systems, however, 
should perform significantly better.  Many of the muons will hit the end-caps 
where a TOF device, one of the \superb options being considered, should 
perform well.  The focusing DIRC option for the barrel PID system should 
also perform significantly better in distinguishing $\mu$ from $\pi$ than 
the \babar DIRC in this momentum range.  Kinematic separation of $\pip\pim$
and $\mu^+\mu^-$ modes is also attainable,
making use of the beam energy constraints used in many CLEO-c analyses.
These issues are under simulation study at this time.  It is conceivable that
running at charm threshold will open up the low $10^{-8}$ or high $10^{-9}$ 
range of sensitivity, a very interesting range for this important branching 
fraction.

Decays of $\Dz\to\epem$ and (LFV) $\Dz\to e^{\mp}\mu^{\pm}$ should also be 
accessible at rates that can be estimated from current \babar and Belle 
results.  The first of these decays is predicted to be $\sim~10^{-18}$, by 
the SM.  The LFV modes can occur in the SM at a rate $\sim 10^{-14}$.  
Observation of these modes at \superb would require explanation beyond the SM.

The best rates so far, ${\cal B}(\Dz\to\epem)<7.9\times 10^{-8}$ and 
${\cal B}(\Dz\to e^{\pm}\mu^{\mp})<2.6\times 10^{-7}$ are preliminary from Belle
\cite{Won:2009eps}.  In the modes with electrons, more background will exist 
from $\gamma$ conversions, but PID would be more reliable than for the 
$\mu^+\mu^-$ mode.  \superb should be able to achieve rates an order of 
magnitude lower than this.

Upper limits on the $\Dz\to\gamma\gamma$ decay rate, of importance in 
estimating long range effects in SM calculations of rare decay modes such as 
$\Dz\to\mu^+\mu^-$, have been published.  The best so far ($<~2.9\times 10^{-5}$)
comes from CLEO 
\cite{Coan:2002te}
using a $13.8~\invfb$ sample taken at $\Upsilon(4S)$.  \babar should publish a 
limit in the region of $2.5\times 10^{-6}$ in the near future using $481~\invfb$.  
Extending this to \superb, it is possible to reach the mid to low $10^{-7}$ 
range at the $\Upsilon(4S)$.

Prospects for a measurement of the $\Dz\to\gamma\gamma$ decay rate at threshold
are good.  A recent CLEO-c thesis
\cite{Asner:2010:cleocthesis}
demonstrates an efficiency for this mode at 5.2\%, with no background events
detected in a $818~\invpb$ sample at $\psi(3770)$.  A reasonable projection
from this is that a limit of a few times $10^{-7}$ can be achieved, a very 
useful clarification of the SM and of the true $\Dz\to\mu^+\mu^-$ rate.

\subsubsection{$D \to l^+l^- X$}

The \superb reach can be estimated from limits in these modes obtained 
by the \babar collaboration
\cite{Aubert:2006aka}
using a $288~\invfb$ sample.  
These are a few parts per million for $\Dp$ modes and a factor three larger 
for $D_s$.  
Best limits for $\Dz$ modes are in the $10^{-3}$ to $10^{-4}$ range
\cite{Freyberger:1996it}
from a $3.85~\invfb$ sample.
One of the largest backgrounds in all cases again comes from semi-leptonic $B$ 
decays.  With a $75~\invab$ sample, this can be modeled quite well from data 
sidebands - more precisely than in \babar.  It is reasonable to expect that 
rates an order of magnitude lower can be achieved, pushing several rates within
the range of SM predictions
\cite{Fajfer:2007qb, Burdman:2003rs}
from long range effects.
At charm threshold, CLEO-c can probe rates for decays of $\Dp$ and for $D_s$ 
in the few times $10^{-6}$ range
\cite{Asner:private},
comparable to rates from \babar.  The projection to a run of $500~\invfb$ at
threshold, therefore, \superb could outperform results from the $\Upsilon(4S)$
by a factor $\sim 2$.

Better estimates for the reach achievable in the modes discussed will require 
more simulation, mostly because PID
devices are an important component of each result.  \superb PID should, in all 
cases, be superior to \babar, so these limits may err on the conservative side.
More information from CLEO-c data, as yet unpublished, should also be forthcoming.
This can be used to improve on estimates of the expected performance from data
at charm threshold.

\subsection{A case for Running at the $D\bar D$ threshold?}

If taken sufficiently early in the \superb run plan, the data sample at 
$\psi(3770)$ could lead to significant discoveries in a much shorter period 
than would be possible without it.  One example is the search for 
$\Dz\to\mu^+\mu^-$ decay.  If this occurs at a rate of a few times 
$10^{-8}$, a sure sign of NP, a signal could be seen in $\Upsilon(4S)$ 
data only after the full 75~\invab sample is analyzed.  Meanwhile, 
the LHCb experiment will have already seen this signal too.  The same 
signal could be seen in the $\psi(3770)$ data that could be taken in a 
few months' running at reduced luminosity
\footnote{A re-configuration of the final focus is required for such a run.  
The machine is designed to make this possible in a period of order 
one month.  It is unlikely that the $\psi(3770)$ run could require as 
much as a whole year to complete.}.
Were this data taken early in the \superb schedule, a more competitive, and 
complementary discovery would be possible much sooner.

There are many good reasons to collect events at (or near) charm threshold.  
For such events, tagging those in which one $D$ meson is identified, the 
other $D$ can be studied with very small background contamination.  These 
can, therefore, be used to search for rare decays and, potentially, 
asymmetries that are especially sensitive to backgrounds.

In several charm studies, particularly those involving leptonic or SL 
decays of charm particles, the CLEO-c collaboration has shown that data 
with an integrated luminosity of 818~\invpb of data taken at the $\psi(3770)$
can provide measurements that are competitive with, or superior to, those 
from approximately three orders more data from \babar and Belle at the 
$\Upsilon(4S)$.  The proposed $500~\invfb$ data sample at $\psi(3770)$, 
almost 1\% of the integrated luminosity anticipated at $\Upsilon(4S)$, 
and the availability of time-dependent information in the decays from 
the (albeit modest) boost, represent an even richer prospect for discovery. 

Running at charm threshold also provides an essential check on any new 
discovery in $\Upsilon(4S)$ data.  A confirmation and possible 
clarification of the result can made more easily in a different charm 
production scenario than in a repetitive study with a significantly enhanced 
sample at $4S$.

Quantum correlations in decays of $D$ pairs from $\psi(3770)$ can lead to 
measurements of their relative strong phases.  It is conceivable that, 
with the \superb boost, any weak phase could, in principal, also be 
studied, though the precision achievable requires further study.

In addition to the aforementioned studies, one could test for possible
\CPT violation in $\Dz\Dzb$ decays, which could be manifest through 
Lorentz violation (For example see 
Refs.~\cite{Kostelecky:1994rn,Colladay:1995qb,Kostelecky:2001ff})
or decoherence effects (For example see Ref.~\cite{Mavromatos:2008bz} 
and references therein) of the correlated wave-function
of the neutral mesons.  This is an interesting area that needs to be 
studied, as \CPT could be violated in different ways in $K$, $D$, and $B$
meson states.

Below, we summarize a few specific areas where $\psi(3770)$ data have 
already been noted to add to the physics reach from that achievable with 
$\Upsilon(4S)$ data.

\begin{itemize}

\item {\em Search for $\Dz\to\mu^+\mu^-$} \\[6pt]
These rare decays are a clean signal for new physics, if seen.  At 
$\Upsilon(4S)$, our upper limit should be in the low to mid $10^{-8}$ 
range when PID efficiency (for mis-identification) will limit our reach.  
At $\psi(3770)$, we would rely on kinematic separation in addition to PID, 
so a lower limit is conceivable, provided that kinematic resolution is
sufficiently good.  This needs further study, but we can expect to achieve a limit similar to that from $\Upsilon(4S)$ data.

\item {\em Improved precision in mixing parameters \xd and \yd} \\[6pt]
Running at $D\bar D$ threshold allows independent measurements of strong
phases $\deltakpi$, $\deltaktwopi$, \etc. for channels that we will use for
mixing measurements at the $\Upsilon(4S)$.  In Table~\ref{tab:babarmix} and
Figs.~\ref{fig:babarmix}(c) and (d), results that might be expected
from inclusion of such measurements of $\deltakpi$ from the 10\invfb threshold
sample expected to come from BES III and also what we would expect from
a 500\invfb at \superb are indicated.  As can be seen, running at threshold
brings a factor 2 in precision of the measurements of \xd and \yd.

A large part of this improvement comes from a model independent strong phase 
measurement over the $\KS h^+h^-$ Dalitz plots.  This improvement will also 
apply to measurements of CKM $\gamma$ that are limited by these models.  
The improvement should be a factor 3 or more than that which will be 
available from BES~III data from $\psi(3770)$.

\item {\em Measurement of $a_{\sst SL}$} \\[6pt]
This parameter cannot be measured well at $\Upsilon(4S)$.  We expect a 
precision of only about $\pm 0.8$ from the small sample of WS $SL$
decays.  Prospects are somewhat better at the $\psi(3770)$, using 
WS $K\pi$ decays to obtain a precision of about $20$\%.  This situation 
could improve to a more useful level if, following studies with CLEO-c 
data yet to be made, we find that we are  able to use events in which 
both $\Dz$'s decay to $SL$ modes.

\item {\em Time-Dependent Measurements} \\[6pt]
A study needs to be made to see if time-dependent correlations in $D$
decays from $\psi(3770)$ can provide any information on weak phases in
the charm sector, a signal for NP.

\item {\em Impact on the measurement of $\gamma$ } \\[6pt]
Measurements of $D$ decays at charm threshold can be used to reduce model 
uncertainties on the extraction of $\gamma$, not only for \superb, but also 
LHCb.  See Section~\ref{sec:bphysics:gamma} for more details.

\end{itemize}

\vspace{5mm}

\vspace{15mm}

\graphicspath{{Polarisation/}{Polarisation/}}
\section{Electroweak neutral current measurements}\label{sec:ew}

The combination of high luminosity and polarized electrons at \superb\ provides a unique
opportunity to measure a number of electroweak neutral current parameters with 
 precisions comparable to those obtained at SLC and LEP but at a $Q^2$ of $(10.58~\gev)^2$.
 The cross-sections for $e^+e^-\rightarrow \mu^+\mu^-$, as for the other  
 final-state fermions, are sensitive to the beam polarization almost entirely
 through $Z-\gamma$ interference. Although the asymmetries are small, the  \superb\
  sample size will be sufficiently large to yield very interesting physics.
 This physics program includes precision $\sin^2 \theta_{W}$ measurements with $\mu^+\mu^-$,
 $\tau^+\tau^-$ and $c \bar{c}$ events as well as measurements of the neutral current
 vector coupling of the $b$. Such measurements are sensitive to a $Z^{\prime}$ and
 can probe neutral current universality at high precision.

 With polarization, \superb\ will make a relatively
  straightforward measurement of the left-right asymmetry of $e^+e^-\rightarrow \mu^+\mu^-$
in a manner identical to that performed by the SLC collaboration~\cite{Abe:1994wx,Schael:2005ema}
 which operated at the Z-pole. SLC measured $\sin^2 \theta_{W}=0.23098\pm0.00026$ where the
error includes a systematic uncertainty component of $\pm 0.00013$  dominated
by the polarization uncertainty of $0.5\%$.
 The ZFITTER software has been used to estimate the level of sensitivity that might be
 reached at \superb\ where the left-right asymmetry is be approximately $-0.0005$.
 A  $e^+e^-\rightarrow \mu^+\mu^-(\gamma)$ selection using \babar\ data
 had a selection efficiency of 53\% for a 99.6\% purity. Such a selection 
 will provide a sample of 46 billion $\mu$-pair events at \superb\
for an integrated luminosity of $75~ab^{-1}$.
 Assuming 80\% polarization can be achieved, 
 the statistical error on the left-right asymmetry will be approximately $5\times 10^{-6}$ 
 which corresponds to a relative error of ${\cal O}(1\%)$. If the polarimeter systematic errors
 can be kept below this level, the uncertainty on  $\sin^2 \theta_{W}$ will be $\sim 0.0002$, which is
 competitive with the SLC measurement but at a much lower $Q^2$. Similar measurements can be made
 with $e^+e^-\rightarrow \tau^+\tau^-(\gamma)$ and with charm, although one would expect the statistical 
 errors to be larger owing to a lower selection efficiency. Nonetheless,
  those measurements will provide the most stringent tests of neutral current universality.
 
\begin{figure}[htb!]
  \begin{center}
    \includegraphics[width=0.45\textwidth]{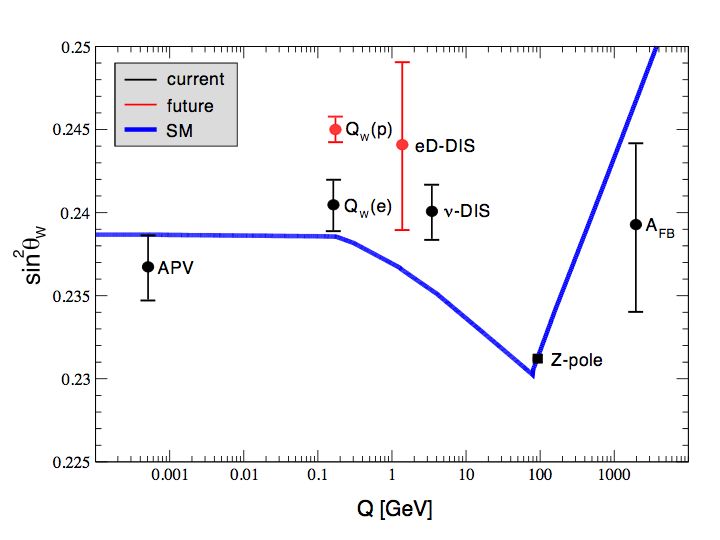}
    \caption{
Summary of experiments that have measured or are proposing to measure $\sin^2 \theta_{W}$
as compiled in \cite{Qweak}.
The standard model  running of $\sin^2 \theta_{W}$ is overlaid on the data points.
\superb\ will provide a point
 at $Q=10.58~\gev$ with an error comparable to that of the measurement at the Z-pole. }
        \label{fig:sin2thetarun}
  \end{center}
 \end{figure}

 These precision measurements are sensitive to the same new physics scenarios, such as a $Z^{\prime}$,
  being probed by
 the QWeak experiment at the Jefferson Laboratory, which will measure 
 $\sin^2 \theta_{W}$ to approximately $0.3\%$ at  $Q^2=(0.16\gev)^2$.  Figure~\ref{fig:sin2thetarun}
 shows the current and planned measurements of $\sin^2 \theta_{W}$.

 As \superb\ will be running on the $\Upsilon(4S)$, the left-right asymmetry for B-mesons will
 be sensitive to the product of the electron neutral current axial coupling and b-quark
 neutral current vector coupling $g^b_V$, as described in a
 proposal for measuring the $g^s_V$ at a $\phi$-factory~\cite{Bernabeu:1995bu}.
 With one billion reconstructed $B\bar{B}$ events from $\Upsilon(4S)$ decays with an 80\% polarized
 beam, \superb\ will provide a measurement of $g^b_V$ that is competitive with the measurement
 from LEP and SLC, $g^b_V = -0.3220\pm0.0077$\cite{:2005ema} but at a lower $Q^2$. 
 In addition to probing new physics, this measurement will shed light on the long-standing
 3$\sigma$ difference between the measurements of $\sin^2 \theta_{W}$ obtained from the forward-backward
 asymmetry of b-quarks and those obtained using leptons.

 We note that other asymmetry measurements at \superb,
 such as the forward-backward left-right asymmetry can provide additional
 information about neutral current couplings.

\graphicspath{{spectroscopy/}{spectroscopy/}}
\section{Spectroscopy}\label{sec:spectroscopy}

\subsection{Introduction}

Although the Standard Model is well-established,
QCD, the fundamental theory of strong interactions, provides a quantitative
comprehension of phenomena at very high energy scales, where perturbation theory is effective due
to asymptotic freedom.
The description of hadron dynamics below the QCD dimensional transmutation scale, in spite
of the success obtained with numerical simulations on the lattice, is not under full theoretical control.

Systems that include heavy quark-antiquark pairs
(quarkonia) are a unique and, in fact, ideal laboratory for probing both the high
energy regimes of QCD, where an expansion in terms of the coupling
constant is possible, and the low energy regimes, where
non-perturbative effects dominate.
For this reason, quarkonia have been studied for decades in great detail.
The detailed level of understanding of the quarkonia mass spectra
is such that a particle mimicking quarkonium properties, but not fitting any quarkonium level,
is most likely to be considered to be of a different nature.

In particular, in the past few years the $B$ Factories and the Tevatron have provided evidence for states that
do not admit the conventional mesonic interpretation and that instead could be
made of a larger number of constituents. While
this possibility has been considered since the
beginning of the quark model~\cite{ GellMann:1964nj}, the actual identification of such states
would represent a major revolution in our understanding of elementary particles. It would also imply the existence
of a large number of additional states that have not yet been observed.

Finally, the study of the strong bound states could be of relevance to
understanding the Higgs boson, if it turns out to be itself
a bound state, as predicted by several technicolor models (with or without extra dimensions)~\cite{Dietrich:2005jn,Contino:2006nn}.

The most likely possible states beyond the mesons and the
baryons are:

\begin{itemize}

\item {\bf hybrids:} bound states of a quark-antiquark pair and a number of
constituent gluons. The lowest-lying state is expected to have quantum numbers
$J^{PC}=0^{+-}$. Since a quarkonium state cannot have
these quantum numbers (see below), this is a unique signature for
hybrids. An additional signature is the preference for a hybrid to
decay into quarkonium and a state that can be produced by the excited
gluons (\eg\ $\pi^+\pi^-$ pairs); see \eg~Ref.~\cite{Kou:2005gt,Close:2005iz}.
\item {\bf molecules:} bound states of two mesons, usually represented as
$[Q\bar{q}][q^{\prime}\bar{Q}]$, where $Q$ is the heavy quark. The
system would be stable if the binding energy were to set the mass of the
states below the sum of the two meson masses.
While this could be the case for when $Q=b$, this does not apply for
$Q=c$, the case for which most of the current experimental data exist. In this case,
the two mesons can be bound by pion exchange. This means that only
states decaying strongly into pions can bind with other mesons (\eg\
there could be $D^*D$ states), but
that the bound state could decay into its constituents~\cite{Braaten:2003he,Close:2003sg,Tornqvist:2004qy,Swanson:2006st,Voloshin:2006wf,Fleming:2007rp,Braaten:2007dw,Braaten:2007ft}.
\item {\bf tetraquarks:} a bound quark pair, neutralizing its color with a bound
antiquark pair, usually represented as $[Qq][\bar{q^{\prime}}\bar{Q}]$. A full
nonet of states is predicted for each spin-parity, \ie, a large number
of states are expected. There is no need for these states to be close to
any threshold~\cite{Maiani:2004vq}.

\end{itemize}

In addition, before the panorama of states is fully clarified, there is always the
lurking possibility that some of the observed states are misinterpretations of
threshold effects: a given amplitude might be enhanced when new hadronic final
 states become energetically possible, even in the absence of resonances.

While there are now several good experimental candidates for unconventional states,
the overall picture is not complete and needs confirmation, as well as
discrimination between the alternative explanations.
A much larger dataset than is currently available is needed, at several energies, to pursue this program; this capability is uniquely within the reach  of 
\superb.

  \subsection{ Light Mesons}
 The problem of the interpretation of the light scalar mesons, namely
$f_0, a_0,\kappa$, and $\sigma$, is one of the oldest problems in hadron physics~\cite{gell}.
For many years the question about the existence of the $\sigma$ meson as a real resonance in $\pi\pi$ scattering has been debated~\cite{Tornqvist:1995ay};  only recently has a 
thorough analysis of $\pi\pi$ scattering amplitudes shown that the $\sigma(500)$ and $\kappa(800)$ can be considered as proper resonances~\cite{Caprini:2005zr,DescotesGenon:2006uk}.

Reconsideration of the $\sigma$ was triggered by the E791 analysis of $D\to 3\pi$ data~\cite{Aitala:2000xu}; a number of papers have commented on those results, 
\eg~Ref.~\cite{Gatto:2000hj, Tornqvist:2000wv,Tornqvist:2000wx,Deandrea:2000yc}.  The role of the scalar mesons in several exclusive $B$ decays could be rather relevant: for example, in the perspective of a high 
precision measurement of the $\alpha$ angle at the \superb factory, the hadronic contributions, like the one of the isoscalar $\sigma$ in $B\to \rho \pi$, must 
be properly controlled~\cite{Deandrea:2000ce,Gardner:2001gc,Bigi:2005fr}.
Also several studies on light and heavy scalar mesons could be
performed analyzing the Dalitz plots of exclusive decays like $B\to KKK$ and $B\to K \pi\pi$. In this respect, having sufficient statistics to clearly
assess the presence of a scalar $\kappa (800)$ resonance, would certainly be a major result for hadron spectroscopy.

Beyond the ``taxonomic'' interest in the classification of scalar mesons, the idea that these mesons could play a key role in our understanding of
aspects of non-perturbative QCD has been raised several times; see for example Ref.~\cite{gribov}.

In what follows we would like to underscore the latter point by observing that:

\begin{itemize}%

\item Light scalar mesons are most likely the lightest particles with an {\it exotic} structure, \ie, they cannot be classified as $q\bar q$ mesons.
\item Their dynamics is tightly connected with instanton physics.  Recent discussions have shown that instanton effects make possible a consistent model for the description of light scalar meson dynamics, under the hypothesis that these particles are diquark-antidiquark mesons.%

\end{itemize}%

Therefore, new modes of aggregation of quark matter could be established by the experimental/theoretical investigation of these particles, further expanding 
the role of instantons in hadron physics.

The idea of four-quark mesons dates back to the pioneering papers by Jaffe~\cite{Jaffe1,Jaffe2,Jaffe:2004ph}, while the discussion of exotic mesons and hadrons in terms of 
diquarks was introduced in Ref.~\cite{Jaffe:2003sg} and then extended in Ref.~\cite{Maiani:2004uc} to the scalar meson sector.

We will assume that the scalar mesons below $1$~GeV are indeed  bound states of  a spin~0 diquark and an anti-diquark (we will often call 
this a tetraquark).  
 A spin~0 diquark field is a color antitriplet $\qq=qq$ bound state (same color of an antiquark).

As in a standard $q\bar q$ meson, the color is neutralized between a diquark and an antidiquark $\qq^\alpha {\bar \qq}_\alpha$. 
Since a spin zero diquark is in a ${\bf \bar 3}$-flavor representation because of Fermi statistics,  
flavor nonets of $\qq\bar \qq$ states are allowed, the so called `crypto-exotic' multiplets. We believe that the sub-GeV scalar mesons most likely represent the lowest tetraquark nonet.

\begin{figure}[!htb]
\psfig{file=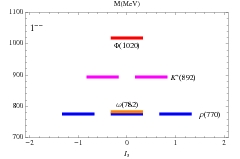,width=4cm}
\includegraphics[width=40mm]{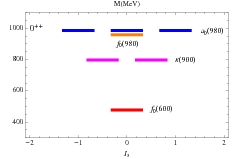}
\caption{\label{fig:i3}
Vector mesons ($q\bar q$ states) and the sub-GeV scalar mesons in the $I_3-m$ plane.}
\end{figure}

The $\qq\bar \qq$ model of light-scalars is very effective at explaining the most striking feature of these particles, namely  their inverted pattern, with 
respect to that of ordinary $q\bar q$ mesons, in the  mass-versus-$I_3$ diagram~\cite{Jaffe1,Jaffe2,Jaffe:2004ph}, as shown in Fig.~\ref{fig:i3}.

Such a pattern cannot be explained in a $q\bar q$ model where, for example, the $f_0(980)$ would be an $s\bar s$ state~\cite{Gatto:2000hj, Tornqvist:2000wv,Tornqvist:2000wx, Deandrea:2000yc} while the
$I=1$, $a_0(980)$, would be a $u\bar u+d\bar d$ state. If this were the case, the degeneracy of the two particles would be rather unnatural.

Besides a correct description of the mass-$I_3$ pattern, the tetraquark model offers the possibility of explaining the decay rates of scalars at a level 
never reached by standard $q\bar q$ descriptions.
The effective decay Lagrangian into two pseudoscalar mesons, \eg\ $\sigma \to \pi\pi$, is written as:
\begin{equation}
{\cal L}_{\rm 1}=c_1 S^i_j  \epsilon^{j t u}\epsilon_{i r s}  \partial_\mu \Pi^r_t  \partial^\mu \Pi^s_u,
\label{exchange}
\end{equation}
where $i,j$ are the flavor labels of $\qq^i$ and $\bar\qq^j$, while $r,s,t,u$ are the flavor labels of the quarks $\bar q^t,\bar q^u$ and  $ q^r, q^s$. $c_1$ 
is an effective coupling and ${S},{\Pi}$ are the scalar and pseudoscalar matrices of meson fields. Observe for example how $\pi^+\pi^-$ are produced by a $[ud][\bar u \bar d]$ tetraquark by setting the right  flavor indices in Eq.~(\ref{exchange}). 

This Lagrangian describes the quark 
exchange amplitude for  the quarks to tunnel out of their diquark shells in $S$ to form ordinary pseudoscalar mesons $\Pi$~\cite{Maiani:2004uc}.  The antisymmetrization in the flavor indices of quarks (${\bf \bar 3}-$flavor representation) is guaranteed by the $\epsilon$ tensors.

Such a mechanism is the straightforward  alternative to the most natural color 
string breaking  $\qq\ \gluon q\bar q\gluon\ \bar \qq\to B\bar B$, {\it i.e.}, a baryon-anti-baryon decay, which happens to be phase-space forbidden to sub-GeV scalar 
mesons. For a discussion about baryonia see~\cite{Cotugno:2009ys}.

The problem with Eq.~(\ref{exchange}) is simply that it is not able to describe the observed decay $f_0\to \pi\pi$, since $f_0\sim[qs][\bar q\bar s]$, with $q=u,d$.
To form a $\pi^+\pi^-$ pair of mesons in the final state one should require to:
$i)$ break the diquarks binding to annihilate the $s$ and the $\bar s$ quarks $ii)$ 
create a $q\bar q$ pair from the vacuum.
Alternatively one could annihilate the diquark and the antidiquark directly into a $q\bar q$ pair via 
a six-fermion interaction, not paying the price of breaking the diquark shells, and hadronize the two light quarks produced into two pions via a quark pair creation. This possibility is provided by six-fermion, instanton induced low energy vertices~\cite{Hooft:2008we}. Such vertices contain a term of the form ${\cal I}=\sum_{i,j} \bar \qq_i \qq_j \bar q_j q_i$, $i,j$ being flavor indices and 
$\qq_{i\alpha}= \epsilon_{ijk}\epsilon_{\alpha\beta\gamma} \bar q^{j\beta}_C\gamma_5 q^{k\gamma}$ being a spin zero diquark.

Alternatively, one can go through a mixing between the two isoscalars $f_0$ and $\sigma$. 
However, as discussed in~\cite{Hooft:2008we}, such mixing is expected to be too small, $<5^\circ$, to account for the structure of the inverted mass pattern (a precise determination of the $\kappa$ mass would be crucial to fix this point).

\begin{figure}[!htb]
\includegraphics[width=80mm]{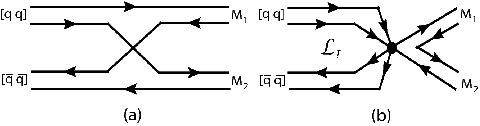}
\caption{\label{fig:qdiags}
Decay of a tetraquark scalar meson $S$ in two $q \bar{q}$ mesons $M_1 M_2$: (a) quark rearrangement (b) instanton-induced process.}
\end{figure}

Thus  in addition to  the quark-exchange diagrams, described at
the effective theory level by the Lagrangian of Eq.~(\ref{exchange}), (see
Fig.~\ref{fig:qdiags}~(a)), we a have six-fermion microscopic interaction of the form ${\cal I}$ (see 
Fig.~\ref{fig:qdiags}~(b)~\footnote{The six-fermion interaction expands to terms of the
  form: $(\bar u^\alpha (1-\gamma_5)u_\alpha)(\bar d^\alpha
  (1-\gamma_5)d_\alpha)(\bar s^\alpha (1-\gamma_5)s_\alpha)$.
  Upon appropriate Fierz rearrangement of, \eg, $(\bar d^\alpha
  (1-\gamma_5)d_\alpha)(\bar s^\alpha (1-\gamma_5)s_\alpha)$, one obtains: $C
  \times (\bar u^\alpha (1-\gamma_5)u_\alpha)
  \qq^{1\gamma}{\bar\qq}_{1\gamma}$, $C$ being a constant factor.})  
which contributes 
to the following effective Lagrangian term:
\begin{equation}
{\cal L}_2=c_2 {\rm Tr} ({{\bf S}  (\partial {\bf \Pi})^2} ),
\label{einst}
\end{equation}
(roughly, introduce a $\bar q_k q_k$ in ${\cal I}$ and call $S_j^i\sim \bar \qq_j \qq^i$, $\Pi_j^i\sim \bar q_j q^i$ respectively).
$c_2$ is an effective coupling expected to be rather smaller than $c_1$ in Eq.~(\ref{exchange}).
Observe that this term is also contained in Eq.~(\ref{exchange}) which actually corresponds to the combination $2{\rm Tr}({\bf S (\partial \Pi)^2})-{\rm Tr}{\bf S}{\rm Tr}(\partial{\bf \Pi})^2$, barring the contribution from the singlet pseudoscalar. The latter term could be described by an `annihilation' diagram at the meson level.  

If on the other hand we assume that the lowest scalar nonet is made up of standard $\bar q q$ mesons, there are no diquarks around, and we expect the instanton contributions to enter only in operators of the kind ${\rm Tr}{\bf S}{\rm Tr}(\partial{\bf \Pi})^2$. Thus the decay Lagrangians to be used to fit data in the $4q$ and $2q$ hypotheses are:
\begin{eqnarray}
&&{\cal L}^{(4q)}= {\cal L}_{\rm 1}(c_1)+{\cal L}_2(c_2),\nonumber\\
&&{\cal L}^{(2q)}= {\cal L}_{\rm 1}(c_2^\prime)+{\cal L}_2(c_1^\prime),\nonumber
\end{eqnarray}
with evident notation. It is expected $|c_1^{(\prime)}|\gg |c_2^{(\prime)}|$.

With such a description of the dynamics one can determine numerical results for the decay amplitudes as reported in Table~\ref{spectroscopy:tab:res} (four-quark fit $|c_1|\simeq 0.02, |c_2|\simeq 0.002$).
\begin{table}[!htb]
\begin{center}
\caption{\label{spectroscopy:tab:res}
\footnotesize Numerical results for amplitudes in GeV. Second and third columns: results obtained with a decay  Lagrangian including or not including instanton 
effects, respectively (Labels $I$ and no-$I$ mean that we add or do not add the instanton contribution).
No $f_0-\sigma$ mixing is assumed in this table. Fourth column: best fit, see text, with instanton effects included. Fifth column: predictions for a $q\bar 
q$ picture of the light scalars.
The $\eta-\eta^\prime$ singlet-octet mixing angle assumed: $\phi_{_{PS}}=-22^\circ$~\cite{Escribano:2005qq,Gerard:2004gx}.
Data for $\sigma$ and $\kappa$ decays are from~\cite{Caprini:2005zr,DescotesGenon:2006uk}, the reported amplitudes correspond to: $\Gamma_{\rm tot}(\sigma) = 272 \pm 6$, $\Gamma_{\rm 
tot}(\kappa) = 557 \pm 24$.
 }
\begin{tabular}{lccccc}
\hline
Proc. &   \multicolumn{3}{c}{ ${\cal A}_{\rm th}([qq][\bar q \bar q])$} &${\cal A}_{\rm th}(q\bar q)$& ${\cal A}_{\rm expt}$  \\
    & $I$ &  no-$ I$ & best fit & $I$ & \\ \hline
$ \sigma   ( \pi^+\pi^-) $ &
{\rm input}  &   {\rm input}  & 1.7 &
{\rm input}  &$2.27 (0.03)$   \\ \hline
$\kappa^+   (K^0 \pi^+)  $ &   $5.0$ &
5.5 & 3.6 & 4.4&$5.2 (0.1) $   \\  \hline
$f_0  (\pi^+\pi^-) $    &    {\rm input} &{\bf 0}&1.6 &
{\rm input}  &$1.4 (0.6)$  \\
$f_0     (K^+ K^-) $    & $4.8$  & 4.5&3.8&  4.4 &$3.8(1.1)$  \\  \hline
$a_0   (\pi^0 \eta)$    &  $4.5$ & 5.4&3.0& 8.9&$2.8 (0.1)$    \\
$a_0  (K^+ K^-)$        &   $3.4 $  & 3.7&2.4 & 3.0 &$2.16(0.04)$  \\  \hline
\end{tabular}  \\ [2pt]
\end{center}
\end{table}
Such a good description of decays is possible {\it only} if the assumption is made that sub-GeV light scalars are
diquark-antidiquark mesons (see Table~\ref{spectroscopy:tab:res}). In the $q\bar  q$ hypothesis, the agreement of $a_0\to\pi^0\eta$ with data appears very poor.

A relative of the lowest lying scalar mesons may have been found very
 recently by \babar: the $Y(2175)$, a particle first observed in the decay
 $Y\to \phi f_0(980)$~\cite{Aubert:2006bu}.  For a discussion see Ref.~\cite{Drenska:2008gr}.

 \subsection{Charmonium}
In the past few years the $B$~Factories have observed several states with clear $c\bar{c}$ content,
which do not behave like standard mesons, and that are therefore an indication of new spectroscopy.

\begin{figure}[!htb]
\epsfig{file=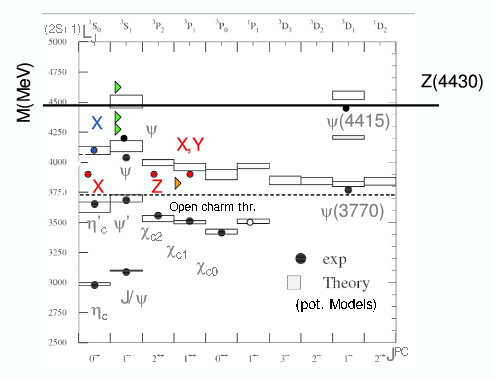, height=5cm}
 \caption{\label{fig:summary}
Measured masses of the newly observed states, positioned
in the spectroscopy according
to their most likely quantum numbers. The charged state ($Z(4430)$) clearly has
no $C$ quantum number.}
\end{figure}

The $X(3872)$ was the first state found not to easily fit into charmonium spectroscopy.
It was initially observed decaying into $J/\psi\pi^+\pi^-$ with a mass just beyond the open charm
threshold~\cite{Choi:2003ue}. The $\pi^+\pi^-$ invariant mass distribution, the observation of the
$X\to J/\psi\gamma$ and the full angular analysis by CDF~\cite{Abulencia:2006ma} and Belle~\cite{Abe:2005iya},
along with the evidence for the $X\to\psi(2S)\gamma$ decay found by \babar\ ~\cite{babar:2008rn},
favor the assignment of $J^{PC}=1^{++}$ for this state, and of $X\to J/\psi\rho$ as its dominant decay.
There are several indications that this is not a (pure) charmonium state: the mass assignment does
not match any prediction of long-verified potential models (see
Fig.~\ref{fig:summary}); the
dominant decay would be isospin-violating; and the state is narrow (less than a few MeV),
despite its mass lying above threshold for the production of two charmed mesons.  At the same time the relative rates to 
$\psi(2S)\gamma$ and $J/\psi\gamma$ are more easily explained in terms of conventional charmonium decays.  
The closeness to the $D^0D^{*0}$ threshold suggests also the hypothesis that it may be  a molecule composed of these two mesons or a threshold effect.

Another aspect of interest is the measurement of the
mass of the $X(3872)$ in the $D^{*0}D^0$ decay mode~\cite{Aubert:2007rva,belle:2008su}, which could differ from the value
measured in the $J/\psi\pi\pi$ decay. The mass difference and the difference in the line-shape in the two modes
could help in discriminating between the many models~\cite{Stapleton:2009ey,Dunwoodie:2007be}. 
If the mass difference is confirmed, it is possible that there are indeed two different states, one decaying to $D^{*0}D^0$ and the other decaying to  
$J/\psi\pi\pi$: the di-quarks with a heavy meson are effectively flavor-triplets, and di-quark pairs would
show the same nonet structure as ordinary mesons, so that it would be natural to expect two states with $S=I_3=0$
very close in mass~\cite{Maiani:2004vq}.

A data sample of ${\cal O}(50\, ab^{-1})$ would yield several (3 to 11) thousand fully reconstructed $B\to X(3872)K$ decays in each of the above-mentioned 
modes.
This would  allow a detailed study of the $X(3872)$ decay dynamics and line-shape, crucial to enlighten possible evidence for non-$q\bar q$ composition.

The $B$~Factories have also found a number of new states with $J^{PC}=1^{--}$ by looking for events where the initial state
radiation brings the $e^+e^-$ center-of-mass energy down to the particle's mass. It was expected that above the open charm threshold
all states would be seen in $R=\sigma_{had}/\sigma_{\mu\mu}$ scans. When the high luminosity at $B$~Factories allowed one to study exclusive final states 
containing a $J/\psi$ or a $\psi(2S)$, at least three new unusual particles were discovered:  the  $Y(4260)$ decaying to $ 
J/\psi\pi^+\pi^-$~\cite{Aubert:2005rm}, the
$Y(4350)$~\cite{Aubert:2006ge} and the $Y(4660)$~\cite{belle:2007ea} decaying to $\psi(2S)\pi^+\pi^-$.

The $\pi^+\pi^-$ invariant mass is a critical observable in discerning the nature of these particles,
which are unlikely to belong to charmonium since there are already other $1^{--}$ known charmonium states, their masses are above the
open-charm threshold, yet they are relatively narrow and are not observed to
decay into two
charmed mesons (the most stringent limit being 
$\BR(Y(4260)\to D\bar{D})/\BR(Y(4260)\to J/\psi\pi^+\pi^-)<1.0$ at $90\%$ CL) Ref.~\cite{Aubert:2007pa}.
Another puzzling feature of these states is the ratio of the partial widths  $\Gamma(J/\psi\pi^+\pi^-)/\Gamma(\psi(2S)\pi^+\pi^-)$, that is small for the 
$Y(4260)$
and large for the $Y(4350)$ and  $Y(4260)$. The current statistics does not allow one
to measure these ratios.

\begin{figure*}[!htb]
\begin{center}
\epsfig{file=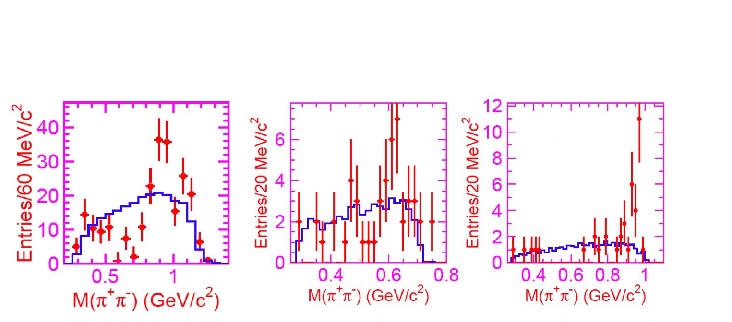,height=6cm}
 \caption{\label{fig:bellepipiinv}
 Di-pion invariant mass distribution in $Y(4260)\to\jpsi \pi^+\pi^-$ (left), $Y(4350)\to \psi(2S) \pi^+\pi^-$ (center),\newline
and $Y(4660)\to \psi(2S) \pi^+\pi^-$ (right) decays.
}

\end{center}
\end{figure*}

Figure~\ref{fig:bellepipiinv}
shows the dipion invariant mass spectra for all
regions in which new resonances have been observed. Only the $Y(4660)$ seems to show a well-defined intermediate
state (most likely an $f_0$), while others have a more complex
structure.

The $Y(4260)$ is currently considered a good hybrid candidate, while the $Y(4350)$ and  $Y(4660)$ are good candidates for
$[cd][\bar{c}\bar{d}]$ and $[cs][\bar{c}\bar{s}]$ tetraquarks, respectively. The latter would prefer to decay to 
$f_0$, while the mass difference is consistent with the hypothesis that the two belong to the same nonet.

An experiment with 50 $ab^{-1}$ of integrated luminosity, yielding samples of 30~K $Y(4260)\to J/\psi\pi^+\pi^-$ and $\approx 3~K$ events each for 
$Y(4350), Y(4660)\to\psi(2S)\pi^+\pi^-$, would allow a detailed study of the line-shape, a measurement of 
$\Gamma(J/\psi\pi^+\pi^-)/\Gamma(\psi(2S)\pi^+\pi^-)$, and a
study of the $\pi^+\pi^-$ invariant mass spectra, as well as of the angular distributions. Furthermore it will be possible to search for other exclusive 
decays to Charmonia such as 
$J/\psi\eta/\pi^0$, $\psi(2S)\eta/\pi^0$, $\chi_{cJ}\pi^+\pi^-$, $\gamma J/\psi$, and $\gamma\psi(2S)$.

\begin{figure}[!htb]
\epsfig{file=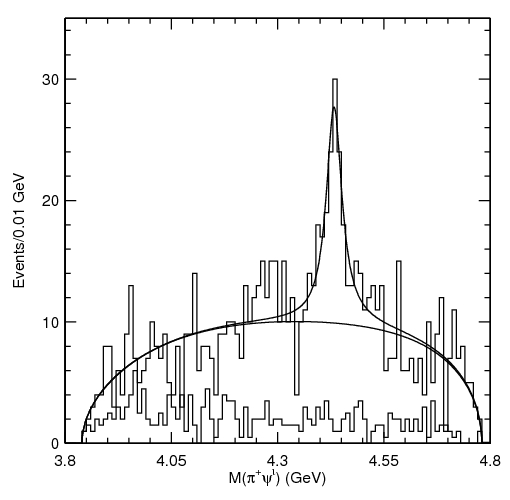,height=6cm}
 \caption{\label{fig:belleZ4430}
 The $\psi(2S)\pi$ invariant mass distribution in
\hbox{$B\to \psi(2S)\pi K$ decays.}}
\end{figure}

The turning point in the query for states beyond charmonium
has been the observation by the Belle Collaboration of a charged
state
decaying into $\psi(2S)\pi^\pm$~\cite{belle:2007wga,belle:2009da}
soon followed by two more charged states, the $Z_1^+(4050)$ and 
the $Z_2^+(4430)$, decaying to $\chi_{c1}\pi^+$\cite{Mizuk:2008me}.
Figure~\ref{fig:belleZ4430}
shows the fit to the  $\psi(2S)\pi$ invariant mass distribution in
$B\to \psi(2S)\pi K$ decays, returning a mass $M=4433\pm4$(stat.)$\pm2$(syst.) MeV/c$^2$
and a width $\Gamma=44^{+18}_{-13}$(stat.)$^{+30}_{-13}$(syst.) MeV.

Such states must
contain a $c$ and a $\bar{c}$, but according to their charge
they must also contain at least an $u$ and a $\bar{d}$. The only possibilities for explaining these state are the tetraquark or the molecule composition, or 
the presence of some threshold effects. The
latter two options are viable for the $Z^+(4430)$ due to the closeness of the $D_1D^*$
threshold.

The analysis is highly complicated by the presence of $K^*$ resonances in the $B\to (c\bar c) \pi^+K$ final state and by the $c\bar c$ polarization. The 
analysis of the full 
 \babar\ data sample did not confirm nor exclude the observation of the $Z^+(4430)$~\cite{babar:2008nk}. No result has yet been presented on the search for 
the $Z_1^+(4050)$ and 
$Z_2^+(4430)$.

It is critical to confirm the existence of these states, and if confirmed to find the corresponding neutral states and/or to observe them in other decay 
modes.
With an integrated luminosity of 50~ab$^{-1}$ we can expect to collect samples of 
100~K to 1.5~M fully reconstructed $B\to J/\psi \pi^+ K$,  
$B\to\psi(2S)\pi^+K$ and $B\to\chi_{cJ}\pi^+K$ events that will allow one 
to establish unambiguously the existence of these states and to determine their 
properties.

In summary, there are several reasons why a run at fifty to a hundred times the 
existing integrated luminosity is decisive to convert these hints into a solid picture:
\begin{itemize}
\item All the new states, apart from the $X(3872)$, have been observed in only
a single decay channel, each with a significance barely above 5$\sigma$.
A hundredfold increase in statistics would allow searches in several other
modes. In particular, it is important to observe both the decay to
charmonium and to $D$-meson pairs and/or $D_s$ meson pairs. Since the
branching fractions of observable final states for the $D$ and especially
for the $D_s$ mesons are particularly small, current experiments do not have
the sensitivity to observe all the decays.
\item Most models predict several other states, such as the neutral
partners of the $Z(4430)$ and the nonet partners, for instance
 $[cd][\bar{c}\bar{s}]$ candidates decaying into a charmonium state and a kaon,
 at a significantly lower rate
(see \eg~Ref.~\cite{Maiani:2007vr}) than the observed modes. Furthermore,
several of these states decay into particles (in particular neutral pions
and kaons) that have a low detection efficiency.
 \end{itemize}

In order to achieve high luminosities the event rate and the machine backgrounds will increase significantly. It is therefore important to estimate the impact 
of the changes in the detector and of this background on the search potentiality. As a first step it has been tested with a fast simulation of the 
$e^+e^-\to Y(4260) \gamma_{ISR}, Y(4260)\to J/\psi\pi\pi$ signal that the detector changes do not affect significantly the efficiency. 
A more comprehensive  study is on the way.

\begin{figure}[!htb]
\epsfig{file=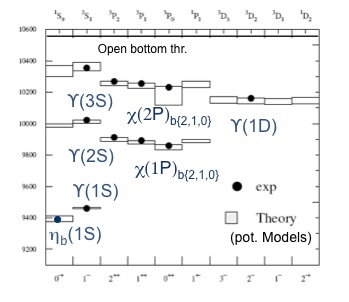, height=5cm}
 \caption{\label{fig:bottomonium}
Measured masses of the bottomonia, positioned
in the spectroscopy according
to their most likely quantum numbers. }
\end{figure}

 \subsection{Bottomonium}
In comparison to charmonium, our knowledge of bottomonium below flavor 
threshold is far from complete: in particular, as shown in Fig.~\ref{fig:bottomonium}, almost all the spectrum of spin 
singlet states (parabottomonia) is still {\it terra incognita}.
Moreover, in the bottomonium system, four narrow D wave states are expected
in the region around 10.16 GeV, and their study \cite{Bonvicini:2004yj}, 
started by CLEO-III, is currently under way in the present generation of B-factories. 
In total, the current generation of B-factories have integrated 
(1.2, 2.6, 1.3)*10$^8$
$\Upsilon$(1, 2, 3S) decays on resonance peak, as shown in table~\ref{tab:datasets}. 

\begin{table}[h]
\small
    \caption{$\Upsilon(nS, n\neq 4)$ datasets after the year 2000 at the \B-factories. }
    \label{tab:datasets}
\begin{center}
\begin{tabular}{l|cccc} \hline
Expt.   & $\Upsilon(1S)$     & $\Upsilon(2S)$ & $\Upsilon(3S)$ & $\Upsilon(5-6S)$ \\
\hline
CLEO    & 20M   &  9M   & 6M    & 0.5 ${\mathrm {fb}}^{-1}$ \\     
Belle   & 98M   &160M   & 11M   & 133 ${\mathrm {fb}}^{-1}$ \\    
\babar   &  -   & 100M   & 122M   & 3.3 ${\mathrm {fb}}^{-1}$ \\ 
\hline
\end{tabular}
\end{center}
\end{table}

Moreover, up to 133 fb$^{-1}$ were accumulated in the $\Upsilon$(5S) region, 
and have started yielding interesting results about transitions to narrow 
states through the open beauty threshold, defying na\"ive expectations.
The analysis of this data is in progress and will probably lead to new 
discoveries in the near future, but it is clear that ten to a hundred times 
the statistics are needed to find all the pieces of the bottomonium puzzle. 

 \subsubsection{Regular bottomonium}
\begin{figure}[h]
  \begin{center}
    \includegraphics[width=.49\textwidth]{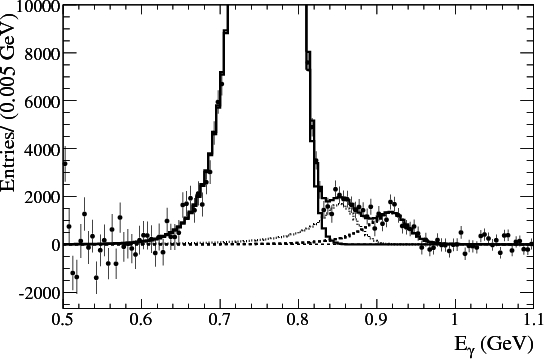}
     \caption{
The inclusive photon spectrum at 3S from \babar, after continuum
subtraction: the peaks from $\chi_{b1,2}\rightarrow\Upsilon(1S)$, ISR production of 
$\Upsilon(1S)$ and  $\Upsilon(3S)\rightarrow\eta_b$ are visible, left to right in the plot.}
    \label{fig:etab}
   \end{center}
\end{figure}

Only recently, the ground state $\eta_b$(1S) has been discovered
by \babar  \cite{Aubert:2008vj,Aubert:2009pz}, as shown in Fig.\ref{fig:etab},
but all other parabottomonia are still missing and surely two of them 
will hardly be within reach of the current generation of B-factories.
Besides the hyperfine splitting, other $\eta_b$ decay properties can be 
predicted with relatively small errors in the NRQCD approximation and deserve 
experimental verification: the total width and the partial 
width to two photons.

 The total width of $\eta_b$ should be measurable by \babar and Belle, at least
 as an upper limit, from the inclusive photon spectra of suppressed transitions. 
A precise measurement 
 (i.e. better than 10\% error) of the $\eta_b$ total width requires 
much higher statistics, which will be available only at a \sff.
Given the large photon background in the low energy part of the spectrum
(i.e. below 100 MeV), the experimentalists are challenged to detect all
the $\eta_b$ decay products on one or more specific channels and try an 
exclusive reconstruction.  
At present date, only few exclusive decay modes have been observed by CLEO
\cite{Asner:2008sx},
with significances above 5$\sigma$, for the $\chi_b$ states, 
which can be reached from 
$\Upsilon$(2, 3S) peaks  via transitions which have branching ratios 
at the 10\% level.
As ratios for direct M1 transitions to $\eta_b$ are expected 
in the 10$^{-4}$ range, at least two orders of magnitude increase in statistics 
is needed. 

In the long term the most important measurement to 
perform is that of the two-photon width, as theory 
predictions on the ratio $\Gamma_{\gamma\gamma}(\eta_b)/\Gamma_{l+l-}(\Upsilon)$ 
are quite insensitive to the renormalization scale \cite{Penin:2004ay}, and yield 
$\Gamma_{\gamma\gamma}(\eta_b)=0.66\pm0.09$ keV. 
If  $\Gamma_{tot}(\eta_b) < 10$ MeV
this would imply a branching ratio at the level of 10$^{-4}$, 
and a cross section 
$\sigma(\Upsilon(2, 3S)\to\gamma\eta_b\to\gamma\gamma\gamma)\sim 0.2$ fb, 
which is by 
far smaller than the cross section for the continuum process 
$\sigma(e^+e^-\to\gamma_{ISR}\to\gamma\gamma\gamma)$. 
The ISR background can actually be avoided by running on the $\Upsilon$3S 
resonance peak and 
using the dipion pair to tag the $\Upsilon$2S decay, and then 
select exclusive $\pi\pi\gamma\gamma\gamma$ events, from the process
$\Upsilon(3S)\to\pi\pi\Upsilon(2S)\to\pi\pi\gamma\eta_b(1S)$. 
In this case, experimentalists are challenged to maximize the efficiency 
on detection and tracking of low momentum dipion pairs.

The discovery of one or more exclusive decay modes of $\eta_b$(1S) 
will also be useful for the search of the analogous direct M1 transitions 
between vector and pseudoscalar 2S and 3S excitations.
For the time being, the current record sample of $\Upsilon(3S)$ decays can 
allow \babar to discover the $\eta_b$(2S) in the inclusive photon spectrum, 
and the $h_b(1P)$ either  via the cascade process 
$\Upsilon(3S)\to \pi^0 h_b(1P)\to\pi^0\gamma\eta_b(1S)$, 
as done by CLEO to  find
the $h_c(1P)$ state in the charmonium system, or via 
$\Upsilon(3S)\to \pi\pi h_b(1P)\to\pi\pi\gamma\eta_b(1S)$, as suggested in 
Ref.~\cite{Voloshin:1985em,Kuang:2006me}.

In order to discover the states $h_b$(2P) and $\eta_b$(3S) we 
probably need a \sff. While $\eta_b$(3S) detection should depend crucially on 
exclusive reconstruction of some decay channel, and it is almost certainly
reachable from the $\Upsilon$(3S), it is not yet clear which transition 
will allow us to reach $h_b$(2P): as the expected mass 
difference $M(\Upsilon(3S))-M(h_b(2P))<M(\pi^0) $, 
detection of $h_b(2P)$ cannot benefit from running on narrow bottomonia.

The recent discovery of unexpectedly large widths for the transitions
 $\Upsilon(4S)\to\eta\Upsilon(1S)$~\cite{Aubert:2008bv} and 
$\Upsilon$(5S)$\to\pi\pi\Upsilon(1S)$~\cite{Abe:2007tk}
may suggest that hadronic transitions to other narrow bottomonia can open new
pathways to these states, \eg\ $\Upsilon$(5S)$\to\eta h_b$(2P).
In the next section, we elaborate on the large physics potential 
of running above the $B_s\bar B_s$ threshold, also for hadron spectroscopy.

  \subsubsection{Exotic bottomonium}

This section discussed recent \babar and Belle scans and the future prospects for
high energy scans at \superb.

Exotic states with two bottom quarks, analogous to those with two charm quarks, could also exist.
In this respect, bottomonium spectroscopy is a
very good test-bench for
speculations advanced to explain the charmonium states.
As a down side, searching for new bottomonium states
is more challenging, since they tend to be broader and there are
more possible decay channels than the charmonium situation. 
This explains why there are still eight unobserved states with masses
below open bottomonium threshold.

Among the known states, there is already one with unusual behavior: there
has been a recent
observation~\cite{Abe:2007tk} of an anomalous enhancement, by two orders
of magnitude, of
the rate of $\FiveS$ decays to the $\OneS$ or a $\TwoS$ and two pions. This
indicates that either the $\FiveS$ itself or a state very close by in
mass has a
decay mechanism that enhances the amplitudes for these processes.

In order to understand whether the exotic state coincides with the
$\FiveS$ or not, a high luminosity
(at least 20 fb$^{-1}$ per point to have a 10\% error) scan of the
resonance region is needed.

In any case, the presence of two
decay channels to other bottomonium states excludes the possibility of
this state being a molecular aggregate, but all other models are possible, and 
would predict a large variety of not yet observed states.

As an example, one can estimate possible resonant states with the tetraquark
model, by assuming that the masses of states with two $b$ quarks can be obtained
from one with two $c$ quarks by adding the mass difference between
the $\OneS$ and the $\jpsi$. Under this assumption, which
works approximately for the known bottomonium states, we could expect three nonets
that could be produced by the $\ThreeS$ and decaying into $\OneS$ and
pions. Assuming that the production and decay rates of these new states
are comparable to the charmonium states, and assuming
a data sample of $\ThreeS$ events comparable in size to the current $\FourS$
sample is needed to clarify the picture, we would need about $10^9\ \ThreeS$
mesons, corresponding to an integrated luminosity of \hbox{0.3 ab$^{-1}$.}

As already mentioned, searching for bottomonium-like states would require
higher statistics than the corresponding charmonium ones; this therefore
represents an even stronger case for \superb.

\subsection{Interplay with other experiments}
\superb is not the only next generation experiment capable of investigating heavy quark spectroscopy.

The LHCb experiment is starting to investigate its potentialities in the field. 
The complementarity of these studies with \superb are evident, considering 
the present 
interplay between B-Factories and the Tevatron: the larger number of mesons 
produced allows detailed studies of the decay modes with final states made of 
charged particles. All other modes are best investigated by $e^+ e^-$ machines.

The only other next generation  experiment at an $e^+ e^-$ machine is BES-III, 
but their current plan is to run below the energies of interest, at the 
$\psi(3770)$~\cite{Asner:2008nq}, where they expect to  collect $5\invfb$ per year. 
Even if a plan to run at the energies of the exotic states were developed, given the lower luminosity the complementarity of \superb and BES-III would be the 
same as the B-Factories and CLEO-c.

A separate mention is deserved by the PANDA experiment at FAIR~\cite{Lutz:2009ff}, a proton-antiproton collider which could produce the exotic resonances at 
threshold (i.e. $e^+e^- \to X,Y$). This innovative production mechanism allows for copious production without the hindrance of fragmentation products. 
Considering the expected characteristics of the antiproton beam and an integrated luminosity of 2 $\invfb$ per year, running at the $J/\psi$ mass would yield 
3.5 $10^9$ $J/\psi$ mesons per year. Considering that $\Gamma_{ee}[Y(4260)]*{\cal{B}}(Y(4260)\to p\bar{p})<0.05 
\Gamma_{ee}[J/\psi]*{\cal{B}}(J/\psi\to p\bar{p})@ 90\%$ C.L.~\cite{Aubert:2005cb} and assuming $\Gamma_{ee}[Y(4260)]=\Gamma_{ee}[J/\psi]$, we could expect 
as many as 30K $Y(4260)\to J/\psi\pi\pi$ with a $J/\psi$ 
decaying leptonically per year. Besides the large uncertainty on the assumption, this estimate can be compared with the 60K events in the same decay chain 
produced in a year at \superb via ISR. The complementarity of the two experiments is guaranteed by the fact that the final states that can be 
studied by the two experiments  are different and that the PANDA experiment can more easily access the narrow states while \superb can study in detail larger 
states if the production mechanism is favorable. Furthermore, in case the center-of-mass-energy of \superb is changed to 
the $Y(4260)$ mass, assuming a factor 10 loss in luminosity with respect to  running at the \FourS, the number of events produced in the decay chain used as 
example would raise to 700K per year: a few weeks scan would then be equivalent to the PANDA dataset.
Finally, PANDA can only reach center-of-mass energies as high as 5 GeV and therefore has no access to bottomonium spectroscopy.

\section{Direct Searches}\label{sec:directsearches}

Bottomonium decays also allow direct searches for physics beyond the SM
in regions of the parameters space that have not been reached by LEP~\cite{Schael:2006cr}:
the possibility of
a rather light non-standard Higgs boson has not been ruled out in
several scenarios beyond the SM
\cite{Dermisek:2005gg,Dermisek:2006py,SanchisLozano:2007wv},
due to the fact that a new scalar may be uncharged under the
gauge symmetries, similar to a sterile neutrino in the fermion
case.  These studies indicate that its mass could be less than twice
the $b$ mass, placing it within the reach of \superb$\!\!$.
Moreover, the
LHC might not be able to unravel a signal from a
light Higgs boson whose mass is
below $B\bar{B}$ threshold, since it will be difficult for the soft decay products
to pass the LHC triggers.  Dark matter may also be
light, evading LEP searches if it does not couple strongly to the
$Z^0$~\cite{Gunion:2005rw,McElrath:2005bp,Fayet:2007ua,Bird:2006jd}.
Finally, the new field of Dark Forces (see Sec.~\ref{sec:DF}) predicts low interacting light particles that couple mostly to photons that can
therefore be produced at a Flavour Factory and that would require a large luminosity to study.

\superb\ will be required in most of these cases to precisely determine
the masses and couplings of any light non-SM particles, and thus will
 play an important discovery role.

\subsection{ Light Higgs}

A Higgs $h$ with $M_h < M_\Upsilon$ can be
produced in $\Upsilon(nS)$ decays via the Wilczek mechanism~\cite{Wilczek:1977zn}
with a branching ratio, at 
 leading-order,
$$ 
\frac{\Gamma(\Upsilon(nS) \to \gamma h)}{\Gamma(\Upsilon(nS) \to \mu
\mu)} =
    \frac{\sqrt{2} G_F m_b^2}{\alpha \pi M_{\NS}} E_\gamma X_d^2,
$$
where $X_d$ is a model-dependent quantity containing the coupling of the
Higgs
to bottom quarks, $m_b$ is the bottom quark mass, $\alpha$ and $G_F$ are
the
electroweak parameters, and $E_\gamma$
is the photon energy.

From a theoretical viewpoint, the existence of a light
pseudoscalar Higgs is not unexpected in many
extensions of the SM. As an example, the
Next-to-Minimal Supersymmetric Standard
Model (NMSSM) has a gauge singlet added to the
MSSM two-doublet Higgs sector~\cite{Ellwanger:2009dp} leading
to seven physical Higgs bosons, five of them neutral, including
two pseudoscalars.
In the limit of either
slightly broken $R$ or Peccei-Quinn (PQ) symmetries,
the lightest CP-odd Higgs boson (denoted by $A_1$)
can be much lighter than the other Higgs bosons, providing
unique signatures at a \sff as discussed in the following.

The $A_1$ coupling to down-type fermions turns out to be proportional to
$X_d = \cos{\theta_A}$ $\tan{\beta}$, where $\tan{\beta}$ denotes
the ratio of the vacuum expectation values of the up- and down-type 
Higgs bosons and $\theta_A$ is the mixing angle of the singlet and 
non-singlet components that constitute the 
physical $A_1$ state \cite{Fullana:2007uq}. 
If $\cos{\theta_A} \sim 0.1-0.5$, present LEP and $B$ physics
bounds can be simultaneously satisfied~\cite{Domingo:2007dx}, while
a light Higgs could still show up
in $\Upsilon$ radiative decays into tauonic pairs: $\NS \to \gamma A_1(\to \tau^+\tau^-)\ ;\ \ \ n=1,2,3. $

As this light Higgs acquires its couplings to SM fermions via mixing
with the SM Higgs, it therefore couples to mass, and will decay to the
heaviest available SM fermion.  In the region $M_{A_1} > 2 M_\tau$, there
are
two measurements which have sensitivity: lepton universality of $\Upsilon$
decays, and searches for a monochromatic photon peak in tauonic $\Upsilon$
decays. 

The measurement of lepton universality
 compares the branching ratios of $\Upsilon$ to $e^+e^-$,
$\mu^+\mu^-$ and $\tau^+\tau^-$
\cite{SanchisLozano:2003ha,SanchisLozano:2006gx},
which should all have identical couplings in the SM, and differ only by 
factors given by phase-space differences.
This inclusive measurement is relevant especially when the monochromatic photon
signal is buried under backgrounds. 
Under reasonable sets of the NMSSM parameters that satisfy all current LEP and $B$ physics bounds, it has 
been shown~\cite{SanchisLozano:2007wv,Domingo:2008rr} 
that $A_1$ bosons with masses between $9$ to $10.5$ GeV 
can give sizeable deviations from the SM if $5 \lesssim \tan\beta \lesssim 20$.

\begin{figure}[!htb]
\begin{center}
\includegraphics[width=16pc]{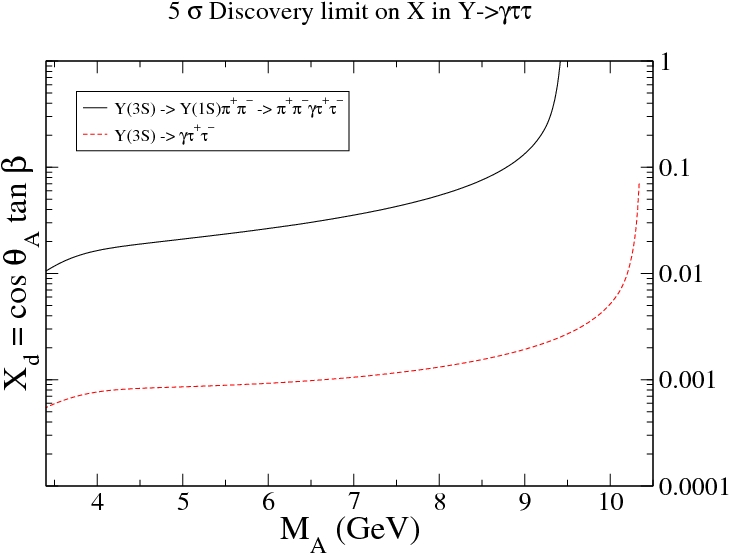}
\caption{\label{fig:ups_gamma_a_mono}
Five $ \sigma$ discovery potential of \superb\
with $\Upsilon(3S)$ data, in the mode $\Upsilon(3S) \to \pi^+ \pi^-
\Upsilon(1S) \to \pi^+ \pi^- \tau^+ \tau^- \gamma$ (solid) and
$\Upsilon(3S) \to \tau^+ \tau^- \gamma$ (dashed). An integrated
luminosity of 1 ab$^{-1}$ has been assumed for this projection.}
\end{center}
\end{figure}

Unfortunately recent measurements of these branching fractions are limited by
systematics and it is hard to conceive of a dramatic improvement below 
the level of $1\%$ precision at \superb.
Alternatively, once can consider the search for monochromatic 
photons~\cite{Dermisek:2006py} where the first relevant decay mode 
is $\ThreeS \to \OneS \pi^+ \pi^-$, which is followed by the 
decay $\OneS \to \gamma \tau^+ \tau^-$.  This only has a
4.5\% branching fraction, but it also has a low background.
The second decay mode is $\ThreeS \to \gamma
\tau^+ \tau^-$, which suffers from a larger background arising 
from $e^+ e^- \to \tau^+\tau^- \gamma$ events, but also has a 
rate that is more than a factor of ten higher than $\ThreeS \to \OneS \pi^+ \pi^-$.
The corresponding exclusion plots expected at \superb are shown in Fig.~\ref{fig:ups_gamma_a_mono}.

Let us finally point out another possible signal for 
detection of a light CP-odd Higgs boson related to 
bottomonium spectroscopy.
As studied in \cite{Domingo:2009tb}, significant $A_1 - \eta_b(nS)$ 
mixing should significantly alter the hyperfine splitting 
$M(\Upsilon(nS))-M(\eta_b(nS))$ compared to SM expectations. 
This kind of search has a great advantage with respect to
the radiative decays of $\Upsilon$ resonances since it
is free of theoretical uncertainties coming from  
QCD and relativistic corrections plaguing the
Wilczek formula. Moreover, from an experimental point of view,
the mixing could spoil a straight forward search for 
narrow peaks in the photon spectrum while 
the measurement of hyperfine splittings could still
yield unexpected results hinting at the 
existence of a light pseudoscalar Higgs~\cite{Domingo:2009tb}.

\subsection{Invisible decays and Dark Matter}

Finally, if Dark Matter is lighter than 5 GeV, it
will require a \sff to determine its properties. 
Generally, in this mass region one needs two
particles, the dark matter particle $\chi$, and a boson
that couples it to the SM U . The most
promising searches are in invisible and radiative decays 
of the $\Upsilon$, which can be measured in the mode
$\Upsilon(3S) \to \pi^+ \pi^- invisible$, which is
sensitive to a vector U~\cite{McElrath:2005bp}. The current best sensitivity
to this process has been achieved by the \babar
Experiment \cite{:2009qd};
however, this result is still an order of magnitude above
the SM prediction. The sensitivity is
limited by the amount of background that needs to
be subtracted, primarily due to undetected leptons
from $\Upsilon(1S) \to \ell^+ \ell^-$ in the final state.
Studies of this background suggest that the only way
to further improve the measurement to the level of the
SM is to employ both far-backward and
far-forward tagging into the design of the detector.
Achieving a $3-5\sigma$ sensitivity to the SM
 will require active background tagging down to
5-10 degrees above the beam-line in both the forward
and backward directions.

The second most promising signature is radiative
decays $\Upsilon(1S) \to \gamma + invisible$.
This is probably the most
favored mode theoretically, and is sensitive to a scalar
or pseudoscalar U. The mediator coupling the SM particles to 
final state $\chi$'s can be a
pseudoscalar Higgs, $U = A_1$, which can be naturally light,
and would appear in this mode [38]. In such models
the Dark Matter can be naturally be a bino-like neutralino. 
Extended detector coverage in the forward and
backward directions is important to reducing the
radiative QED backgrounds which dominate this final state.

It is expected that improving detector coverage with
active coverage for tagging low-angle
or missing-particle backgrounds
will also improve the sensitivity in flagship
measurements of \superb, including $B \to K \nu \bar{\nu }$
and $B \to \ell \nu$.

\subsection{Dark Forces} 
\label{sec:DF}

Recent cosmic ray measurements of the electron and positron flux from
ATIC~\cite{atic}, FERMI~\cite{Abdo:2009zk}, and PAMELA~\cite{Adriani:2008zr} have
spectra which are not well described by galactic cosmic ray models
such as GALPROP~\cite{Moskalenko:2005xu}.  For instance,  PAMELA shows an
increase in the positron/electron fraction with increasing energy.  No
corresponding increase in the antiproton spectrum is observed.  There
have been two main approaches attempting to explain these features:  
astrophysical sources (particularly from undetected, nearby  pulsars)~\cite{Hooper:2008kg}
 and annihilating or decaying dark matter.  

Arkani-Hamed \ea~\cite{ArkaniHamed:2008qn} and Pospelov \ea~\cite{Pospelov:2008jd} have introduced a class of
theories containing a new ``dark force'' and a light, hidden sector.
In this model, the ATIC and PAMELA signals are due to dark matter
particles with mass $\sim400-800\gevcc$ annihilating into the gauge
boson force carrier with mass $\sim 1\gevcc$, dubbed the $A^\prime$, which subsequently 
decays to SM particles.  
If the $A^\prime$ mass is below twice the proton mass, decays to $p\overline{p}$ are kinematically 
forbidden allowing only decays to 
states like $\epem$, $\mu^+\mu^-$, and $\pi\pi$.  If the dark force is non-Abelian, 
this theory can also accommodate the 511 keV 
signal found by the INTEGRAL satellite \cite{Jean:2003ci} and the DAMA modulation data \cite{Bernabei:2008yi}.

\begin{figure}[!htb]
\begin{center}
\epsfig{file=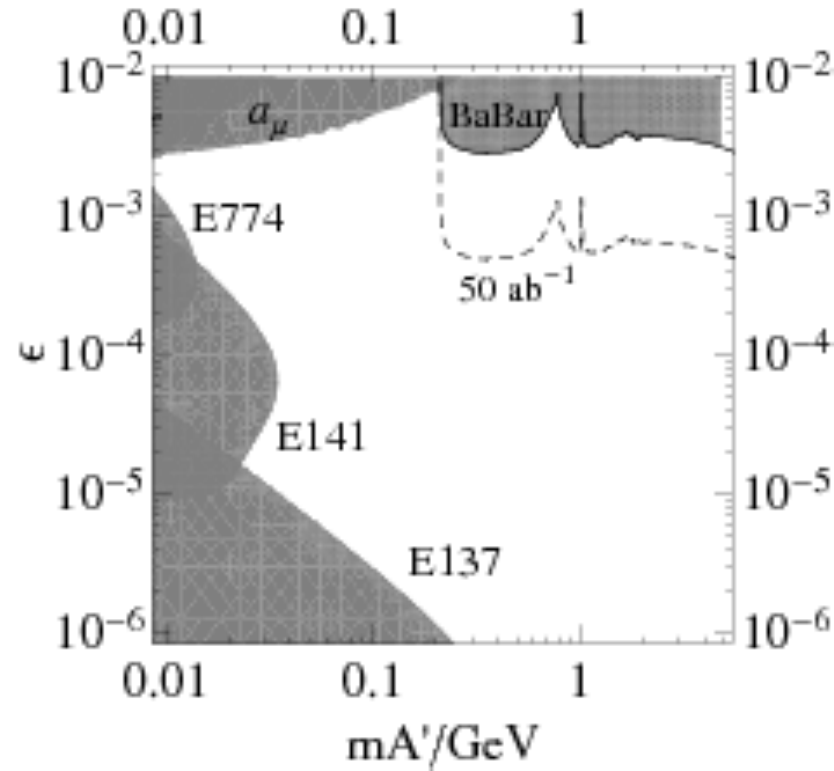,height=6cm}
 \caption{\label{fig:epsilon}
 Shaded: The current constraints on the kinetic mixing parameter $\epsilon$ as a function of dark photon mass.  Dashed line:  the expected constraint from \superb with $50ab^{-1}$ of data.   }
\end{center}
\end{figure}

The dark sector couples to the SM through kinetic mixing with the photon (hence we
call the  $A^\prime$ the ``dark photon'') with 
a mixing strength $\epsilon$.  The current limits on $\epsilon$ from various experiments 
are shown on Figure \ref{fig:epsilon}.  
Low-energy, high luminosity $\epem$ experiments like the \B-Factories are in excellent position to 
probe these theories, as pointed out in papers by Batell \ea~\cite{Batell:2009yf} and Essig \ea~\cite{Essig:2009nc}.
Broadly speaking, there are three categories for dark force searches at \superb:  direct production,
 rare B-decays, and rare decays of other mesons.  

The most general searches for dark forces are in direct $\epem$ production.  The primary 
model independent signature is $\epem \to \gamma A^\prime\to\gamma l^+ l^-$.  While
these channels are the cleanest  theoretically, they suffer from large irreducible QED backgrounds.  Searches 
for narrow resonances in $\epem\to\gamma\mu\mu$ and $\epem\to\gamma\tau\tau$ have been 
carried out by CLEO~\cite{:2008hs} and \babar~\cite{Aubert:2009cp}.  The limit on $\epsilon$ obtained from the \babar
 $\epem\to\gamma\mu\mu$ analysis using 32$fb^{-1}$ is shown on Figure~\ref{fig:epsilon}.  With 
the increased luminosity, \superb should be sensitive to values of $\epsilon$ down to $5\times 10^{-4}$.
Since the gauge symmetry of the dark sector is by construction broken, there is also at least one
``dark Higgs'' ($h^\prime$) in the model.  Therefore there can also be 
interactions like $\epem \to A^\prime h^\prime \to  3(l^+ l^-)$.  While this channel is suppressed 
with respect to $\epem \to \gamma A^\prime$, the final state of 6-leptons (with possibly all three pairs
giving a narrow resonance) should be much cleaner with a small irreducible QED background.  There 
are a number of other, more model dependent searches we can do at \superb.  For instance, if the 
dark force is non-Abelian there can be  final states with 4-,8-, or even 12- or more leptons with many
pairs forming a narrow resonance.  While these final states are harder to use to extract $\epsilon$ limits, 
any evidence of a narrow resonance in them would be evidence for new physics.

Searches can also be performed in very rare decays of the \B meson.  Generally speaking, and decay 
involving a photon can be used to search for a dark photon.  We can search in the $ l^+l^-$
mass spectrum in modes such as $B\to K l^+l^-$ for a narrow resonance, although there will be a large 
background from the normal SM process.  
In addition, loop dominated modes such as $B^0\to l^+l^- l^+l^-$ or $B\to K l^+l^- l^+l^-$ can enhanced 
by a ``Higgs$^\prime$-strahlung'  from the top quark in the loop~\cite{Batell:2009jf}.  If these modes are observed, We can look in the di-lepton mass spectrum for a resonance.  

Finally, we can search for dark forces in rare meson decays~\cite{Reece:2009un}.  
The \superb experiment will not be just a \B meson factory, it will also 
produce huge samples of other mesons such as $\pi^0$, $\eta$, $K$, $\phi$, and 
$J\psi$.   For instance, there are roughly $10^{10} \pi^0/ab^{-1}$ and 
$10^9 \eta/ab^{-1}$ produced which can be used to search for the channel 
$\pi^0/\eta\to\gamma A^\prime\to\gamma l^+ l^-$.  Searching the huge meson samples for rare decays such as these should give limits on $\epsilon$ that are competitive to other measurements.

\graphicspath{{Lattice/}{Lattice/}}
\section{Role of Lattice QCD}\label{sec:lattice}

This section describes the role of lattice QCD in the physics case of \superb.

\begin{figure}[!htb]
\begin{center}
\includegraphics[width=0.5\textwidth]{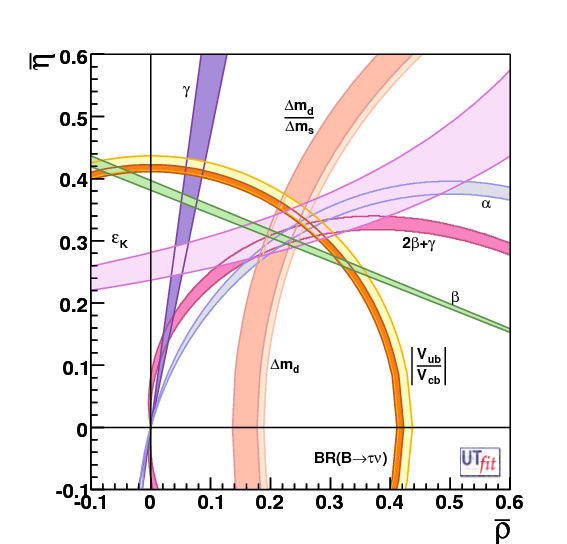}
\caption{\label{fig:ut2015}
Unitarity triangle fit within the SM extrapolated using expected results at
\superb and future lattice QCD calculations~\cite{Bona:2007qt}. Central values of
the constraints are chosen from the present UT fit. The bands show the 95\%
probability regions selected by the single constraints.}
\end{center}
\end{figure}

While there are some flavour observables, like the angles of unitarity triangle,
which can be determined with rather small or even negligible theoretical
uncertainties, in other cases the extraction of physical results also relies on
theoretical inputs, mainly on lattice QCD calculations. This is the case, for
example, of several among the constraints entering the unitarity triangle
analysis, for which an extrapolation at the \superb is illustrated in
Fig.~\ref{fig:ut2015}~\cite{Bona:2007qt}.
In this analysis, in order to convert into constraints in the $(\bar \rho, \bar
\eta)$-plane the measurements of leptonic ($B \to \tau \nu$) and semi-leptonic
($B \to \pi(\rho) l \nu/B \to D(D^*) l \nu$) \B decay rates and of $K^0 - \bar
K^0$ and $B^0_{d/s} - \bar B^0_{d/s}$ mixing amplitudes ($\varepsilon_K$,
$\Delta m_d$ and $\Delta m_d/\Delta m_s$), a determination of the corresponding
hadronic matrix elements is required. These matrix elements are expressed in
terms of decay constants ($f_B$), form factors (${\cal F}^{B\to D/D^*}$,
$f_+^{B\pi}, \ldots$) and bag parameters ($f_{Bs}\sqrt{B_{B_s}}$, $\xi$).

For the physics case of \superb, in order to exploit the full power of flavour
physics for NP searches and, even more, NP characterization,
improved theoretical predictions are essential. Hadronic uncertainties in
particular need to be controlled with an unprecedented accuracy, comparable
to the one achieved by the experimental measurements. For most of the hadronic
parameters, the precision necessary to fulfill such a requirement is at the
level of few percent or better.

Lattice QCD is the theoretical tool of choice to compute hadronic quantities.
Being only based on first principles, it does not introduce additional free
parameters besides the fundamental couplings of QCD, namely the strong coupling
constant and the quark masses. In addition, all systematic uncertainties
affecting the results of lattice calculations can be systematically reduced in
time, with the continuously increasing availability of computing power. The
development of new algorithms and of new theoretical techniques further speeds
up the process of improving precision.

The important issue of whether the precision of lattice QCD calculations will
succeed in competing with the experimental one at the time when a \superb factory
could be running has been addressed in a dedicated study~\cite{lubicz@superbIV}
reported in the \superb CDR~\cite{Bona:2007qt}. The result of this study was
promising: in order to reach the few percent accuracy required in the
determination of the most relevant hadronic parameters, supercomputers
performing in the 1-10 PFlops range are required. This computing power is just
in the ball park of what is expected to be available to lattice QCD
collaborations in $\sim 2015$, when a \superb factory could be running and
producing results.

In the study of Ref.~\cite{Bona:2007qt}, the estimate of the precision expected
to be reached by lattice QCD calculations covered a temporal extension of about
10 years (2006-2015). Such an estimate is unavoidably affected by some
uncertainties. The dominant sources of errors in lattice QCD calculations have
systematic origin, so that the accuracy of the lattice results does not improve
in time by following simple scaling laws (at variance with the computing power,
which increases instead according to a rather predictable exponential
behavior). Therefore, predictions in this context are necessarily based also on
educated guesses, and their reliability decreases the more we attempt to go
further in time.

\setlength{\tabcolsep}{2pt}  
\begin{table*}[!t]
\caption{Prediction of the accuracy on the lattice QCD determinations of
various hadronic parameters from Ref.~\cite{Bona:2007qt}. The $5^{th}$ column has
been added for the present work.}
\label{tab:lattice}
\begin{center}
\begin{tabular}{lccc||c||cc}  
\hline
Measurement & 
\begin{tabular}{c} Hadronic \\ Parameter \end{tabular} &  
\begin{tabular}{c} Status \\ End 2006 \end{tabular} & 
\begin{tabular}{c} 6 TFlops \\ (Year 2009) \end{tabular}&
\begin{tabular}{c} {\bf Status } \\ {\bf End 2009} \end{tabular}&
\begin{tabular}{c} 60 TFlops \\ (Year 2011) \end{tabular}&
\begin{tabular}{c} 1-10 PFlops \\ (Year 2015) \end{tabular} \\ \hline
$K \to \pi\, l\, \nu$   & $f_+^{K\pi}(0)$
& 0.9\,\% & 0.7\,\% & 0.5\,\% & 0.4\,\% & $<0.1\,\%$ \\
$\varepsilon_K$ & $\hat B_K$
& 11\,\%  &  5\,\%  &   5\,\%  & 3\,\%  & 1\,\% \\
$B \to l\, \nu$  & $f_B$
& 14\,\% & 3.5-4.5\,\% &  5\,\%  & 2.5-4.0\,\% &  1.0-1.5\,\% \\
$\Delta m_d$ & $f_{Bs}\sqrt{B_{B_s}}$ 
& 13\,\% & 4-5\,\% &  5\,\%  & 3-4\,\% & 1-1.5\,\% \\
$\Delta m_d/\Delta m_s$ & $\xi $
& 5\,\% & 3\,\% &   2\,\%  & 1.5-2\,\% & 0.5-0.8\,\% \\
$B\to D/D^*\,l\,\nu$& ${\cal F}^{B\to D/D^*}$
& 4\,\%  &  2\,\%  &  2\,\%  &  1.2\,\% & 0.5\,\% \\
$B\to \pi/\rho \,l\,\nu$ & $f_+^{B\pi},\ldots$
& 11\,\% & 5.5-6.5\,\% &  11\,\%  & 4-5\,\% & 2-3\,\% \\
$B\to K^*/\rho \,(\gamma, l^+l^-)$ & $T_1^{B\to K^*/\rho}$
& 13\,\% & ------ &   13\,\%  & ------ & 3-4\,\% \\
\hline
\end{tabular}
\end{center}
\end{table*}
After three years from the presentation of Ref.~\cite{Bona:2007qt}, we are now
in the position of start verifying whether the improvements predicted for
lattice QCD calculations were accurate. This is already
a non-trivial check. Indeed, while for many years lattice calculations have been
plagued by the use of the quenched approximation, so that the typical lattice
uncertainties at the time of Ref.~\cite{Bona:2007qt} were at the level of
10-15\%, in the last few years extensive unquenched lattice QCD simulations have
been performed, by various lattice collaborations and using different approaches
(i.e. different lattice actions, renormalization techniques, etc.). For this
reason, for several hadronic parameters, the typical uncertainties are now
significantly reduced with respect to three years ago, by a factor 2 or 3.

A summary of lattice uncertainties and predictions for the future is presented
in Table~\ref{tab:lattice}, which is reported from Ref.~\cite{Bona:2007qt}
except for the $5^{th}$ column, which is new. A representative set of measurements
relevant for flavour physics and corresponding hadronic parameters is listed in
the Table. The corresponding lattice uncertainty, as it was quoted at the end of
2006, is given in the 3rd column. In the $4^{th}$, $6^{th}$ and $7^{th}$ columns, the accuracy
predicted for the future is presented, assuming the availability of a computing
power of about 6 TFlops, 60 TFlops and 1-10 PFlops respectively. These
performances are those expected for supercomputers typically available to
lattice QCD collaborations in the years 2009, 2011 and 2015 respectively. Thus,
the last column of the Table predicts in particular the accuracy that is
expected to be reached by lattice QCD calculations at the time of the \superb.
This prediction indicates that, for most of the relevant quantities, a precision
at the level of 1\% should be reached.

In Table~\ref{tab:averages} we collect a set of current lattice averages for the
same hadronic parameters listed in Table~\ref{tab:lattice}.
\begin{table}[!ht]
\caption{Lattice averages for various hadronic parameters. The result for the
semi-leptonic form factor $f_+^{B\pi}$ has been already converted into the
corresponding exclusive determination of $|V_{ub}|$.}
\label{tab:averages}
\begin{center}
\begin{tabular}{ccc}  
\hline
\begin{tabular}{c} Hadronic \\ Parameter \end{tabular} &  
\begin{tabular}{c} Lattice \\ average \end{tabular} & Ref. \\ \hline
$f_+^{K\pi}(0)$ & 0.962(3)(4) & \cite{lubicz@latt09} \\
$\hat B_K$ & 0.731(7)(35) & \cite{lubicz@latt09} \\
$f_B$ (MeV) & 192.8(9.9) & \cite{Laiho:2009eu} \\
$f_{Bs}\sqrt{\hat B_{B_s}}$ (MeV) & 275(13) & \cite{Laiho:2009eu} \\
$\xi$ & 1.243(28) & \cite{Laiho:2009eu} \\
${\cal F}^{B\to D^*}(1)$ & 0.924(22) & \cite{Lubicz:2008am} \\
${\cal G}^{B\to D}(1)$ & 1.060(35) & \cite{Lubicz:2008am} \\
$|V_{ub}|_{excl.}$ &  35(4)$\,10^{-4}$ & \cite{Lubicz:2008am} \\
\hline
\end{tabular}
\end{center}
\end{table}
Central values and errors are quoted from Ref.~\cite{lubicz@latt09} for the kaon
observables ($f_+^{K\pi}(0)$ and $\hat B_K$), Ref.~\cite{Laiho:2009eu} for the
\B physics parameters ($f_B$, $f_{Bs} \sqrt{B_{B_s}}$, $\xi$) and
Ref.~\cite{Lubicz:2008am} for the semi-leptonic form factors (${\cal F}^{B\to
D^*}$, ${\cal G}^{B\to D}$ and $f_+^{B\pi}$). On the basis of these results
we have compiled the new ($5^{th}$) column of Table~\ref{tab:lattice} summarizing the
status of lattice calculations at the end of 2009.

The main conclusion which can be drawn from this analysis is that there is quite
a good agreement between the predictions for the year 2009 (4th column of
Table~\ref{tab:lattice}) and the accuracy actually reached by lattice
calculations ($5^{th}$ column of Table~\ref{tab:lattice}). Even though the prediction
was made only 3 years ago, thus a relatively short time with respect to the
whole time interval of about 10 years considered in Table~\ref{tab:lattice}, it
should be also noted that we have witnessed significant changes in the last 3
years. Realistic unquenched lattice simulations have been performed, and for
most of the hadronic quantities listed in Table~\ref{tab:lattice} the achieved
accuracy has improved by a factor two or more (compare the $3^{rd}$ and $5^{th}$ columns of the
Table). This improvement has been quite precisely predicted, at a quantitative
level, by the dedicated studies of Refs.~\cite{lubicz@superbIV,Bona:2007qt}.

\begin{figure}[!htb]
\begin{center}
\includegraphics[width=0.5\textwidth]{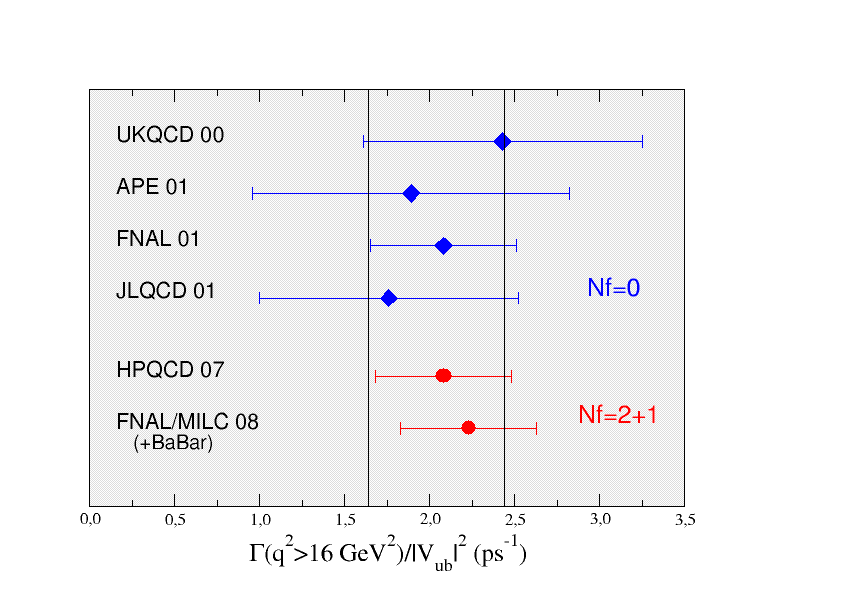}
\caption{\label{fig:vubexcl}
Lattice QCD results for $\Gamma(B \to \pi\,l\,\nu; q^2 > 16~GeV^2)/|V_{ub}|^2$.
The estimated average is shown by the vertical band. The plot is an update from
Ref.~\cite{Lubicz:2008am}.}
\end{center}
\end{figure}

There is one notable exception to the previous conclusion. It is represented by
the lattice determination of the form factor controlling the exclusive $B\to
\pi \,l\,\nu$ semi-leptonic decays. The relative uncertainty on this form factor
was about 11\% at the end of 2006. It was predicted to decrease by approximately
a factor 2 within by the end of 2009, but it is actually unchanged with respect to
three years ago. One possible reason for that is the following: at variance with
the other quantities listed in Table~\ref{tab:lattice}, only two modern lattice
studies~\cite{Dalgic:2006dt,Bailey:2008wp} of the exclusive $B\to \pi\,l\,\nu$
decays have been performed so far (both based on the same staggered gauge
configurations generated by the MILC collaboration). The results of these
studies are in agreement within each other and also with older quenched results,
see Fig.~\ref{fig:vubexcl}.
The accuracy on the determination of the relevant form factor,
$f_+^{B\pi}(q^2)$, in the two modern calculations, however, is still at the
level of 10\%, which is the uncertainty quoted in Tables~\ref{tab:lattice}
and~\ref{tab:averages}. Thus, a larger number of (possibly) more accurate
lattice studies of $B\to \pi\,l\,\nu$ decays would be welcome in order to
improve the theoretical prediction also in this case.

For all other quantities listed in Table~\ref{tab:lattice}, the agreement
between the prediction of Refs.~\cite{lubicz@superbIV,Bona:2007qt} and the
accuracy actually reached at present by lattice calculations is encouraging. It
shows that lattice calculations are rapidly improving in the last few years, and
to expectations. It also supports the prediction for the time in which a
\superb factory could be running, indicating that the percent accuracy could be
actually reached by lattice QCD. If this will be the case, then the next 
generation of flavour physics experiments will not be limited by the hadronic 
uncertainties, and the theoretical accuracy will succeed in competing with the 
experimental one.

\graphicspath{{interplay/}{interplay/}}
\section{Interplay between measurements}\label{sec:interplay}

Numerous studies of flavour and \CP violating observables in New Physics models have been
performed over the past years. In the following we give a brief summary of the results
obtained within the MSSM with various realizations of flavour, in the SM with
a 4th generation of quarks and leptons (SM4), in Randall-Sundrum models with bulk fields,
both with the SM bulk gauge group (minimal RS) and with a protective custodial symmetry (RSc),
and in the Littlest Higgs model with T-parity (LHT model). Clearly our focus lies on those
observables that can be measured at a \sff with high precision and thus provide a
powerful tool to discriminate among various scenarios.

\subsection{MSSM}
MSSM contains large numbers of new parameters which can potentially produce the 
flavour and \CP violating phenomena beyond SM.  
The search for SUSY via flavour physics is particularly important since many of the
parameters mentioned above are related to the SUSY breaking mechanism, which occurs at
higher energies than that accessible at the energy frontier experiments.
For this reason, various scenarios to search for signals of MSSM at \superb
have been devised.

Since the SUSY breaking term contains a huge number of free parameters, how to
interpret the New Physics signal in the framework of MSSM is non trivial. With
regard to this aspect, the large number of observables measurable at \superb is advantageous as
the correlations of those different observables play a central role in constraining the
SUSY parameters. In the following we will illustrate different kinds of
``interplay'' (correlation among the observables) between observables
measurable at \superb.

\subsubsection{Minimal flavour model: interplay to the LHC direct search}
\begin{figure}[h]
\begin{center}
\includegraphics[width=8.cm]{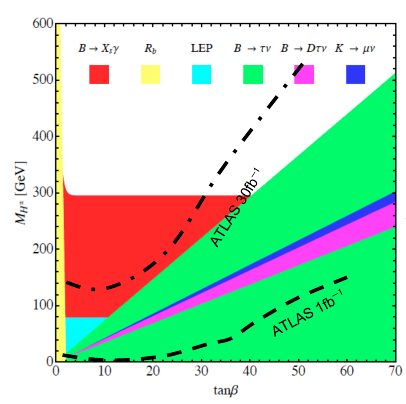}
\end{center}
\caption{The excluded region at 95\% confidence level for the charged Higgs mass
versus $\tan\beta$ from Ref.~\cite{Haisch:2008ar}. The branching ratio of $B\to \tau \nu$ as well as
$B\to D\tau \nu$ will be significantly improved at \superb. The ATLAS
potential expected for the excluded region obtained using 1 and 30\invfb of running at a center of mass
energy of 14 TeV is also shown.}
\label{fig_kou_2}
\end{figure}

Even if we assume the SUSY breaking is flavour blind (no additional flavour
violation beyond that introduced via the CKM matrix), there are still 
SUSY effects observable at \superb. The effects are expected for example when one
chooses a large value for $\tan\beta$ and/or the split Higgs mass. 
Since most of the SUSY particle searches at LHC are aiming to investigate this
class of models one can study the complementarity of \superb and the
energy frontier experiments such as ATLAS and CMS. Figure~\ref{fig_kou_2} (taken 
from Ref.~\cite{Haisch:2008ar})  shows the constraints that can be obtained for the charged
Higgs mass for given value of $\tan\beta$ from different observables, which
include the \superb golden channels, $B\to \tau\nu$ and $B\to D\tau\nu$.  We have
superimposed the constraint expected to be achieved from direct 
searches at ATLAS~\cite{Aad:2009wy}~\footnote{Additional constraints on these parameters
are expected to come from the measurement of $\BR(B_s\to\mu^+\mu^-)$ at LHCb.}.
In Table~\ref{tab_kou_1}, we show the expected deviation from the SM for various
\superb observables for the so-called LHC benchmark points. 

\begin{table}[th]
\caption{ Predictions of flavor observables based on expected measurements
 from the LHC in mSUGRA at SPS1a, SPS4, SPS5 benchmark points. Quantities denoted
 ${\cal R}$ are the ratios of the branching fractions relative to their Standard Model values.
 Quoted uncertainties (when available) come from the errors on the measurement of the
 New Physics parameters at LHC. Uncertainties on the Standard Model predictions of flavor
 observables are not included. For the SPS4 benchmark point the sensitivity
 study at LHC are not available. The SPS parameters are as follows
 (see~\cite{Hitlin:2008gf} for more details).  
 SPS 1a: ($m_0 = 100$  GeV, $m_{1/2}= 250$ GeV, $A_0 = -100$  GeV, $\tan\beta =
 10$, $\mu > 0$), \ \ SPS 4: ($m_0 = 400$  GeV, $m_{1/2} = 300$ GeV, $A_0 = 0$, 
 $\tan\beta = 50$, $\mu > 0$), and SPS 5: ($m_0 = 150$  GeV, $m_{1/2} = 300$ 
 GeV, $ A_0 = -1000$, $\tan\beta = 5$, $\mu > 0$).}
\label{tab_kou_1}
  \begin{center}
   \begin{tabular}{lccc} 
    \hline %
                                                &      SPS1a             &  SPS4   &  SPS5              \\
    \hline
    ${\cal R}(B \to X_s\gamma)$                 &    0.919  $\pm$ 0.038  &  0.248  & 0.848  $\pm$ 0.081 \\
    ${\cal R}(B \to \tau\nu)$                   &    0.968  $\pm$ 0.007  &  0.436  & 0.997  $\pm$ 0.003 \\
    ${\cal R}(B \to X_s l^+l^-)$                &    0.916  $\pm$ 0.004  &  0.917  & 0.995  $\pm$ 0.002 \\
    ${\cal R}(B \to K \nu \overline{\nu})$      &    0.967  $\pm$ 0.001  &  0.972  & 0.994  $\pm$ 0.001 \\
    ${\cal B}(B_d \to \mu^+\mu^-)/10^{-10}$     &    1.631  $\pm$ 0.038  &  16.9   & 1.979  $\pm$ 0.012 \\
    ${\cal R}(\Delta m_s)$                      &    1.050  $\pm$ 0.001  &  1.029  & 1.029  $\pm$ 0.001 \\
    ${\cal B}(B_s \to \mu^+\mu^-)/10^{-9}$      &    2.824  $\pm$ 0.063  &   29.3  & 3.427  $\pm$ 0.018 \\
    ${\cal R}(K \to \pi^0 \nu \overline{\nu})$  &    0.973  $\pm$ 0.001  &  0.977  & 0.994  $\pm$ 0.001 \\
    \hline %
  \end{tabular}
\end{center}
\end{table}

\subsubsection{Model independent analysis: interplay among the similar flavour transitions}
\begin{figure*}[!h]
\resizebox{16cm}{!}{
\includegraphics{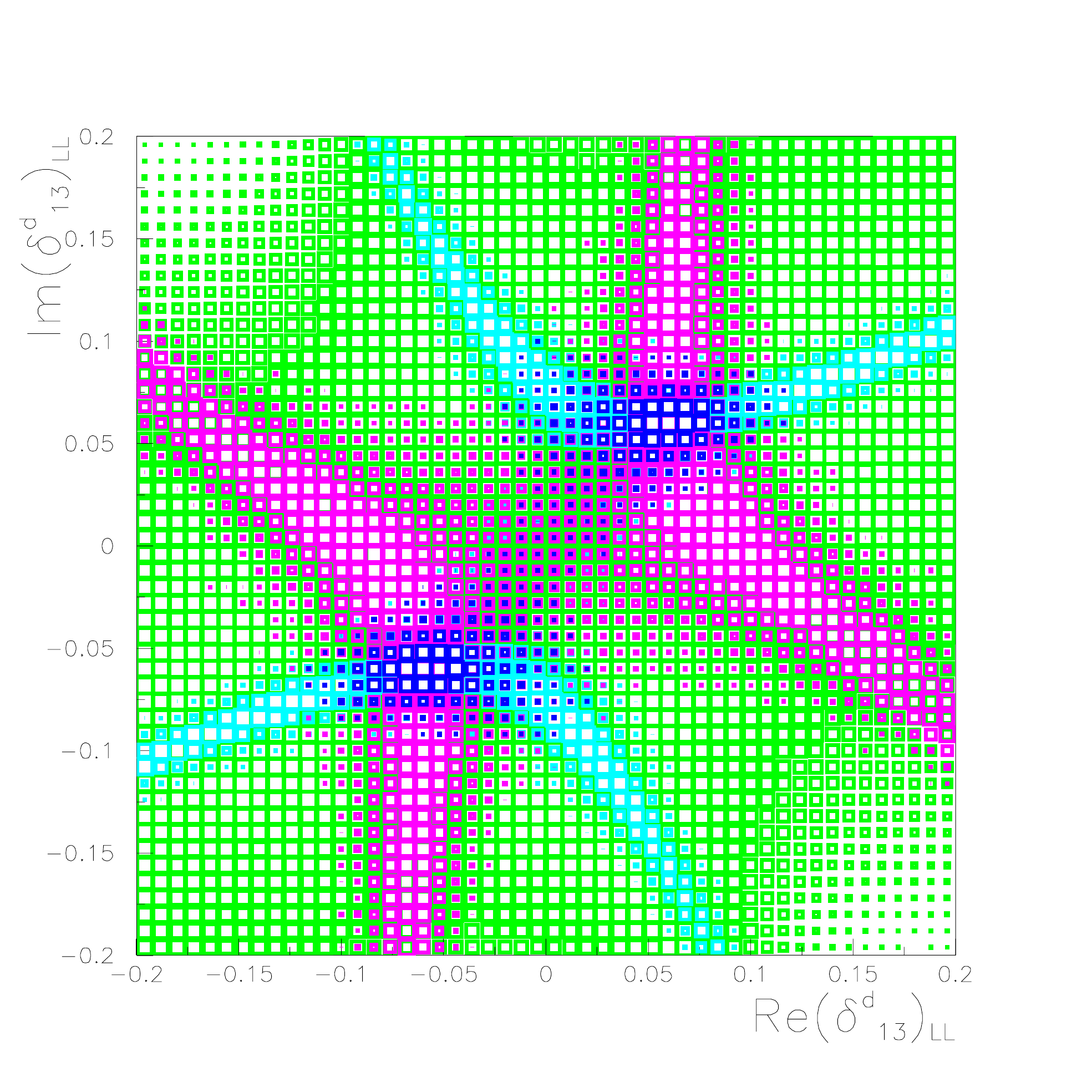}
\includegraphics{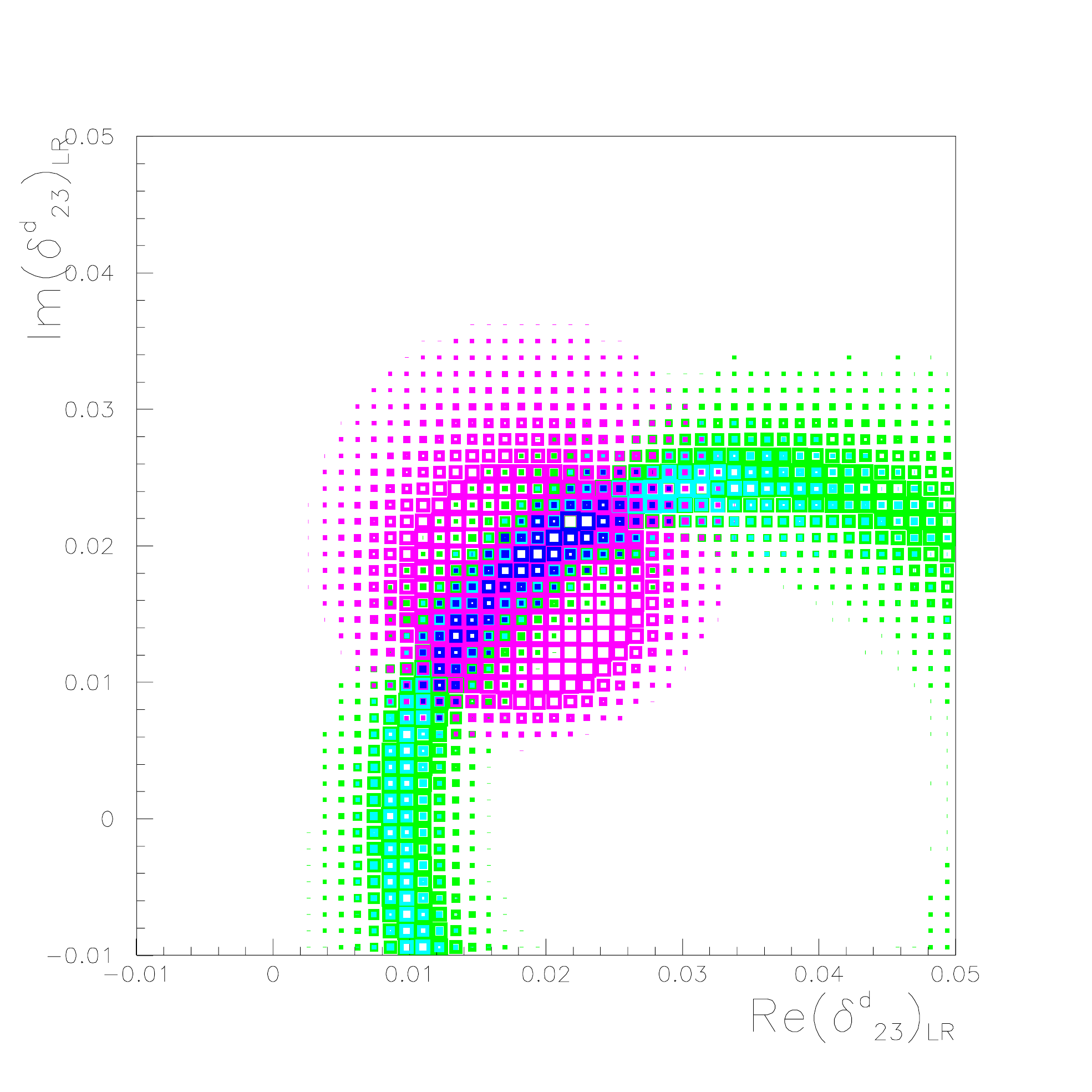}
}
\caption{Left: Density plot of the selected region in the
$Re(\delta^d_{13})_{LL}-Im(\delta^d_{13})_{LL}$ for
$m_{\tilde{q}}=m_{\tilde{g}}=1$ TeV and $(\delta^d_{13})_{LL}=0.085e^{i\pi/4}$
using \superb measurements (namely, 1-3 generation transitions). Different
colors correspond to different constraints: $A_{\rm SL}^d$ (green), $\beta$
(cyan), $\Delta m_d$ (magenta), all together (blue). Right: 
Density plot of the selected region in the
$Re(\delta^d_{23})_{LR}-Im(\delta^d_{23})_{LR}$ for
$m_{\tilde{q}}=m_{\tilde{g}}=1$ TeV and $(\delta^d_{23})_{LR}=0.028e^{i\pi/4}$
using \superb measurements (namely, 2-3 generation transitions). Different
colors correspond to different constraints: ${\mathcal{B}}(B\to X_s\gamma)$
(green), ${\mathcal{B}}(B\to X_sl^+l^-)$ (cyan), $A_{CP}(B\to X_s\gamma)$
(magenta), all together (blue).} 
\label{fig_kou_3}
\end{figure*}

Contrary to the previous type of study, one can also consider the SUSY breaking
parameters as free and use the \superb observables to constrain them. Such study
has been intensively carried out in the \superb framework using the so-called
Mass Insertion Approximation (MIA) (see e.g.~\cite{Bona:2007qt}). In this approximation,
a certain level of degeneracy in the squark mass is assumed in order to
guarantee the suppression of the unwanted large flavour violation effects
(i.e. Super GIM mechanism) to the previously observed flavour phenomena, which
are in good agreement with the SM predictions. Since the mass insertion
parameters are defined at the electroweak scale, the relation between the SUSY
parameters and the observables are quite simple. As a result, one can readily
study the effect of the same mass insertion contribution to the different \superb
observables  (e.g. mass insertion $(\delta^d)_{13/23}$ to various type of $b\to
d/s$ transitions) as shown in Fig.~\ref{fig_kou_3}.  Such interplay is extremely
useful to distinguish between different types of SUSY.

\subsubsection{Model dependent analysis: interplay among different type of flavour observables} 

\begin{figure*}[th]
\includegraphics[width=0.9\textwidth]{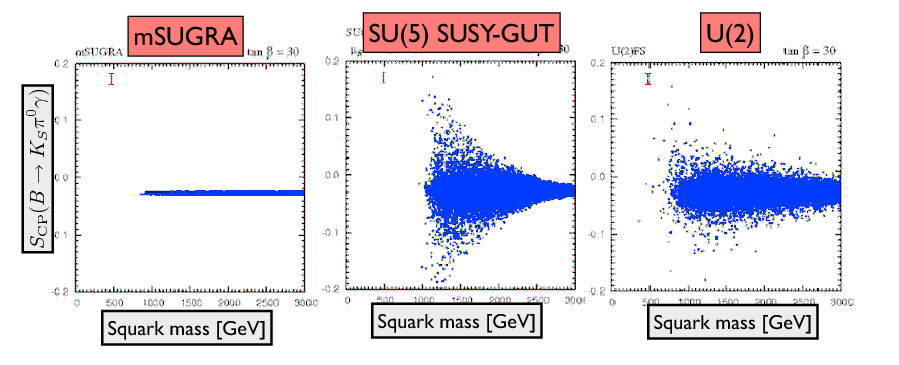}
\includegraphics[width=0.9\textwidth]{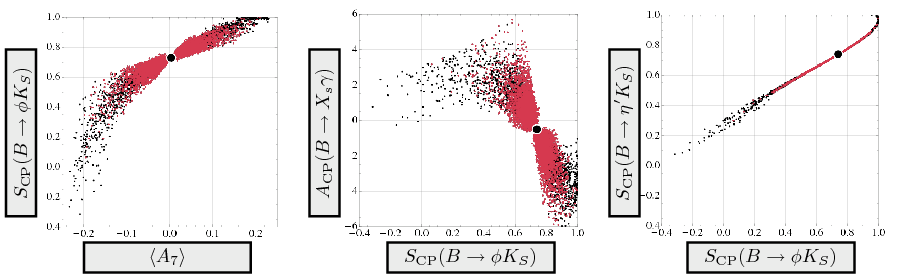}
\caption{The top figures are scatter plots of the time-dependent \CP
asymmetry for $B\to K_S\pi^0\gamma$ in terms of the averaged squark mass. The
varied parameters are those given in the specific flavour models, mSUGRA
(top-left), SU(5) SUSY-GUT (top-middle), U(2) (top-right) (see~\cite{Goto:2007ee} for
more details). The bottom figures are the result for the various observables for
the so-called $\delta LL$ mass insertion model~\cite{Hall:1995es}. Large
correlations can be observed between various \superb observables
(see~\cite{Altmannshofer:2009ne} for more details). }
\label{fig_kou_4}
\end{figure*}

Another approach to tackle the large number of SUSY parameters is to use a
theoretically motivated flavour symmetry for the SUSY parameters at a high
energy (at the SUSY breaking scale, GUT scale etc.). There have been various
attractive proposals for such symmetry. One such example is
a Grand Unification Theory (GUT) in which the quark and the
lepton sectors are unified at the GUT scale. In this class of approach,
non-trivial correlations can appear. One of such examples contains an apparent
relation between the the $2-3$ generation transition of quark and lepton sectors
(such as  $b\to s$ transitions and $\tau\to\mu$ transition)  in SUSY-GUT models.
Some examples are shown in Fig.~\ref{fig_kou_4} from Ref.~\cite{Goto:2007ee}.

In Ref.~\cite{Altmannshofer:2009ne},  various kinds of flavour models are studied. 
A brief summary of their results are shown in Table \ref{tab:DNA}, which indicates the possible size of
effects in various $B$ physics observables, in $D^0-\bar D^0$ mixing and in the
$\tau\to \mu\gamma$ decay. Finding for instance large NP effects in the latter
decay or in the \CP asymmetry $S_{\phi K_S}$ would rule out the AKM model \cite{Antusch:2007re}
while favoring the other models analyzed. Similarly observing significant \CP violating
effects in $D-\bar D$ mixing would disfavor all models analyzed except the
AC~\cite{Agashe:2003rj} model~\cite{Altmannshofer:2010ad}.

\newcommand{\three}{{\color{red}$\bigstar\bigstar\bigstar$}}
\newcommand{\two}{{\color{blue}$\bigstar\bigstar$}}
\newcommand{\one}{{\color{black}$\bigstar$}}
%
\begin{table}[t]
\caption{
``DNA'' of flavour physics effects for the most interesting observables in a selection of SUSY
 models from Ref.~\cite{Altmannshofer:2009ne}. \three\ signals large effects, \two\ visible but small
 effects and \one\ implies that the given model does not predict sizable effects in that
 observable.
\label{tab:DNA}}
\addtolength{\arraycolsep}{4pt}
\renewcommand{\arraystretch}{1.5}
\centering
\begin{tabular}{|l|c|c|c|c|c|}
\hline
&  AC & RVV2 & AKM  & $\delta$LL & FBMSSM 
\\
\hline
$D^0-\bar D^0$& \three & \one & \one & \one & \one 
\\
\hline
$ S_{\psi\phi}$ & \three & \three & \three & \one & \one  
\\
\hline\hline
$S_{\phi K_S}$ & \three & \two & \one & \three & \three  \\
\hline
$A_{\rm CP}\left(B\rightarrow X_s\gamma\right)$ & \one & \one & \one & \three & \three 
\\
\hline
$A_{7,8}(B\to K^*\mu^+\mu^-)$ & \one & \one & \one & \three & \three 
\\
\hline
$A_{9}(B\to K^*\mu^+\mu^-)$ & \one & \one & \one & \one & \one 
\\
\hline
$B\to K^{(*)}\nu\bar\nu$  & \one & \one & \one & \one & \one 
\\
\hline
$B_s\rightarrow\mu^+\mu^-$ & \three & \three & \three & \three & \three 
\\
\hline
$\tau\rightarrow \mu\gamma$ & \three & \three & \one & \three & \three  \\
\hline
\end{tabular}
\renewcommand{\arraystretch}{1}
\end{table}

\begin{figure}[ht!]
\begin{center}
\includegraphics[width=.45\textwidth]{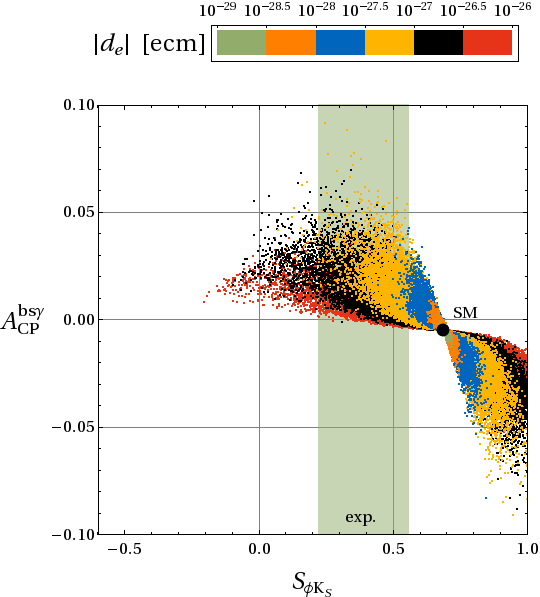}
\caption{Correlation between the CP asymmetries $A_\text{CP}^{b\to s\gamma}$ and
$S_{\phi K_S}$ in the FBMSSM \cite{Altmannshofer:2008hc}. The various colors indicate
the predicted lower bound on the electron EDM. \label{fig:FBMSSM}
}
\end{center}
\end{figure}

In the same article, it is also pointed out that even the flavour blind MSSM (FBMSSM) analyzed in \cite{Altmannshofer:2008hc} can account for large effects in various $B$ physics observables. Of particular interest in this case are \CP violating observables like $A_\text{CP}^{b\to s\gamma}$ and $S_{\phi K_S}$ which, due to the minimal flavour structure of the model, are highly correlated with electric dipole moments (EDMs). In Fig.\ \ref{fig:FBMSSM} we show $A_\text{CP}^{b\to s\gamma}$ as a
function of $S_{\phi K_S}$. Due to the strong correlation between these two asymmetries, the
aim to address the present tension in $S_{\phi K_S}$ unambiguously predicts large NP effects
in the \CP asymmetry in $b\to s\gamma$, which even changes sign with respect to the SM prediction.

\subsection{Fourth generation of quarks and leptons}

\begin{figure}[ht!]
\begin{center}
\includegraphics[width=.45\textwidth]{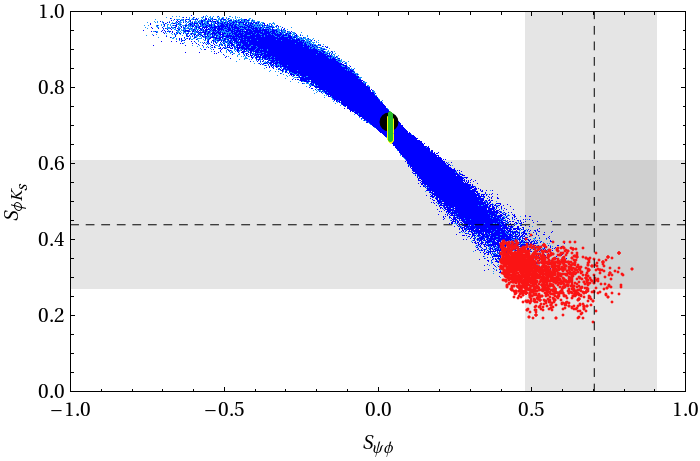}
\caption{Correlation between the \CP asymmetries $S_{\phi K_S}$ and $S_{\psi\phi}$ in the
SM4 \cite{Buras:2010pi}.\label{fig:SM4-phiKS}}
\end{center}
\end{figure}

\begin{figure}
\begin{center}
\includegraphics[width=.45\textwidth]{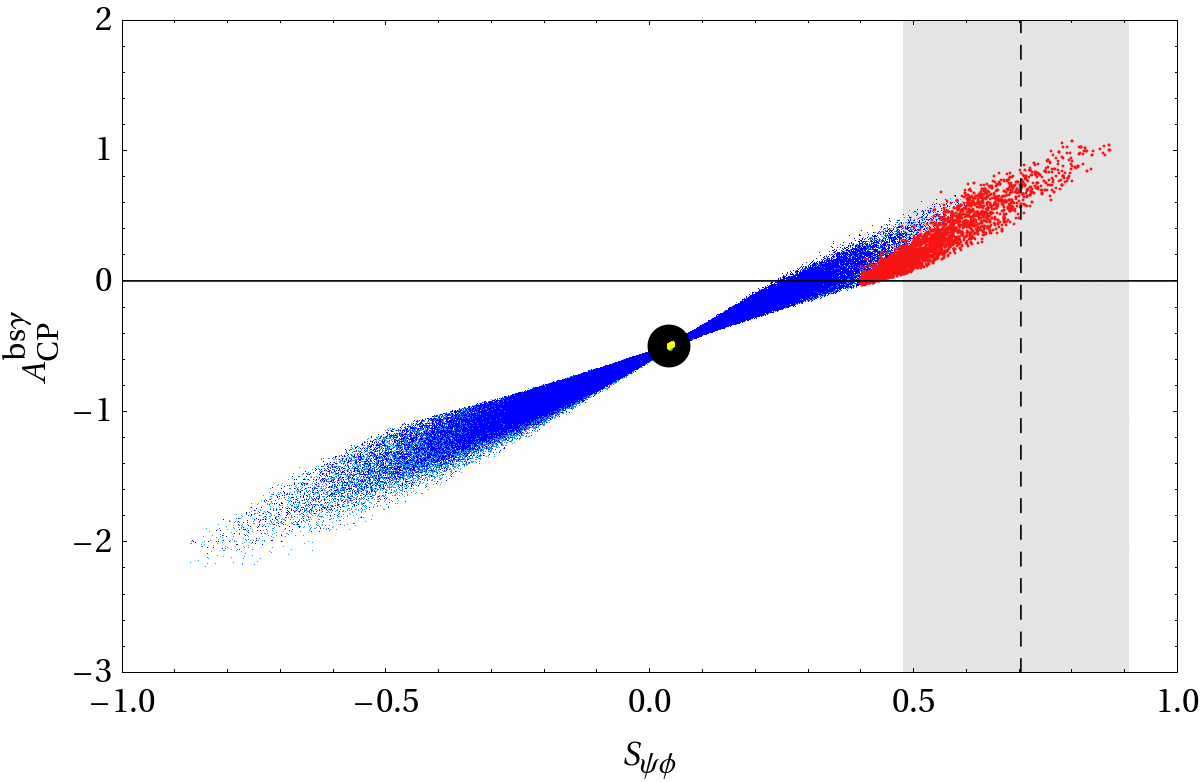}
\caption{Correlation between the \CP asymmetries $A_\text{CP}^{b\to s\gamma}$ and
$S_{\psi\phi}$ in the SM4 \cite{Buras:2010pi}.\label{fig:SM4-bsg}}
\end{center}
\end{figure}

Recently the implications on flavour physics observables from extending the SM by adding a fourth
generation of quarks and leptons (SM4) have received a lot of attention, see
e.\,g.~\cite{Hou:2005yb,Hou:2005yb,Hou:2006mx,Herrera:2008yf,Soni:2010xh,Buras:2010pi}. The guidelines of how to extract the new parameters
of the CKM4 matrix from future data has been presented in \cite{Buras:2010pi} and will not
be repeated here. Instead we show in Figs.\ \ref{fig:SM4-phiKS} and \ref{fig:SM4-bsg} the \CP
asymmetries $S_{\phi K_S}$ and $A_\text{CP}^{b\to s\gamma}$, respectively, as functions of
$S_{\psi\phi}$. In both cases a strong correlation can be observed. Therefore, if the present
deviation from the SM prediction in $S_{\psi\phi}$ will be confirmed in the future more accurate
experiments, the SM4 unambiguously predicts large effects in $S_{\phi K_S}$ and
$A_\text{CP}^{b\to s\gamma}$. Together with the possible direct observation of a 4th
generation at the LHC, these effects can be used to tighten the allowed SM4 parameter space.

\subsection{Minimal and custodially extended RS models}

A theoretically appealing approach to the SM flavour puzzle is given by
Randall-Sundrum models with bulk fermions \cite{Randall:1999ee}. In this scenario the observed
hierarchies in quark masses and CKM mixings are naturally obtained from the different
localization of fermions along the 5D bulk. Implications for low energy flavour violating
observables have been studied extensively in the literature, see
e.\,g.\ \cite{Agashe:2004ay,Blanke:2008zb,Bauer:2009cf}.

\begin{figure}
\begin{center}
\includegraphics[width=.45\textwidth]{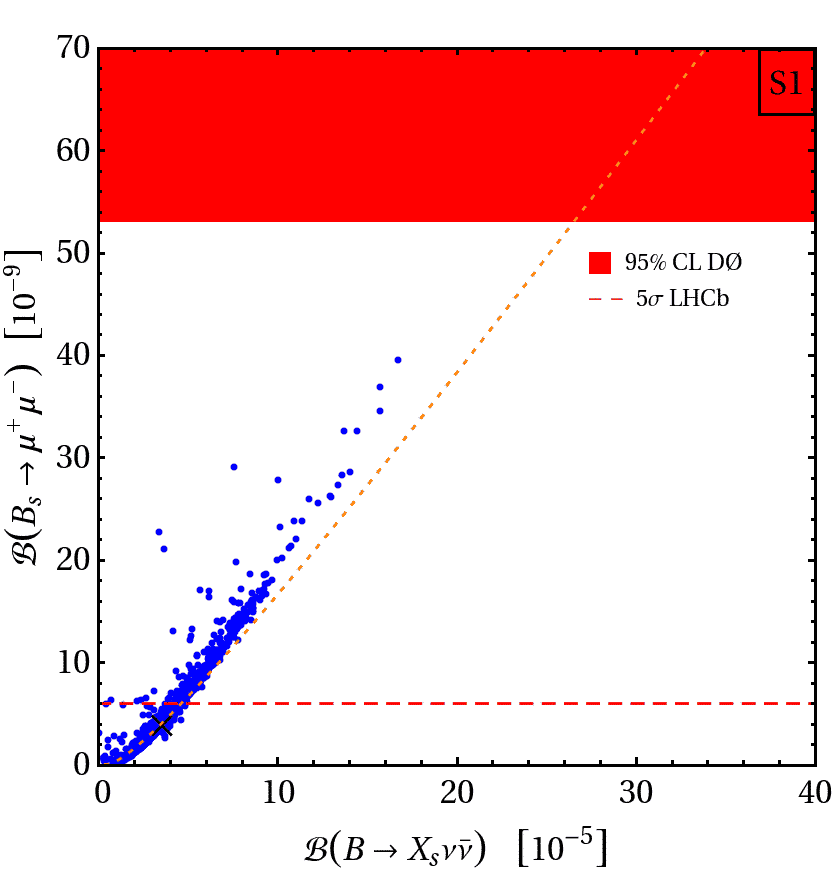}
\caption{Correlation between the branching ratios for $B_s\to \mu^+\mu^-$ and
$B\to X_s\nu\bar\nu$ in the minimal RS model \cite{Bauer:2009cf}.\label{fig:minRS}}
\end{center}
\end{figure}

Interestingly the observed pattern of effects depends crucially on the realization of the model.
In the minimal scenario with only the SM gauge group in the bulk, the NP contributions to rare
decays are dominantly left-handed. Consequently large effects could be expected in both $B$
and $K$ decays \cite{Agashe:2004ay,Blanke:2008zb,Bauer:2009cf}. As an example
Fig.\ \ref{fig:minRS} shows the correlation between $Br(B_s\to\mu^+\mu^-)$ and
$Br(B\to X_s\nu\bar\nu)$ in the minimal RS model. The latter branching ratio can reach values
larger than $10^{-4}$, which necessarily coincide with large NP effects also in the former
channel.

\begin{figure}
\begin{center}
\includegraphics[width=.45\textwidth]{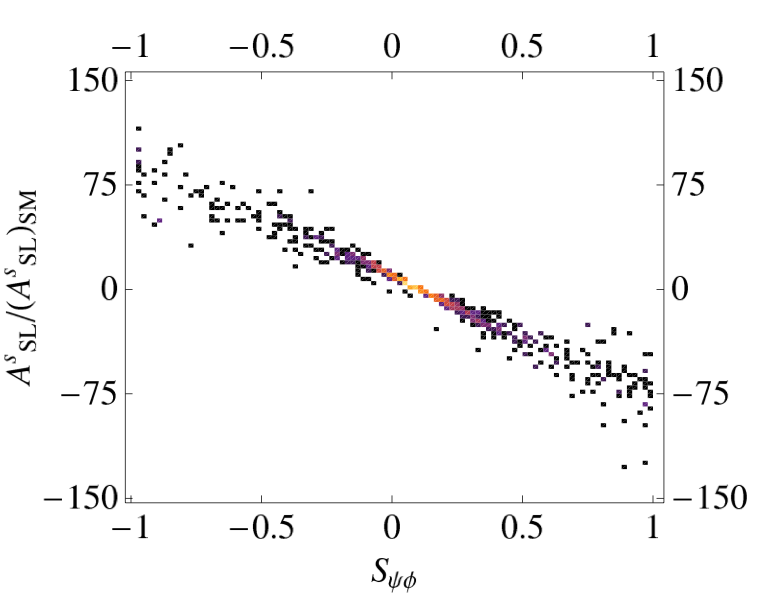}
\caption{Correlation between the \CP asymmetries $A^s_\text{SL}$ and $S_{\psi\phi}$ in the
RSc model \cite{Blanke:2008zb}.\label{fig:RSc}}
\end{center}
\end{figure}

The situation is completely different in the case of a custodially extended bulk gauge
symmetry \cite{Blanke:2008zb}. Due to the suppression of left-handed flavour changing $Z$
couplings, rare decays in this case are dominated by right-handed currents. Consequently,
while large NP effects can appear in the kaon sector, the effects in rare $B$ decays are
predicted to be small and therefore difficult to disentangle from the SM. The situation is
however different in the $\Delta F=2$ sector, where a large new phase in $B_s - \bar B_s$
mixing can be generated (see Fig.\ \ref{fig:RSc}).

\subsection{Littlest Higgs model with T-parity}

The detailed FCNC studies in the Littlest Higgs model with T-parity (LHT) performed 
in 2006--2007 \cite{Blanke:2006sb} have recently been updated \cite{Blanke:2009am} in light
of an additional LHT contribution to the $Z$ penguin pointed out in \cite{Goto:2008fj} and
of new input from experiments and lattice calculations. While the additional contribution
affected the size of some of the possible effects, the main conclusions from \cite{Blanke:2006sb}
remained intact:
\begin{itemize}
\item
Large NP effects are possible in \CP asymmetries related to $B_s - \bar B_s$ mixing and in
rare $K$ decays.
\item
The effects in rare $B$ decays are small and therefore difficult to measure.
\item
Large effects can be expected in LFV $\mu$ and $\tau$ decays, as summarized in
Table \ref{tab:bounds}.
\item
Ratios of LFV branching ratios turn out to be very different from the MSSM predictions and can
therefore serve as a clean tool to distinguish between these two models (see
Table \ref{tab:ratios}).
\end{itemize}

\begin{table}[h!]
\caption{{Maximal values} on LFV $\tau$ decay branching ratios in the LHT model, for two
different values of the scale $f$, after imposing the constraints on $\mu\to e\gamma$ and
$\mu^-\to e^-e^+e^-$ \cite{Blanke:2009am}.  \label{tab:bounds}}
{
\begin{center}
\begin{tabular}{|c|c|c|c|}
\hline
decay & $f=1000\gev$ & $f=500\gev$ & Super$B$ \\
 & & & sensitivity \\\hline
$\tau\to e\gamma$ & $8\cdot 10^{-10}$  & ${2\cdot 10^{-8}}$ & ${2\cdot10^{-9}}$  \\
$\tau\to \mu\gamma$ & $8\cdot 10^{-10}$  &$2\cdot 10^{-8}$   &${2\cdot10^{-9}}$ \\
$\tau^-\to e^-e^+e^-$ & $1\cdot10^{-10}$  & ${2\cdot10^{-8}}$   & $2\cdot10^{-10}$ \\
$\tau^-\to \mu^-\mu^+\mu^-$ & $1\cdot10^{-10}$  & ${2\cdot10^{-8}}$   & $2\cdot10^{-10}$ \\
$\tau^-\to e^-\mu^+\mu^-$ & $1\cdot10^{-10}$ & ${2\cdot10^{-8}}$   & \\
$\tau^-\to \mu^-e^+e^-$ & $1\cdot10^{-10}$ & ${2\cdot10^{-8}}$  & \\
$\tau^-\to \mu^-e^+\mu^-$ & $6\cdot10^{-14}$ & ${1\cdot10^{-13}}$ & \\
$\tau^-\to e^-\mu^+e^-$ & $6\cdot10^{-14}$ &${1\cdot10^{-13}}$   &  \\
$\tau\to\mu\pi$ & $4\cdot10^{-10} $  & ${5\cdot10^{-8}}$  & \\
$\tau\to e\pi$ & $4\cdot10^{-10} $ & ${5\cdot10^{-8}}$   & \\
$\tau\to\mu\eta$ & $2\cdot10^{-10}$  & ${2\cdot10^{-8}}$  & $4\cdot10^{-10}$ \\
$\tau\to e\eta$ & $2\cdot10^{-10}$  & ${2\cdot10^{-8}}$  & $6\cdot10^{-10}$ \\
$\tau\to \mu\eta'$ & $1\cdot10^{-10}$ & ${2\cdot10^{-8}}$  & \\
$\tau\to e\eta'$ & $1\cdot10^{-10}$ & ${2\cdot10^{-8}}$   & \\\hline
\end{tabular}
\end{center}
}
\end{table}

\begin{table}[h!]
\caption{Comparison of various ratios of branching ratios in the LHT model
($f=1\tev$) \cite{Blanke:2009am} and in the MSSM without \cite{Ellis:2002fe} and
with \cite{Paradisi:2005tk} significant Higgs contributions.\label{tab:ratios}}
{\renewcommand{\arraystretch}{1.5}
\begin{center}
\begin{tabular}{|c|c|c|c|}
\hline
ratio & LHT  & MSSM     & MSSM \\
      &      & (dipole) & (Higgs)\\\hline\hline
$\frac{Br(\tau^-\to e^-e^+e^-)}{Br(\tau\to e\gamma)}$   & 0.04\dots0.4     &$\sim1\cdot10^{-2}$ & ${\sim1\cdot10^{-2}}$\\
$\frac{Br(\tau^-\to \mu^-\mu^+\mu^-)}{Br(\tau\to \mu\gamma)}$  &0.04\dots0.4     &$\sim2\cdot10^{-3}$ & $0.06\dots0.1$ \\\hline
$\frac{Br(\tau^-\to e^-\mu^+\mu^-)}{Br(\tau\to e\gamma)}$  & 0.04\dots0.3     &$\sim2\cdot10^{-3}$ & $0.02\dots0.04$ \\
$\frac{Br(\tau^-\to \mu^-e^+e^-)}{Br(\tau\to \mu\gamma)}$  & 0.04\dots0.3    &$\sim1\cdot10^{-2}$ & ${\sim1\cdot10^{-2}}$\\
$\frac{Br(\tau^-\to e^-e^+e^-)}{Br(\tau^-\to e^-\mu^+\mu^-)}$     & 0.8\dots2.0   &$\sim5$ & 0.3\dots0.5\\
$\frac{Br(\tau^-\to \mu^-\mu^+\mu^-)}{Br(\tau^-\to \mu^-e^+e^-)}$   & 0.7\dots1.6    &$\sim0.2$ & 5\dots10 \\\hline
\end{tabular}
\end{center}\renewcommand{\arraystretch}{1.0}
}
\end{table}

\begin{figure}[h!]
\begin{center}
\includegraphics[width=.45\textwidth]{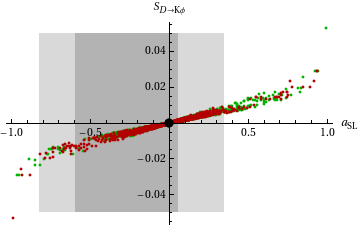}
\caption{Correlation between the \CP asymmetries $a_\text{SL}$ and $S_{K_S \phi}$ in the
LHT model \cite{Bigi:2009df}.\label{fig:LHT-D}}
\end{center}
\end{figure}

A detailed study of $D^0 -\bar D^0$ mixing in the LHT model has been performed
in \cite{Bigi:2009df}. While in case of the \CP conserving observables $x$ and $y$ a possible
NP contribution is difficult to disentangle due to the poor knowledge of the SM long-distance
contributions, an observation of \CP violation in the $D$ system would be an unambiguous sign
of NP. Figure~\ref{fig:LHT-D} shows the correlation between the semi-leptonic \CP asymmetry
$a_\text{SL}$ and the asymmetry in $D\to K_S\phi$ decays. We observe that in both observables
LHT physics can lead to spectacular deviations from the tiny SM prediction. A deviation from
the correlation in Fig.\ \ref{fig:LHT-D} would be a clear sign of direct \CP violation in the
$D \to K_S \phi$ channel.

\subsection{Precision CKM constraints.}\label{sec:interplay:precisionckm}

The CKM ansatz has been tested at the 10\% level by \babar and \belle.  One significant 
consequence of recording 75\invab of data at the \FourS is that it will be possible to 
push the precision of this global CKM test down to the percent level.  It is worth recalling that 
there are direct (i.e. measurements of the angles of the unitarity triangle) and 
indirect ways to test the CKM mechanism. One advantage of a \sff compared to other flavour 
experiments is that it will be able to perform a wide array of measurements of both the direct and indirect 
constraints.  The consequence
of this is that \superb will be able to perform a self-consistent over-constraint of
the description of quark mixing in the SM, and as is shown in Figure~\ref{fig:ut2015}, if one
extrapolates measurements from today, to the era of \superb 
we could find the dream scenario where constraints do not converge on a single point indicating
that new physics modifies our understanding of quark mixing. The alternative so-called nightmare 
scenario would be that once again the SM description is a good enough description of the experimental 
picture of nature that we have built up since the pivotal work of Cabibbo from 1963 on quark mixing~\cite{cabibbo}.

The nightmare scenario is by no means the end of the road.  In fact in many situations, this
would signify the beginning of a number of new physics searches, some of which would be possible at \superb,
but there would also be a number of new physics searches possible at other experiments.  One example is that of
the measurement of the the $K\to\pi\nu\overline{\nu}$ branching fraction.  The theoretical uncertainty on the branching fraction
of the charged and neutral modes is dominated by knowledge of the CKM matrix.  In the case
of $K^+\to\pi^+\nu\overline{\nu}$ this is 33\%, and for $K^0_L\to\pi^0\nu\overline{\nu}$ this is 52\%~\cite{Haisch:2007pd}.
With the current knowledge of the CKM mechanism a measurement of either of these modes
would provide another constraint on the SM, however with a precision over-constraint of 
the CKM mechanism from \superb, then these kaon branching fractions would
be sensitive probes of new physics via loop amplitudes in analogy with the discussions in 
Section~\ref{sec:bphysics} for the study of rare $B$ decays.  Thus the role of \superb
in elucidating flavour physics and searching for new physics, transcends the limitations
of measurements possible at a \sff and has potentially important consequences for 
the interpretation of the results from other proposed or existing flavour physics experiments.

\graphicspath{{conclusion/}{conclusion/}}
\section{Conclusions}\label{sec:conclusion}

\superb is a next generation high luminosity $e^+e^-$ collider that will accumulate 
a data sample of $75ab^{-1}$ within five years of nominal data taking.  This experiment 
could start running as early as 2015, by which time the LHC will have accumulated a 
significant sample of data, and would be reporting the results of searches for 
or direct measurements of new physics.  Those results are limited in that they
measure only flavour diagonal processes. In order to fully understand the 
nature of new physics, one also has to measure the off-diagonal terms, in 
analogy to the CKM and PMNS mixing matrices.  The new physics capability of 
the \superb experiment is completely complementary to the direct searches 
that are now underway at the LHC.  There are many measurements that could 
provide an unequivocal signal for new physics, and with hind-sight it would be
possible to decode the more subtle nature of new physics by comparing the results
of many measurements against theoretical predictions.  The interplay between
measurements made at \superb\ and those possible at other experiments is 
discussed in detail in Section~\ref{sec:interplay} where a strategy for 
elucidating the nature of new physics is outlined.  This strategy is only feasible
through a combination of direct and indirect searches, where most of the latter 
are only possible at a \sff like \superb.

The new physics sensitive measurements possible at \superb are discussed in detail 
throughout this paper.  Some of the golden channels that we aim to measure
are discussed in the following summary.  In terms of Higgs physics, one can 
combine information from rare $B$ decays in order to precisely measure 
$\tan\beta$ or the coupling $A$ in CMSSM.  In addition to learning about the 
couplings and structure of the Higgs sector beyond the Standard Model, one 
can indirectly search for charged Higgs particles to a level that exceeds 
the LHC direct search capabilities by a factor of $3-5$ over the full range of 
$\tan\beta$.  CP violation parameters in \B and \D decays are also sensitive 
probes of Higgs and SUSY particles, and these will be studied to the fullest 
extent possible.  

Again, using rare \B\ decays measured at \superb it is possible to probe the 
structure of SUSY.  For example, two thirds of the MSSM parameters are flavour 
couplings, and with rare decay measurements from a \sff it would be possible 
measure the real and imaginary parts of a number of flavour couplings of SUSY 
models to a few percent.  

In should be noted that exclusive decays of the rare processes $b\to s \ell\ell$ 
and $b\to s \gamma$ will be measured with high statistics at LHCb.  At 
\superb however one can also perform both inclusive and exclusive measurements 
of these decays.  Inclusive decays provide important additional constraints 
as the theoretical uncertainties on the exclusive processes are much larger 
than the inclusive ones.  These sets of measurements at \superb will be limited 
by theoretical uncertainties as discussed in Section~\ref{sec:bphysics}.

\superb has by far the greatest sensitivity for studies of Lepton Flavour Violating 
$\tau$ decays and will be able to search down to branching fractions of the 
level of $2\times 10^{-10}$.  SUSY GUT models, using constraints from the current 
$B_s$ mixing and phase measurements from the Tevatron predict that such $\tau$ 
LFV channels could exist with branching fractions of a few $10^{-8}$.  In addition to 
LFV studies in $\tau$ decays, \superb will search for LFV in di-lepton decays of 
light mesons.  Such decays are sensitive to light Higgs or Dark Matter particles
that would be difficult or impossible to detect in high energy machines.  Similarly,
the detailed study of the decays of light mesons could elucidate, or exclude large
parts of the parameter space for the dark sector which is commonly referred to as
`Dark Forces' in the literature.

There are a number of deviations from the Standard Model at the level of 
$2-3\sigma$ at the existing B-factories.  If any of these were a manifestation 
of new physics, the increased precision obtainable at \superb would be able to 
convert these hints into discoveries of new physics. One such deviation is that
of a CPT test using di-lepton decays of $B$ mesons.  A precision test of 
CPT could be performed at \superb which in turn could probe new physics near 
the plank scale for some quantum gravity scenarios.  The current discrepancy 
between measurement at the Standard Model is at the level of $2.7\sigma$.  

If the recent Tevatron claim of evidence for non-Standard Model 
physics~\cite{Abazov:2010hv} is confirmed, then the texture of 
flavour physics beyond the Standard Model will be rich. The
strategy for elucidating nature outlined in Section~\ref{sec:interplay} 
can be expected to provide at least a partial reconstruction of 
the new physics Lagrangian, and does not depend on confirmation 
of the D0 result as a manifestation of new physics.

If no new physics were found at the LHC, the flavour problem
could become the flavour opportunity.
Nowadays, $B$ data already point to a new physics scale exceeding
the TeV. Using the clean environment at \superb,
particle physicists would be able to indirectly probe the energy range 
$10-100$ TeV.

Likewise, if the LHC fails to find the Standard Model Higgs, it would be 
possible to combine information from \superb with measurements of $g-2$, 
and $\Omega_{CDM}$ to 
improve the indirect constrains on the Higgs mass in CMSSM.  The currently 
preferred mass for the Higgs from this method is compatible with the results
of precision electroweak fits.

In addition to the aforementioned new physics search capabilities, \superb will 
be able to perform precision tests of the Standard Model, which in turn could 
reduce theoretical uncertainties sufficiently to pave the way for additional 
new physics searches.  One highlight of the Standard Model measurements is the 
possibility to perform a precision measurement of $\sin^2\theta_{\mathrm{W}}$.
Charm mixing has been firmly established in recent years, and a detailed study of possible \CP
violation effects in charm would be performed at \superb.  These would include
the study of quantum-correlated $D^0\overline{D}^0$ decays, which would give
access to additional experimental observables beyond those studied in charm
decays at the \FourS.  Any large manifestation of \CP violation in charm
would be a clear indication of new physics.  In addition to the aforementioned
studies, data from \superb could be used to measure a number of benchmark 
parameters, such as meson masses and decay constants, which in turn could be 
used to further validate Lattice QCD and hone our understanding of theory.
The current $B$ factories provided the community with a rich harvest
in terms of meson spectroscopy.  They range from the
the discovery of the ground state $\eta_b$ that had been sought after for
many decades to the unexpected observation of several candidate for bound
states with a quark structure never observed before. The limited
statistics of the B-Factories has only opened the question of the
definition of this new spectroscopy, but the hundred times larger samples
of \superb are needed to see the complete the picture.

In summary the \superb experiment would be able to perform precision measurements
of a wide array of new physics sensitive and Standard Model observables.  By
interpreting the resulting pattern of measurements and deviations from the
Standard Model this experiment would be able to elucidate details of the nature
of new physics to energy scales up to 100 TeV.  This broad physics programme
is complementary to the direct search programme at the LHC.

\bibliography{wp}{}

\end{document}